\documentclass{msuphddissertation}
\usepackage{amsmath,amssymb,amsthm,paralist,graphicx,epstopdf,dcolumn,braket,caption,titlesec}
\author{Luke Titus} 
\title{Effects of Nonlocality on Transfer Reactions} 
\unit{Physics -- Doctor of Philosophy} 

\begin{document}
\maketitlepage 

\begin{abstract}
Nuclear reactions play a key role in the study of nuclei away from stability. Single-nucleon transfer reactions involving deuterons provide an exceptional tool to study the single-particle structure of nuclei. Theoretically, these reactions are attractive as they can be cast into a three-body problem composed of a neutron, proton, and the target nucleus.
Optical potentials are a common ingredient in reactions studies. Traditionally, nucleon-nucleus optical potentials are made local for convenience. The effects of nonlocal potentials have historically been included approximately by applying a correction factor to the solution of the corresponding equation for the local equivalent interaction. This is usually referred to as the Perey correction factor. In this thesis, we have systematically investigated the effects of nonlocality on $(p,d)$ and $(d,p)$ transfer reactions, and the validity of the Perey correction factor. We implemented a method to solve the single channel nonlocal equation for both bound and scattering states. We also developed an improved formalism for nonlocal interactions that includes deuteron breakup in transfer reactions. This new formalism, the nonlocal adiabatic distorted wave approximation, was used to study the effects of including nonlocality consistently in $(d,p)$ transfer reactions.  

For the $(p,d)$ transfer reactions, we solved the nonlocal scattering and bound state equations using the Perey-Buck type interaction, and compared to local equivalent calculations. Using the distorted wave Born approximation we construct the T-matrix for $(p,d)$ transfer on $^{17}$O, $^{41}$Ca, $^{49}$Ca, $^{127}$Sn, $^{133}$Sn, and $^{209}$Pb at $20$ and $50$ MeV. Additionally we studied $(p,d)$ reactions on $^{40}$Ca using the the nonlocal dispersive optical model. We have also included nonlocality consistently into the adiabatic distorted wave approximation and have investigated the effects of nonlocality on on $(d,p)$ transfer reactions for deuterons impinged on $^{16}$O, $^{40}$Ca, $^{48}$Ca, $^{126}$Sn, $^{132}$Sn, $^{208}$Pb at $10$, $20$, and $50$ MeV. 

We found that for bound states the Perry corrected wave functions resulting from the local equation agreed well with that from the nonlocal equation in the interior region, but discrepancies were found in the surface and peripheral regions. Overall, the Perey correction factor was adequate for scattering states, with the exception for a few partial waves. Nonlocality in the proton scattering state reduced the amplitude of the wave function in the nuclear interior. The same was seen for nonlocality in the deuteron scattering state, but the wave function was also shifted outward. In distorted wave Born approximation studies of $(p,d)$ reactions using the Perey-Buck potential, we found that transfer distributions at the first peak differed by $15-35\%$ as compared to the distribution resulting from local potentials. When using the dispersive optical model, this discrepancies grew to $\approx 30-50\%$. When nonlocality was included consistently within the adiabatic distorted wave approximation, the disagreement was found to be $\sim 40\%$. 

If only local optical potentials are used in the analysis of experimental $(p,d)$ or $(d,p)$ transfer cross sections, the extracted spectroscopic factors may be incorrect by up to $50\%$ in some cases due to the local approximation. This highlights the necessity to pursue reaction formalisms that include nonlocality exactly. \end{abstract}





\begin{acknowledgment}
I would like to thank my advisor Prof. Filomena Nunes whose support and guidance has helped me both professionally and personally.  I am grateful for her continued support throughout my journey in nuclear physics, and her encouragement in my exploration of other fields and careers. Her consistent positive attitude is a model for me to follow, and has made the difficult task of completing my degree a pleasure. Undoubtedly I would not be where I am in life without her. I am very fortunate to have her as an advisor.

I would like to express my thanks to my guidance committee, Prof. Mark Voit, Prof. Phil Duxbury, Prof. Morten Hjorth-Jensen, and Prof. Remco Zegers for their invaluable advice and suggestions along the way. I'd like to thank again Prof. Mark Voit for taking me under his wing in astronomy for a semester, and for the countless hours of discussions we had about galaxy clusters and the universe. I would also like to thank Prof. Ian Thompson, Prof. Ron Johnson, Prof. Pierre Capel, Prof. Jeff Tostevin, and Dr. Gregory Potel, without whom, this work would not have been possible.

I owe a debt of gratitude to the Department of Physics and Astronomy at Michigan State University, the National Superconducting Cyclotron Laboratory, and the theory group for providing financial, academic, and technical support. I would also like thank the National Science Foundation and the Department of Energy for their financial support.

I thank my colleagues and friends at Michigan State University, and especially my current and past group members: Bich Nguyen, Neelam Upadhyay, Muslema Pervin, Amy Lovell, Alaina Ross, Terri Poxon-Pearson, Gregory Potel, Jimmy Rotureau, and Ivan Brida. 

Last but not least, I thank my parents, Carrie Gossen and Joe Titus, my brother Blake Titus, and the rest of my family for love and support throughout my entire life, and their constant encouragement as I developed my interest in science. Without them, I would never have made it this far. 
\end{acknowledgment}


\TOC 
\LOT
\LOF


\newpage
\pagenumbering{arabic}
\begin{doublespace}


\chapter{Introduction}
\label{Intro}

Since the dawn of nuclear physics, reaction studies have been performed to investigate the properties of the nucleus. One of the many reasons these studies have been carried out is to address the overarching goal of nuclear physics. This is to understand where all the matter in the universe came from and how it was formed. To solve this problem, we not only need to understand the environments in which nuclear reactions occur, but we also need to understand the nature of the nuclei undergoing the reactions. This is a daunting task with hundreds of stable nuclei, and thousands of unstable nuclei known to exist \cite{Erler_n2012}. 

\begin{figure}[h]
\begin{center}
\includegraphics[width=1.0\textwidth]{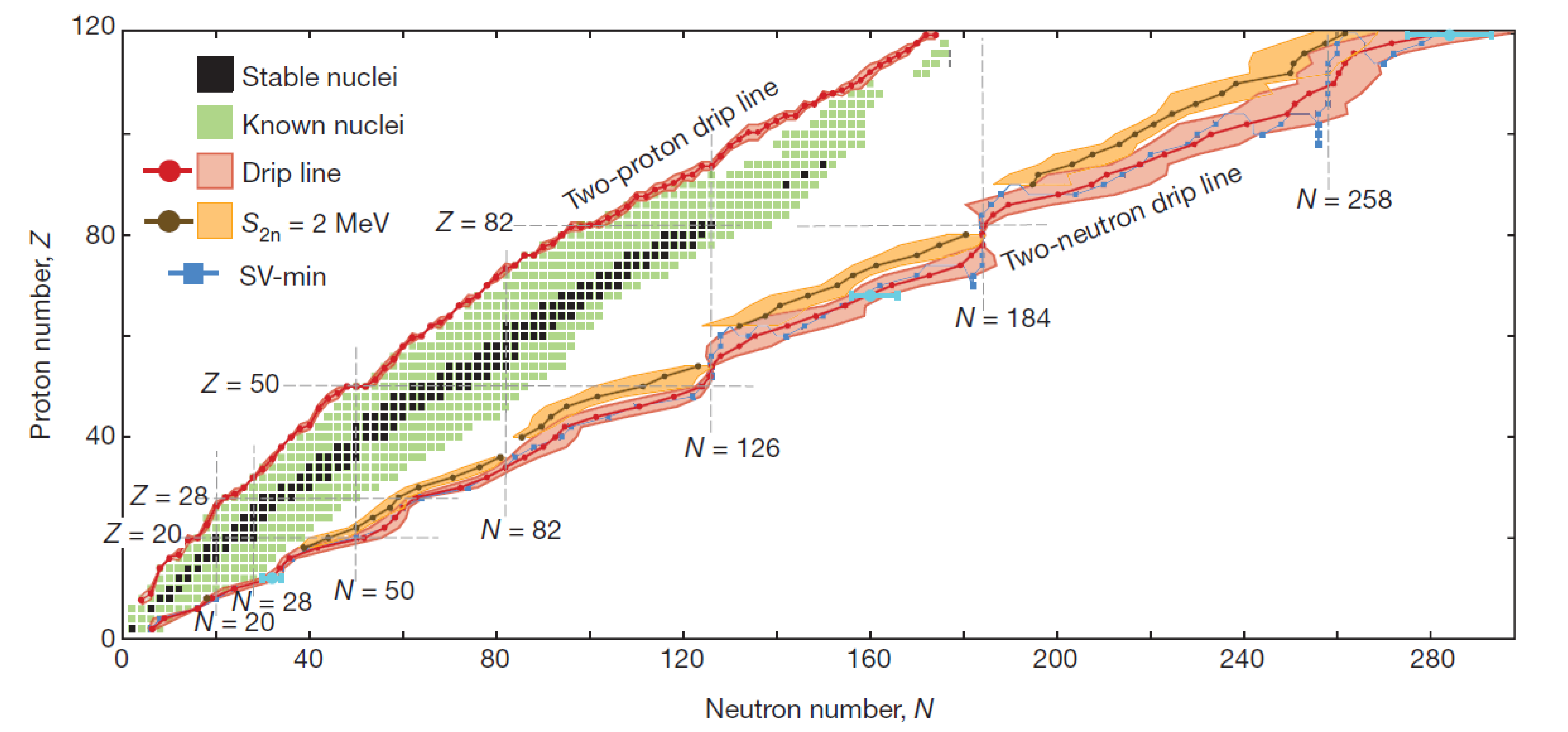}
\end{center}
\caption{The chart of the nuclides. The proton drip line is indicated by the line above the stable nuclei, and the neutron drip line is indicated below the stable nuclei. The proton (neutron) drip line indicates where the addition of a single proton (neutron) will make the resulting nucleus unbound. Figure reprinted from \cite{Erler_n2012} with permission.}
\label{fig:Nuclear_Chart}
\end{figure}

In Fig. \ref{fig:Nuclear_Chart} the chart of the nuclides is shown with the corresponding proton and neutron drip lines. The drip line is the point that separates bound from unbound nuclei. The neutron drip line, for example, defines the point where the addition of a single neutron will make the resulting nucleus unbound.  While an extraordinary amount of progress has been made in experimentally measuring unstable nuclei, it is remarkable how far the neutron drip line is expected to extend, and how many nuclei are yet to be discovered.

For many decades, intense experimental and theoretical effort has been put into studying stable nuclei. While experiments aimed at studying stable isotopes are still performed, the focus in modern times has shifted towards the study of exotic nuclei. In the context of understanding the origin of the matter in the universe, exotic nuclei play a crucial role. While exotic nuclei live for a very short period of time, reactions on exotic nuclei are essential to creating heavy elements \cite{Thompson_book}. In certain astrophysical environments, nuclei rapidly capture protons or neutrons, pushing them towards the drip line. These unstable nuclei then $\beta$ decay back to the valley of stability. To fully understand the path the nucleosynthesis takes, and the elements that are produced, we must understand the properties of the exotic nuclei very far from stability, and the reaction mechanisms of neutrons, protons, or heavier elements on those exotic nuclei.

For many nuclear reaction experiments, a good reaction theory is required to extract reliable information. The same can be said about the potentials we put into our theories. In fact, the two work hand in hand. An excellent model can be held back by the use of poor interactions, while the best interaction available will provide little insight when used in a poor model.

\begin{figure}[h!]
\begin{center}
\includegraphics[scale=0.6]{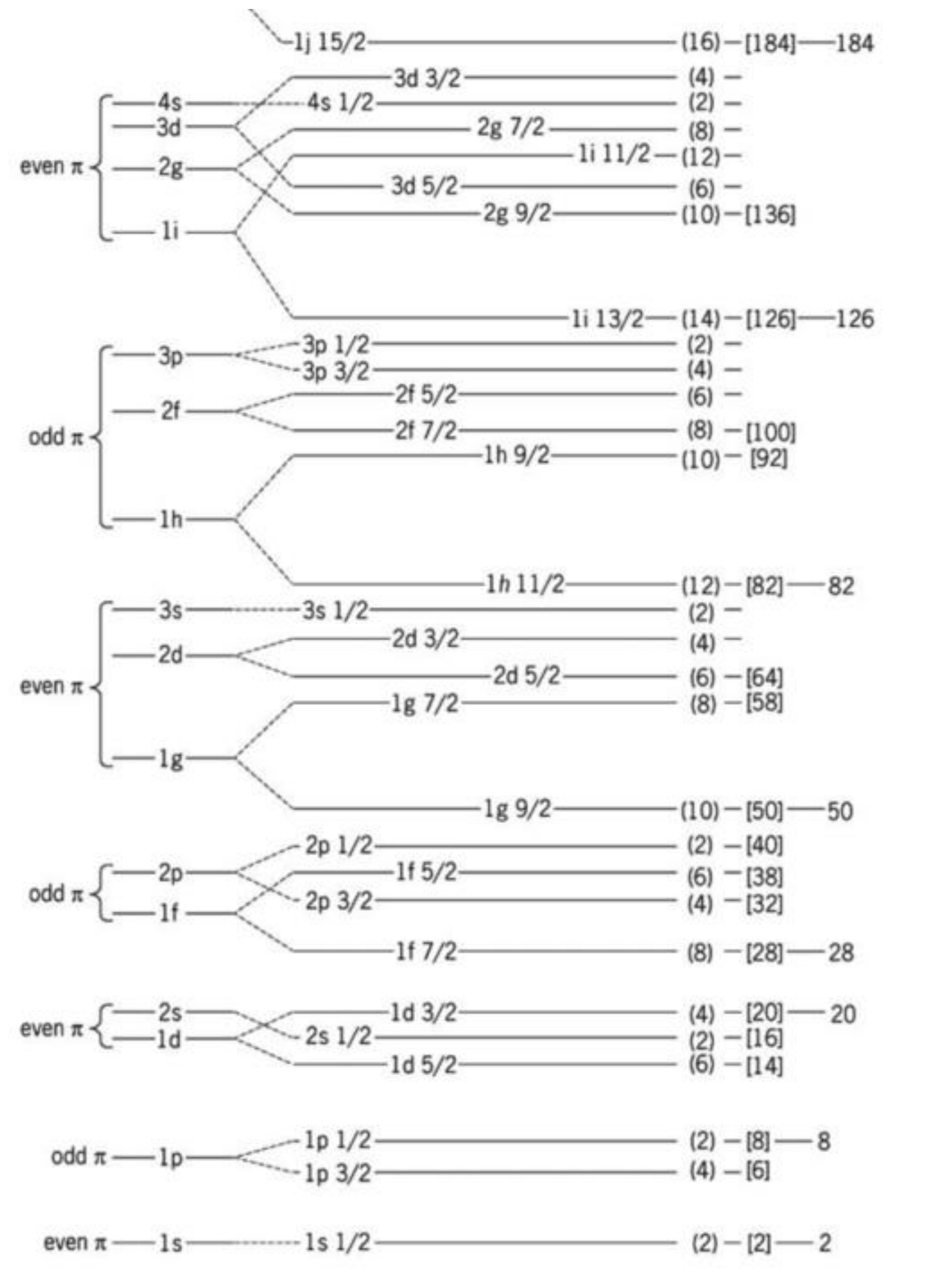}
\end{center}
\caption{Typical Nuclear Shell Structure.}
\label{fig:Energy_Levels}
\end{figure}

An important part of understanding the properties of nuclei is knowing the spin and parity assignments of the various energy levels. Single nucleon transfer reactions are an excellent tool for understanding these properties. The protons and neutrons inside a nucleus arrange themselves in an organized way, roughly following the way levels organize themselves in a harmonic oscillator potential with a spin-orbit interaction. Filling a shell provides additional stability. Indicated in Fig. \ref{fig:Energy_Levels} (right) are the magic numbers corresponding to the number of neutrons or protons needed to fill in a shell. The ordering shown in Fig. \ref{fig:Energy_Levels} provides a guide to assigning energy levels. As one moves away from stability there is shell reordering and different magic numbers emerge.

The use of single nucleon transfer reactions such as $(d,p)$ or $(p,d)$ as a probe to study nuclear structure began in the early 1950s. Butler realized that the spins and parities of nuclear energy levels can be obtained from angular distributions, without the need to know properties of excited states \cite{Butler_pr1950}. This fact was reiterated by Huby \cite{Huby_n1950, Huby_pmjs1951}, and later followed up with theoretical calculations by Bhatia and collaborators \cite{Bhatia_pmjs1952}. While these early studies relied on the very simple plane wave Born approximation, it drew considerable attention to $(d,p)$ reactions as a means to study nuclear structure through the analysis of angular distributions of transfer reactions.

Since these pioneering studies, the shell structure of nuclei has been studied with the aid of single nucleon transfer reactions. Of particular interest for this work are the stripping $(d,p)$ or pickup $(p,d)$ reactions. These types of reactions are an excellent tool for measuring the energy levels of nuclei, as well as the spin and parity assignments of the corresponding energy levels. It is transfer reactions such as these which provided much of the structure information of stable isotopes in the early days of nuclear physics \cite{Lee_pr1964, Bingham_prc1973, Schiffer_pr1967, Lee_pr1967, Bingham_prc1973-2, Kolata_prc1972}.

\begin{figure}[h!]
\begin{center}
\includegraphics[scale=0.6]{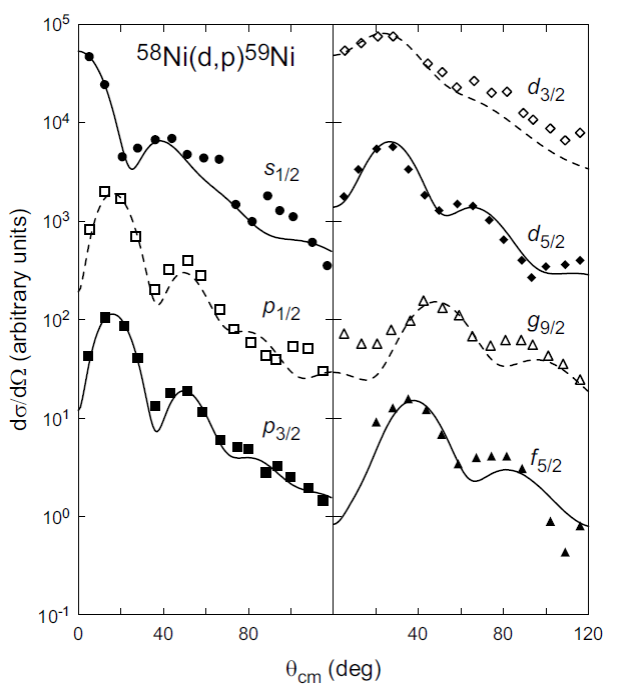}
\end{center}
\caption{Dependence of the transfer angular distribution on the transferred angular momentum for $^{58}$Ni$(d,p)^{59}$Ni at $8$ MeV, with data from \cite{Chowdhury_npa1973}. Reprinted from \cite{Thompson_book} with permission.}
\label{fig:Transfer_Diff_L}
\end{figure}

In Fig. \ref{fig:Transfer_Diff_L} we show the dependence of the transfer angular distribution on the transferred angular momentum for $^{58}$Ni$(d,p)^{59}$Ni at $10$ MeV. The transferred angular momentum has an influence on the shape of the transfer distribution, as well as the location of the peak of the transfer distribution. It is seen that for the $s_{1/2}$ state the peak occurs at $0^{\circ}$, and for increasing angular momentum transfer the first peak gets shifted to increasing angles. The oscillations of the transfer distribution can be understood in terms of a diffraction pattern, analogous to that of a single slit diffraction pattern. With increasing energy the diffraction pattern is found to have more oscillations. Also, as the beam energy increases, the transfer distribution gets shifted to more forward angles. The magnitude of the cross section is related to the $Q$-value, or energy mismatch, of the reaction. The magnitude of the cross section is largest when $Q=0$, and decreases as energy mismatch increases.

Modern reaction theories have progressed greatly since the $1950$s, allowing for more reliable nuclear structure information to be extracted from experimental data. The theoretical advances of reaction theory coupled with advances in experimental techniques have made the use of transfer reactions to study exotic nuclei feasible. In the early days of nuclear physics, transfer reactions were performed by making a target using stable nuclei, and impinging protons, deuterons, $^3$He, or other nuclei on the target to initiate the transfer process. When studying unstable nuclei, these reactions are done in inverse kinematics \cite{Jones_prc2011, Jones_prc2004, Winfield_npa2001, Catford_jpg2005, Fernandez-Dominguez_prc2011, Thomas_prc2007}. Since exotic nuclei are too short lived to make a target, a deuterated target, for example, is sometimes used, and a beam of exotic nuclei is impinged on the target to perform the experiment.

As $(d,p)$ or $(d,n)$ transfer reactions are a useful tool for studying the overlap function of the final nucleus, these reactions are also a preferred method to extract the normalization of the tail of the overlap function. This quantity is known as the asymptotic normalization coefficient (ANC), and is defined in Eq.(\ref{eq:ANC}). At very low energies, the transfer cross section is dominated by the amplitude of the overlap function in the asymptotic region. Thus, a $(d,n)$ transfer reaction can provide information on the proton bound state of the final nucleus. The ANC can be used to determine astrophysically important $(p,\gamma)$ reaction rates at energies unobtainable experimentally via the ANC method \cite{Xu_prl1994}.

Making use of the ANC method, transfer reactions have also become a common tool to extract information relevant in the understanding of astrophysically important processes \cite{Burjan_jpg2013, Fukui_prc2015}. Sometimes, the ANC for the system of interest is not accessible directly, while the mirror system is. When this is the case, charge symmetry of the nuclear force can be exploited to derive a model independent quantity relating the ratio of ANCs of the two systems \cite{Timofeyuk_prl2003}. This has been shown to be a reliable method to indirectly extract an ANC \cite{Timofeyuk_prc2005, Titus_prc2011}, and has been used in practice \cite{Guo_npa2005, Guo_prc2006}. 

Whereas ANCs calculated theoretically can be very different depending on the model that is used, the idea behind the method proposed in \cite{Timofeyuk_prl2003} suggests that the ratio of ANCs of mirror pairs is model independent.  This method is very useful to extract the ANC of the proton state, useful in $(p,\gamma)$ reactions important for astrophysics, by measuring the mirror partner. In the early stage of my graduate work, we performed a study to test the model independence of the ratio of ANCs of mirror pairs, and the validity of the analytic formula derived in \cite{Timofeyuk_prl2003}. This project is discussed in Appendix \ref{Mirror_Symmetry}.




\section{Nuclear Interactions}

The elastic scattering of a nucleon off of a nucleus is a complicated quantum many-body problem. To solve the problem exactly would require the fully anti-symmetrized many-body wave function that includes the couplings of the elastic channel to all the other non-elastic channels available (transfer, inelastic scattering, charge exchange, fusion, fission, etc.). This is a very difficult problem to solve, and in practice, the scattering process is not solved in this manner. However, the elastic scattering of a particle from some arbitrary potential, $U(R)$, is well understood \cite{Thompson_book, Goldberger_book}. Assuming that the complicated interaction between some particle and the nucleus can be represented by a complex mean-field is the basis of the optical model. 

\begin{figure}[h!]
\begin{center}
\includegraphics[width=0.9\textwidth]{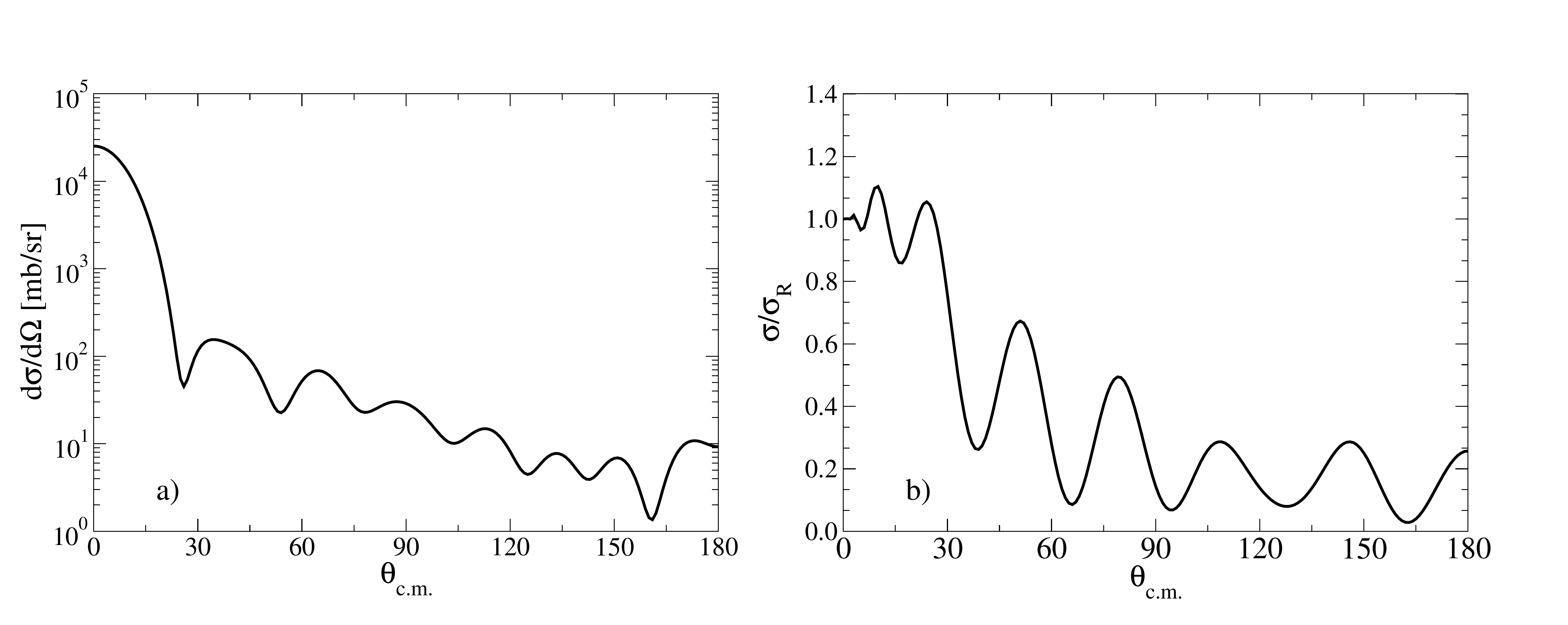}
\end{center}
\caption{Angular distributions for elastic scattering of nucleons off $^{208}$Pb at $25$ MeV. (a) $n+^{208}$Pb (b) $p+^{208}$Pb with differential cross section normalized to Rutherford.}
\label{fig:Elastic_Scattering}
\end{figure}

In Fig. \ref{fig:Elastic_Scattering} we show the angular distributions for elastic scattering of nucleons off $^{208}$Pb at $25$ MeV. In panel (a) is $n+^{208}$Pb, and in panel (b) is $p+^{208}$Pb. Due to the Coulomb potential, proton elastic scattering is usually normalized to Rutherford, which is the point-Coulomb cross section, and always goes to unity at $0^{\circ}$. When this is done, the angular distributions for proton elastic scattering are unitless. The oscillations result from a diffraction pattern which can be understood qualitatively in a similar way as single slit diffraction. For a larger target or lower energy, there will be fewer oscillations between $0^{\circ}$ and $180^{\circ}$, and there will be more oscillations for a smaller target or a higher energy. 

In the optical model, elastic scattering data are fit by varying potential parameters in an assumed form for $U(R)$. This complex interaction, referred to as the optical potential, is used to describe the elastic scattering process of the particle off the nucleus, with the imaginary part taking into account loss of flux to non-elastic channels. Once the optical potential is defined, it can then be used as an input to a model that describes some other process with the goal of obtaining an observable other than elastic scattering, such as transfer cross sections.

Elastic scattering data for the desired target and energy are often times not available. To remedy this problem, optical potentials are constructed through simultaneous fits to large data sets of elastic scattering. These are referred to as global optical potentials. The energy, target, and projectile dependent parameters are varied to produce a best fit to the entire data set. The purpose of using a global potential is that one can easily interpolate in order to obtain a potential for a nucleus in which there is no experimental data available. Obtaining a potential, and therefore predictions on observables, of un-measured nuclei is a very attractive feature of using a global potential and is a credit to their success over the decades. It is for this reason that considerable effort has been put into creating many different global optical potentials over the years which have received widespread use \cite{Perey_adndt1974, Varner_pr1991, Koning_np2003}. 

Global potentials are a very useful tool for studying nuclear reactions and predicting observables. However, the way they are constructed leaves out a considerable amount of physics. Elastic scattering only constrains the normalization of the scattering wave function outside the range of the interaction. It is not sensitive to the short-range properties of the wave function. Therefore, the short-range physics is not constrained at all by elastic scattering. Also, much of the elastic scattering data that exists is for stable nuclei. With the increasing interest of the study of rare isotopes, the extrapolations to exotic nuclei may not be reliable. It is for this reason that a more physically motivated form for the optical potential should be pursued.

All widely used global optical potentials are local. However, when derived from a many-body theory, the resulting optical potential is nonlocal. The strong energy dependence of global potentials is assumed to account for the nonlocality that is neglected. With increasing interest in microscopically derived optical potentials, it is becoming necessary to investigate the validity of the local assumption, and develop methods to incorporate nonlocal potentials into modern reaction theories.


\section{Nonlocality}
\label{Sec:Nonlocality}

It has long been known that the optical potential is nonlocal \cite{Bell_prl1959}. In the Hartree-Fock theory, the existence of an exchange term introduces an explicit nonlocal potential \cite{Dickhoff_Book}. For scattering, the complicated coupling of the elastic channel to all other non-elastic channels accounts for another significant source of nonlocality \cite{Feshbach_ap1958, Feshbach_ap1962}. These two sources of nonlocality, anti-symmetrization and channel couplings, have been known and studied for decades (e.g. \cite{Fraser_epja2008}). 

\begin{figure}[h!]
\begin{center}
\includegraphics[scale=0.12]{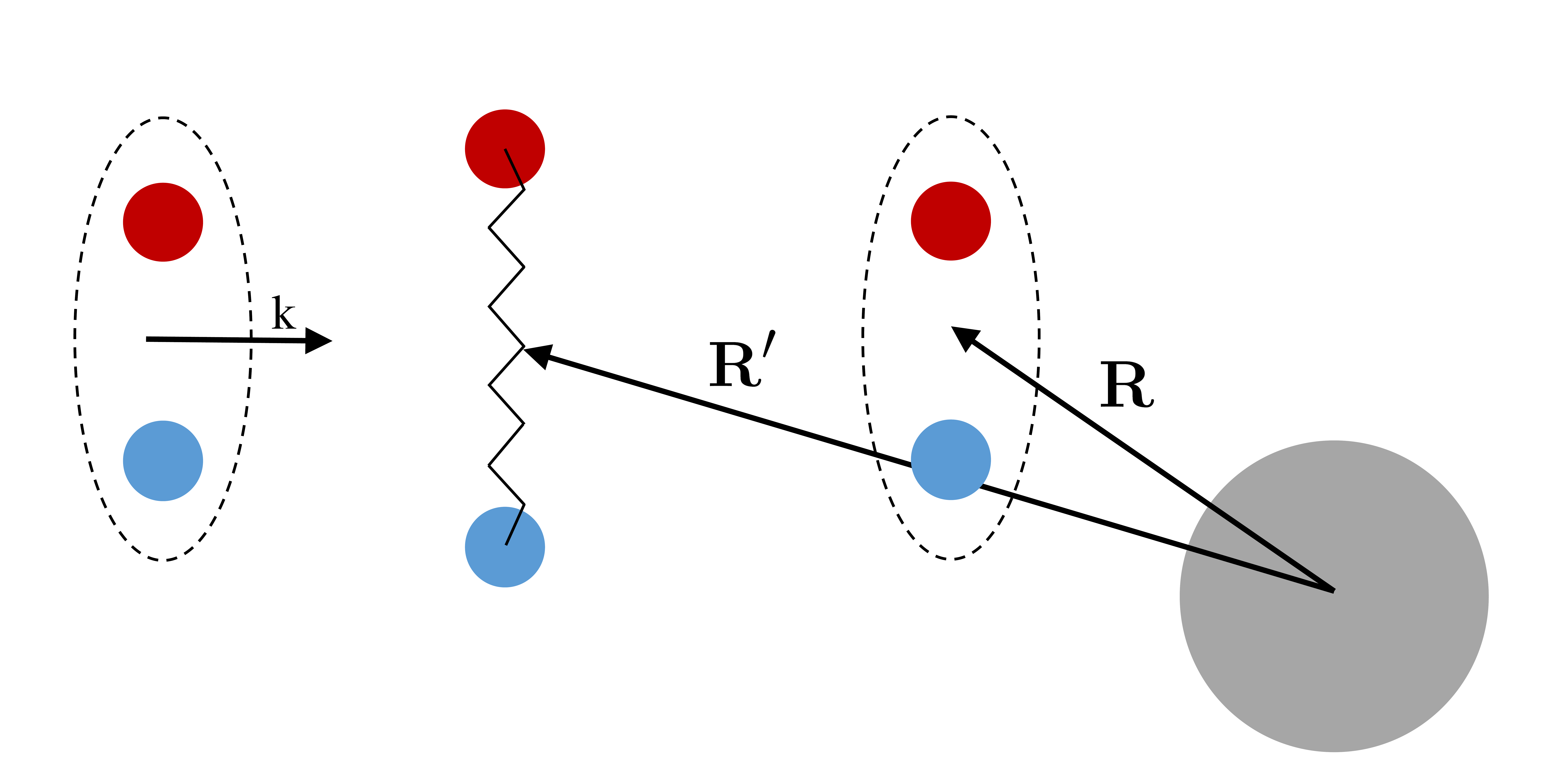}
\end{center}
\caption{An example of a channel coupling nonlocality. In this case, the deuteron is impinged on some target. The channel coupling nonlocality results from the deuteron breaking up as it approaches the nucleus, propagating through space in its broken up state, and then recombining to form the deuteron again.}
\label{fig:NL_Example}
\end{figure}

As a physical example, consider a deuteron impinging on a target, and let \textbf{R} and \textbf{R}' locate the center of the deuteron relative to the center of the target. Let's say that the deuteron breaks up at $\textbf{R}'$ as it approaches the target. The deuteron can then propagate through space in its broken up state, then recombine to form the deuteron again at $\textbf{R}$. This process is depicted in Fig. \ref{fig:NL_Example}. Such a process would constitute a channel coupling nonlocality. This will result in a potential of the form $V(\textbf{R},\textbf{R}')$ since the interaction at a given point is dependent on the value of the potential and the scattering wave function at all other points in space.

As another example, consider a single nucleon scattering off a nucleus. Since the system wave function is a fully anti-symmetric many-body wave function, it is not guaranteed that the projectile in the incident channel is the same particle as the one in the exit. The Pauli principle also plays a role when the projectile is propagating through the nuclear medium, and most notably has the effect of reducing the amplitude of the wave function in the nuclear interior. All of these effects will manifest in a potential of the form $V(\textbf{R},\textbf{R}')$.

\subsection{Microscopic Optical Potentials}

This work is not concerned with constructing a microscopic optical potential, but rather with using current phenomenological nonlocal optical potentials, and studying the effects of nonlocality on transfer observables. However, it is important to understand the considerable amount of effort that has been put forth in recent decades to construct optical potentials from microscopic theories. In this thesis we will demonstrate that nonlocality is an important feature of the nuclear potential that must be considered explicitly. Moving forward, the development of ab-initio many-body theories offer the promise of realistic microscopic optical potentials. The methods outlined here will be the tools for future studies.

Several studies have been made to construct a microscopically based optical potential. In the pioneering work of Watson \cite{Watson_pr1953, Francis_pr1953}, and later refined by Kerman, McManus, and Thaler (KMT) \cite{Kerman_ap1959}, the theory of multiple scattering was developed, where the optical potential to describe elastic scattering is constructed in terms of the amplitudes for the scattering of the incident particle by the individual neutrons and protons in the target nucleus. This theory for constructing the optical potential is limited to relatively high energies ($>100$ MeV). Deriving the multiple scattering expansion of the KMT optical potential is a complicated task, but has been done successfully, such as for $^{16}$O \cite{Crespo_prc1992}. 

The optical potential can also be identified with the self-energy, as first indicated by Bell and Squires \cite{Bell_prl1959}. H\"ufner and Mahaux studied the optical potential in great detail through use of a systematic expansion of the self-energy within the Greens function approach to the many-body problem \cite{Hufner_ap1972}. This approach is analogous to the Bethe-Brueckner expansion for the calculation of the binding energy \cite{Day_rmp1967}. This formulation of the optical potential in terms of the self-energy is attractive as it is suitable for both intermediate and high energy scattering, and reduces to the expressions of multiple scattering theory at high energies. 

It is through the connection to the self-energy that Jeukenne, Lejeune, Mahaux (JLM) formulated their optical model potential for infinite nuclear matter \cite{Jeukenne_prc1974}. In infinite nuclear matter, the concept of a projectile and target lose their meaning. Instead, a potential energy and a lifetime for a quasiparticle state obtained by creating a particle or hole with momentum $k$ above the correlated ground state is defined. Later, the JLM approach was extended to finite nuclei using a local density approximation \cite{Jeukenne_prc1977}. 

The link between the self-energy and the optical potential was further explored by Mahaux and Sartor \cite{Mahaux_anp1991}. This implementation is known as the dispersive optical model (DOM). The advantage of this method is that it provides a link between nuclear reactions and nuclear structure through a dispersion relation. In recent years, a local version of the DOM was introduced for Calcium isotopes \cite{Charity_prc2007}, and a nonlocal DOM was subsequently developed for $^{40}$Ca \cite{Dickhoff_prc2010, Mahzoon_prl2014}. Transfer reaction studies have shown that the local DOM is able to describe transfer cross sections as well as or better than global potentials \cite{Nguyen_prc2011}, and that the nonlocal DOM can significantly modify the shell occupancy, or spectroscopic factor, of the states populated in transfer reactions \cite{Ross_prc2015}. 

Various other techniques exist which construct an optical potential through the self-energy using modern advances in nuclear theory. Making use of the progress that has been achieved, Holt and collaborators constructed a microscopic optical potential from the self-energy for nucleons in a medium of infinite isospin-symmetric nuclear matter within the framework of chiral effective field theory \cite{Holt_prc2013}. 

The two sources of nonlocality, channel coupling and anti-symmetrization, have been studied over the years by numerous authors \cite{Fraser_epja2008, Rawitscher_prc1994, Rawitscher_npa1987} to name only a few. Many of these studies derive the nonlocal potential using some microscopic theory, then compare the potential obtained to commonly used phenomenological nonlocal potentials. Such was done in \cite{Fraser_epja2008}  where the multichannel algebraic scattering (MCAS) method \cite{Amos_npa2003} was used to obtain the nonlocal potential resulting from channel coupling. The resulting nonlocal potential was found to be very different from the simple Gaussian nonlocalities assumed in phenomenological potentials. However, the MCAS method is only suitable for very low energy projectiles, where just a few excited states are relevant to the coupling, and thus, can be explicitly coupled together to generate the channel coupling nonlocal potential. 

\subsection{Phenomenological Nonlocal Optical Potentials}

The formalism to develop a microscopic optical potential is complicated, and requires considerable computation time to implement. However, constructing a nonlocal potential phenomenologically provides a practical alternative to construct a nonlocal potential applicable for widespread use. The seminal work of Perey and Buck, \cite{Perey_np1962}, was the first attempt to constrain the parameters of a nonlocal potential through fits to elastic scattering data. This work was done in the sixties, but it is still the most commonly referenced nonlocal optical potential. In the late seventies, Giannini and Ricco constructed a phenomenological nonlocal optical potential, \cite{Giannini_ap1976, Giannini_ap1980}. In that work, the potential parameters were constrained with fits to a local form, then a transformation formula was used to obtain the nonlocal potential. Very recently, Tian, Pang, and Ma (TPM) introduced a third nonlocal global optical potential, \cite{Tian_ijmpe2015}. These three works are to our knowledge the only attempts to construct a phenomenologial nonlocal global optical potential. 

A common feature of using a nonlocal potential is that the amplitude of the wave function is reduced in the nuclear interior as compared to the wave function resulting from using a local potential. Numerous studies have been performed to investigate this effect, and to find ways to correct for it \cite{Austern_pr1965, Austern_Book, Fiedeldey_np1966}. These studies were focused on potentials of the form of the phenomenological Perey-Buck nonlocal potential. A local equivalent potential to the nonlocal potential should formally exist. Attempts have been made to find this local equivalent potential \cite{Fiedeldey_npa1967, Coz_ap1970}. In nearly all these cases, the Perey-Buck form for the nonlocal potential was assumed. 

\subsection{Solving Nonlocal Equations}

While the theoretical foundation for constructing nonlocal potentials has been around for many decades, the broad application of nonlocal potentials in the field of nuclear reactions has never come to fruition. With a nonlocal potential, the Schr\"odinger equation transforms from a differential equation to an integro-differential equation. Therefore, the most straightforward way to solve the equation is through iterative methods, which dramatically increases the computational cost. 

Since the knowledge of nonlocality dates back to the 1950s when computer power was much more limited than today, the preferred method was to include nonlocality approximately through a correction factor \cite{Austern_pr1965, Austern_Book, Fiedeldey_np1966}. However, several methods now exist that improve the efficiency of the basic iteration scheme. Kim and Udagawa have presented a rapid method using the Lanczos technique \cite{Kim_prc1990, Kim_cpc1992}. A method by Rawitscher uses either Chebyshev or Sturmian functions as a basis to expand the scattering wave function \cite{Rawitscher_npa2012}. Also, an improved iterative method has been proposed by Michel \cite{Michel_epja2009}. 

Computation time is no longer an issue. In this work, we used an iterative method outlined in Appendix \ref{SolvingEquation} to solve the integro-differential equation. This is, by far, the easiest, but definitely not the most efficient way to solve the equation. Since the increase in computation time is minimal, pursuing a faster way was not a priority and will be pursued at a later time. If one desired to construct their own global nonlocal potential by fitting large amounts of elastic scattering data, it would be advantageous to further optimize our technique.


\section{Motivation for present work}

In this work, we would like to describe single nucleon transfer reactions involving deuterons while using nonlocal optical potentials. Ever since the early days of nuclear physics, right up to the modern day, the distorted wave Born approximation (DWBA) has been a common theory used to analyze data from transfer reaction experiments \cite{Lee_pr1964-2,Kanungo_plb2010}. In the DWBA, the transfer process is assumed to occur in a one-step process, and an optical potential fitted to deuteron elastic scattering is used to describe the deuteron scattering state. The shortcoming of the DWBA is that the deuteron is loosely bound, so it is likely that the deuteron will breakup as it approaches the nucleus. Not taking deuteron breakup into account explicitly can have a significant effect on transfer cross sections. In all known implementations of the DWBA to describe transfer cross sections, local deuteron optical potentials have been used. These deuteron optical potentials were obtained either by fitting a single elastic scattering angular distribution, or using a global parameterization such as that from Daehnick \cite{Daehnick_prc1980}. 

In order to include deuteron break up explicitly, it is necessary to include the $n-p$ degrees of freedom. This then requires solving the $n+p+A$ three-body problem. A three-body approach was introduced in the zero range approximation by Johnson and Soper \cite{Johnson_prc1970}, and later extended to include finite range effects by Johnson and Tandy \cite{Johnson_npa1974}. This is known as the adiabatic distorted wave approximation (ADWA). A recent systematic study of $(d,p)$ reactions within the formalism of \cite{Johnson_npa1974} shows the importance of finite range effects \cite{Nguyen_prc2010}. In these theories the deuteron scattering state is treated as a three-body problem, composed of $n+p+A$. The breakup of the deuteron is included explicitly, and the input potentials are neutron and proton optical potentials, which are much better constrained than deuteron optical potentials. In this sense, ADWA is a more advanced theory than the DWBA with the added advantage that nucleon optical potentials exist in a nonlocal form. Therefore, in this work the explicit inclusion of nonlocality in single nucleon transfer reactions within the ADWA will be pursued.

As mentioned before, nonlocality in $(d,p)$ transfer reactions has traditionally been included approximately through a correction factor. This is the method exploited in commonly used transfer reaction codes such as TWOFNR \cite{twofnr}. The bound and scattering wave functions are calculated using a suitable local potential, normally a global potential for elastic scattering and a mean field reproducing the experimental binding energy for the bound state. The correction factor used implies that the nonlocality assumed is of the Perey-Buck form. From microscopic calculations, it is known that a single Gaussian is not sufficient to take into account the complex nature of nonlocality \cite{Fraser_epja2008}. Therefore, not only is this method of including nonlocality not accurate, but it is limited to a form for the nonlocality that may not adequately represent the true nonlocality in the nuclear potential.

Recently, some attempts have been made to include nonlocality within the adiabatic model by introducing an energy shift to the optical potentials used to calculate the scattering wave functions \cite{Timofeyuk_prl2013, Timofeyuk_prc2013}. This method is very attractive as all local codes which calculate $(d,p)$ transfer can still be used without modification. However, the adequacy of this energy shift to take nonlocality into account must be quantified. Another limitation of this method is that it relies on energy independent nonlocal nucleon optical potentials assumed to have the Perey-Buck form.

While the existence of nonlocality in the optical model has been known for many decades, not many calculations of transfer reactions with the explicit inclusion of nonlocality have ever been performed. While the approximate ways to correct for nonlocality are common, it is not known if these approximate methods are sufficient. The method of constructing local optical potentials through fits to elastic scattering data has been practical and useful, but since elastic scattering does not constrain the short range nonlocalities present in the nuclear potential, it is unlikely this approach to constructing the optical potential will be reliable when moving towards exotic nuclei. Also, it must be understood how other observables are affected due to the way in which the optical potentials are constructed. 

The goal of this thesis is to study the explicit inclusion of nonlocality on single nucleon transfer reactions involving deuterons. Since nonlocality has either been ignored or included approximately in nearly all reaction calculations for over half a century, the effect of neglecting nonlocaly on reaction observables must be quantified. Also, the quality of the commonly used approximate techniques need to be assessed. For this purpose we extend the formalism of the ADWA to include nonlocality. Finally, with renewed interest in microscopic optical potentials, the formalism must be kept general so that nonlocal potentials of any form can be used.

In this thesis, we will first test the concept of the correction factor using the Perey-Buck potential. This will be done by performing DWBA calculations of $(p,d)$ reactions on a wide range of nuclei and energies. The correction factor will be applied to the proton scattering state, and the neutron bound state in the entrance channel. We will then include nonlocality explicitly in the entrance channel in order to quantify the adequacy of the correction factor to account for nonlocality. For this part of the study, a local deuteron optical potential will be used to describe the deuteron scattering state within the DWBA.

Since it is well known that deuteron breakup plays an important role in describing the reaction dynamics, it is crucial to incorporate nonlocality into a reaction theory that explicitly includes deuteron breakup. Thus, we chose to extend the formalism of the ADWA to include nonlocal potentials. Also, since the Perey-Buck form for the nonlocality is not consistent with microscopic calculations, the formalism was kept general so that it can be used with a nonlocal potential of any form that may result from a microscopic calculation.

Finally, through a systematic study, the effect of ignoring nonlocality in the optical potential on transfer observables can be quantified. We will choose a range of nuclei and energies, and perform calculations of $(d,p)$ transfer reactions using nonlocal potentials in both the entrance and exit channels. The resulting cross sections will be compared to cross sections generated from local phase equivalent potentials in order to quantify the effect of neglecting nonlocality in the optical potential.


\section{Outline}

This thesis is organized in the following way. In chapter \ref{Theory} we will present the necessary theory. We will begin with a discussion of elastic scattering, and the two-body T-matrix. We will extend the two-body T-matrix to three-bodies. Then we will introduce the adiabatic distorted wave approximation, and finally extend this theory to include nonlocal potentials. In chapter \ref{Potentials} we will discuss optical potentials. First we will introduce the concept of a global optical potential, then turn our attention to nonlocal potentials. We will introduce Perey-Buck type potentials, and the corresponding correction factor. We will then describe the Giannini-Ricco potential and the DOM nonlocal potential. Last there will be a discussion of local equivalent potentials. In Chapter \ref{Results} we will present our results beginning in Sec. \ref{PB_Transfer} with a discussion of $(p,d)$ reactions using the Perey-Buck potential in the entrance channel within the DWBA. In Sec. \ref{DOM_Transfer} we compare the effects of including the DOM potential and the Perey-Buck potential in the entrance channel of $(p,d)$ reactions using the DWBA. Lastly, in Sec. \ref{NL_ADWA} we study $(d,p)$ transfer reactions within the ADWA while including nonlocality consistently. Finally, in Chapter \ref{Conclusions} we will draw our conclusions and discuss the outlook for future work.

Some of the work developed during this thesis, while critical, is too technical to present in the main body. We have thus collected that information in the following appendices. In Appendix \ref{SolvingEquation} we discuss the method by which we solve the scattering and bound state nonlocal equations. In Appendix \ref{Correction_Factor} we derive the correction factor that is applied to wave functions resulting from a local potential in order to account for the neglect of nonlocality. In Appendix \ref{Nonlocal_Adiabatic} we derive the nonlocal adiabatic potential, and in Appendix \ref{tmatrix} we derive the partial wave decomposition of the T-matrix used to calculate transfer reaction cross sections. In Appendix \ref{Checks} we go over some checks to ensure the accuracy of the code I developed to compute transfer cross sections, NLAT (NonLocal Adiabatic Transfer). In Appendix \ref{Mirror_Symmetry}, we discuss a method to extract astrophysically relevant ANCs using the concept of mirror symmetry. While Appendix \ref{Mirror_Symmetry} is a research project of relevance to the field that stands on its own \cite{Titus_prc2011}, it does not fit the theme of the thesis. Therefore, we chose to include it as a separate appendix.


\chapter{Reaction Theory for the Transfer of Nucleons}
\label{Theory}

Elastic scattering is the anchor of many reaction theories since elastic scattering wave functions are often times inputs to these theories, and are used to calculate quantities such as transfer cross sections. Elastic scattering is also the primary means by which we construct the nuclear potential. Therefore, for reaction theory to make useful predictions, we must have a good understanding of elastic scattering.

The theoretical study of transfer reactions commonly uses the distorted-wave Born approximation (DWBA). In this theory, the transfer process is assumed to be a single step, and the breakup of the deuteron is included implicitly through the deuteron optical potential. The deuteron is loosely bound, and is likely to breakup during the course of the reaction. Therefore, not including the breakup of the deuteron explicitly is known to be inaccurate \cite{Schmitt_prl2012}. Despite breakup not being included explicitly, the DWBA theory is still commonly used to describe transfer reactions due to its simplicity and the legacy of codes available.

 Modern reaction theories that incorporate breakup begin with the three-body picture of the process. The three bodies are the neutron and the proton making up the incident deuteron, and the target nucleus. A practical method for including deuteron breakup was introduced by Johnson and Tandy \cite{Johnson_npa1974}. This method is usually referred to as the adiabatic distorted-wave approximation (ADWA). The ADWA has been benchmarked with more advanced techniques \cite{Nunes_prc2011, Upadhyay_prc2012}, and shown to be competitive.  In \cite{Nunes_prc2011}, $(d,p)$ angular distributions for the ADWA and the exact Faddeev method are compared. It was found that the results from the ADWA are within 10\% of the full solution at forward angles, demonstrating that the ADWA is a reliable and practical method for calculating angular distributions of transfer reactions. The ADWA theory will be the focus of this work. 

An attractive feature of the ADWA is that it includes breakup explicitly, and also relies on nucleon optical potentials, which are much better constrained than the deuteron optical potentials used in the DWBA. In all known uses of the ADWA, local nucleon optical potentials were used. However, recent studies have shown that the nonlocality of the nuclear potential can have a significant impact on transfer cross sections \cite{Deltuva_prc2009, Titus_prc2014, Ross_prc2015}. Thus, it has become necessary to extend the ADWA formalism to include nonlocal potentials \cite{Titus_prc2015}.


\section{Elastic Scattering}
\label{Sec:Elastic_Scattering}

To describe elastic scattering distributions, we begin by solving the partial wave decomposed Schr\"odinger equation

\begin{eqnarray}\label{radial-eqn}
\left[-\frac{\hbar^2}{2\mu}\left(\frac{\partial^2}{\partial R^2}-\frac{L(L+1)}{R^2} \right)+U_N(R)+V_C(R)-E \right]\psi_{\alpha}(R)=0,
\end{eqnarray}

\noindent with $U_N(R)$ being some short-range nuclear potential, $V_C$ the Coulomb potential, $\mu$ the reduced mass of the projectile target system, and $E$ the projectile kinetic energy in the center of mass frame. Here, $\alpha=\{L I_p J_p I_t\}$ is a set of quantum numbers that define each partial wave, where $L$ is the orbital angular momentum between the projectile and the target, $I_p$ and $I_t$ are the spin of the projectile and target respectively, and $J_p$ is the angular momentum resulting from coupling the orbital angular momentum with the spin of the projectile. In the asymptotic limit where the nuclear potential goes to zero, the scattering wave function takes the form

\begin{eqnarray}\label{scat-asymptotic}
\psi_{\alpha}(R)=\frac{i}{2}\left[H^-_{L}(\eta_{L},kR)-\textbf{S}_{\alpha}H^+_L(\eta_{L},kR) \right],
\end{eqnarray}

\noindent where $\eta=Z_1Z_2e^2\mu/\hbar^2k$ is the Sommerfeld parameter, k is the wave number, $\textbf{S}_{\alpha}$ is the scattering matrix element (S-matrix), and $H^-$ and $H^+$ are the incoming and outgoing Hankel functions \cite{Abramowitz_Book}, respectively. For neutrons, $\eta=0$. The theoretical scattering amplitude for elastic scattering is related to the S-Matrix by

\begin{eqnarray}
f_{\mu_p \mu_t \mu_{p_i} \mu_{t_i}}(\theta)&=&\delta_{\mu_p \mu_{p_i}}\delta_{\mu_t \mu_{t_i}}f_c(\theta) +\frac{2\pi i}{k_i}\sum_{L_i L J_{p_i} J_p M_{p_i} M_{p} M_i J_{T}}C_{L_i M_i I_{p_i} \mu_{p_i}}^{J_{p_i} M_{p_i}}C_{J_{p_i}M_{p_i} I_{t_i} \mu_{t_i}}^{J_{tot} M_{tot}} \nonumber \\
&\phantom{=}& \times \ C_{L M I_p \mu_p}^{J_p M_p}C_{J_p M_p I_t \mu_t}^{J_{tot} M_{tot}}Y_{L M}(\hat{k})Y^*_{L_i M_i}(\hat{k}_i) \nonumber \\
&\phantom{=}& \times \ \left(1-\textbf{S}_{\alpha} \right)e^{i\left(\sigma_{L}(\eta_\alpha)+\sigma_{L_i}(\eta_{\alpha_i})\right)}
\end{eqnarray}

\noindent with $\mu_{p_i}$ and $\mu_{t_i}$ being the projections of the spin of the projectile and target, respectively, before the scattering process, while $\mu_{p}$ and $\mu_{t}$ are the spin projections after the scattering process.  In this equation, $f_c$ is the point Coulomb scattering amplitude:

\begin{eqnarray}
f_c(\theta)=-\frac{\eta}{2k\sin^2(\theta/2)}\exp\left[-i\eta\ln(\sin^2(\theta/2))+2i\sigma_0(\eta) \right],
\end{eqnarray}

\noindent with the Coulomb phase given by $\sigma_L(\eta)=\textrm{arg}\Gamma(1+L+i\eta)$. 

The $S_{\alpha}$ are determined by matching a numerical solution of Eq.(\ref{radial-eqn}) to the known asymptotic form (\ref{scat-asymptotic}). This is done by constructing the $R$-Matrix, which is simply an inverse logarithmic derivative. 

\begin{eqnarray}
\textbf{R}_{\alpha}=\frac{1}{R_{match}}\frac{H^-_{L}-\textbf{S}_{\alpha}H^+_{L}}{{H^-_{L}}'-\textbf{S}_{\alpha}{H^+_{L}}'}
\end{eqnarray}

\noindent with the primes indicating derivatives with respect to $R$. The $R$-Matrix is evaluated at some matching point outside the range of the nuclear interaction, denoted by $R_{match}$. The $R$-matrix uniquely determines the $S$-matrix by 

\begin{eqnarray}
\textbf{S}_{\alpha}=\frac{H^-_L-R_{match}\textbf{R}_{\alpha}{H^-_L}'}{H^+_L-R_{match}\textbf{R}_{\alpha}{H^+_L}'}.
\end{eqnarray}

Once the $S$-matrix for each partial wave is calculated, the theoretical differential cross section, which is the quantity that is compared with experiment, is obtained by summing the squared magnitude of the scattering amplitude over the final $m$-states, and averaging over the initial states:

\begin{eqnarray}
\frac{d\sigma}{d\Omega}&=&\frac{1}{\hat{I}_{p_i}\hat{I}_{t_i}}\sum_{\mu_p \mu_t \mu_{p_i} \mu_{t_i}}\left|f_{\mu_p \mu_t, \mu_{p_i} \mu_{t_i}}(\theta) \right|^2
\end{eqnarray}


\section{Two-Body T-Matrix}

We would like to find the transition amplitude (T-matrix) for a $(d,p)$ transfer reaction. Before we get to transfer reactions, let us first consider the T-matrix for two-body scattering, such as elastic scattering. The discussion of Sec. \ref{Sec:Elastic_Scattering} formulated elastic scattering in terms of an S-matrix. This is the way most codes solve elastic scattering. Another way of formulating elastic scattering is in terms of the T-matrix, and leads to a natural generalization to three-body scattering, which is the case for $d+A$ reactions.

We begin with a partial wave decomposed two-body coupled channel equation \cite{Thompson_book}

\begin{eqnarray}
\left[-\frac{\hbar^2}{2\mu}\left(\frac{d^2}{dR^2}-\frac{L(L+1)}{R^2}\right)+V_c(R)-E \right]\psi_{\alpha}(R)=-\sum_{\alpha'}\langle \alpha|V|\alpha'\rangle \psi_{\alpha'}(R').
\end{eqnarray}

\noindent The T-matrix is an important quantity as it gives the amplitude of the outgoing wave after scattering. In Eq.(\ref{scat-asymptotic}) we wrote the asymptotic form of the scattering wave function in terms of the S-matrix. We can write an equivalent expression for the asymptotic form of the wave function in terms of the T-matrix

\begin{eqnarray}
\psi_{\alpha \alpha_i}(R)\rightarrow \delta_{\alpha \alpha_i}F_{L_i}(\eta_L,kR)+\textbf{T}_{\alpha \alpha_i}H_{L}^{+}(\eta_L,kR),
\end{eqnarray}

\noindent where $F_{\alpha}(R)$ is the regular Coulomb function, and again, $H^+$ is the out going Hankel function. If $U=0$ then $\textbf{T}=0$. The goal is thus to find an expression for the T-matrix. Using Green's function techniques, the T-matrix for two-body scattering is given by \cite{Thompson_book}

\begin{eqnarray}
\textbf{T}_{\alpha \alpha_i}=-\frac{2\mu}{\hbar^2k}\langle \phi^{(-)}|V|\Psi\rangle,
\end{eqnarray}

\noindent where $\phi$ is the homogeneous solution when no coupling potentials are present, $\mu$ is the reduced mass of the two-body system, and $k$ is the wave number. The $^{(-)}$ superscript indicates that $\phi^{(-)}$ has incoming spherical waves as the boundary condition. $\phi^{(-)}$ is thus the time reverse of $\phi$. The complex conjugation implied in the bra-ket notation cancels the complex conjugation implied in the $^{(-)}$. 

Often times, we can decompose $V$ into two parts so that $V=U_1+U_2$. We would like to calculate the T-matrix for the transition when two potentials are present, and derive the \textit{two-potential formula}. We begin by writing the T-matrix substituting in the separated expression for $V$

\begin{eqnarray}
-\frac{\hbar^2k}{2\mu}\textbf{T}^{1+2}=\int \phi \left(U_1+U_2\right)\psi dR.
\end{eqnarray}

Using these two potentials, we can define various functions. $\phi$ is the free field solution, $\chi$ is the solution distorted by $U_1$ only, and $\psi$ is the full solution. These are related to each other through the relations

\begin{eqnarray}\label{ImplicitForms}
[E-T]\phi&=&0 \nonumber \\
\chi&=&\phi+\hat{G}_0U_1 \chi \nonumber \\
\psi&=&\phi+\hat{G}_0(U_1+U_2)\psi \nonumber \\
&=&\chi+\hat{G}_1U_2\psi,
\end{eqnarray}

\noindent with the two Green's functions given by

\begin{eqnarray}
\hat{G}_0&=&[E-T]^{-1} \nonumber \\
\hat{G}_1&=&\left[E-T-U_1 \right]^{-1}.
\end{eqnarray}

\noindent Using these relations, we can rewrite the T-matrix as

\begin{eqnarray}\label{TwoPotFormula}
-\frac{\hbar^2k}{2\mu}\textbf{T}^{1+2}_{\alpha \alpha_i}&=&\int \left[\chi(U_1+U_2)\psi-(\hat{G}_0U_1\chi)(U_1+U_2)\psi \right]dR \nonumber \\
&=&\int [\phi U_1 \chi+\chi U_2 \psi]dR \nonumber \\
&=&\langle \phi^{(-)}|U_1|\chi\rangle+\langle \chi^{(-)}|U_2|\psi \rangle. \nonumber \\
\end{eqnarray}

Consider the elastic scattering of protons as an illustrative example. In this case, $U_1$ could be the Coulomb potential, and $U_2$ could be the nuclear potential. The first term would be the Coulomb scattering amplitude, $f_c(\theta)$, and the second term would be the Coulomb-distorted nuclear amplitude $f_n(\theta)$. Thus, the nuclear scattering amplitude when Coulomb is present is not simply the amplitude due to the short-ranged nuclear forces alone, but from the effect of Coulomb on top of nuclear. From these scattering amplitudes we obtain the differential elastic cross section by calculating $|f_c(\theta)+f_n(\theta)|^2$. This is used in our studies for computing elastic scattering of charged particles.

\subsection{Born Series}
\label{Sec:BornSeries}

Using Eq.(\ref{TwoPotFormula}), and the implicit form for $\psi$ in Eq.(\ref{ImplicitForms}), we can, by iteration, form what is known as the Born series:

\begin{eqnarray}
\textbf{T}^{(1+2)}_{\alpha \alpha_i}&=&\textbf{T}^{(1)}+\textbf{T}^{2(1)} \nonumber \\
&=&\textbf{T}^{(1)}-\frac{2\mu}{\hbar^2k}\left[\langle \chi^{(-)}|U_2|\chi\rangle+\langle \chi^{(-)}|U_2\hat{G}_1U_2|\chi\rangle+ \dots \right].
\end{eqnarray}

\noindent Truncating the series after the first term is known as the first-order distorted-wave Born approximation (DWBA). 

\begin{figure}[h]
\begin{center}
\includegraphics[scale=0.3]{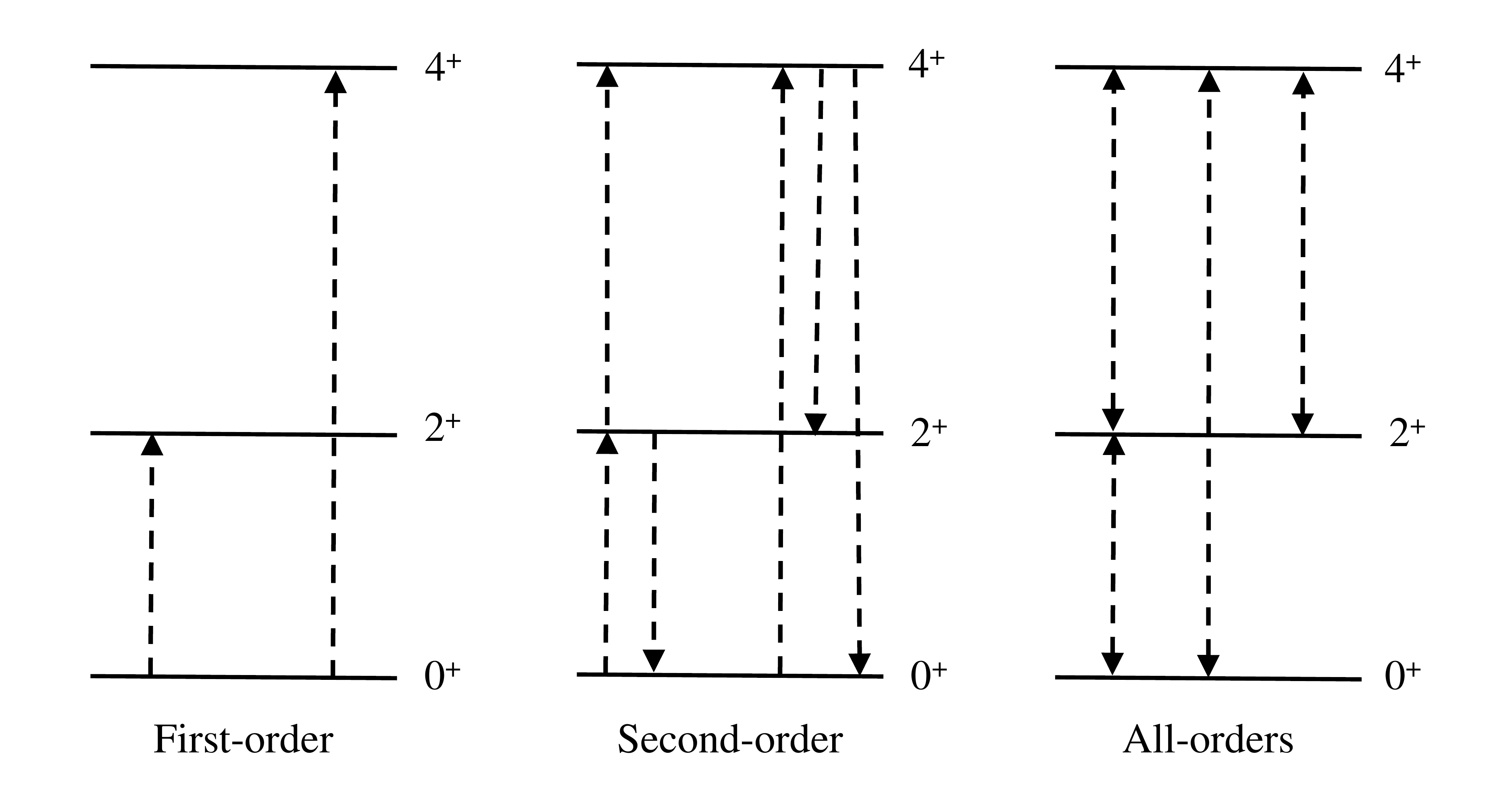}
\end{center}
\caption{First, second, and all-order couplings within a set of $0^+$, $2^+$, and $4^+$ nuclear levels, starting from the ground state. }
\label{fig:DWBA}
\end{figure}

The DWBA is particularly useful when we are describing some kind of transition. If $U_1$ is a central optical potential for all non-elastic channels, it cannot cause the transition since central potentials are not able to change the quantum numbers of the scattered particle, or change their energy. When this is the case, $\textbf{T}^{(1)}=0$, and we get an expression for the T-matrix to describe the transition from an incoming channel $\alpha_i$ to an exit channel $\alpha \ne \alpha_i$.

\begin{eqnarray}
\textbf{T}_{\alpha \alpha_i}^{DWBA}=-\frac{2\mu_\alpha}{\hbar^2k_{\alpha}}\langle \chi_{\alpha}^{(-)}|U_2|\psi_{\alpha_i}\rangle
\end{eqnarray}

\noindent Let us consider, as an example, inelastic excitation of a rotational band in a nucleus. Fig. \ref{fig:DWBA} illustrates first, second, and all order couplings between the $0^+$ ground state, and the $2^+$, and $4^+$ excited states. The first-order DWBA can be thought of as a one step process, where the ground state couples to either the $2^+$ state or the $4^+$ state. Similarly, the second-order DWBA is a two-step process where, for example, the ground state can couple to the $2^+$ state, and then the $2^+$ state can either couple to the $4^+$ state or the $0^+$. For a part of the transfer reaction studies in this thesis, the first-order DWBA was used. From here on out, the first-order DWBA will simply be referred to as the DWBA.


\section{Three-Body T-Matrix}
\label{Sec:Three_Body_Tmatrix}

\begin{figure}[h]
\begin{center}
\includegraphics[scale=0.2]{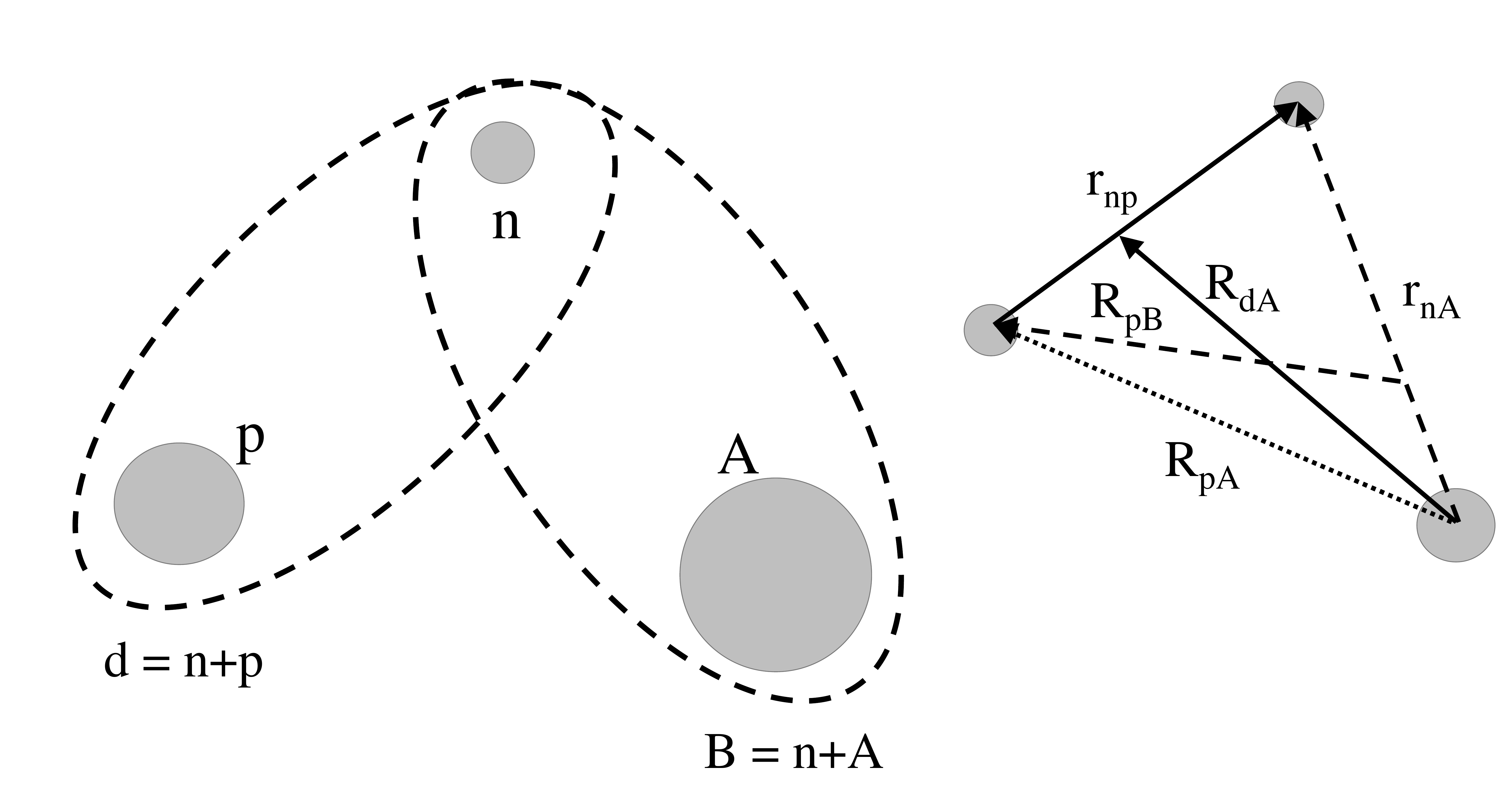}
\end{center}
\caption{The coordinates used in a one particle transfer reaction. }
\label{fig:Transfer_Coordinates}
\end{figure}

We can generalize the above discussion to a three-body system. Consider the collection of the three bodies $n+p+A$, with the coordinates appropriate for $A(d,p)B$ given in Fig. \ref{fig:Transfer_Coordinates}. The coordinates $\textbf{r}_{np}$ and $\textbf{R}_{dA}$ refer to the configuration before the transfer occurs, and the coordinates $\textbf{r}_{nA}$ and $\textbf{R}_{pB}$ are for immediately after the transfer. The Hamiltonian for the three bodies is given by

\begin{eqnarray}
H=T_{\textbf{r}_{np}}+T_{\textbf{R}_{dA}}+V_p(\textbf{r}_{np})+V_t(\textbf{r}_{nA})+U_{pA}(\textbf{R}_{pA}),
\end{eqnarray}

\noindent where $U_{pA}(\textbf{R}_{pA})$ is the core-core optical potential. We can equivalently express the two kinetic energy terms as $T_{\textbf{r}_{np}}+T_{\textbf{R}_{dA}}=T_{\textbf{r}_{nA}}+T_{\textbf{R}_{pB}}$. This allows us to write two different internal Hamiltonians for the bound states, $H_d=T_{\textbf{r}_{np}}+V_p(\textbf{r}_{np})$ and $H_B=T_{\textbf{r}_{nA}}+V_t(\textbf{r}_{nA})$. Thus, we can write the Hamiltonian in two ways, called the post and the prior form

\begin{eqnarray}\label{Hpost}
H&=&H_{prior}=T_{\textbf{R}_{dA}}+U_i(R_{dA})+H_p(\textbf{r}_{np})+\mathcal{V}_i \nonumber \\
&=&H_{post}=T_{\textbf{R}_{pB}}+U_f(R_{pB})+H_t(\textbf{r}_{nA})+\mathcal{V}_f,
\end{eqnarray}

\noindent where $U_{i,f}$ are the entrance and exit channel optical potentials, respectively, and the $\mathcal{V}_{i,f}$ interaction terms are given by

\begin{eqnarray}
\mathcal{V}_i&=&V_t(\textbf{r}_{nA})+U_{pA}(\textbf{R}_{pA})-U_i(R_{dA}) \nonumber \\
\mathcal{V}_f&=&V_p(\textbf{r}_{np})+U_{pA}(\textbf{R}_{pA})-U_f(R_{pB}).
\end{eqnarray}

\noindent For $(d,p)$ reactions it is advantageous to work in the post form. In such a case, we see that $U_{pA}(\textbf{R}_{pA})-U_{pB}(R_{pB})\approx 0$. This term is called the remnant term and approximately cancels for all but light targets. This is because the optical potentials between $p+A$ and $p+(A+1)$ are not likely to be significantly different. We demonstrate that the remnant can be neglected in Sec. \ref{Remnant}.

Just like in the case of elastic scattering, the differential cross section for an $A(d,p)B$ reaction is found by summing the squared magnitude of the scattering amplitude over the final m-states, and averaging over initial states. The T-matrix is related to the scattering amplitude by

\begin{eqnarray}\label{eq:scat-amp}
f_{\mu_A M_d \mu_p M_B}(\textbf{k}_f,\textbf{k}_i)=-\frac{\mu_f}{2\pi\hbar^2}\sqrt{\frac{v_f}{v_i}}T_{\mu_A M_d \mu_p M_B}(\textbf{k}_f,\textbf{k}_i),
\end{eqnarray}

\noindent where the subscript $i (f)$ represents the initial (final) state, $\mu_f$ is the reduced mass, and $v$ is the velocity of the projectile. Here, $\mu_A$, $M_d$, $\mu_p$, and $M_B$ are the projection of the spin of the target in the entrance channel, the deuteron, the proton, and the target in the exit channel, respectively.

We would like to find the T-matrix for $A(d,p)B$ reactions. Using the post representation for the Hamiltonian, Eq.(\ref{Hpost}), and the two-potential formula, Eq.(\ref{TwoPotFormula}), we can identify $U_1=U_f(R_{pB})+V_t(\textbf{r}_{nA})$ and $U_2=\mathcal{V}_f$. Since $U_f(R_{pB})+V_t(\textbf{r}_{nA})$ produces the elastic scattering state of $p+B$, it cannot cause the transfer transition. Therefore, $\textbf{T}^{(1)}=0$. As a result, $\textbf{T}^{2(1)}$ is the only non-zero term. In our notation, the exact T-matrix for a given projection of angular momentum in the post form is:

\begin{eqnarray}\label{Tmatrix}
T_{\mu_A M_d \mu_p M_B}^{post}(\textbf{k}_f,\textbf{k}_i)=\langle \Psi_{\textbf{k}_f}^{\mu_p M_B}|V_{np}+\Delta|\Psi_{\textbf{k}_i}^{\mu_A M_d}\rangle.
\end{eqnarray}

\noindent The remnant term $\Delta=U_{pA}-U_{pB}$ is negligible for all but light targets. The ket in Eq. (\ref{Tmatrix}) for the T-matrix is the full three-body wave function for $n+p+A$, while the bra is the product of a proton distorted wave and the $n+A$ bound state wave function. As a first approximation, we can approximate the ket as a product of a deuteron bound state and a deuteron distorted wave. This is the well known distorted wave Born approximation (DWBA). In this case, the ket is given by

\begin{eqnarray}
|\Psi_{\textbf{k}_i}^{M_d \mu_A}\rangle &=&\Xi_{I_A \mu_A}(\xi_A)\phi_{j_i}(r_{np})\chi_i^{(+)}(\textbf{k}_i,\textbf{r}_{np},\textbf{R}_{dA},\xi_p,\xi_n),
\end{eqnarray}

\noindent where $\Xi_{I_A \mu_A}(\xi_A)$ is the spin function for the target, with spin $I_A$ and projection $\mu_A$. $\phi_{j_i}(r_{np})$ is the radial wave function for the bound state, which in this case is the deuteron, and $j_i$ is the angular momentum resulting from coupling the spin of the fragment in the bound state (the neutron) to the orbital angular momentum between the fragment and the core. The distorted wave, $\chi^{(+)}_i$, is given by

\begin{eqnarray}\label{eq:Deuteron-Distorted-Wave}
&\phantom{=}&\chi_i^{(+)}(\textbf{k}_i,\textbf{r}_{np},\textbf{R}_{dA},\xi_p,\xi_n)=\frac{4\pi}{k_i}\sum_{L_i J_{P_i}}i^{L_i}e^{i\sigma_{L_i}}\frac{\hat{J}_{P_i}}{\hat{J}_d}\frac{\chi_{L_i J_{p_i}}(R_{dA})}{R_{dA}} \\
&\phantom{=}& \times \ \left\{\tilde{Y}_{L_i}(\hat{k}_i)\otimes \left\{\left\{\Xi_{I_p}(\xi_p) \otimes \left\{\tilde{Y}_{\ell_i}(\hat{r}_{np}) \otimes \Xi_{I_n}(\xi_n)\right\}_{j_i} \right\}_{J_d}\otimes \tilde{Y}_{L_i}(\hat{R}_{dA})\right\}_{J_{P_i}}\right\}_{J_d M_d}, \nonumber
\end{eqnarray}

\noindent where $\Xi_{I_p}(\xi_p)$ and $\Xi_{I_n}(\xi_n)$ are the spin functions for the proton and neutron respectively, with spin $I_p=I_n=\tfrac{1}{2}$. The spin of the deuteron is given by $J_d=1$, and the spin of the deuteron coupled to the orbital angular momentum between the deuteron and the target, $L_i$, gives the total projectile angular momentum, $J_{P_i}$. The spherical harmonics, $\tilde{Y}_{L}$, are defined with the phase convention that has a built in factor of $i^L$. Therefore, $\tilde{Y}_L=i^LY_L$ with $Y_L$ defined on p.133 of the book \cite{Varshalovich_Book}. The hatted quantities are given by $\hat{J}=\sqrt{2J+1}$.  The function $\chi_{L_i J_{P_i}}(R_{dA})$ satisfies the equation

\begin{eqnarray}
\left[-\frac{\hbar^2}{2\mu_i}\left(\frac{\partial^2}{\partial R_{dA}^2}-\frac{L_i(L_i+1)}{R_{dA}^2} \right)+U^{dA}+V^{SO}_{1 L_i J_{P_i}}+V_C(R_{dA})-E_d \right]\chi_{L_i J_{P_i}}(R_{dA})=0, \nonumber \\
\end{eqnarray}

\noindent where $V^{SO}$ is the spin-orbit potential, and $V_C$ is the Coulomb potential. $U^{dA}$ is a deuteron optical potential. For the bra we have

\begin{eqnarray}
\langle \Psi_{\textbf{k}_f}^{\mu_p M_B}|&=&\left\{\Xi_{I_A}(\xi_A)\otimes \left\{ \tilde{Y}_{\ell_f}(\hat{r}_{nA})\otimes \Xi_{I_n}(\xi_n) \right\}_{j_f}\right\}^*_{J_B M_B}\phi_{j_f}(r_{nA})\chi_f^{(-)*}(\textbf{k}_f,\textbf{R}_{pB}) \nonumber \\
\end{eqnarray}

\noindent where $\phi_{j_f}(r)$ is the $n+A$ bound state radial wave function, $j_f$ is the angular momentum of the bound state resulting from coupling the spin of the neutron to the orbital angular momentum of the bound state, $\ell_f$, while $J_B$ is the total angular momentum of the final nucleus, $B$. The exit channel distorted wave is given by

\begin{eqnarray}
\chi_f^{(-)*}(\textbf{k}_f,\textbf{R}_{pB},\xi_p)&=&\frac{4\pi}{k_f\hat{I}_p}\sum_{L_f J_{P_f}}i^{-L_f}e^{i\sigma_{L_f}}\hat{J}_{P_f} \frac{\chi_{L_f J_{P_f}}(R_{pB})}{R_{pB}} \nonumber \\
&\phantom{=}& \times \ \left\{\tilde{Y}_{L_f}(\hat{k}_f)\otimes \left\{\Xi_{I_p}(\xi_p) \otimes \tilde{Y}_{L_f}(\hat{R}_{pB}) \right\}_{J_{P_f}} \right\}_{I_p \mu_p}
\end{eqnarray}

\noindent and the function $\chi_{L_f J_{P_f}}(R_{pB})$ satisfies

\begin{eqnarray}
\left[-\frac{\hbar^2}{2\mu_f}\left(\frac{\partial^2}{\partial R_{pB}^2}-\frac{L_f(L_f+1)}{R_{pB}^2} \right)+U^{pB}+V^{SO}_{I_p L_f J_{P_f}}+V_C(R_{pB})-E_p \right]\chi_{L_f J_{P_f}}(R_{pB})=0 \nonumber \\
\end{eqnarray}

\noindent with $U^{pB}$ being the proton optical potential in the exit channel.

\begin{figure}[h!]
\begin{center}
\includegraphics[scale=0.5]{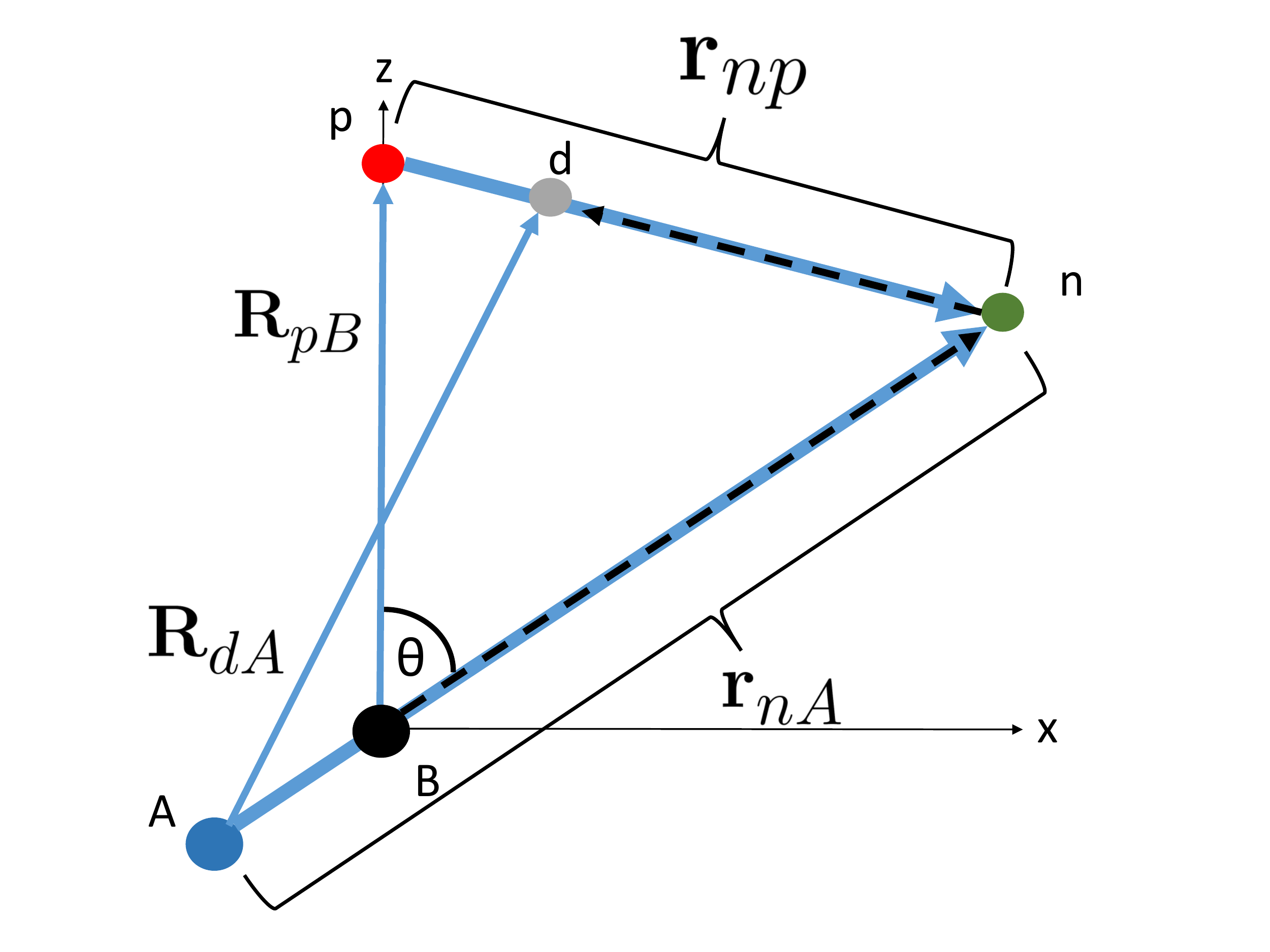}
\end{center}
\caption{The coordinates used to calculate the T-matrix for $(d,p)$ transfer.}
\label{fig:Tmatrix_Coordinates}
\end{figure}

We need to do a partial wave decomposition of the T-matrix, Eq.(\ref{Tmatrix}), so that we can calculate the scattering amplitude, Eq.(\ref{eq:scat-amp}), and hence, the cross section, in a numerically efficient way. We show in Appendix \ref{tmatrix} that for a general $\ell_i$ and $\ell_f$ relative orbital angular momentum in the initial and final bound states, the partial wave decomposition of the T-matrix is given by

\begin{eqnarray}\label{TmatrixSum}
T_{Q M_Q m_f}&=&\mathcal{C}\sum_{K}\sum_{L_i J_{P_i}}\sum_{L_f J_{P_f}}\mathcal{A}^{K,L_i J_{P_i} L_f J_{P_f}}_{Q M_Q m_f}(\hat{k}_f)\mathcal{I}^{K,L_i J_{P_i} L_f J_{P_f}},
\end{eqnarray}

\noindent where phase and statistical factors are collected in

\begin{eqnarray}
\mathcal{C}&=& \frac{32\pi^3\hat{I}_n}{\hat{I}_pk_ik_f}\frac{(-)^{3I_p+j_i+J_d+2j_f}}{\hat{j}_i\hat{j}_f},
\end{eqnarray}

\noindent angular momentum couplings are mostly put in,

\begin{eqnarray}
\mathcal{A}^{K,L_i J_{P_i} L_f J_{P_f}}_{Q M_Q m_f}(\hat{k}_f) &=& \frac{(-)^K}{\hat{K}}\langle \ell_f \ell_i (K) I_n I_n (0) K M| \ell_f I_n (j_f) \ell_i I_n (j_i) K M \rangle  \nonumber \\
&\phantom{=}& \times \ \sum_{L_i J_{P_i}}\sum_{L_f J_{P_f}}i^{3L_i+L_f+\ell_f+\ell_i}e^{i(\sigma_{L_i}+\sigma_{L_f})}\hat{L}_i\hat{L}_f\hat{J}_{P_i}\hat{J}_{P_f} \nonumber \\
&\phantom{=}& \times \langle I_p J_d (j_i) L_f L_i (K) j_f m_f|I_p L_f (J_{P_f}) J_d L_i (J_{P_i}) j_f m_f \rangle \nonumber \\
&\phantom{=}& \times \ \sum_{g} \langle L_f L_i (g) J_{P_f} J_{P_i} (j_f) Q M_Q|L_f J_{P_f} (I_p) L_i J_{P_i} (J_d) Q M_Q \rangle  \nonumber \\
&\phantom{=}& \times \ \sum_{m_g}C_{g m_g j_f m_f}^{Q M_Q}C_{L_f m_g L_i 0}^{g m_g}Y_{L_f m_g}(\hat{k}_f), \nonumber \\ 
\end{eqnarray}

\noindent and the radial integrals are contained in

\begin{eqnarray}
\mathcal{I}^{K,L_i J_{P_i} L_f J_{P_f}} &=& \sum_{M_K}(-)^{M_K}C_{L_f 0 L_i, -M_K}^{K,-M_K}\sum_{\tilde{m}_f\tilde{m}_i}C_{\ell_f \tilde{m}_f \ell_i \tilde{m}_i}^{K M_K}  \nonumber \\
&\phantom{=}& \times \ \int \phi_{j_f}(r_{nA})\chi_{L_f J_{P_f}}(R_{pB})V(r_{np})\phi_{j_i}(r_{np})\chi_{L_i J_{p_i}}(R_{dA}) \frac{R_{pB}r^2_{nA}}{R_{dA}}     \nonumber \\
&\phantom{=}& \times \   Y_{L_i, -M_K}(\hat{R}_{dA})Y_{\ell_f \tilde{m}_f}(\hat{r}_{nA})Y_{\ell_i \tilde{m}_i}(\hat{r}_{np})\sin\theta dR_{pB}dr_{nA}d\theta. \nonumber \\
\end{eqnarray}

\noindent The 9j symbol, $\langle j_1 j_2 (j_{12}) j_3 j_4 (j_{34}) jm| j_1 j_3 (j_{13})j_2 j_4 (j_{24}) j' m' \rangle$, is given on p.334 of \cite{Varshalovich_Book}, while the $C_{j_1 m_1 j_2 m_2}^{j_3 m_3}$ are the Clebsh-Gordan coefficients. The coordinates used to calculate the integral in the equation above are given in Fig. \ref{fig:Tmatrix_Coordinates}. 

With this partial wave decomposition, the differential cross section is given by

\begin{eqnarray}\label{TransferDiffCS}
\frac{d\sigma}{d\Omega}&=&\frac{k_f}{k_i}\frac{\mu_i \mu_f}{4\pi^2\hbar^4}\frac{\hat{J}_B^2}{\hat{J}^2_d \hat{J}^2_A \hat{j}_f^2}\sum_{m_f Q M_Q}T_{Q M_Q m_f}T^*_{Q M_Q m_f}.
\end{eqnarray}

\noindent Introducing Eq.(\ref{TmatrixSum}) into Eq.(\ref{TransferDiffCS}) we obtain the form for the transfer cross section used in this work.


\section{Three-Body Models}

We are interested in describing the reaction $A(d,p)B$ where the final nucleus $B=A+n$ is a bound state. In principle, the scattering state for the deuteron can be modeled as a $d+A$ two-body problem. This is often done where the $d+A$ optical potential is taken from fits to deuteron elastic scattering. However, due to the loosely bound nature of the deuteron, it is important to consider deuteron breakup explicitly. Thus, we begin with a three-body Hamiltonian for the $n+p+A$ system,

\begin{eqnarray}\label{3body-Hamiltonian}
\mathcal{H}_{3B}=T_R+T_r+U_{nA}+U_{pA}+V_{np}.
\end{eqnarray}

Here $T_R$ and $T_r$ are the kinetic energy operators for the center of mass motion and the $n-p$ relative motion, respectively. $V_{np}$ is the neutron-proton interaction, while $U_{pA}$ and $U_{nA}$ are the proton-target and neutron-target interactions. The wave function $\Psi(\textbf{r},\textbf{R})$ describes a deuteron incident on a nucleus $A$ and is a solution to the equation $\mathcal{H}_{3B}\Psi=E\Psi$.  

A variety of methods exist to solve the three-body problem. The Faddeev approach offers an exact method to solve the three-body problem for a particular Hamiltonian \cite{Deltuva_prc2013}, such as the Hamiltonian given in Eq.(\ref{3body-Hamiltonian}). Faddeev methods are computationally expensive, and so far current implementations have difficulties with handling heavy systems due to the Coulomb potential. The Continuum Discretized Coupled Channel (CDCC) method offers another means of solving the three-body problem \cite{Austern_pr1987}. However, this method too is computationally expensive. The ADWA can provide a reliable description of transfer cross sections while requiring minimal computation costs. Studies have benchmarked these three methods and have shown that the ADWA can reliably reproduce transfer cross sections when compared to the other two more advanced methods in the energy ranges relevant for this study \cite{Nunes_prc2011,Upadhyay_prc2012}. 


\section{Adiabatic Distorted Wave Approximation}
\label{Section-ADWA}

Consider the three-body wave function describing the deuteron scattering state. A formal expansion of this wave function is given by

\begin{eqnarray}\label{3BodyWF}
\Psi(\textbf{r},\textbf{R})=\Phi_d(\textbf{r})X_d(\textbf{R})+\int d\textbf{k}\Phi_{\textbf{k}}(\textbf{r})X_{\textbf{K}}(\textbf{R}),
\end{eqnarray}

\noindent where $\Phi_d(\textbf{r})$ is the deuteron bound state wave function, and $X_d(\textbf{R})$ is the elastic deuteron center of mass scattering wave function. $\Phi_{\textbf{k}}(\textbf{r})$ describes the relative motion of an $n-p$ pair, and the continuum components $X_{\textbf{K}}(\textbf{R})$ describe the motion of the center of mass of this $n-p$ pair scattered with relative energy $\epsilon_{\textbf{K}}$. 

In the DWBA, $\Psi(\textbf{r},\textbf{R})=\Phi_d(\textbf{r})X_d(\textbf{R})$, so breakup is not included since the second term in Eq.(\ref{3BodyWF}) is neglected, which contains all the breakup components. While it is known that breakup is important to the dynamics of deuteron induced transfer reactions, calculating the second term in Eq.(\ref{3BodyWF}) to all orders accurately is difficult. 

In formulating the ADWA, Johnson and Tandy \cite{Johnson_npa1974} realized that to calculate transfer cross sections, we need to know the three-body wave function only in the combination $V_{np}|\Psi\rangle$, as is seen in Eq.(\ref{Tmatrix}) with the remnant term neglected. Therefore, an alternative expansion should be sought that accurately represents the three-body wave function within the range of $V_{np}$. The essence of the ADWA method \cite{Johnson_npa1974} is to expand the three-body wave function in a discrete set of Weinberg states,

\begin{eqnarray}\label{Weinberg-Expansion}
\Psi(\textbf{r},\textbf{R})=\sum_{i=0}^{\infty}\Phi_i(\textbf{r})X_{i}(\textbf{R}).
\end{eqnarray}

\noindent The Weinberg states are a complete set of states within the range of the $V_{np}$ interaction, and are given by

\begin{eqnarray}
\left[T_{\textbf{r}}+\alpha_iV_{np}(\textbf{r})+\epsilon_d \right]\Phi_{i}(\textbf{r})=0,
\end{eqnarray}

\noindent where $\epsilon_d$ is the deuteron binding energy, and ach state is orthogonal by the relation

\begin{eqnarray}\label{eq:Weinberg-Normalization}
\langle \Phi_i|V_{np}|\Phi_j\rangle = -\delta_{ij}.
\end{eqnarray}

\begin{figure}[h]
\begin{center}
\includegraphics[scale=0.4]{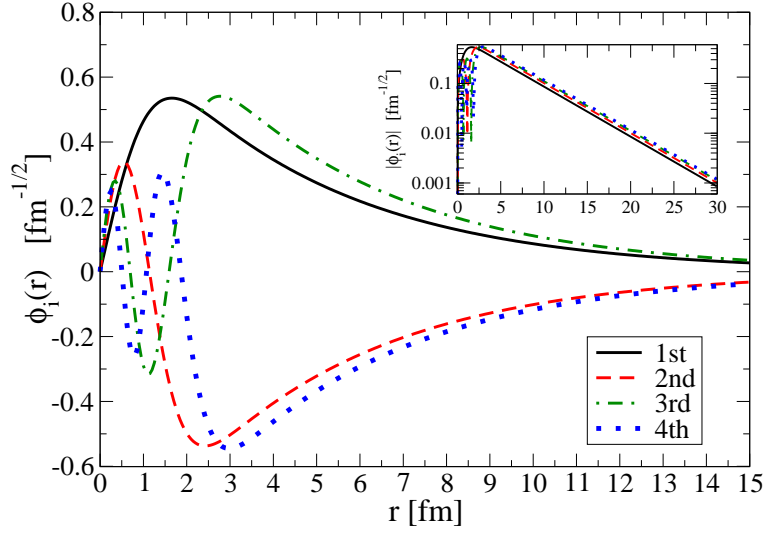}
\end{center}
\caption{The first four Weinberg States when using a central Gaussian which reproduces the binding energy and radius of the deuteron ground state. The inset shows the asymptotic properties of each state.}
\label{fig:WeinbergStates}
\end{figure}

The first Weinberg component $\Phi_0(\textbf{r})$ occurs when $\alpha_0=1$. Therefore, the first component is simply the deuteron ground state wave function with a different normalization condition. Each successive Weinberg component will contain an additional node.  Since each Weinberg state has the same binding energy, the asymptotic properties of each Weinberg state will be identical. In Fig. \ref{fig:WeinbergStates} we show the first four Weinberg states when using a central Gaussian which reproduces the binding energy and radius of the deuteron ground state, as in \cite{Moro_prc2009}. For the first four states, $\alpha_i=\{1,5.2,12.7,23.4 \}$. The inset shows the asymptotic properties of each state. Since each Weinberg state has the same binding energy, they decay with the same rate outside the range of the interaction.

Since we are only interested in describing the short-ranged properties of the three-body wave function, having the wrong asymptotics is not a concern. For an effective expansion, only a finite number of terms should be necessary for an adequate description of the wave function with the inclusion of breakup. Keeping all the terms in the expansion of Eq.(\ref{Weinberg-Expansion}) results in a complicated coupled channel set of equations to describe the scattering process. To eliminate this complication, the typical procedure is to keep only the first term of the expansion. This has been shown to be an excellent approximation \cite{Pang_prc2013}.

To derive the adiabatic potential, we insert the expansion of the three-body wave function, Eq.(\ref{Weinberg-Expansion}), into the Schr\"odinger equation using our particular three-body Hamiltonian, Eq.(\ref{3body-Hamiltonian}). Since we are keeping only the first term, we will write the wave function as $\Psi(\textbf{r},\textbf{R})\approx \Phi_0(\textbf{r})X_{AD}(\textbf{R})$. This gives us

\begin{eqnarray}
\left[T_R+T_r+U_{nA}(\textbf{R}_n)+U_{pA}(\textbf{R}_p)+V_{np}-E\right]\Phi_0(\textbf{r})X_{AD}(\textbf{R})=0
\end{eqnarray}

\noindent Here, $E$ is the total system energy given by $E=E_d-\epsilon_d$, where $E_d$ is the incident deuteron kinetic energy in the center of mass frame, and $\epsilon_d$ is the deuteron binding energy. Since $(T_{\textbf{r}}+V_{np}(r))\Phi_0(\textbf{r})=-\epsilon_d\Phi_0(\textbf{r})$, we can make this replacement giving us

\begin{eqnarray}
\left[T_R+U_{nA}(\textbf{R}_n)+U_{pA}(\textbf{R}_p)-E_d\right]\Phi_0(\textbf{r})X_{AD}(\textbf{R})=0
\end{eqnarray}

\noindent We now multiply by $\langle \Phi_0|V_{np}$ and use the orthogonality properties of the Weinberg states to obtain

\begin{eqnarray}
\left[T_R+U_{AD}^{Loc}(\textbf{R})-E_d\right]X_{AD}(\textbf{R})=0,
\end{eqnarray}

\noindent where the local adiabatic potential, $U_{AD}^{Loc}(\textbf{R})$ is given by

\begin{eqnarray}\label{LocalAdPot}
U_{AD}^{Loc}(\textbf{R})=-\langle \Phi_0|V_{np}(U_{nA}(\textbf{R}_{n})+U_{pA}(\textbf{R}_p))|\Phi_{0}\rangle.
\end{eqnarray}

It is important to note that $X_{AD}(\textbf{R})$ is not the same as the elastic scattering wave function $X_d(\textbf{R})$ in the plane wave basis of Eq.(\ref{3BodyWF}). $X_d(\textbf{R})$ describes elastic scattering, and the potential used to generate $X_d(\textbf{R})$ would be a deuteron optical potential obtained by fitting elastic scattering data. On the other hand, $U_{AD}$ does not describe deuteron elastic scattering. In fact, the adiabatic potential is only of use to describe transfer reactions. However, the input optical potentials, $U_{nA}$ and $U_{pA}$, do describe elastic scattering, and are obtain by fits to nucleon data. This is an advantageous feature of the ADWA as nucleon optical potentials are much better constrained than deuteron optical potentials.


\section{Nonlocal Adiabatic Distorted Wave Approximation}
\label{Sec:Nonlocal_ADWA}

We would like to consider the adiabatic potential in Eq.(\ref{LocalAdPot}) when we are using nonlocal nucleon optical potentials. A detailed derivation is presented in Appendix \ref{Nonlocal_Adiabatic}. Here we will give an overview of the derivation. As an example, consider first the neutron nonlocal operator acting on the three-body wave function:

\begin{eqnarray}\label{Neutron-Nonlocal-Integral}
\hat{U}_{nA}\Psi(\textbf{r},\textbf{R})&=&\int U_{nA}(\textbf{R}_n,\textbf{R}'_n)\Psi(\textbf{R}'_n,\textbf{R}'_p)\delta(\textbf{R}'_p-\textbf{R}_p)d\textbf{R}'_nd\textbf{R}'_p \nonumber \\
&=&8\int U_{nA}\left(\textbf{R}-\frac{\textbf{r}}{2},2\textbf{R}'-\textbf{R}-\frac{\textbf{r}}{2}\right)\Psi(\textbf{r}-2(\textbf{R}'-\textbf{R}),\textbf{R}')d\textbf{R}'.
\end{eqnarray}

\begin{figure}[h]
\begin{center}
\includegraphics[scale=0.25]{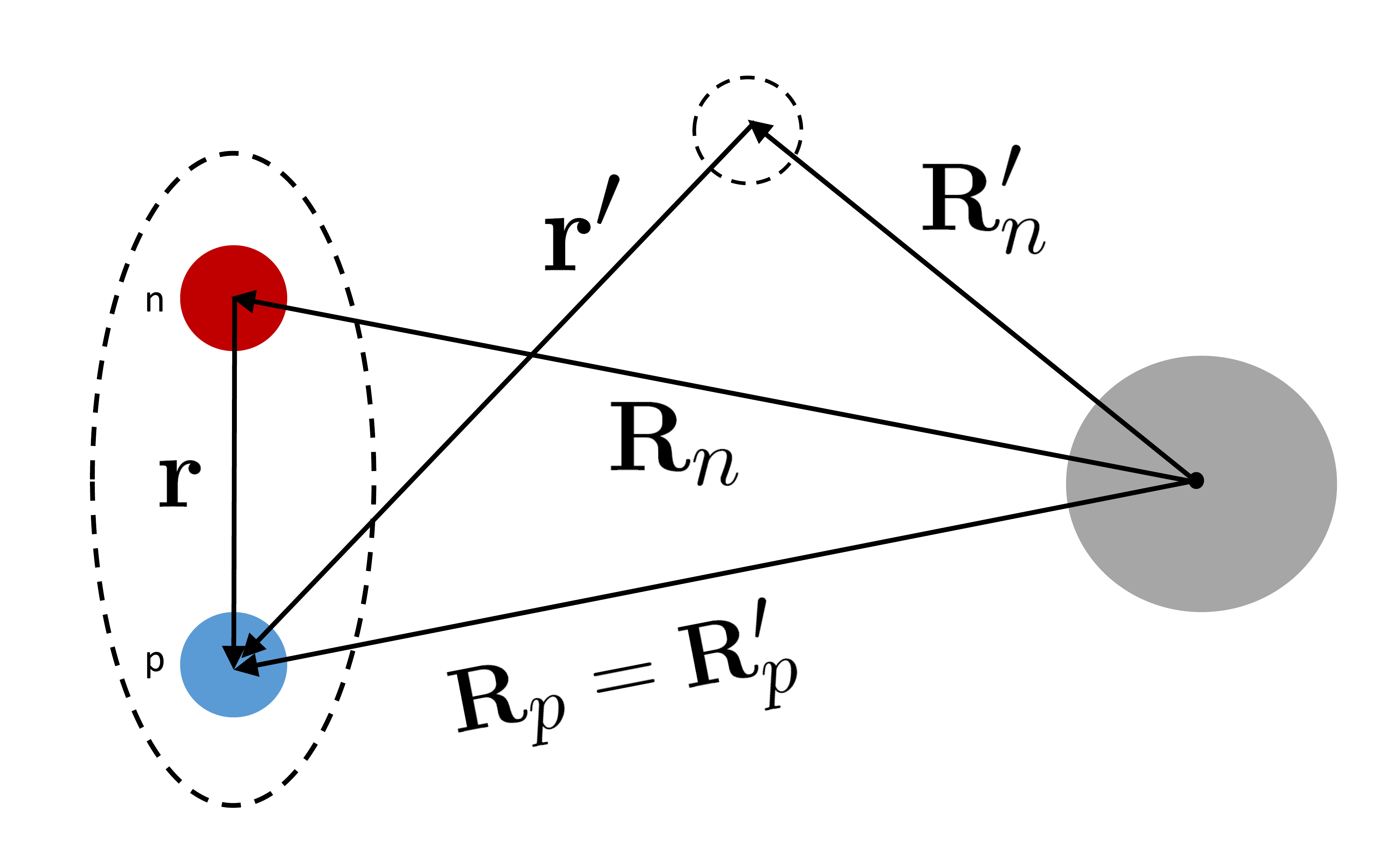}
\end{center}
\caption{The coordinates used for constructing the neutron nonlocal potential. The open dashed circle represents the neutron in a different point in space to account for nonlocality.}
\label{fig:Neutron-Nonlocal-Coordinates}
\end{figure}

The coordinates used for calculating the neutron nonlocal potential are shown in Fig. \ref{fig:Neutron-Nonlocal-Coordinates}, where the open dashed circle represents the neutron in a different point in space to account for nonlocality. Since we are calculating the optical potential for the neutron interacting with the target, the proton remains stationary when integrating the neutron coordinate over all space. Hence, the reason for the delta function in Eq.(\ref{Neutron-Nonlocal-Integral}). 

In Eq.(\ref{Neutron-Nonlocal-Integral}), the Jacobian for the coordinate transformation is unity, and we integrated over $d\textbf{r}'$ to eliminate the delta function. We used the vector definitions $\textbf{R}_{p,n}=\textbf{R}\pm\frac{\textbf{r}}{2}$, where $\textbf{R}_p$ uses the ``$+$'' sign and $\textbf{R}_n$ uses the ``$-$'' sign. A similar expression is found for the proton nonlocal operator. 

Since we are using only the first Weinberg state, we will drop the ``$0$'' subscript on the wave functions, and write the expansion of the three-body wave function as $\Psi(\textbf{r},\textbf{R})\approx \Phi(\textbf{r})X(\textbf{R})$. Thus, the general nucleon nonlocal operator is

\begin{eqnarray}
&\phantom{=}&\hat{U}_{NA}\Phi(\textbf{r})X(\textbf{R})=8\int U_{NA}\left(\textbf{R}\pm\frac{\textbf{r}}{2} ,2\textbf{R}'-\textbf{R}\pm\frac{\textbf{r}}{2}\right)\Phi(\textbf{r}\pm2(\textbf{R}'-\textbf{R}))X(\textbf{R}')d\textbf{R}'. \nonumber \\
\end{eqnarray}

\noindent Adding and subtracting $\textbf{R}$ in the second argument of $U_{NA}$ and making the definition $\textbf{s}=\textbf{R}'-\textbf{R}$, we can rewrite the nucleon nonlocal operator as

\begin{eqnarray}
\hat{U}_{NA}\Phi(\textbf{r})X(\textbf{R})&=&8\int U_{NA}\left(\textbf{R}_{p,n},\textbf{R}_{p,n}+2\textbf{s}\right)\Phi(\textbf{r}\pm 2\textbf{s})X(\textbf{R}+\textbf{s})d\textbf{s}.
\end{eqnarray}

In Eq.(\ref{eq:Deuteron-Distorted-Wave}) we gave the deuteron distorted wave for each projection of angular momentum of the deuteron and target. Now we need the deuteron wave function for relative motion between $d$ and $A$ for each value and projection of total angular momentum, $J_TM_T$. This is given by

\begin{eqnarray}\label{WFexpansion}
\Psi(\textbf{r},\textbf{R})&\approx& \Phi(\textbf{r})X(\textbf{R})=\sum_{\ell L J_p}\phi_{\ell}(r)\frac{\chi_{L J_p}^{J_T M_T}(R)}{R} \\
&\phantom{=}& \times \ \left\{ \left\{ \left\{     \left\{ \Xi_{1/2}(\xi_n)\otimes \Xi_{1/2}(\xi_p) \right\}_{1} \otimes \tilde{Y}_{\ell}(\hat{r})\right\}_{1} \otimes \tilde{Y}_{L}(\hat{R})\right\}_{J_p} \otimes \Xi_{I_t}(\xi_t) \right\}_{J_T M_T}. \nonumber
\end{eqnarray}

\begin{figure}[h]
\begin{center}
\includegraphics[scale=0.25]{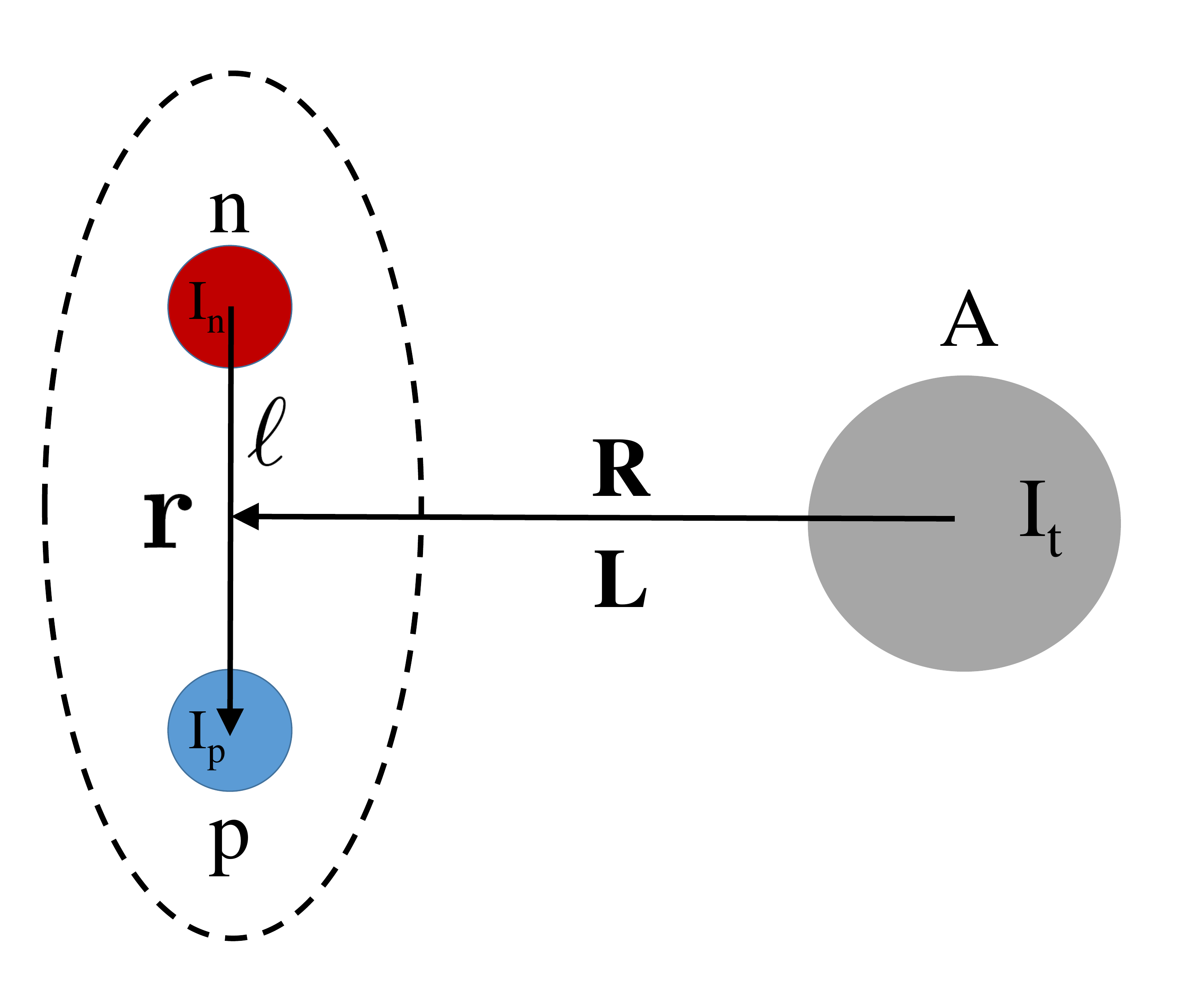}
\end{center}
\caption{The coordinates used for constructing the system wave function for the $d+A$ deuteron scattering state.}
\label{fig:Deuteron-System-WF}
\end{figure}

\noindent The description of each term is given after Eq.(\ref{eq:Deuteron-Distorted-Wave}). The coordinates for constructing the system wave function for the deuteron scattering state are given in Fig. \ref{fig:Deuteron-System-WF}. 

We would like to find the partial wave decomposition of 

\begin{eqnarray}\label{eqn:NL-Adiabatic}
\left[\hat{T}_{\textbf{R}}+V_C(R)+U_{so}(R)-E_d \right]\Phi(\textbf{r})X(\textbf{R})=-\left(\hat{U}_{nA}+\hat{U}_{pA} \right)\Phi(\textbf{r})X(\textbf{R}), \nonumber \\
\end{eqnarray}

\noindent where $U_{so}(R)$ is the sum of the neutron and proton spin-orbit potentials. To begin the partial wave decomposition, multiply Eq.(\ref{eqn:NL-Adiabatic}) by

\begin{eqnarray}
\sum_{\ell'}\phi_{\ell'}(r)V_{np}(r)\left\{ \left\{ \left\{     \left\{ \Xi_{1/2}(\xi_n)\otimes \Xi_{1/2}(\xi_p) \right\}_{1}    \otimes \tilde{Y}_{\ell'}(\hat{r})\right\}_{1} \otimes \tilde{Y}_{L'}(\hat{R})\right\}_{J'_p} \otimes \Xi_{I_t}(\xi_t) \right\}_{J_T M_T}^* \nonumber
\end{eqnarray}

\noindent and integrate over $d\textbf{r}$, $d\Omega_R$, $d\xi_{n}$, $d\xi_{p}$ and $d\xi_t$. The $lhs$ of the equation becomes

\begin{eqnarray}\label{lhs}
\frac{1}{R}\left[\frac{\hbar^2}{2\mu}\left(\frac{\partial^2}{\partial R^2}-\frac{L(L+1)}{R^2} \right)+E_d \right]\chi_{L J_p}^{J_T M_T}(R).
\end{eqnarray}

\noindent As we only considered $\ell=0$ deuterons in our calculations, let us make this assumption right at the beginning of our partial wave decomposition of the $rhs$.  Therefore, the two $\tilde{Y}_{\ell}(\hat{r})$ terms give $1/4\pi$, and the partial wave decomposition of the $rhs$ of Eq. (\ref{eqn:NL-Adiabatic}) is

\begin{eqnarray}
&\phantom{=}& -\frac{8}{4\pi}\sum_{L' J'_p} \int \phi_{0}(r)V_{np}(r) U_{NA}\left(\textbf{R}_{p,n},\textbf{R}_{p,n}+2\textbf{s}  \right)\phi_{0 }(|\textbf{r}\pm 2\textbf{s}|)\frac{\chi_{L' J'_p}^{J_T M_T}\left(\left|\textbf{R}+\textbf{s} \right|\right)}{\left|\textbf{R}+\textbf{s} \right|}  \\
&\times& \left\{ \left\{ \Xi_{1}(\xi_{np}) \otimes \tilde{Y}_{L}(\hat{R})\right\}_{J_p} \otimes \Xi_{I_t}(\xi_t) \right\}_{J_T M_T}^*\left\{ \left\{ \Xi_{1}(\xi_{np})\otimes \tilde{Y}_{L'}(\widehat{R+s})\right\}_{J'_p} \otimes \Xi_{I_t}(\xi_t) \right\}_{J_T M_T} \nonumber \\
&\times& d\textbf{s} d\textbf{r} d\Omega_R d\xi_t d\xi_{n}d\xi_{p}. \nonumber
\end{eqnarray}

Our goal is to couple the integrand up to zero angular momentum. This will be spherically symmetric so we can use symmetry to reduce the dimensionality of the integral. After several additional steps of algebra we arrive at:

\begin{eqnarray}\label{NL-Adiabatic}
&\phantom{=}&\left[\frac{\hbar^2}{2\mu}\left(\frac{\partial^2}{\partial R^2}-\frac{L(L+1)}{R^2} \right)-V_C(R)-U_{so}(R)+E_d \right]\chi_{L J_p}^{J_T M_T}(R) \\
&=&-\frac{8R \sqrt{\pi}}{\hat{L}}\int \phi_0(r)V_{np}(r)\frac{\chi_{L J_p}^{J_T M_T}\left(\left|\textbf{R}+\textbf{s} \right|\right)}{\left|\textbf{R}+\textbf{s} \right|} Y_{L0}(\widehat{R+s})  \nonumber \\
&\phantom{=}& \times \ \left[ U_{nA}\left(\textbf{R}_{n},\textbf{R}_{n}+2\textbf{s}  \right)\phi_{0}(|\textbf{r}- 2\textbf{s}|) + U_{pA}\left(\textbf{R}_{p},\textbf{R}_{p}+2\textbf{s}  \right)\phi_{0}(|\textbf{r}+ 2\textbf{s}|) \right] r^2 \sin\theta_{r} dr d\theta_{r}d\textbf{s}. \nonumber
\end{eqnarray}

\noindent This is ultimately the nonlocal equation we solve to obtain the adiabatic wave that represents $d+A$ initial scattering to be introduced into the T-matrix, Eq.(\ref{Tmatrix}). More details of the derivation are given in Appendix \ref{Nonlocal_Adiabatic}. 


\section{Spectroscopic Factors}

Transfer reactions are performed not only to extract spin and parity assignments of energy levels, but also to extract spectroscopic factors. As an example to understand the concept of a spectroscopic factor, let us consider the $^{17}$O nucleus, which can be modeled as a $^{16}$O core plus a valence neutron. Let us assume that $^{16}$O contains only a $0^+$ ground state and a $2^+$ excited state. The ground state of $^{17}$O is a $5/2^{+}$ state. Due to the possible excited states of the core, the ground state of $^{17}$O can occur in various configurations. Here we consider only two for simplicity:

\begin{eqnarray}\label{eq:configuration}
|^{17}\textrm{O}_{g.s.}\rangle = \alpha_1\left[^{16}\textrm{O}(0^+) \otimes n_{1d_{5/2}} \right]_{5/2^+}+ \alpha_2\left[^{16}\textrm{O}(2^+) \otimes n_{2s_{1/2}} \right]_{5/2^{+}}
\end{eqnarray}

\noindent These two configurations for $^{17}$O are: the ground state of the $^{16}$O core coupled to the valence nucleon in a $1d_{5/2}$ orbital, and the $^{16}$O core in its excited $2^+$ state coupled to the valence neutron in a $2s_{1/2}$ orbital. Both configurations must correspond to the ground state energy, which means that the available energy for the neutron in the $1d_{5/2}$ orbital is different than that for the neutron in the $2s_{1/2}$ orbital due to core excitation. The spectroscopic factor tells us how probable it is to find the valence neutron in $^{17}$O$_{g.s.}$ in a $1d_{5/2}$ configuration with $^{16}$O$_{g.s.}$, and is given by:

\begin{eqnarray}
S_{1d_{5/2}}=|\langle ^{16}\textrm{O}(0^+)|^{17}\textrm{O}_{g.s.}\rangle|^2=\alpha_1^2
\end{eqnarray}

The spectroscopic factor for the configuration with $^{16}$O in the ground state can often be cleanly extracted from the $^{16}$O$(d,p)^{17}$O$_{g.s.}$. The reason being the fast radial fall off for the other configurations due to the additional binding of the neutron caused by core excitation. In a simple theoretical DWBA analysis, only the first configuration of Eq.(\ref{eq:configuration}) is included in the calculation. The peak of the transfer distribution corresponds to impact parameters for the deuteron grazing the surface. Therefore, one expects the transfer process to adequately be described as a one-step process. Since we left out all of the other configurations, our theory assumed that $\alpha_1^{Theory}=1$, so most often it will over-predict the transfer cross section at the peak. By normalizing the theoretical transfer distribution at the first peak to the experimental distribution at the first peak, we can extract the physical $|\alpha_1|^2$ value. It is for this reason that we are interested in the magnitude of the transfer cross section at the first peak throughout this thesis.


\chapter{Optical Potentials}
\label{Potentials}

Effective potentials describing the scattering process are needed when doing calculations of reactions, and an accurate theoretical description of these reactions is required for the reliable extraction of desired quantities. Optical potentials have been obtained phenomenologically, primarily from elastic scattering data, but sometimes from absorption cross sections and polarization observables \cite{Becchetti_pr1969, Varner_pr1991, Koning_np2003, Weppner_prc2009, Perey_adndt1974}. In all optical potentials, the nuclear potential is assumed to be complex, where the imaginary part takes into account loss of flux to non-elastic channels. 

In all commonly used global optical potentials, the interaction is assumed to be local. As a consequence, these potentials all have a strong energy dependence. Inherent in the local assumption of the potential is a factoring out of the many-body degrees of freedom. Therefore, the anti-symmetrization of the many-body wave function, and the coupling to all the non-elastic channels, is not explicitly taken into account, and must be introduced effectively into the local potential through an energy dependence of the parameters.


\section{Global Optical Potentials}

Global optical potentials are often used in the analysis of nuclear reactions. Global potentials are very convenient as they can easily be extrapolated to regions of the nuclear chart where data is not available, or they can be used at energies where data has not been taken. However, such extrapolations should always be done carefully. A global optical potential constructed from fits to stable nuclei may not give sensible results when extrapolated to exotic nuclei. Nonetheless, using a global potential is sometimes the only option available when making theoretical predictions of experiments on exotic nuclei.

Global optical potentials attempt to describe the nuclear potential across some range of mass and energy. To do this, some kind of form for the complex mean field must be assumed. In most constructions of global optical potentials, the real and imaginary parts contain combinations of Volume (v), Surface (d), and Spin-Orbit (so) terms given by

\begin{eqnarray}
U_v(R)&=&-V_vf(R,r_v,a_v) \nonumber \\
U_d(R)&=&4a_dV_d\frac{d}{dR}f(R,r_d,a_d) \nonumber \\
U_{so}(R)&=&\left(\frac{\hbar}{m_\pi c} \right)^2V_{so}\frac{1}{R}\frac{d}{dR}f(R,r_{so},a_{so})2\textbf{L}\cdot\textbf{s},
\end{eqnarray}

\noindent where 

\begin{eqnarray}
f(R,r,a)=\left[1+\exp\left(\frac{R-rA^{1/3}}{a} \right) \right]^{-1}.
\end{eqnarray}

\noindent The Coulomb potential is taken to be that of a homogeneous sphere of charge 

\begin{eqnarray}
V_C(R)=
\begin{cases}
\frac{Z_1 Z_2 e^2}{2}\left(3-\frac{R^2}{R_C^2} \right) & \textrm{if} \ R < R_c \\
\frac{Z_1 Z_2 e^2}{R} & \textrm{if} \  R \geq R_c, 
\end{cases}
\end{eqnarray}

\noindent where the Coulomb radius is given by $R_c=r_cA^{1/3}$. Given this definition of the optical potential there are, in principle, $19$ free parameters: $3$ parameters per term, $6$ terms assuming the volume, surface, and spin-orbit terms are all complex, and the Coulomb radius. 

The real part usually comes from the density distribution of the nucleus, $\rho(r)$, which is typically of a Woods-Saxon form. This justifies the real volume term, and if $\rho(r)$ has surface ripples, then one would need a real surface potential as well. The imaginary volume term is responsible for loss of flux from the elastic channel occurring somewhere inside the nucleus. This term is sometimes used in global optical potentials, and becomes more important at higher energies. The imaginary surface term is responsible for removing flux due to non-elastic events occurring at the nuclear surface. This is a very important term and is included in all global optical potentials because most reactions occur at the surface. The spin-orbit term is the interaction between the spin of the projectile and its orbital angular momentum with the target. A real spin-orbit interaction is always included in global optical potentials. An imaginary spin-orbit term is sometimes included, but the depth of the imaginary part is often times small. 

Normally, 19 parameters are too many to constrain from just elastic scattering, so one needs to further constrain the form of the global optical potential. Phenomenology has guided us towards the basic form of a real volume, an imaginary surface, and a real spin-orbit term. Some global optical potentials, such as \cite{Koning_np2003}, include an imaginary volume and imaginary spin-orbit term as well. These terms are normally needed when higher energy reactions ($>50$ MeV) are considered in the fit. 

Once a functional form is chosen, the free parameters are varied to obtain a best fit to a large amount of elastic scattering data. In these fits, the depth, and sometimes the radius and diffuseness, of the various terms can be energy and mass dependent. When using the potential, the target mass, charge, and the projectile energy must be specified. Then, the value for the depth, radius, and diffuseness of each term is calculated. There are several global optical potentials on the market, some of the common ones are discussed here \cite{Koning_np2003, Varner_pr1991, Perey_adndt1974}.


\section{Motivating Nonlocal Potentials}

We already discussed in Sec. \ref{Sec:Nonlocality} the sources of nonlocality for the effective NA interaction. Here we provide additional perspective based on Feshbach's work \cite{Feshbach_ap1958,Feshbach_ap1962}. When derived from the many-body problem, the single particle Schr\"odinger equation describing the motion of nucleons in nuclei is nonlocal. In the projection operator theory of Feshbach, a formal equation for the single particle motion can be derived. The formalism of Feshbach uses the projection operator $\mathcal{P}$ to project the many-body wave function onto the channels that are considered explicitly, and the projection operator $\mathcal{Q}$ projects onto all channels left out from the model space. Consider the case when $\mathcal{P}$ projects the many-body wave function, $\Psi$, onto the elastic channel. When this is the case, the operators are defined by

\begin{eqnarray}
\mathcal{P}=|\Psi_{gs}\rangle \langle \Psi_{gs} |; \quad \quad \mathcal{Q}=1-\mathcal{P}; \quad \quad \mathcal{Q}|\Psi_{gs}\rangle = 0,
\end{eqnarray}

\noindent where $|\Psi_{gs}\rangle$ gives the elastic scattering channel where the target remains in its ground state, $|\Phi_{gs}\rangle$ and the projectile undergoes elastic scattering, $|X_{el}\rangle$:

\begin{eqnarray}
|\Psi_{gs}\rangle=|X_{el}\rangle|\Phi_{gs}\rangle.
\end{eqnarray}

\noindent Using this projection operator formalism, Feshbach showed that a formal expression for the Schr\"odinger equation to describe elastic scattering is given by

\begin{eqnarray}
\left(E-T_{\textbf{R}}-\langle \Phi_{gs}|V|\Phi_{gs}\rangle-\langle\Phi_{gs}|VQ\frac{1}{E-QHQ}QV|\Phi_{gs}\rangle \right)X_{el}(\textbf{R})=0,
\end{eqnarray}

\noindent with $V$ being the bare projectile-target interaction. From this equation, we can identify the optical potential as 

\begin{eqnarray}\label{eq:Formal-NL-Pot}
U=V+VQ\frac{1}{E-QHQ}QV.
\end{eqnarray}

\noindent The first term appears local while the second is inherently nonlocal. If we allow for anti-symmetriztion between the projectile and all the nucleons of the target, whereby the incident nucleon may not the be same as the exiting nucleon, even the first term becomes nonlocal. In the Hartree-Fock theory, used for bound state calculations, the naturally arising exchange term is a direct result of anti-symmetrization.

While Eq.(\ref{eq:Formal-NL-Pot}) is a formal equation, it gives some physical insight into the nature of nonlocality. The nucleon begins in the space of elastic scattering, $\mathcal{P}$-space. The system then couples to some non-elastic channel and propagates through that space, $\mathcal{Q}$-space, before returning back to the elastic channel at some later location in space. This also gives a physical justification for the need to have a potential with two arguments, $U(\textbf{R},\textbf{R}')$. Flux leaves $\mathcal{P}$-space and goes into $\mathcal{Q}$-space at $\textbf{R}'$. The flux propagates through $\mathcal{Q}$-space, before being deposited back into $\mathcal{P}$-space at $\textbf{R}$. 

With a nonlocal potential, the Schr\"odinger equation, Eq.(\ref{radial-eqn}), gets transformed into an integro-differential equation

\begin{eqnarray}
\frac{\hbar^2}{2\mu}\nabla^2 \Psi(\textbf{R})+E\Psi(\textbf{R})=U_o(\textbf{R})\Psi(\textbf{R})+\int U^{NL}(\textbf{R},\textbf{R}')\Psi(\textbf{R}')d\textbf{R}'.
\end{eqnarray}

\noindent To describe the physics of flux leaving $\mathcal{P}$-space at $\textbf{R}'$, propagating through $\mathcal{Q}$-space, and returning to $\mathcal{P}$-space at $\textbf{R}$, it becomes natural to describe the potential at the point $\textbf{R}$ to be dependent on the overlap of the wave function and the potential at all other points in space, hence, the need for the integral.


\section{Perey-Buck Type}

Due to the success of local global optical potentials, it would be natural to assume that similar global parameterizations have been made to nonlocal potentials. Unfortunately, this is not the case. To our knowledge, there are only three global nonlocal optical potentials that are constructed phenomenologically from elastic scattering. The seminal paper of Perey and Buck in 1962 was the first attempt to make a parameter set for a nonlocal model \cite{Perey_np1962}. In the late 70s Giannini and Ricco constructed their potential by fitting a large amount of data to a local form. They then used an approximate transformation formula to obtain the nonlocal parameters \cite{Giannini_ap1976,Giannini_ap1980}. Finally, in 2015, Tian, Pang, and Ma (TPM) constructed their potential through fits to elastic scattering and analyzing powers \cite{Tian_ijmpe2015}. 

While the existence of nonlocality in the nuclear potential has long been known, there has historically been great difficulty in specifying the exact form for the nonlocal nuclear potential. A simple form was first proposed by Frahn and Lemmer \cite{Frahn_nc1957}, and later developed and implemented by Perey and Buck \cite{Perey_np1962}. The Perey-Buck potential is the most commonly referred to phenomenological nonlocal optical potential due to its simplicity. The Perey-Buck potential is given by 

\begin{eqnarray}\label{PB-Form}
U^{NL}_{PB}(\textbf{R},\textbf{R}')&=&U\left(\left|\frac{\textbf{R}+\textbf{R}'}{2} \right|\right)H\left(\left|\textbf{R}-\textbf{R}'\right|, \beta \right),
\end{eqnarray}

\noindent where the function $U\left(\left|\frac{\textbf{R}+\textbf{R}'}{2} \right|\right)$ is of a Woods-Saxon form, and the function $H(\left|\textbf{R}-\textbf{R}'\right|, \beta )$ is chosen to be a normalized Gaussian function,
 
\begin{equation}
H\left(\left|\textbf{R}-\textbf{R}'\right|, \beta \right)=\frac{\textrm{exp}\left({-\left|\frac{\textbf{R}-\textbf{R}'}{\beta}\right|^2}\right)}{\pi^{\frac{3}{2}}\beta^3}.
\end{equation}

\noindent Making the definition $p=\tfrac{R+R'}{2}$, $U(p)$ has a form similar to those in local optical model calculations. For the Perey-Buck nonlocal potential, $U(p)$ consists of a nonlocal real volume, nonlocal imaginary surface, and a local real spin-orbit potential.

The parameter that defines the range of the nonlocality is $\beta$. As a physical example to understand this parameter, consider anti-symmetrization, which as we already discussed is a source of nonlocality. Since the true many-body wave function is anti-symmetric, it is possible for the incident nucleon to not be the same as the scattered nucleon. For the incident nucleon to ``switch places'' with one of the nucleons within the target, it is reasonable to assume that the two nucleons must be relatively close to each other for this to occur. Typically, nonlocality ranges are of the order of the size of the nucleon. For the Perey-Buck potential, $\beta$ is fixed at $0.85$ fm. For other nonlocal potentials, such as the TPM, $\beta$ is an additional parameter in their fit. The resulting value for $\beta$ in neutron and proton versions of the TPM are very similar to that of the Perey-Buck potential.

As the Perey-Buck potential is phenomenological, the parameters involved are obtained by fitting elastic scattering. Two data sets were used: $n+^{208}$Pb at $7.0$ and $14.5$ MeV. Perey and Buck assumed that the parameters were energy and mass independent. Therefore, a single parameter set completely defines the nonlocal potential of Perey and Buck. The parameter set for the Perey-Buck potential is given in Table \ref{PB-Form-Parameters}. 

The work of Tian, Pang, and Ma (TPM) was the first modern attempt to find a parameter set for a nonlocal potential \cite{Tian_ijmpe2015}. In their fit, a multitude of data was considered, spanning energy and mass. This is a great improvement over the two data sets Perey and Buck used in their fit. A separate potential for protons and neutrons was found for the TPM potential, unlike Perey and Buck where no protons were used in the fit. As with Perey and Buck, the parameters in the TPM potential are assumed energy and mass independent. For higher energy reactions, the TPM potential was found to provide a better $\chi^2$ than the Perey-Buck potential, while at lower energies the two potentials are comparable. The parameter set for the TPM potential is given in Table \ref{PB-Form-Parameters}. 

The TPM potential was published after much of the work for this study was completed. Hence, for this reason, and due to the popularity and widespread use of the Perey-Buck potential, the potential we used to assess the effects of nonlocality was that of Perey and Buck. In this study, we are interested in differences between nonlocal and local equivalent calculations, and not so much on the quality of the nonlocal calculations themselves. When it becomes necessary to use nonlocal potentials to extract information from experiments, the improved TPM potential is the better choice.

\begin{table}
\begin{center}
  \begin{tabular}{|c||c|c|c|}
     \hline
              & Perey-Buck &    TPM     & TPM      \\
              &            &  Neutrons  & Protons  \\ \hline \hline
     $V_v$    & $71.00$    & $70.00$    & $70.95$  \\ \hline
     $r_v$    & $1.22$     & $1.25$     & $1.29$   \\ \hline
     $a_v$    & $0.65$     & $0.61$     & $0.58$   \\ \hline
     $W_v$    &   ---      & $1.39$     & $9.03$   \\ \hline
     $r_{w_v}$ &   ---      & $1.17$     & $1.24$   \\ \hline
     $a_{w_v}$ &   ---      & $0.55$     & $0.50$   \\ \hline
     $W_{d}$   &  $15.00$   & $21.11$    &  $15.74$  \\ \hline
     $r_{w_d}$ &  $1.22$    & $1.15$     & $1.20$   \\ \hline
     $a_{w_d}$ &  $0.47$    & $0.46$     & $0.45$   \\ \hline
     $V_{so}$  &  $7.18$    & $9.00$     & $8.13$   \\ \hline
     $r_{so}$  &  $1.22$    & $1.10$     & $1.02$   \\ \hline
     $a_{so}$  &  $0.65$    & $0.59$     & $0.59$   \\ \hline
     $r_c$    &  $1.22$     & ---        &  $1.34$   \\ \hline
     $\beta$  &  $0.85$     & $0.90$     &  $0.88$   \\ \hline
  \end{tabular}
  \caption{Potential parameters for the Perey-Buck \cite{Perey_np1962} and TPM \cite{Tian_ijmpe2015} nonlocal potentials.}
  \label{PB-Form-Parameters}
\end{center}
\end{table}

\begin{figure}[h!]
\begin{center}
\includegraphics[width=0.45\textwidth]{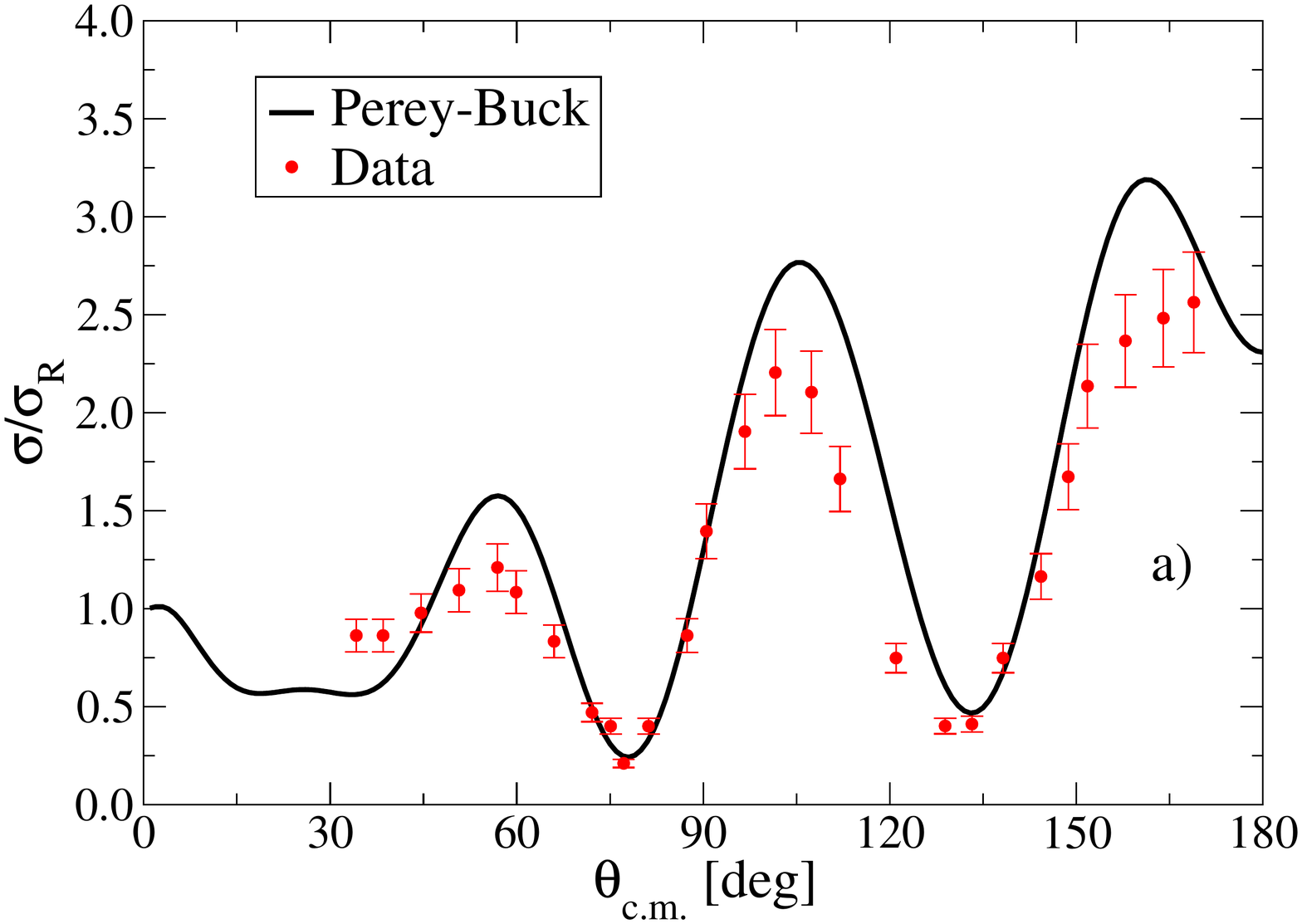}
\includegraphics[width=0.45\textwidth]{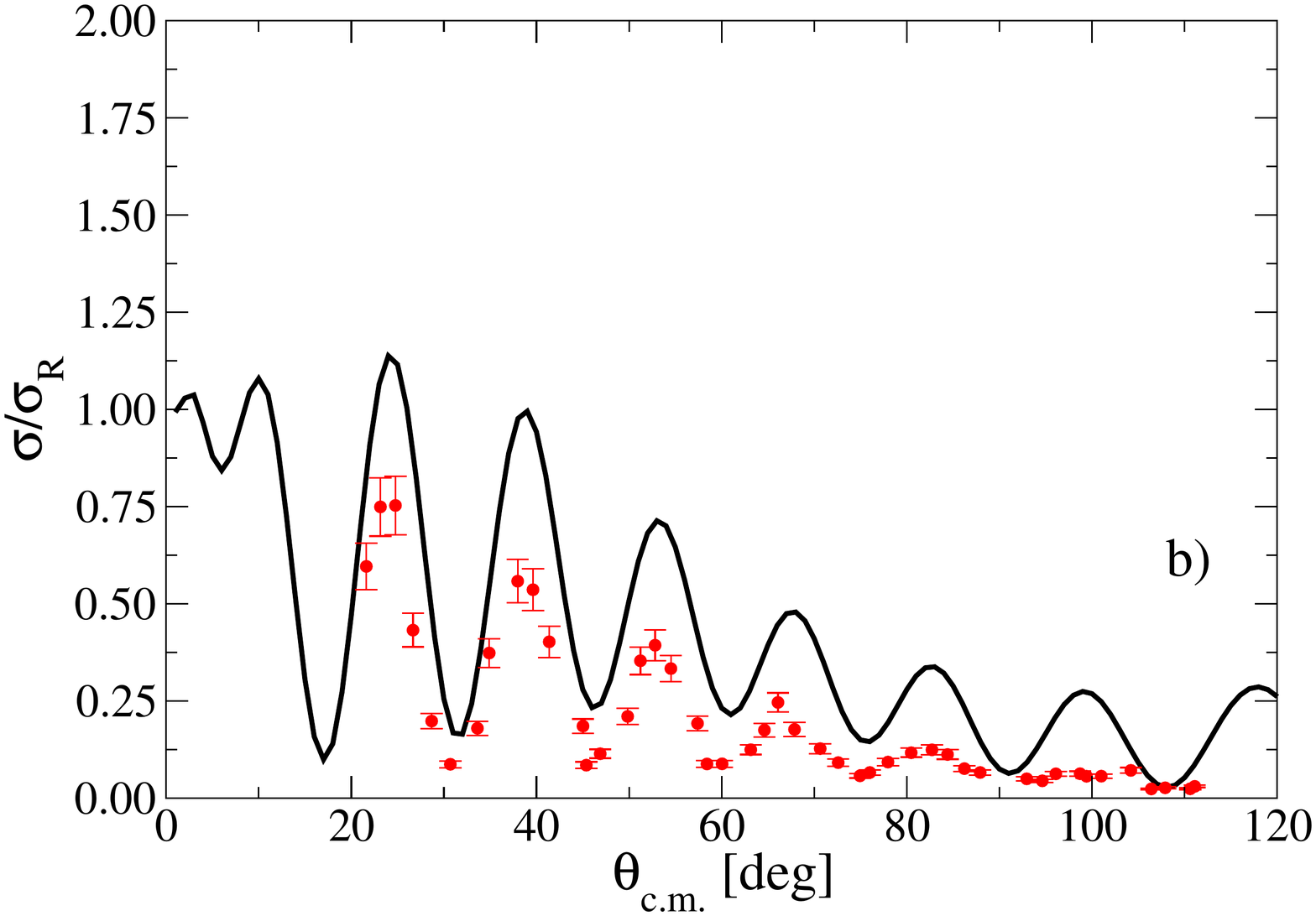}
\end{center}
\caption{Differential elastic scattering relative to Rutherford as a function of scattering angle. (a) $^{48}$Ca$(p,p)^{48}$Ca at $15.63$ MeV with data from \cite{Lombardi_npa1972} (b) $^{208}$Pb$(p,p)^{208}$Pb at $61.4$ MeV with data from \cite{Fulmer_pr1969}.}
\label{fig:PereyBuck_vs_Data}
\end{figure}

With local optical potentials, hundreds of elastic scattering data sets using both protons and neutrons scattering off of a range of nuclei at a range of energies are used to constrain the parameters of the potential. Therefore, it would be expected that the two data sets Perey and Buck used to constrain their parameters would not be sufficient to reproduce elastic scattering over a wide range of nuclei and energies. Nonetheless, reasonable agreement with data is seen despite the simplistic way in which the potential parameters are constrained, as is seen in Fig. \ref{fig:PereyBuck_vs_Data}. 

In order to solve the nonlocal equation, we first need to do a partial wave expansion of the nonlocal potential, 

\begin{eqnarray}
U^{NL}_{PB}(\textbf{R},\textbf{R}')&=&\sum_{L M}\frac{g_{L}(R,R')}{RR'}Y_{L M}(\hat{R})Y^*_{L M}(\hat{R}') \nonumber \\
&=&\sum_L\frac{2L+1}{4\pi} \frac{g_L(R,R')}{RR'}P_L(\cos \theta),
\end{eqnarray}

\noindent where we defined $\theta$ as the angle between \textbf{$R$} and \textbf{$R'$}. Multiplying both sides by $P_{L}(\cos \theta)$, integrating over all angles, using the orthogonality of the Legendre polynomials, and solving for $g_L(R,R')$, we find that

\begin{equation}
g_{L}(R,R')=2{\pi}RR'\int_{-1}^{1}U^{NL}_{PB}(\textbf{R},\textbf{R}')P_{L}(\cos\theta)d(\cos\theta).
\end{equation}

\noindent Now, inserting the Perey-Buck form for the nonlocal potential, replacing $\frac{1}{2}\left| \textbf{R}+\textbf{R}' \right|$ with $\frac{1}{2}(R+R')$, and doing a few lines of algebra outlined in Appendix \ref{SolvingEquation}, we arrive at

\begin{eqnarray}
g_{L}(R,R')&=&\frac{2i^{L}z}{\pi^{\frac{1}{2}}\beta}j_{L}(-iz)\textrm{exp}\left(-\frac{R^2+R'^2}{\beta^2}\right)U\left(\frac{1}{2}(R+R')\right) 
\end{eqnarray}

\noindent with z=$\frac{2RR'}{\beta^2}$, and $j_L$ being spherical Bessel functions. We now have a partial wave equation in terms of $g_L(R,R')$ for each function $\chi_{L}(r)$:

\begin{equation}\label{NLeqn}
\frac{\hbar^2}{2\mu}\left[\frac{d^2}{dr^2}-\frac{L(L+1)}{R^2}\right]\chi_L(R)+E\chi_L(R)=\int g_{L}(R,R')\chi_L(R')dR'.
\end{equation}


\subsection{Correction Factor}
\label{Section-PCF}

Ever since  Perey and Buck introduced their potential in 1962, nearly all analytic work involving approximations to nonlocal potentials, or corrections to wave functions due to a nonlocal potential, assumed the Perey-Buck form for the nonlocality. That is, there is only one nonlocality parameter, $\beta$, and the nonlocal part of the potential takes the form Eq.(\ref{PB-Form}). This is not true for the DOM or the Giannini-Ricco nonlocal potential, discussed later, where there are several terms with a different nonlocality parameter.

Accounting for the nonlocality through the energy dependence of a local optical potential is known to be insufficient. One key feature of a nonlocal potential is that it reduces the amplitude of the wave function in the nuclear interior compared to the wave function from an equivalent local potential. This is the so-called Perey effect \cite{Perey_pl1964}. Physically, the reduction of the wave function can be understood to result from the repulsion due to the Pauli principle. 

Since it wasn't practical to solve the integro-differential equation with a nonlocal potential in the 1960s, there was great interest in finding a way to account for this reduction of amplitude while still keeping the simplicity of solving a local equation. This was first accomplished by Austern, who studied the wave functions of nonlocal potentials and demonstrated the Perey effect in one dimension \cite{Austern_pr1965}. Later, Fiedeldey did a similar study for the three dimensional case \cite{Fiedeldey_np1966}. Using a different method, Austern presented a way to relate wave functions obtained from a nonlocal and a local potential in the three-dimensional case \cite{Austern_Book}. Since then, nonlocal calculations have been avoided using the Perey correction factor (PCF). 

The Perey correction factor is derived in detail in Appendix \ref{Correction_Factor}. Here we simply outline the derivation. To derive the PCF, we begin with the three dimensional Schr\"odinger equation

\begin{eqnarray}\label{NL-Sch-Eq}
\frac{\hbar^2}{2\mu}\nabla^2 \Psi^{NL}(\textbf{R})+E\Psi^{NL}(\textbf{R})=U_o(\textbf{R})\Psi^{NL}(\textbf{R})+\int U^{NL}(\textbf{R},\textbf{R}')\Psi^{NL}(\textbf{R}')d\textbf{R}',
\end{eqnarray}

\noindent where $U_o(\textbf{R})$ is the local part of the potential, and typically contains spin-orbit and Coulomb terms. Let us define a function, $F(\textbf{R})$ that connects the local wave function $\Psi^{loc}(\textbf{R})$ resulting from the potential $U^{LE}(\textbf{R})$ with the wave function resulting from a nonlocal potential, $\Psi^{NL}(\textbf{R})$:

\begin{eqnarray}\label{PCF-Def}
\Psi^{NL}(\textbf{R}) \equiv F(\textbf{R})\Psi^{loc}(\textbf{R}).
\end{eqnarray}

\noindent The potential $U^{LE}(\textbf{R})$ is defined such that it reproduces the exact same elastic scattering as the nonlocal potential. Since the local and nonlocal equations describe the same elastic scattering, the wave functions should be identical outside the nuclear interior. Thus, $F(\textbf{R}) \rightarrow 1$ as $R \rightarrow \infty$. The local equation that $\Psi^{loc}$ satisfies is:

\begin{eqnarray}\label{Loc-Sch-Eq}
\frac{\hbar^2}{2\mu}\nabla^2 \Psi^{loc}(\textbf{R})+E\Psi^{loc}(\textbf{R})=U^{LE}(\textbf{R})\Psi^{loc}(\textbf{R}).
\end{eqnarray}

\noindent Combining Eq.(\ref{NL-Sch-Eq}) and Eq.(\ref{Loc-Sch-Eq}) with the assumption of Eq.(\ref{PCF-Def}) we obtain:

\begin{eqnarray}\label{CorrectionFactor}
F(r)=\left(1-\frac{\mu \beta^2}{2\hbar^2}\left[U^{LE}(R)-U_o(R) \right] \right)^{-1/2}.
\end{eqnarray}

It should be noted that the PCF is only valid for nonlocal potentials of the Perey-Buck form. However, there is no reason to expect that the full nonlocality in the optical potential will look anything like the Perey-Buck form. On physical grounds, the optical potential must be energy dependent due to nonlocalities arising from channel couplings. While the specific form chosen for the Perey-Buck potential is convenient for numerical calculations, a single Gaussian term mocking up all energy-independent nonlocal effects is likely to be an oversimplification.


\section{Giannini-Ricco Nonlocal Potential}

The Perey-Buck potential remained the only widely known and used nonlocal potential available for the following 15 years after its development. An attempt by Giannini and Ricco was made to construct a nonlocal potential of a similar form but with more data constraining the parameters \cite{Giannini_ap1976, Giannini_ap1980}. Their first work focused on $N=Z$ spherical nuclei, while their second work made an extension to $N\ne Z$ nuclei. Unfortunately, in doing the fits, no nonlocal calculations were performed. Instead, the fits were done using a purely local optical potential, and a transformation formula was used which related nonlocal and local form factors. This transformation formula is derived in Appendix \ref{Correction_Factor}.

To construct their potential, Giannini and Ricco first derived a general expression of the nonlocal potential in the framework of Watson multiple-scattering theory \cite{Watson_pr1953}. The form of the derived nonlocal potential is a guide for the parametrization of the phenomenological optical potential, whose parameter values are fitted to both elastic scattering and bound state properties. To do the fit, a local form for the optical potential was chosen. The parameters were varied to obtain a best fit of the available data, and then the transformation formulas were used to get the nonlocal Giannini-Ricco potential for $N=Z$ nuclei (GR76) \cite{Giannini_ap1976}, and later for $N \ne Z$ nuclei (GR80) \cite{Giannini_ap1980}. We chose not to use this potential since the fits to data were done using local potentials, and an unreliable transformation formula was used to get the nonlocal potential.


\section{Nonlocal Dispersive Optical Model Potential}

An alternative method for obtaining the optical potential is through the self-energy, which can be calculated microscopically using modern day structure theory. This is the method by which the Dispersive Optical Model (DOM) is constructed. The DOM makes use of the Kramers-Kronig dispersion relation that links the imaginary and real parts of the nucleon self-energy \cite{Mahaux_anp1991, Dickhoff_prc2010}. The optical potential is constrained by this dispersion relation. This method was first introduced by Mahaux and Sartor \cite{Mahaux_prl1986}. The nuclear mean field is a function of energy, where for $E<0$ it is the shell-model potential that describes single-particle states, while for $E>0$ it is the optical model potential that describes scattering cross sections. 

\begin{figure}[h!]
\begin{center}
\includegraphics[scale=0.30]{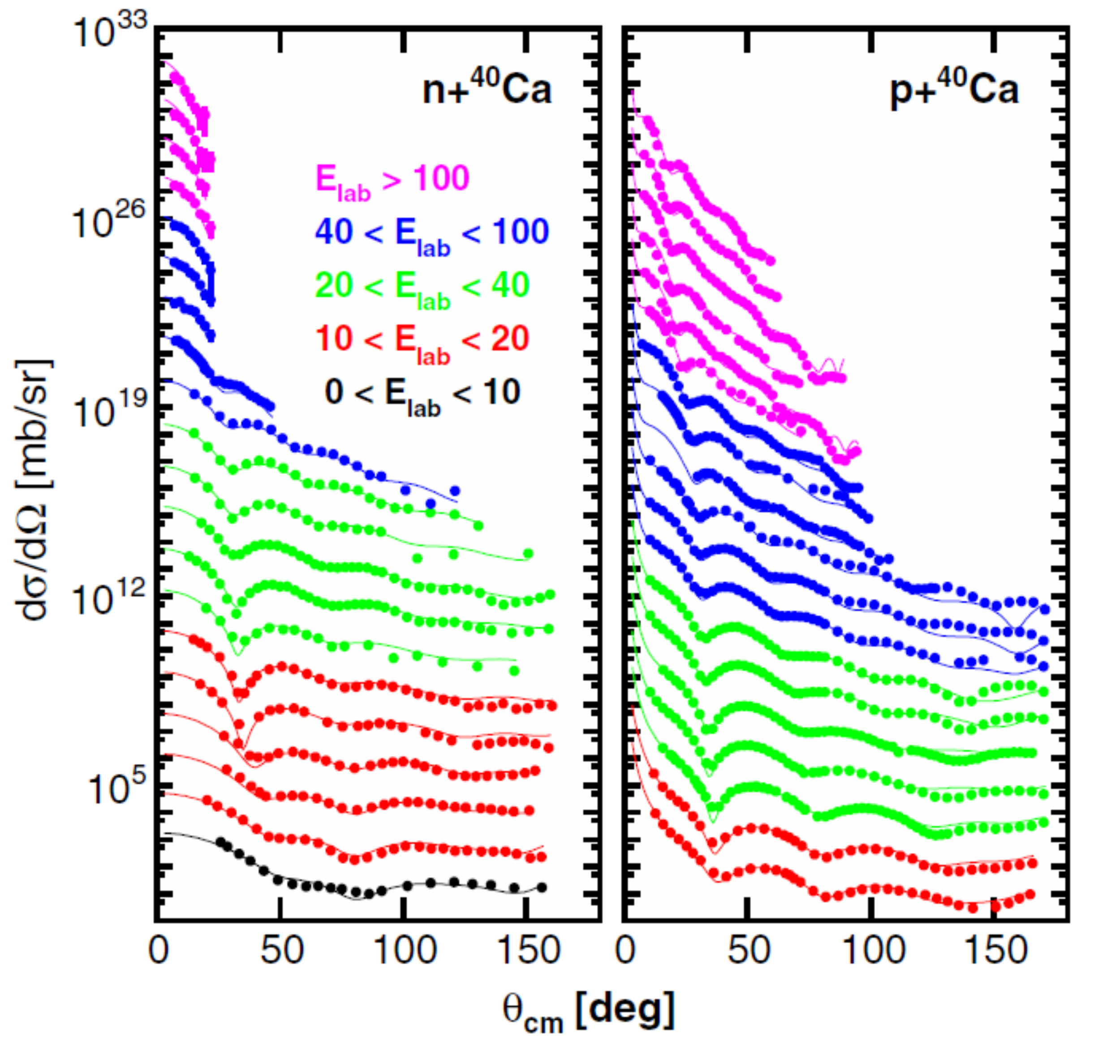}
\end{center}
\caption{Calculated and experimental elastic scattering angular distributions using the nonlocal DOM potential. Data for each energy are offset for clarity with the lowest energy at the bottom and highest at the top. Data references in \cite{Mahzoon_prl2014}. Figure reprinted from \cite{Mahzoon_prl2014} with permission.}
\label{fig:DOM_vs_Data}
\end{figure}

While the nuclear mean field is a continuous function of energy, its behavior as the energy changes sign is not simple due to the coupling between elastic and inelastic channels. It is this coupling that gives rise to the energy dependence and the imaginary component in the optical potential. Through the dispersive relation, the scattering and bound state parts of the nuclear mean field can be linked. The scattering parameters can be constrained by use of fitting elastic scattering, and the bound state parameters can be constrained by comparisons to single particle energies and $(e,e'p)$ observables. As seen in Fig. \ref{fig:DOM_vs_Data}, the nonlocal DOM potential reproduces experimental elastic scattering data across a wide range of energies.

Using this dispersive relation, a local version of the DOM has been developed \cite{Mueller_prc2011}. The local DOM was subsequently used to in the analysis of $(d,p)$ transfer reactions on closed shell nuclei and shown to describe transfer angular distributions with similar adequacy as some of the local global optical potentials on the market \cite{Nguyen_prc2011}. 

Recently, the dispersive optical model formalism has been extended to explicitly include nonlocality, specifically for $^{40}$Ca \cite{Mahzoon_prl2014}. As compared to the Perey-Buck potential, the nonlocal DOM has very different ranges for the nonlocality, and a different value for the nonlocal range in the volume and surface absorption terms. The different ranges of nonlocality for each term in the potential makes the application of a correction factor difficult. However, as the adequacy of the correction factor to take into account nonlocal effects has been put into question \cite{Titus_prc2014}, a correction factor for the DOM potential should not be sought.


\section{Local Equivalent Potentials}
\label{Section-LPE}

To assess the effects of nonlocal potentials, a local phase equivalent (LPE) potential  needs to be found. A local potential is considered 'phase equivalent' to a nonlocal potential if it reproduces the same elastic scattering. This definition is chosen since optical potentials are constructed through fits to elastic scattering data. Therefore, if two potentials are able to generate the same elastic scattering distribution, then the two potentials are indistinguishable at the level of elastic scattering, regardless of their form. 

The downside of this definition is that the short-ranged nonlocal effects are not constrained through elastic scattering. To find a LPE potential, we first assume some form for the LPE potential. This form is normally chosen to mimic the shape of the nonlocal potential. As an example, the Perey-Buck nonlocal potential has real volume, real spin-orbit, and imaginary surface terms. Therefore, the LPE potential was chosen to have the same terms. We calculate the elastic scattering distribution generated from the nonlocal potential, then vary the parameters of our LPE potential to obtain a best fit to the elastic scattering distribution. This was done with the code \textsc{SFRESCO} \cite{fresco} which performs a $\chi^2$ minimization. 

Another method to obtain a LPE potential is through S-matrix inversion \cite{Mackintosh_pl1982, Ioannides_np1985}. This has the advantage over elastic scattering fits since the resulting potential will exactly reproduce the S-matrix elements you started with \cite{Kukulin_jpg2004}. However, it is important to note that the S-matrix is not an observable, so one cannot extract an S-matrix for each partial wave from elastic scattering data.

Exactly reproducing the S-matrix elements from the nonlocal calculation when doing the local fit was one difficulty we encountered in this study. While the fits visually looked very good, there were some very minor differences between the S-matrix elements generated with the nonlocal potential and the fitted LPE potential. There was always particular difficulty for surface partial waves. Nonetheless, when there were differences, the differences were small, and not noticeable in elastic angular distribution.

Finding a LPE potential is also an attractive way to make very sophisticated calculations of the optical potential practical, and to assess their validity. Such has been done with the nonlocal optical potential generated from multiple scattering \cite{Crespo_prc1994} using the S-matrix inversion technique. Here, the PCF was calculated by taking the ratio of the wave function generated from the nonlocal potential with that from the local potential. A similar procedure was done using the g-folding model for $p+^{12}$C scattering at various energies \cite{Lovell_prc2000}. In that study, they investigated the energy dependence of the equivalent local potential, showing that this energy dependence does not take into account the full nonlocality, and that the nonlocality itself must be energy dependent. 

While the S-matrix inversion technique was useful to find a LPE potential in these studies, it may not always be the most attractive way to obtain a LPE potential in practice. A fit to elastic scattering is a much more practical and natural way to obtain a local potential, since it is based on an observable for which one may have data. In practice, when an optical potential is desired, there may sometimes be elastic scattering data available on the nucleus of interest at the correct energy. When this is the case, a common procedure may be to fit the elastic scattering data directly, rather than rely on the extrapolations of some global potential. It is this philosophy we wanted to follow when obtaining local potentials that are phase equivalent to a given nonlocal potential. However, rather than fitting theory to data, we fit theory to theory.

\begin{figure}[h]
\begin{center}
\includegraphics[scale=0.35]{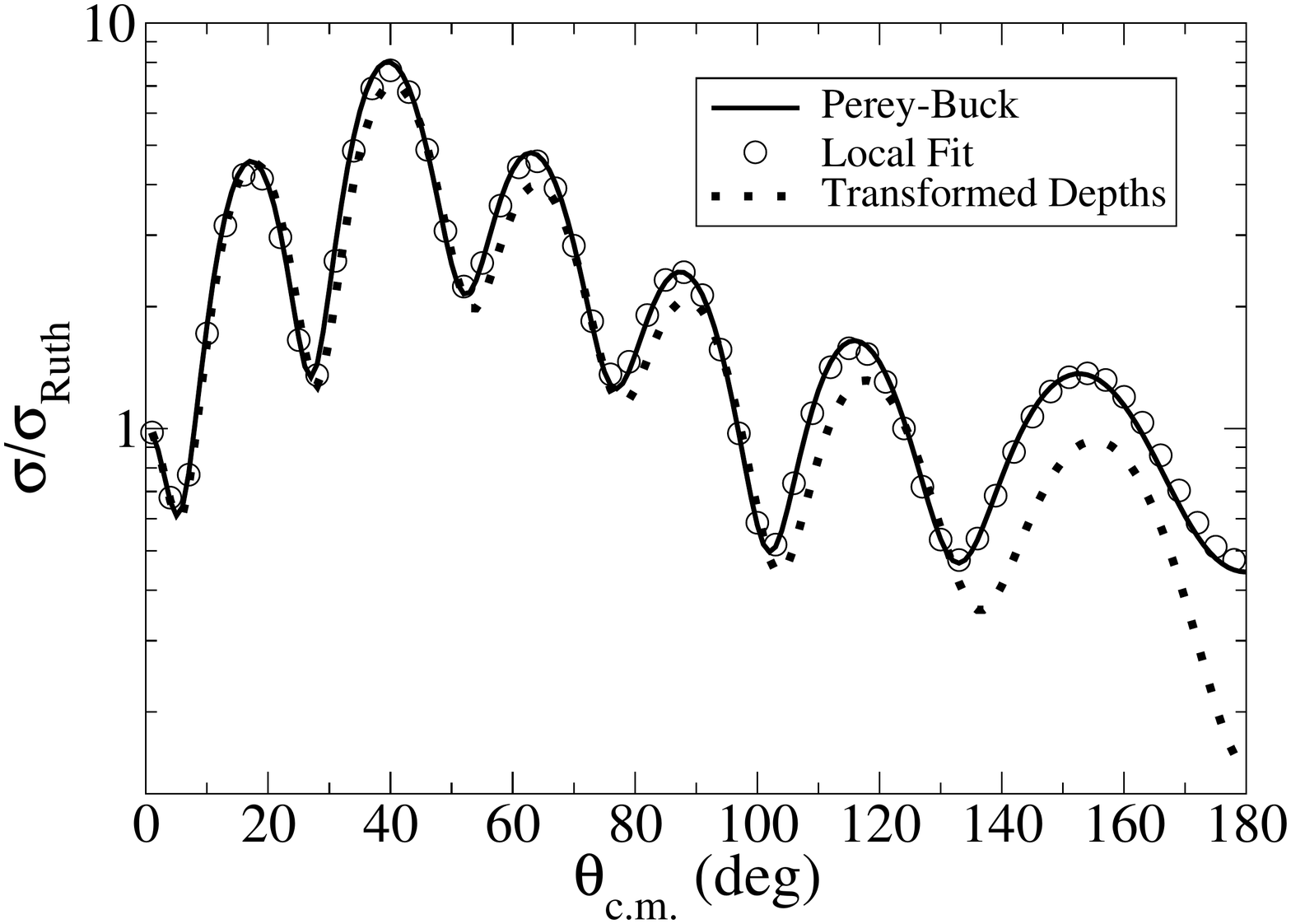}
\end{center}
\caption{$^{49}$Ca$(p,p)^{49}$Ca at 50.0 MeV: The solid line is obtained from using the Perey-Buck nonlocal potential, the open circles are a fit to the nonlocal solution, and the dotted line is obtained by transforming the depths of the volume and surface potentials according to Eq. (\ref{Trans}). Figure reprinted from \cite{Titus_prc2014} with permission.}
\label{fig:p49Ca_50-0}
\end{figure}

An example of one such fit is shown in Fig. \ref{fig:p49Ca_50-0}. We show the differential cross section over the Rutherford cross section as a function of the scattering angle. The solid line is the elastic scattering distribution generated using the Perey-Buck nonlocal potential. The open circles are a fit to the nonlocal solution, and the dotted line is obtained by transforming the depths of the volume and surface potentials \cite{Titus_prc2014}. Notice that the local fit is essentially exact all the way out to 180$^{\circ}$. The transformation formulas relied on by Giannini and Ricco to construct their potential represents the dotted line. The inadequacy of the transformation formula to reproduce the solution with the nonlocal potential is why the Giannini-Ricco potential was not favored in this study.

\chapter{Results}
\label{Results}

For many years nonlocality has been effectively included in calculations by use of the Perey correction factor (PCF) \cite{Austern_pr1965, Fiedeldey_np1966}, as discussed in Sec. \ref{Section-PCF}. However, the PCF is only suitable for use with potentials of the Perey-Buck form, and thus not of use for the DOM potential or a microscopically derived optical potential. In addition, the quality of the PCF has never been rigorously tested. Therefore, the first part of this study was to investigate if the correction factor was adequately able to account for the impact of nonlocality on the wave functions.

The reduction of the wave function can be understood physically in terms of the repulsion between fermions due to the Pauli principle.  Since one major source of nonlocality is due to anti-symmetrization, this repulsion will naturally have the effect of pushing some of the wave function out of the interior as compared to an interaction that doesn't take anti-symmetrization into account.

The deuteron scattering state is also affected by nonlocality. When using the DWBA, it is possible to apply a correction factor to the deuteron scattering wave function. However, a nonlocal global deuteron optical potential does not exist for the purpose of comparison. Also, as we have discussed, the DWBA does not take deuteron breakup into account explicitly. Therefore, we would like to use the more advanced ADWA which does consider breakup, and relies on better constrained nucleon optical potentials, of which nonlocal global potentials exist (i.e. Perey-Buck). 

The numerical details of the calculations performed in this thesis are presented in Sec. \ref{Sec:Numerical_Details}. The results of this thesis will be presented in three parts. The first part in Sec. \ref{PB_Transfer} investigates $(p,d)$ transfer reactions on $^{17}$O, $^{41}$Ca, $^{49}$Ca, $^{127}$Sn, $^{133}$Sn, and $^{209}$Pb at proton energies of $E_p=20$ and $50$ MeV. The transfer cross sections were calculated within the DWBA, and nonlocality in the deuteron channel is not included. In this study, we investigated the effect of nonlocality on the proton scattering wave function and the neutron bound state wave function. We also examined the validity of the commonly used PCF to effectively include nonlocality.

Next, in Sec. \ref{DOM_Transfer}, we studied $(p,d)$ transfer reactions on $^{40}$Ca at proton energies of $E_p=20$, $35$, and $50$ MeV using the nonlocal DOM potential, as well as the Perey-Buck potential. Once again, the transfer cross section was calculated within the DWBA, and nonlocality in the deuteron channel is not included.  Here we studied hole states rather than single particle states, as in the previous study. Hence, the goal of this study was to understand if the effects of nonlocality seen in the previous study could be generalized to hole states, and to see if the same conclusions can be drawn when using a different form for the nonlocal potential.

Finally, in Sec. \ref{NL_ADWA}, we studied $(d,p)$ reactions on $^{16}$O, $^{40}$Ca, $^{48}$Ca, $^{126}$Sn, $^{132}$Sn, and $^{208}$Pb at deuteron energies of $E_d=10$, $20$ and $50$ MeV. For these cases, nonlocality was included explicitly in the deuteron scattering state within the ADWA, as well as in the proton channel. In all wave functions, the Perey-Buck nonlocal potential was used. This study sought to quantify the effect of nonlocality when included consistently in calculations of single nucleon transfer reactions including deuteron breakup.  

It is important to note that the purpose of this work is not to describe the data. We do not expect that the Perey-Buck potential, developed in the sixties for $n+^{208}$Pb at intermediate energies using two data sets, will do well for a wide range of targets and energies. The focus should be on the differences between the nonlocal and the local calculations under the constraint of the same physical input, namely, that both the nonlocal and local optical potentials introduced reproduce the exact same elastic scattering.


\section{Numerical Details}
\label{Sec:Numerical_Details}

To compare the results of the nonlocal calculations, we must compare our results to calculations using local potentials with the same constraints. Therefore, to constrain the local nucleon-target optical potentials, we require that they reproduce the same elastic scattering obtained when using the Perey-Buck or the DOM potential at the relevant energies. For the proton scattering states, we calculate $(p,p)$ elastic scattering at the relevant energy using the Perey-Buck or DOM potential, then fit the resulting distribution to a local form. The fitting of these local phase equivalent (LPE) potentials was performed using the code \textsc{SFRESCO} \cite{fresco}. 

For the deuteron scattering states, the procedure is somewhat different. In Secs. \ref{PB_Transfer} and \ref{DOM_Transfer} we use the local global deuteron optical potential of Daehnick \cite{Daehnick_prc1980} evaluated at the relevant energy.  In Sec. \ref{NL_ADWA}, we calculate $(n,n)$ and $(p,p)$ elastic scattering at half the deuteron energy using the Perey-Buck potential, and again found LPE potentials for the elastic scattering distributions. The local adiabatic potential is then calculated with the proton and neutron LPE potentials. 

For the neutron bound states, we calculated the nonlocal equation using a real Woods-Saxon form with a nonlocality range of $\beta=0.85$ fm. We also used a local spin-orbit interaction with a depth fixed at $6$ MeV. For each term we used a radius of $r=1.25$ fm and $a=0.65$ fm. The depth of the nonlocal real Woods-Saxon form was then adjusted to reproduce the physical binding energy. Also, in Sec. \ref{DOM_Transfer} we used the nonlocal DOM potential to calculate the neutron bound state. The corresponding bound state resulting from local potentials was obtained by setting $\beta=0$ and adjusting the local real Woods-Saxon depth to reproduce the binding energy. 

In Secs. \ref{PB_Transfer} and \ref{DOM_Transfer}, the calculated wave functions were read into the code \textsc{FRESCO} to calculate the $(p,d)$ transfer cross sections. We used the Reid soft core interaction \cite{Reid_ap1968} in the $(p,d)$ T-matrix and to calculate the deuteron bound state. In Sec. \ref{NL_ADWA}, the bound and scattering states that are calculated are inserted into the $(d,p)$ T-matrix Eq.(\ref{Tmatrix}). This was implemented in the code NLAT (NonLocal Adiabatic Transfer). The NN interaction in this case was a central Gaussian which reproduces the binding energy and radius of the deuteron ground state, as in \cite{Moro_prc2009}. 

In Secs. \ref{PB_Transfer} and \ref{DOM_Transfer} the scattering wave functions were solved by using a $0.05$ fm radial step size with a matching radius of $40$ fm. For the bound states solutions, we used a radial step size of $0.02$ fm. The matching radius was half the radius of the nucleus under consideration, and the maximum radius was $30$ fm. The cross sections contain contributions of partial waves up to $J=30$. 

In Sec. \ref{NL_ADWA} the scattering wave functions were calculated in steps of $0.01$ fm with a matching radius of $30$ fm. The nonlocal adiabatic potential was obtained on a radial grid of step $0.05$ fm. We used linear interpolation to calculate the nonlocal adiabatic potential in steps of $0.01$ fm in order to calculate the adiabatic deuteron wave function with the same step size. The bound state wave functions were also calculated in steps of $0.01$ fm with a maximum radius of $30$ fm and a matching radius of half the radius of the nucleus under consideration. Again, converged cross sections contain partial waves up to $J=30$. 

\subsection{Effects of Neglecting Remnant}
\label{Remnant}

To get an idea of the significance of the remnant term, we show in Table \ref{tab:Rem_vs_NoRem} the percent difference at the first peak of the $(d,p)$ transfer cross section for a calculation with the remnant term relative to a calculation without the remnant term for a wide range of targets. These DWBA calculations used the deuteron global optical potential of Daehnick \cite{Daehnick_prc1980} to describe the deuteron scattering state, and the LPE potentials to the Perey-Buck potential for the proton scattering state, a central Gaussian for the deuteron bound state, and a real Woods-Saxon form for the neutron bound state that reproduces the experimental binding energy.

\begin{table}
\begin{center}
  \begin{tabular}{|c||c|c|c|c|c|c|}
     \hline
               &  $^{16}$O$(d,p)$   &  $^{40}$Ca$(d,p)$  & $^{48}$Ca$(d,p)$   & $^{126}$Sn$(d,p)$  & $^{132}$Sn$(d,p)$ & $^{208}$Pb$(d,p)$   \\ \hline \hline
   $10$ MeV    & $-1.92\%$         & $-2.69\%$   & $-0.39\%$  & $0.32\%$   & $0.73\%$   & $1.48\%$     \\ \hline
   $20$ MeV    & $-1.87\%$         & $-2.26\%$   & $-0.34\%$  & $-0.65\%$  & $0.17\%$   & $0.11\%$     \\ \hline
   $50$ MeV    & $-5.57\%$         & $0.07\%$    & $-2.59\%$  & $-0.41\%$  & $-0.08\%$  & $-0.38\%$    \\ \hline
  \end{tabular}
  \caption{Percent difference of the $(d,p)$ transfer cross section at the first peak for a calculation including the remnant term relative to a calculation without the remnant term.}
  \label{tab:Rem_vs_NoRem}
\end{center}
\end{table}


\section{Distorted Wave Born Approximation with the Perey-Buck Potential}
\label{PB_Transfer}

The first part of this study was to investigate the effect of the Perey-Buck potential on the entrance channel of $(p,d)$ transfer reactions. For this study, nonlocality was included explicitly in the proton scattering state and the neutron bound state. Using the wave functions generated with the nonlocal potentials, $(p,d)$ transfer reactions were calculated. These cross sections were compared to those generated with LPE potentials, discussed in Sec. \ref{Section-LPE}. Also, wave functions were modified with the PCF, and the corresponding transfer cross sections were calculated. The goal in this study was to assess the effect of nonlocality on transfer cross sections when compared to cross sections generated with LPE potentials, as well as to determine the quality of the PCF and its ability to reproduce the effects of nonlocality.

\subsection{Proton Scattering State}
\label{PB_Proton_Scat}

When doing calculations of $(p,d)$ or $(d,p)$ reactions using the T-matrix formalism of Eq.(\ref{Tmatrix}), it is required to calculate a proton elastic scattering wave function in either the entrance or exit channel. Because some codes, such as \textsc{TWOFNR} \cite{twofnr}, allow for nonlocality to be included through the PCF, this approach has become common practice. However, until recently, the accuracy of this approach was not understood. Using the Perey-Buck potential, this methodology to include nonlocality has been tested. To do this check, a LPE potential needed to be found. The open circles in Fig. \ref{fig:p49Ca_50-0} are one such example of a LPE potential. The LPE potential found from the fit is the $U^{LE}$ term in Eq.(\ref{CorrectionFactor}).

\begin{figure}[h!]
\begin{center}
\includegraphics[scale=0.35]{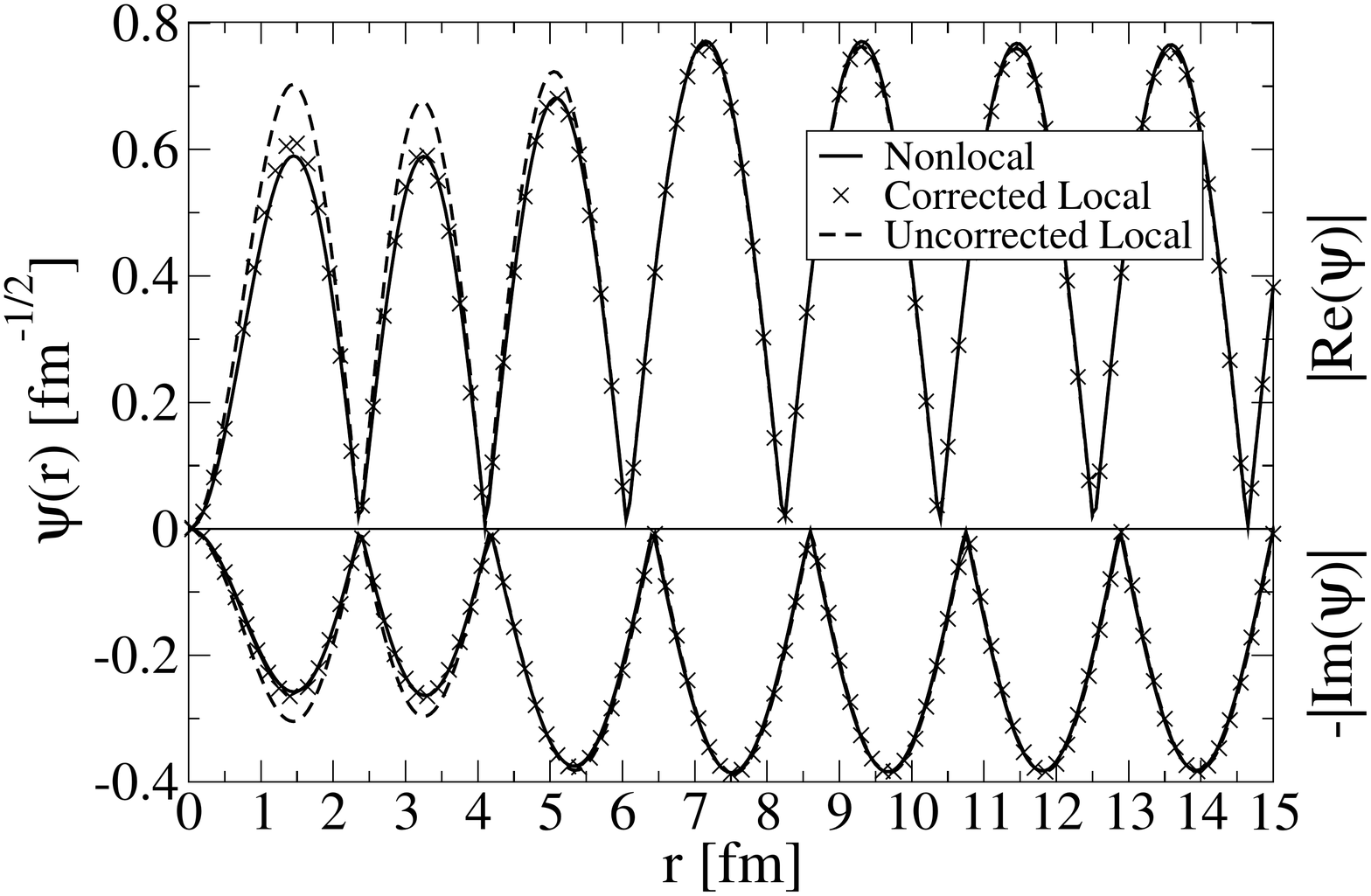}
\end{center}
\caption{Real and imaginary parts of the $J^{\pi}=1/2^-$ partial wave of the scattering wave function for the reaction $^{49}$Ca$(p,p)^{49}$Ca at 50.0 MeV: $\psi^{NL}$ (solid line), $\psi^{PCF}$ (crosses), and $\psi^{loc}$ (dashed line). Top (bottom) panel: absolute value of the real (imaginary) part of the scattering wave function. Figure reprinted from \cite{Titus_prc2014} with permission.}
\label{fig:p49Ca_50-0_001_001}
\end{figure}

\begin{figure}[h!]
\begin{center}
\includegraphics[scale=0.35]{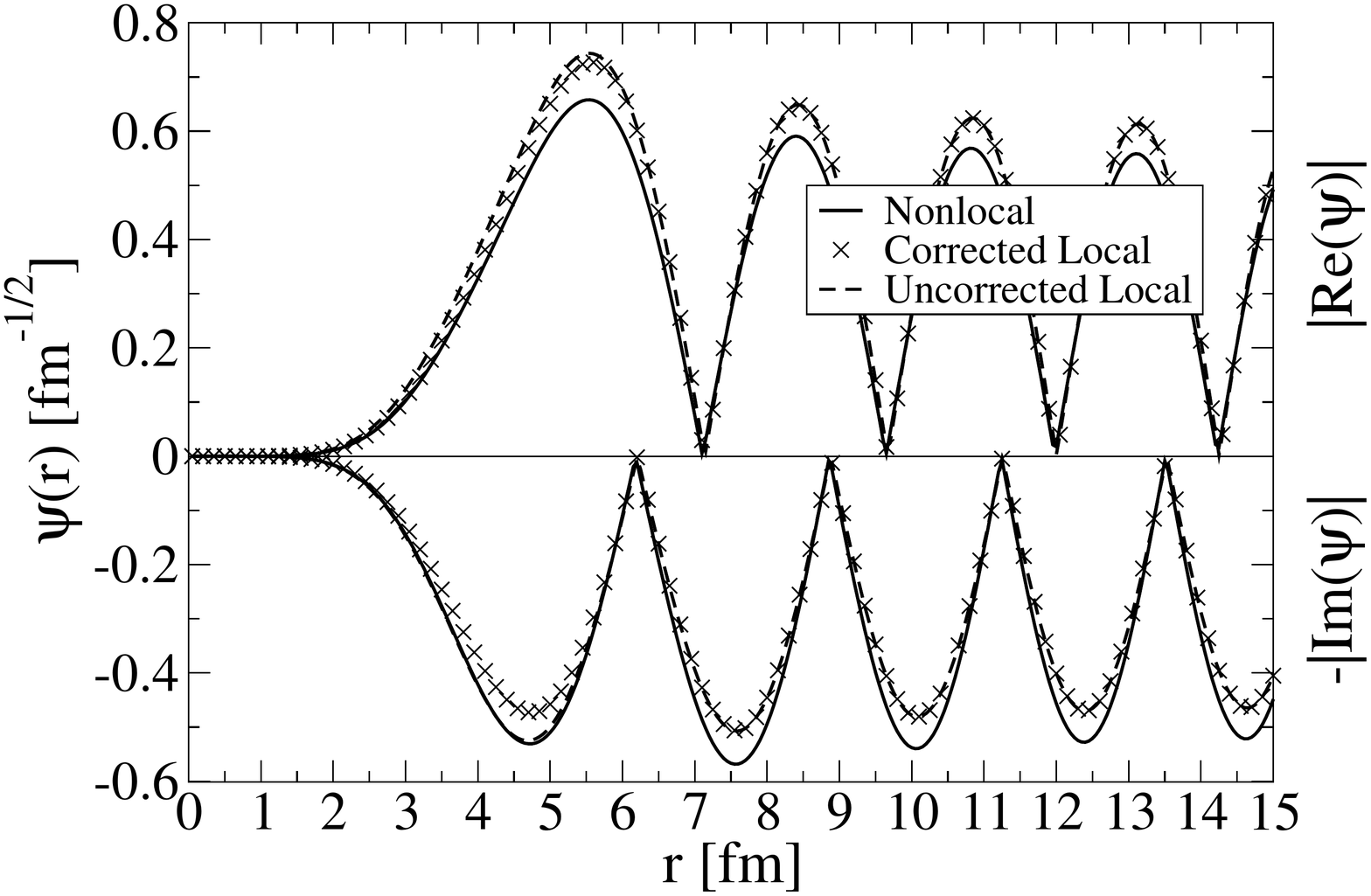}
\end{center}
\caption{Real and imaginary parts of the $J^{\pi}=11/2^+$ partial wave of the scattering wave function for the reaction $^{49}$Ca$(p,p)^{49}$Ca at 50.0 MeV. See caption of Fig. \ref{fig:p49Ca_50-0_001_001}. Figure reprinted from \cite{Titus_prc2014} with permission.}
\label{fig:p49Ca_50-0_006_011}
\end{figure}

As an example, we will use the LPE potential from Fig. \ref{fig:p49Ca_50-0} and consider the scattering wave function for the reaction $^{49}$Ca$(p,p)^{49}$Ca at $50$ MeV. The $J^{\pi}=1/2^-$ partial wave is shown in Fig. \ref{fig:p49Ca_50-0_001_001}. As is seen in the figure, the reduction of the wave function resulting from the nonlocal potential (solid line) relative to the wave function from the LPE potential (dashed line) is apparent. Also seen is that the wave functions from the nonlocal potential and the local potential with the PCF applied (crosses) are in good agreement. This was a general result for most partial waves. However, in all cases that were studied, problems arose for partial waves corresponding to impact parameters around the surface region, shown in Fig. \ref{fig:p49Ca_50-0_006_011}. Since transfer cross sections tend to be most sensitive to the surface region, the differences for these angular momenta are particularly relevant. We will see how the inadequacy of the PCF for surface partial wave affects the resulting transfer cross sections in Sec. \ref{pd_Transfer_PB}. 

This inability of the PCF to correct surface partial waves is partly due to the way in which it was derived. When deriving the PCF, terms related to $\nabla^2F$ were neglected, such as the one in Eq.(\ref{FullCorrectionFactor}). This term only contributes around the nuclear surface. In addition, when performing the local fit, we occasionally found slight differences in the S-matrix elements for a particular partial wave. Since the scattering wave functions are normalized according to Eq.(\ref{scat-asymptotic}), these small changes in the S-matrix will result in different amplitudes for the real and imaginary parts of the scattering wave function in the asymptotic region.  

\subsection{Neutron Bound State}
\label{PB_Neutron_Bound}

We now turn our attention to the neutron bound state that exists in the entrance channel of $(p,d)$ reactions. In order to investigate the effects of nonlocality on the bound state wave functions, and the adequacy of the PCF to correct for nonlocality, the PCF was applied to the local bound state wave function, and the resulting wave function was renormalized to unity. 

\begin{figure}[h!]
\begin{center}
\includegraphics[scale=0.4]{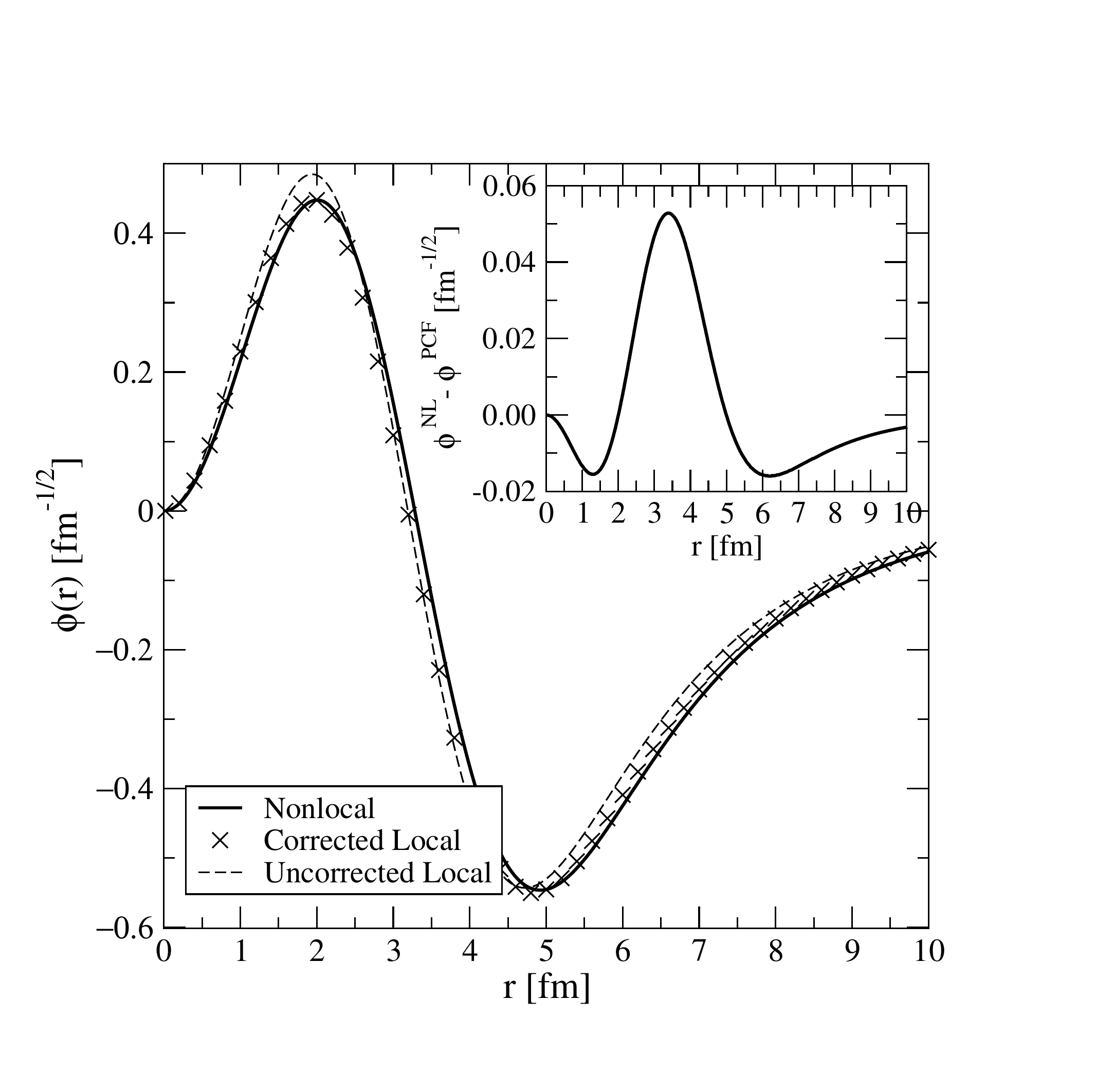}
\end{center}
\caption{Ground state, $2p_{3/2}$, bound wave function for $n+^{48}$Ca. $\phi^{NL}$ (solid line), $\phi^{PCF}$ (crosses), and $\phi^{loc}$ (dashed line). The inset shows the difference $\phi^{NL}-\phi^{PCF}$. Figure reprinted from from \cite{Titus_prc2014} with permission.}
\label{fig:n48Ca_Bound}
\end{figure}

To illustrate, the $2p_{3/2}$ ground state wave function for $n+^{48}$Ca is shown in Fig. \ref{fig:n48Ca_Bound}. Visually, the correction factor does an excellent job correcting for nonlocality in the bound state. However, it is important to notice that in the surface region ($2-5$ fm), the PCF does very little to bring the wave function resulting from the local equivalent potential into agreement with the wave function resulting from the nonlocal potential. The inset, which shows the difference between $\phi^{NL}$ and $\phi^{PCF}$, emphasizes this fact. 

As stated in Sec. \ref{PB_Proton_Scat}, the reason for the inadequacy of the PCF in the surface region goes back to the way in which the PCF was derived. In this case, the bound wave function has a large slope around the surface resulting in large differences between the wave function generated using nonlocal interactions and the one generated from local interactions.

Another important point to note is that nonlocality has the effect of increasing the normalization of the asymptotic properties of the wave function (the ANC). Since nonlocality reduces the amplitude of the wave function in the nuclear interior, and the wave function is always normalized to unity, the ANC must increase. Therefore, the ANC of the bound wave function resulting from nonlocal potentials was found to always be larger than the ANC from local potentials, and the ANC of the corrected wave function was somewhere in between. 

\subsection{$(p,d)$ Transfer Cross Sections - Distorted Wave Born Approximation}
\label{pd_Transfer_PB}

Now that we have studied the effect of nonlocality on the scattering and bound state wave functions, we can investigate the effect nonlocality has on $(p,d)$ transfer reactions when nonlocality is included explicitly in the entrance channel.

\begin{figure}[h!]
\begin{center}
\includegraphics[scale=0.35]{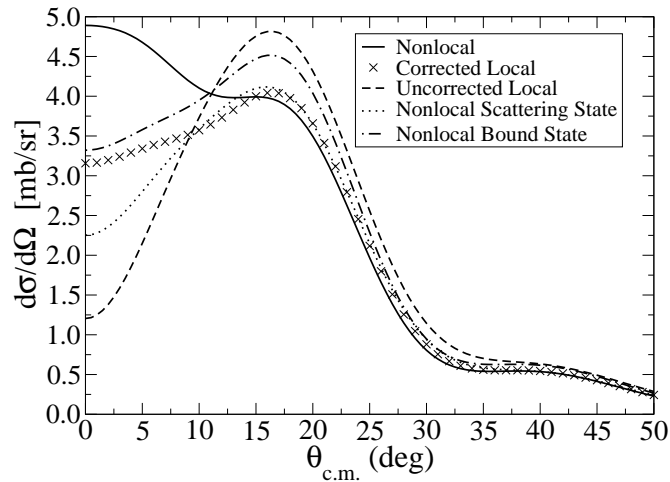}
\end{center}
\caption{Angular distributions for $^{49}$Ca$(p,d)^{48}$Ca at $50$ MeV: Inclusion of nonlocality in both the proton scattering state and the neutron bound state (solid), using LPE potentials, then applying the correction factor to both the scattering and bound states (crosses), using the LPE potentials without applying any corrections (dashed line), including nonlocality only in the proton scattering state (dotted line) and including nonlocality only in the neutron bound state (dot-dashed line). Figure reprinted from \cite{Titus_prc2014} with permission.}
\label{fig:49Ca_50-0pd}
\end{figure}

As a first example, consider the transfer reaction corresponding to the wave functions we have been studying in Secs. \ref{PB_Proton_Scat} and \ref{PB_Neutron_Bound}, $^{49}$Ca$(p,d)^{48}$Ca at a proton energy of $E_p=50$ MeV in the laboratory frame. The separate and combined effects of nonlocality in the bound and scattering states are shown in Fig. \ref{fig:49Ca_50-0pd}. The solid line corresponds to when nonlocality is included in both the proton scattering state and the neutron bound state, the dashed line corresponds to the distribution obtained when only local equivalent potentials are used, the crosses correspond to the cross section obtained when the proton scattering state and neutron bound state wave functions are both corrected with the PCF. Also shown with the dotted line is the cross section when nonlocality was only added to the scattering state, and the dot-dashed line when nonlocality is only added to the neutron bound state.

The results of Fig. \ref{fig:49Ca_50-0pd} are unique for the cases we considered in that the shape of the distribution was significantly changed. The reason for the significant changes around zero degrees can be seen from an analysis of the scattering and bound wave functions. The bound wave function has a node which occurs at a radius corresponding to the surface region for $^{49}$Ca. Since the bound wave function has a large slope in this region, the percent difference between the nonlocal and local wave functions can be quite large. For this case, the nonlocal bound wave function is smaller than the local wave functions in this region, reducing the cross section at the peak around $20$ degrees. On the other hand, the magnitude of the bound wave function is large in the asymptotic region, which increases the cross section at zero degrees.

For the scattering wave function, the most significant differences were for partial waves corresponding to the surface region. Also, the asymptotics of the scattering wave functions were different due to small differences in the S-matrix, again mostly for surface partial waves. There is an interplay between the real and imaginary parts of the scattering wave function which influences the cross section at forward angles. Then, the complex combination of all these effects produces the interesting behavior of the transfer cross section at forward angles.

\begin{figure}[h!]
\begin{center}
\includegraphics[scale=0.35]{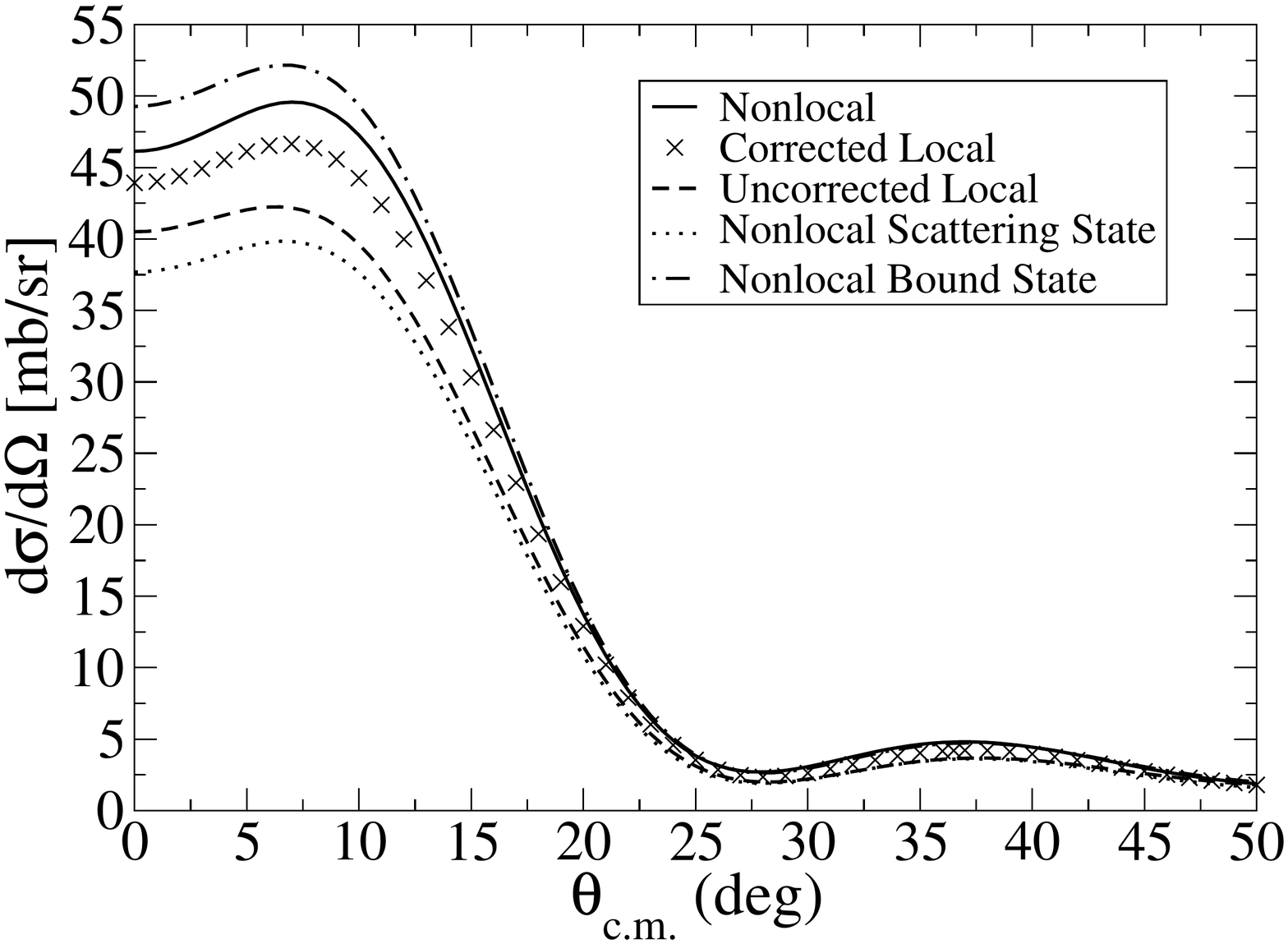}
\end{center}
\caption{Same as in Fig. \ref{fig:49Ca_50-0pd} but for $^{49}$Ca$(p,d)^{48}$Ca at $E_p=20$ MeV. Figure reprinted from \cite{Titus_prc2014} with permission.}
\label{fig:49Ca_20-0pd}
\end{figure}

\begin{figure}[h!]
\begin{center}
\includegraphics[scale=0.35]{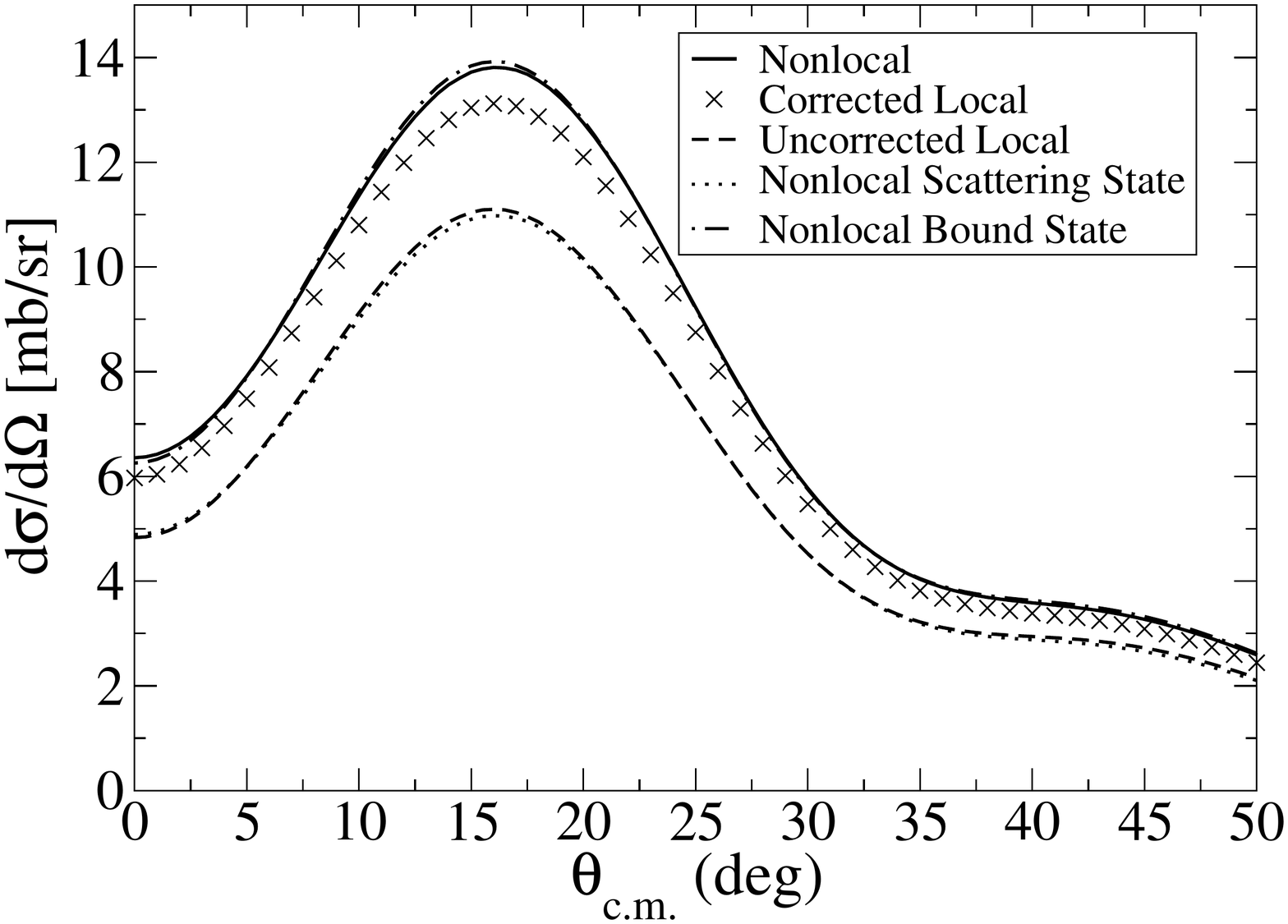}
\end{center}
\caption{Same as in Fig. \ref{fig:49Ca_50-0pd} but for $^{133}$Sn$(p,d)^{132}$Sn at $E_p=20$ MeV. Figure reprinted from \cite{Titus_prc2014} with permission.}
\label{fig:133Sn_20-0pd}
\end{figure}

Now consider the same reaction but at a lower energy. In Fig. \ref{fig:49Ca_20-0pd} we show $^{49}$Ca$(p,d)^{48}$Ca at a proton energy of $E_p=20$ MeV in the laboratory frame. This case is more representative of the general features we saw in this systematic study. The nonlocality in the scattering state had the effect of reducing the transfer cross section due to the reduction of the scattering wave function, while the nonlocality in the bound state had the effect of enhancing the cross section due to the increase of the wave function in the asymptotic region. At this lower energy, the overall effect was an enhancement of the transfer cross section at the first peak. In addition, it is seen that the PCF moves the transfer distribution generated with local potentials in the direction of that generated with nonlocal potentials. However, the PCF was not able to fully take the effects of nonlocality into account.

\begin{figure}[h!]
\begin{center}
\includegraphics[scale=0.35]{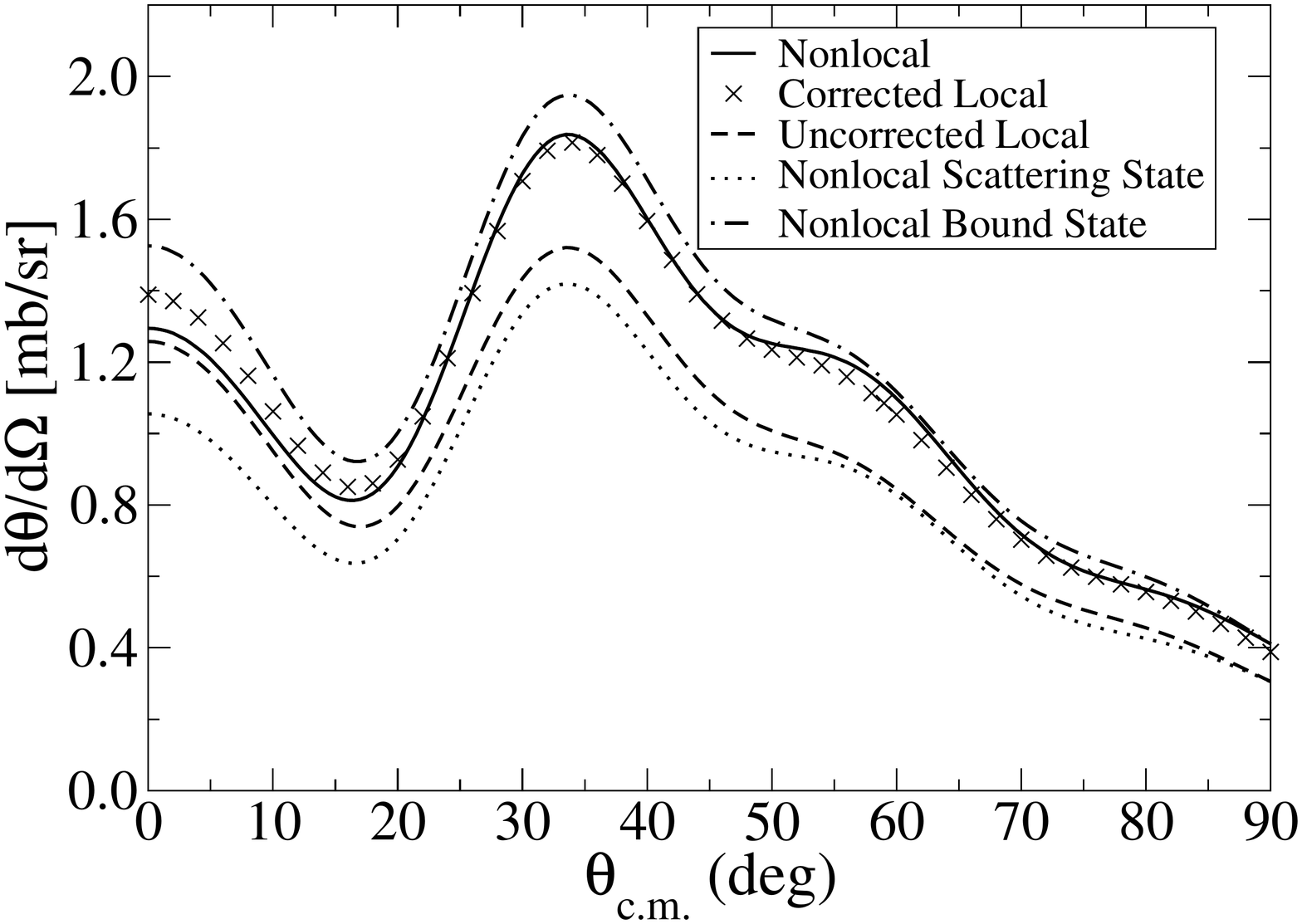}
\end{center}
\caption{Same as in Fig. \ref{fig:49Ca_50-0pd} but for $^{209}$Pb$(p,d)^{208}$Pb at $E_p=20$ MeV. Figure reprinted from \cite{Titus_prc2014} with permission.}
\label{fig:209Pb_20-0pd}
\end{figure}

Next we consider heavier targets, such as $^{133}$Sn and $^{209}$Pb, at $E_p=20$ MeV. In both of these cases the inclusion of nonlocality in the scattering state decreased the cross section by a smaller amount than was seen before. This is due to the low energy of the proton and the high charge of the target. Nonlocality in the proton scattering state reduced the magnitude of the wave function in the nuclear interior, but the energy of the proton was not high enough to penetrate deeply due to the large Coulomb barrier. On the other hand, nonlocality in the bound state is very significant. Since the projectile energy was low, and the charge of the target was high, the reaction is dominated by the asymptotitcs of the bound wave function, which is enhanced in the nonlocal case.

For $^{133}$Sn$(p,d)^{132}$Sn at $E_p=20$ MeV, Fig. \ref{fig:133Sn_20-0pd}, the PCF does a reasonable job taking nonlocality into account, but there are still discrepancies between the full nonlocal and corrected local solutions. For $^{209}$Pb$(p,d)^{208}$Pb at $E_p=20$ MeV, Fig. \ref{fig:209Pb_20-0pd}, there are discrepancies at forward angles, but the distributions resulting from nonlocal and local potentials coincidentally agree at the peak. This agreement is accidental, and comes from the nonlocal effect in the bound state canceling that in the scattering state. 

\begin{table}[h!]
\centering
\begin{tabular}{|c|r|r|}
\hline
& Corrected         & Nonlocal          \\
$E_{lab}=20$ MeV & Relative to Local & Relative to Local  \\
\hline
$^{17}$O$(1d_{5/2})(p,d)$ & $7.1\%$  & $18.8\%$  \\ 
$^{17}$O$(2s_{1/2})(p,d)$ & $20.1\%$  & $26.5\%$ \\ 
$^{41}$Ca$(p,d)$ &$11.4\%$ & $21.9\%$     \\ 
$^{49}$Ca$(p,d)$ & $10.4\%$  & $17.3\%$   \\ 
$^{127}$Sn$(p,d)$  & $17.5\%$  & $17.3\%$ \\ 
$^{133}$Sn$(p,d)$ & $18.2\%$  & $24.4\%$  \\ 
$^{209}$Pb$(p,d)$ & $19.4\%$  & $20.8\%$  \\ 
\hline
\end{tabular}
\caption{Percent difference of the $(p,d)$ transfer cross sections at the first peak when using the PCF (2nd column), or a nonlocal potential (3rd column), relative to the local calculation with the LPE potential, for a number of reactions occurring at 
$20$ MeV.}
\label{Tab:Percent_Difference_20}
\end{table}

\begin{table}[h!]
\centering
\begin{tabular}{|c|r|r|}
\hline
& Corrected         & Nonlocal  \\
$E_{lab}=50$ MeV & Relative to Local & Relative to Local \\
\hline
$^{17}$O$(1d_{5/2})(p,d)$  & $17.0\%$  & $35.4\%$  \\ 
$^{17}$O$(2s_{1/2})(p,d)$  & $0.2\%$  & $12.7\%$  \\ 
$^{41}$Ca$(p,d)$ & $2.9\%$  & $5.8\%$  \\ 
$^{49}$Ca$(p,d)$ & $-16.0\%$  & $-17.1\%$  \\ 
$^{127}$Sn$(p,d)$  & $10.1\%$  & $4.5\%$  \\  
$^{133}$Sn$(p,d)$ & $-6.7\%$  & $-16.9\%$  \\ 
$^{209}$Pb$(p,d)$ & $8.6\%$  & $8.6\%$  \\ 
\hline
\end{tabular}
\caption{Percent difference of the $(p,d)$ transfer cross sections at the first peak when using the PCF (2nd column), or a nonlocal potential (3rd column), relative to the local calculation with the LPE potential, for a number of reactions occurring at 
$50$ MeV.}
\label{Tab:Percent_Difference_50}
\end{table}

The percent difference at the first peak of the transfer distributions for all the cases that were studied are summarized in Tables \ref{Tab:Percent_Difference_20} and \ref{Tab:Percent_Difference_50} for the $(p,d)$ reactions at $20$ and $50$ MeV. For the lower energy cases, it is seen that the inclusion of nonlocality provided a general enhancement to the transfer cross section at the first peak. This is due to the enhancement of the bound state wave function in the asymptotic region playing a more significant role in the magnitude of the transfer cross section at low energies. At higher energies, there is a competition between the enhancement due to the bound state, and the reduction due to the scattering state. In most cases, there is still an enhancement of the cross section, but the overall effect is less significant than for the lower energy cases.

\subsection{Summary}

In this study, the long established Perey correction factor (PCF) and the effects of nonlocality on the entrance channel of $(p,d)$ reactions were studied. The integro-differential equation containing the Perey-Buck nonlocal potential was solved numerically for single channel scattering and bound states. A local phase equivalent (LPE) potential was obtained by fitting the elastic distribution generated by the Perey-Buck potential. The PCF was applied to the wave functions generated with the LPE potentials or the local equivalent binding potentials, and the scattering and bound state wave functions were then used in a finite-range DWBA calculation in order to obtain $(p,d)$ transfer cross sections. 

We found that the explicit inclusion of nonlocality in the entrance channel increased the transfer distribution at the first peak by $15-35\%$. In most cases, the transfer distribution from using a nonlocal potential increased relative to the distribution from the local potential. In all cases, the PCF moved the transfer distribution in the direction of the distribution which included nonlocality explicitly. However, nonlocality was never fully taken into account with the PCF. The PCF can be improved by including the surface terms that were neglected, and not assuming that the local momentum approximation is valid. Such additional corrections were not pursued since the full nonlocal solution can be calculated.


\section{Distorted Wave Born Approximation with the Dispersive Optical Model Potential}
\label{DOM_Transfer}

The results of the previous work by Titus and Nunes \cite{Titus_prc2014}, covered in Sec. \ref{PB_Transfer}, demonstrated that nonlocality is significant in the study of transfer reactions, and that the PCF is not able to fully reproduce the complex effects of nonlocality. However, the previous study only considered the Perey-Buck nonlocal potential. Recently, the nonlocal DOM potential for $^{40}$Ca was developed \cite{Mahzoon_prl2014}. We wanted to see if nonlocality remains an important ingredient to transfer reactions when a different nonlocal potential is used. Since the DOM was constructed only for $^{40}$Ca, we were only able to consider $^{40}$Ca$(p,d)^{39}$Ca reactions. For this study, we investigated laboratory proton energies of $E_p=20$, $35$, and $50$ MeV.

\begin{figure}[h!]
\begin{center}
\includegraphics[scale=0.35]{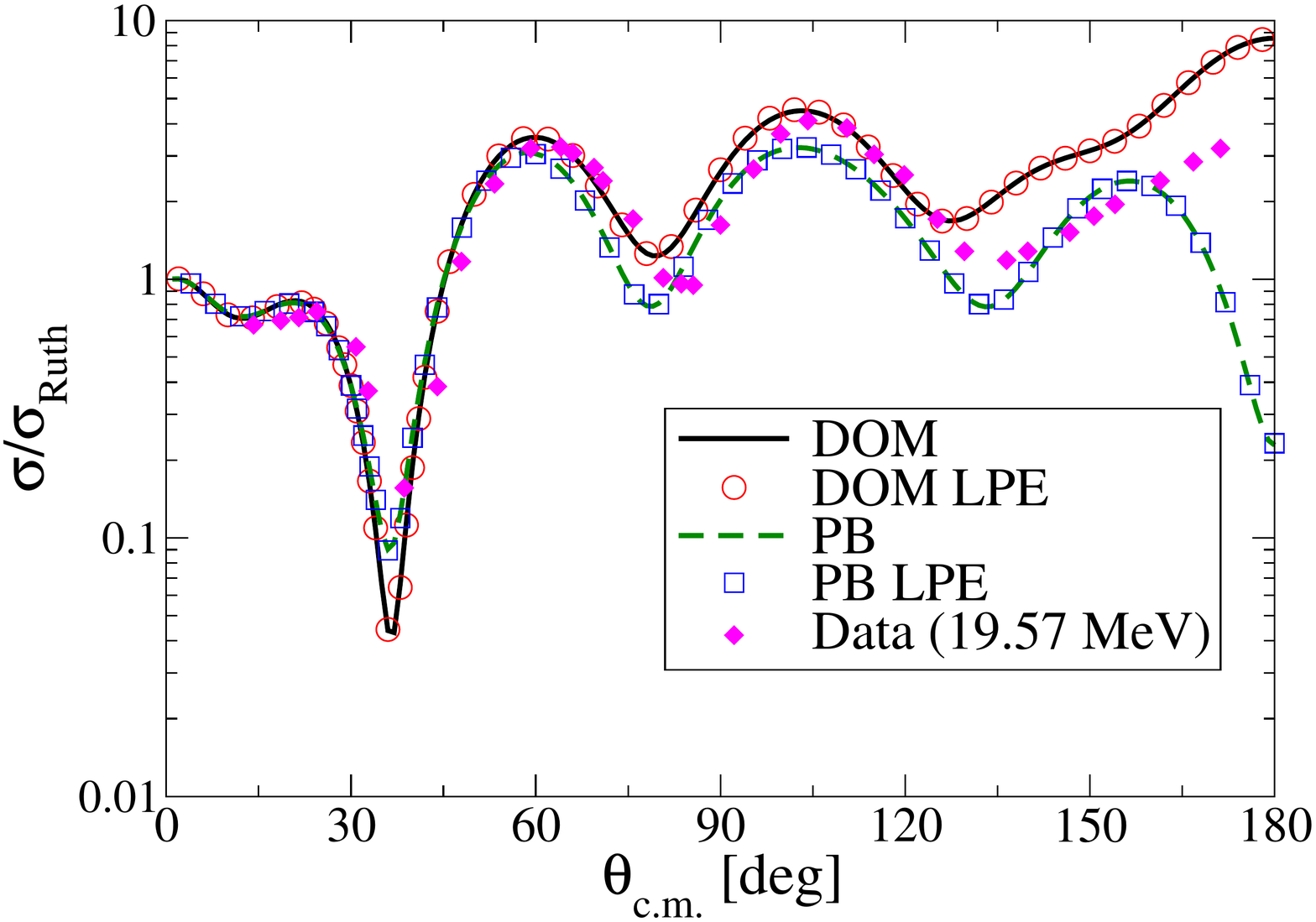}
\end{center}
\caption{Angular distributions for elastic scattering normalized to Rutherford for protons on $^{40}$Ca at $E_p=20$ MeV. The elastic scattering with the DOM potential (solid line), the DOM LPE potential (open circles), the Perey-Buck interaction (dashed line), and the Perey-Buck LPE potential (open squares). The data (closed diamonds) from \cite{Dicello_prc1971}. Figure reprinted from \cite{Ross_prc2015} with permission.}
\label{fig:Elastic_20_DOM}
\end{figure}

\begin{figure}[h!]
\begin{center}
\includegraphics[scale=0.35]{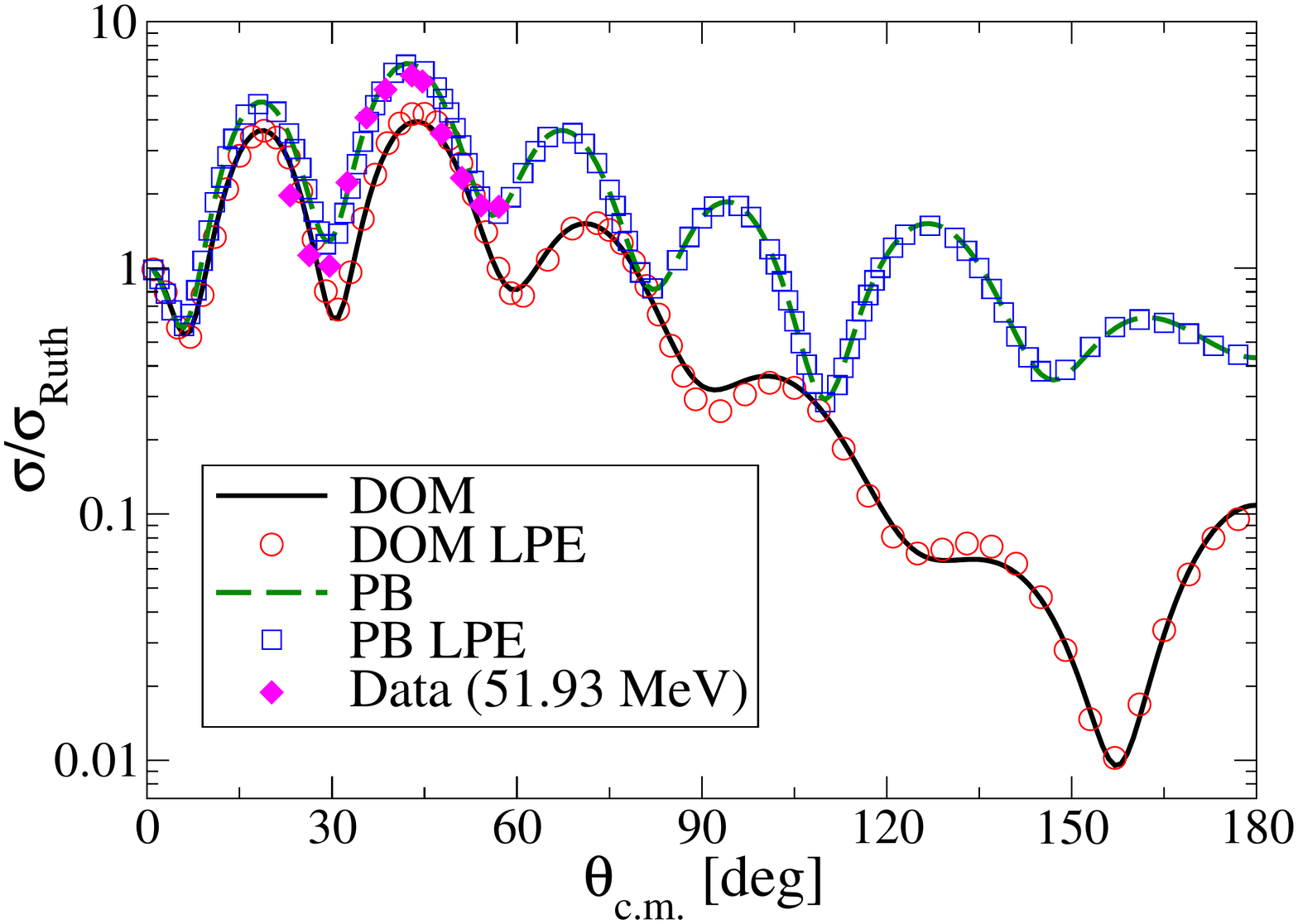}
\end{center}
\caption{Same as in Fig. \ref{fig:Elastic_20_DOM} but for $E_p=50$ MeV. Data from \cite{Ohnuma_jpsj1980}. Figure reprinted from \cite{Ross_prc2015} with permission.}
\label{fig:Elastic_50_DOM}
\end{figure}

\subsection{Proton Scattering State}

In investigating the effect of nonlocality when using the DOM potential, we no longer considered the PCF since there is no easy generalization of the PCF to the DOM potential. This is because each term of the DOM potential has a different value for the nonlocality parameter, $\beta$. While it would be possible to construct a PCF for the DOM potential, the results of the previous work by Titus and Nunes \cite{Titus_prc2014} made pursuing a PCF for the DOM potential irrelevant.

\begin{figure}[h!]
\begin{center}
\includegraphics[scale=0.35]{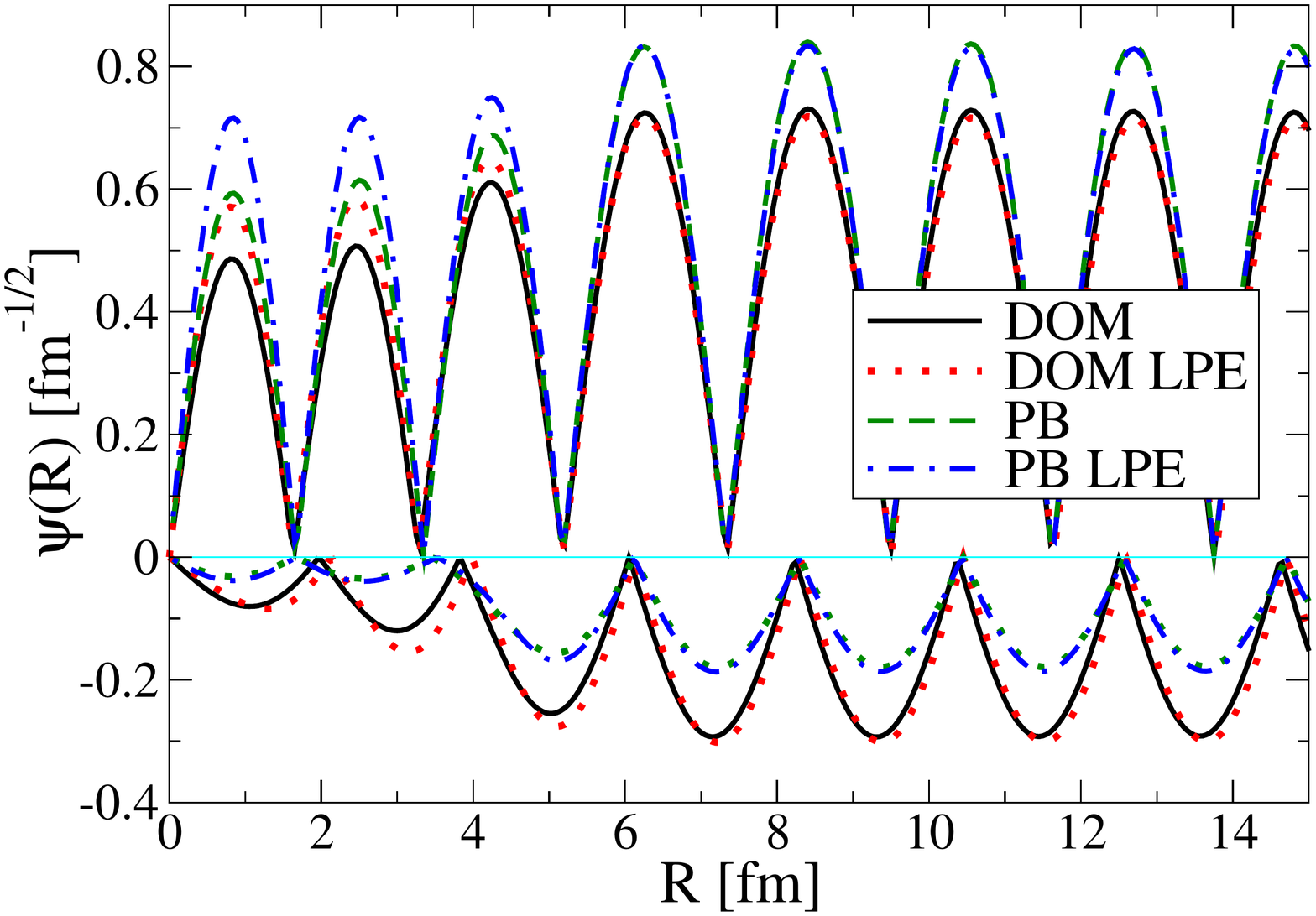}
\end{center}
\caption{The real and imaginary parts of the $J^{\pi}=1/2^+$ partial wave of the scattering wave function for the reaction $^{40}$Ca$(p,p)^{40}$Ca at $E_p=50$ MeV. This shows the wave function resulting from the DOM potential (solid line) and its LPE potential (dotted line), the Perey-Buck potential (dashed line) and its LPE potential (dot-dashed line). The top (bottom) panel shows the absolute value of the real (imaginary) part of the scattering wave function. Figure reprinted from \cite{Ross_prc2015} with permission.}
\label{fig:WF_50_Leq0_DOM}
\end{figure}

In Figs. \ref{fig:Elastic_20_DOM} and \ref{fig:Elastic_50_DOM} we show the elastic distributions generated from the DOM and the Perey-Buck nonlocal potentials along with the corresponding LPE potentials. We also show the corresponding elastic scattering data at the closest relevant energy. It is seen from the distributions that the DOM potential is superior when it comes to describing the data. This should not be a surprise as the DOM potential was constructed from fits to nucleon elastic scattering data on $^{40}$Ca while the Perey-Buck potential was constructed from neutron elastic scattering on $^{208}$Pb at low energy. Nonetheless, the Perey-Buck potential does a reasonably good job at describing the elastic scattering data for the energy and angular range that the data is available.

To investigate the scattering wave functions, we consider the $J^{\pi}=1/2^+$ partial wave for scattering at $E_p=50$ MeV in Fig. \ref{fig:WF_50_Leq0_DOM}. For both the DOM and the Perey-Buck nonlocal potential, we see the reduction of the scattering wave function relative to the wave function from the LPE potential, which is consistent with earlier studies \cite{Titus_prc2014, Austern_pr1965, Fiedeldey_np1966}. Since the two nonlocal potentials describe different elastic scattering distributions, they will have different S-matrix elements for each partial wave, and hence, the different normalizations in the asymptotic region is expected.

\begin{figure}[h!]
\begin{center}
\includegraphics[scale=0.35]{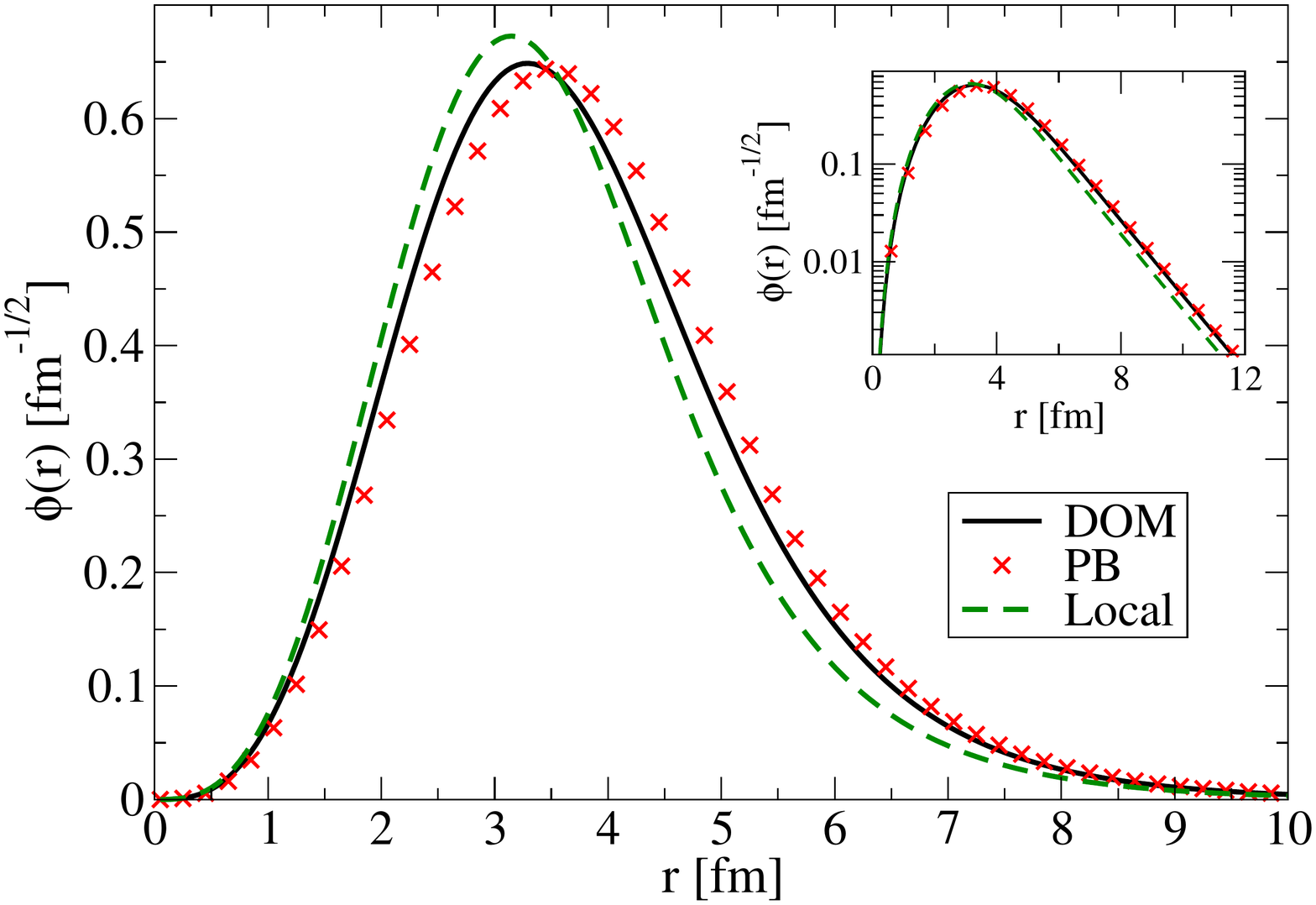}
\end{center}
\caption{The neutron ground state $1d_{3/2}$ bound wave function for $n+^{39}$Ca. Shown is the wave function obtained using the DOM potential (solid line), the Perey-Buck potential (crosses) and the local interaction (dashed line). The inset shows the asymptotic properties of each wave function. Figure reprinted from \cite{Ross_prc2015} with permission.}
\label{fig:Bound_State_DOM}
\end{figure}

\subsection{Neutron Bound State}

The neutron $1d_{3/2}$ bound state wave functions using the various potentials are shown in Fig. \ref{fig:Bound_State_DOM}. The DOM bound wave function was found using the potential defined in \cite{Mahzoon_prl2014}. The same general features of the bound wave functions in \cite{Titus_prc2014} are seen here. Nonlocality reduces the amplitude of the bound wave function and thus pushes the wave function outward. 

\subsection{$(p,d)$ Transfer Cross Sections - Distorted Wave Born Approximation}

\begin{figure}[h!]
\begin{center}
\includegraphics[scale=0.28]{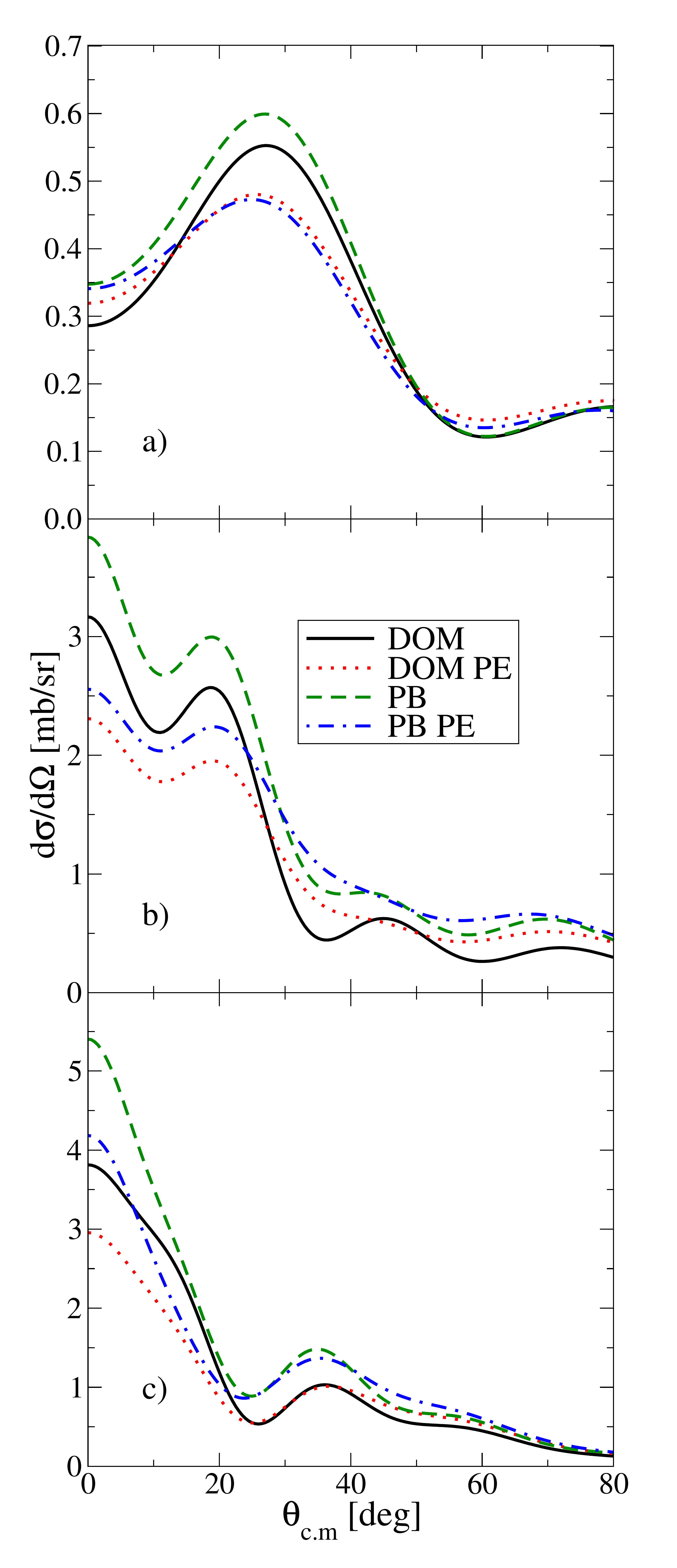}
\end{center}
\caption{Angular distributions for the $^{40}$Ca$(p,d)^{39}$Ca reaction at (a) $E_p=20$ MeV, (b) $E_p=35$ MeV, and (c) $E_p=50$ MeV. In this figure is the transfer distribution resulting from using the nonlocal DOM (solid line) and its LPE potential (dotted line), the Perey-Buck potential (dashed line) and the Perey-Buck LPE potential (dot-dashed line). Figure reprinted from \cite{Ross_prc2015} with permission.}
\label{fig:All_Transfer_DOM}
\end{figure}

In Fig. \ref{fig:All_Transfer_DOM} we show the transfer distributions for the three energies calculated. Shown is the transfer distribution resulting from the DOM nonlocal potential and its LPE potential, as well as the Perey-Buck nonlocal potential and its LPE potential. In general we see that nonlocality for both potentials provides an enhancement of the cross section at the first peak. This is consistent with the conclusions of \cite{Titus_prc2014}. However, at higher energies, there is not as much cancellation between the scattering and bound states so that the full nonlocal calculation still resulted in a fairly significant increase in the cross section. The key difference is that the neutron was bound by $15.6$ MeV in this study whereas the neutron was always bound less than $10$ MeV in all cases in the study of Sec. \ref{PB_Transfer}. 

\begin{table}[h]
\centering
\begin{tabular}{|c|r|r|r|}
\hline
 $E_{p}$ (MeV)       & bound state      & scattering state   & full nonlocal \\
\hline
20& 27 \% &-14 \%  & 15 \%\\
35& 31 \% &10 \%  & 52 \%\\
50& 31 \% &-3 \%  & 29 \%\\
\hline
\end{tabular}
\caption{Percent differences of the $(p,d)$ transfer cross sections at the first peak at the listed beam energies using the DOM potential relative to the calculations with the phase-equivalent potential. Results are listed  separately for the effects of nonlocality on the bound state, the scattering state, and the total.}
\label{tab:diff-dom}
\end{table}

\begin{table}[h]
\centering
\begin{tabular}{|c|r|r|r|}
\hline
 $E_{p}$ (MeV)       & bound state      & scattering state  & full nonlocal  \\
\hline
20& 42 \%&-15 \% & 27 \% \\
35& 55 \%&-8 \%  & 52 \% \\
50& 42 \%&-11 \% & 29 \% \\
\hline
\end{tabular}
\caption{Same as Table \ref{tab:diff-dom}, but now for the Perey-Buck potential.}

\label{tab:diff-pb}
\end{table}

Shown in Table \ref{tab:diff-dom} and \ref{tab:diff-pb} are the percent difference of the $(p,d)$ transfer cross sections at the first peak when using the DOM and Perey-Buck potentials, respectively. In the tables we show the separate effects of the neutron bound state and the proton scattering state, as well as the percent difference for the full nonlocal calculation.

In most cases nonlocality in the scattering state had the effect of reducing the transfer cross section. One exception was for $E_p=35$ MeV when using the DOM potential. The increase was due to the shape of the scattering wave function near the surface region. In this particular case, obtaining and exact fit to the nonlocal distribution was much more difficult than in the other cases. All other cases reduced the cross section by a similar amount. Since the Coulomb barrier is not large for $^{40}$Ca, there was no suppression of the nonlocal effects in the scattering state as we seen in the previous study for heavier systems, such as $^{133}$Sn or $^{209}$Pb. 

The effect of nonlocality in the bound state at all proton energies was to increase the cross section. This is because nonlocality shifts the bound wave function towards the surface region where these transfer reactions are more sensitive. We also note that the nonlocal effects for the Perey-Buck interaction are generally larger than for the previous study \cite{Titus_prc2014}. This is because in the previous study we were studying single particle states in closed shell nuclei, while here we focus on hole states in $^{40}$Ca.

\subsection{Summary}

In this work we studied the effects of adding non-locality in the entrance channel of transfer reactions using a nonlocal potential obtained from the dispersive optical model (DOM) and comparing it to the results from the older Perey-Buck interaction. Our studies focus on the $^{40}$Ca$(p,d)^{39}$Ca reaction at $E_p=20$, $35$ and $50$ MeV. We consider the nonlocality in the proton channel, and solve the integral-differential equation to obtain the proton scattering and neutron bound state solutions for both nonlocal potentials. We then computed the transfer matrix element in the DWBA, ignoring nonlocality in the deuteron channel. 

Our results show that, irrespective of the details of the potential, nonlocality reduces the strength of the wave function in the nuclear interior, an effect most noticeable in the bound states, but also significant in scattering states. Due to the normalization condition, nonlocality in the bound state also shifts the wave function to the periphery region, causing an increase in the transfer cross sections. Typically, nonlocality in the scattering state acts in the opposite direction, reducing the overall effect. When nonlocality is included in both the bound and scattering states, the transfer cross sections are increased by $\approx 15-50$\% for the DOM potential, in contrast with $\approx 30-50$\% obtained with the Perey-Buck interaction. 



\section{Nonlocal Adiabatic Distorted Wave Approximation with the Perey-Buck Potential}
\label{NL_ADWA}

The previous two studies (Titus and Nunes \cite{Titus_prc2014} discussed in Sec. \ref{PB_Transfer}, and Ross, Titus, Nunes, \textit{et. al.} \cite{Ross_prc2015} discussed in Sec. \ref{DOM_Transfer}) have demonstrated that the explicit inclusion of nonlocality, at least in the entrance channel of $(p,d)$ reactions, is very important to take into account explicitly. They have also shown that commonly used correction factors are not sufficient to effectively include nonlocality, and that nonlocality is important regardless of the form chosen for the nonlocal potential. 

In both of these studies, a local deuteron optical potential was used to describe the deuteron scattering state. We will now turn our attention to studying transfer reactions within the ADWA, discussed in Sec. \ref{Section-ADWA}, which includes deuteron breakup explicitly. As the ADWA is based on a three-body Hamiltonian, we included nonlocality consistently in all nucleon-target interactions. For this study, we will focus on $(d,p)$ transfer reactions. It should be noted that due to time reversal invariance, the cross sections for $(d,p)$ and $(p,d)$ transfer reactions differ only by a statistical constant, assuming that the initial and final states are the same. The statistical constant can be determined by detailed balance \cite{Thompson_book}. Therefore, even though we are considering a different reaction, we are building on the learning from the previous studies.


\subsection{The Source Term}
\label{Sec:Source_Term}

In order to compare the effect of nonlocality on the adiabatic potential, we define the $rhs$ of Eq. (\ref{NL-Adiabatic}) to be $S_{AD}^{NL}(R)$. It is difficult to compare the nonlocal adiabatic potential directly to the local adiabatic potential since $S_{AD}^{NL}(R)$ has the scattering wave function built into it. However, we can compare $S_{AD}^{NL}(R)$ with the local corresponding quantity, $S_{AD}^{Loc}(R)=U_{AD}^{Loc}(R)X_{AD}^{Loc}(R)$. After a partial wave decomposition, the source terms become functions dependent only on the scalar $R$ for each $LJ$ combination of the deuteron scattering state. 

\begin{figure}[h!]
\begin{center}
\includegraphics[scale=0.3]{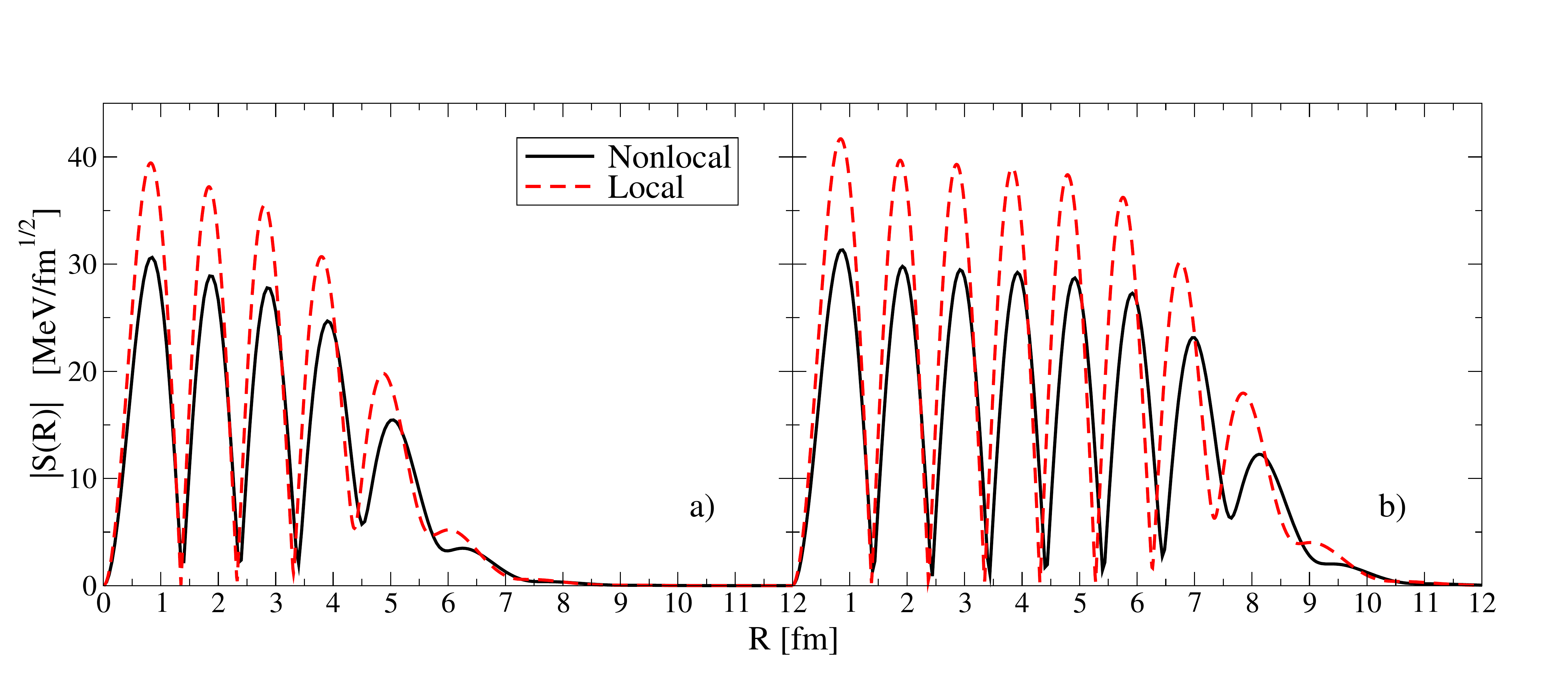}
\end{center}
\caption{Absolutes value of the $d+A$ source term when nonlocal and local potentials are used. (a) $d+^{48}$Ca at $E_d=50$ MeV. (b) $d+^{208}$Pb at $E_d=50$ MeV. Both are for the $L=1$ and $J=0$ partial wave. Figure reprinted from \cite{Titus_prc2015} with permission.}
\label{fig:Source_Comparison_Leq1_Jeq0}
\end{figure}

\begin{figure}[h!]
\begin{center}
\includegraphics[scale=0.3]{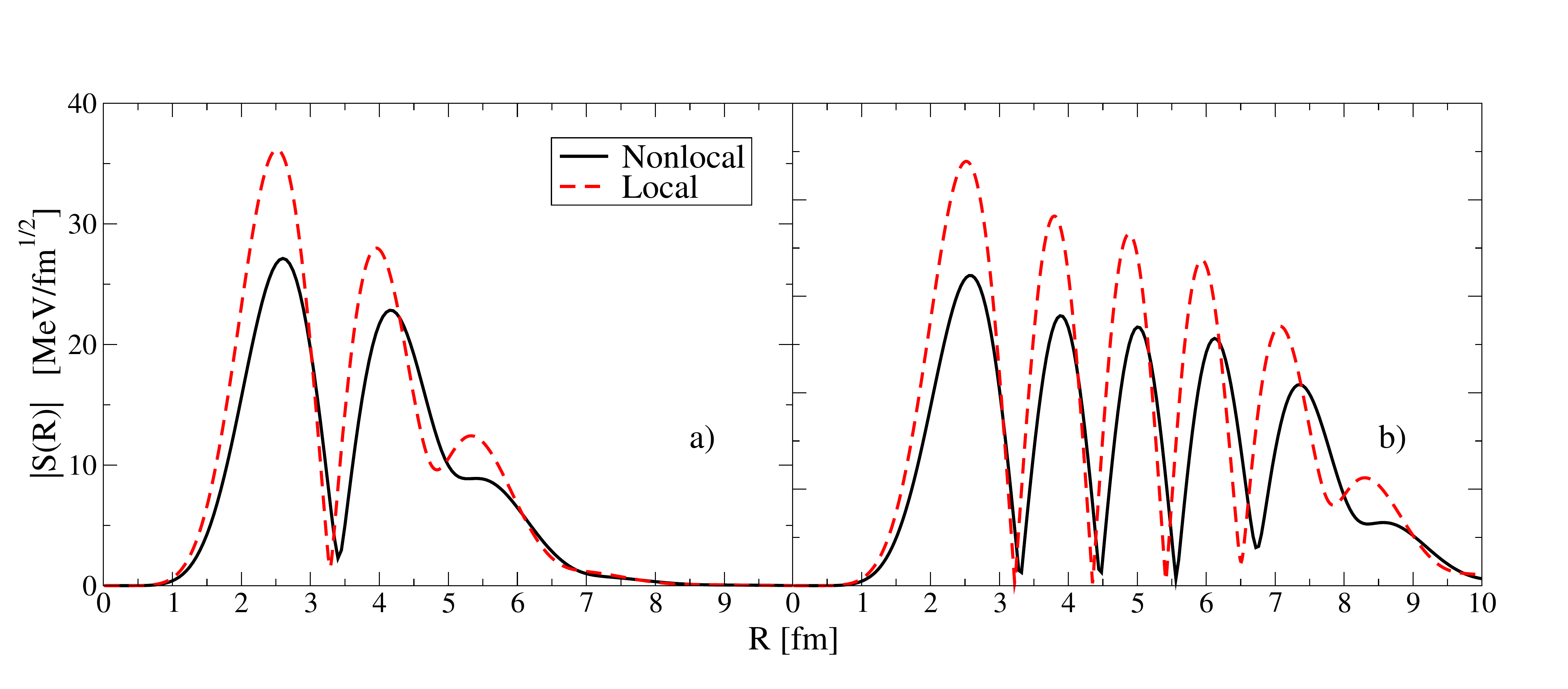}
\end{center}
\caption{Absolute value of the $d+A$ source term when nonlocal and local potentials are used. (a) $d+^{48}$Ca at $E_d=50$ MeV. (b) $d+^{208}$Pb at $E_d=50$ MeV. Both are for the $L=6$ and $J=5$ partial wave. Figure reprinted from \cite{Titus_prc2015} with permission.}
\label{fig:Source_Comparison_Leq6_Jeq5}
\end{figure}

To get an idea of the effect of nonlocality on the adiabatic potential, we make a comparison in Fig. \ref{fig:Source_Comparison_Leq1_Jeq0} of this radial source term for the angular momentum values of $L=1$ and $J=0$ of the $d+^{48}$Ca and $d+^{208}$Pb wave function, both at a beam energy of $E_d=50$ MeV. In Fig. \ref{fig:Source_Comparison_Leq6_Jeq5} we make the same comparison but for the $L=6$ and $J=5$ partial wave. In both figures, the solid line corresponds to the nonlocal source term, while the dashed line is its local equivalent. The magnitude of the nonlocal source term is reduced compared to the local source term. It is also seen that the source term in the nonlocal case gets shifted outward relative to the local case. Both these effects imprint themselves on the adiabatic deuteron wave function, as we will see in Sec. \ref{Sec:Deuteron_Scattering_State}. 

\subsection{Deuteron Scattering State}
\label{Sec:Deuteron_Scattering_State}

The necessary formalism for the local implementation of the ADWA and the nonlocal extension of the ADWA has been addressed in Chapter \ref{Theory}. The radial equation that must be solved for each partial wave is given by Eq.(\ref{NL-Adiabatic}). The $rhs$ of this equation acts as a source term, and the differences between the nonlocal source term and the corresponding local source term was compared in Sec. \ref{Sec:Source_Term}. 

\begin{figure}[h!]
\begin{center}
\includegraphics[scale=0.3]{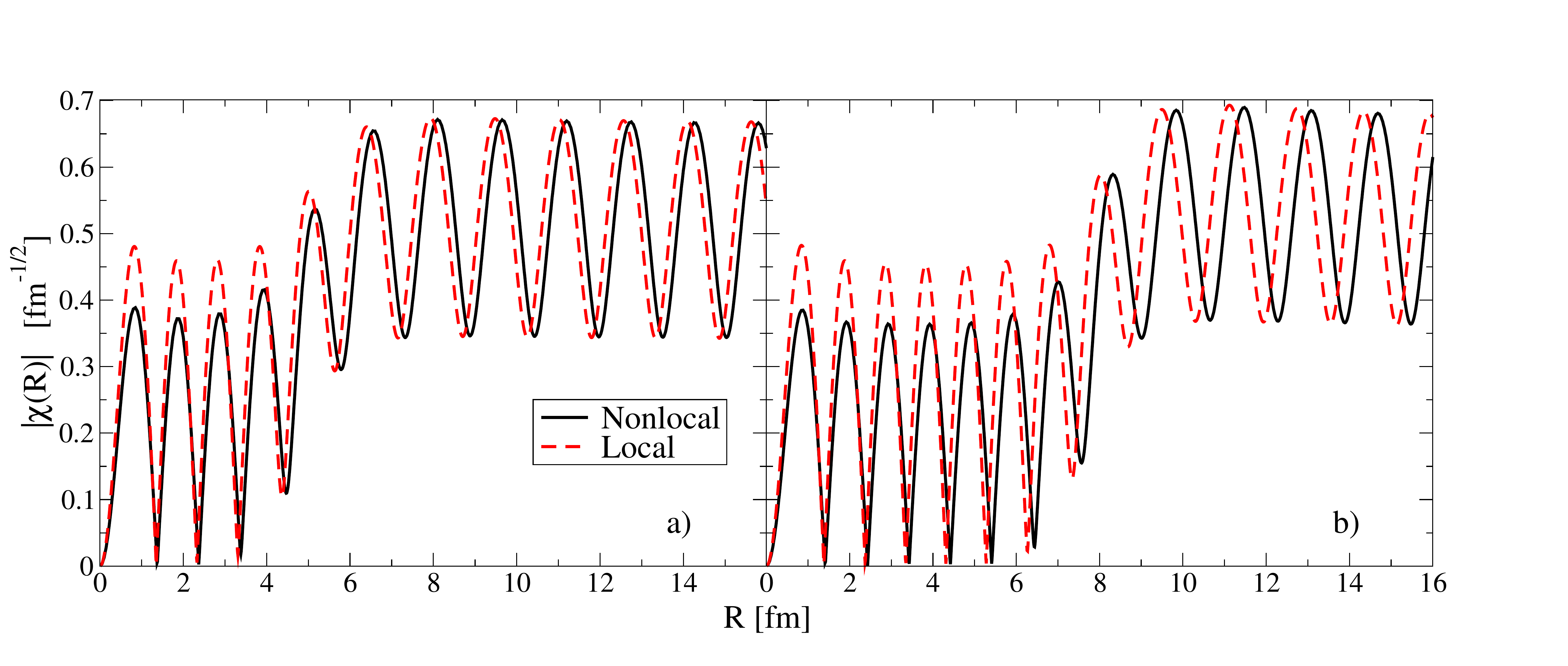}
\end{center}
\caption{Absolute value of the $d+A$ scattering wave function using the ADWA theory when nonlocal and local potentials are used. (a) $d+^{48}$Ca and (b) $d+^{208}$Pb. Both for the $L=1$ and $J=0$ partial wave at $E_d=50$ MeV in the laboratory frame. Figure reprinted from \cite{Titus_prc2015} with permission. }
\label{fig:dWF_Comparison_Leq1_Jeq0}
\end{figure}

\begin{figure}[h!]
\begin{center}
\includegraphics[scale=0.3]{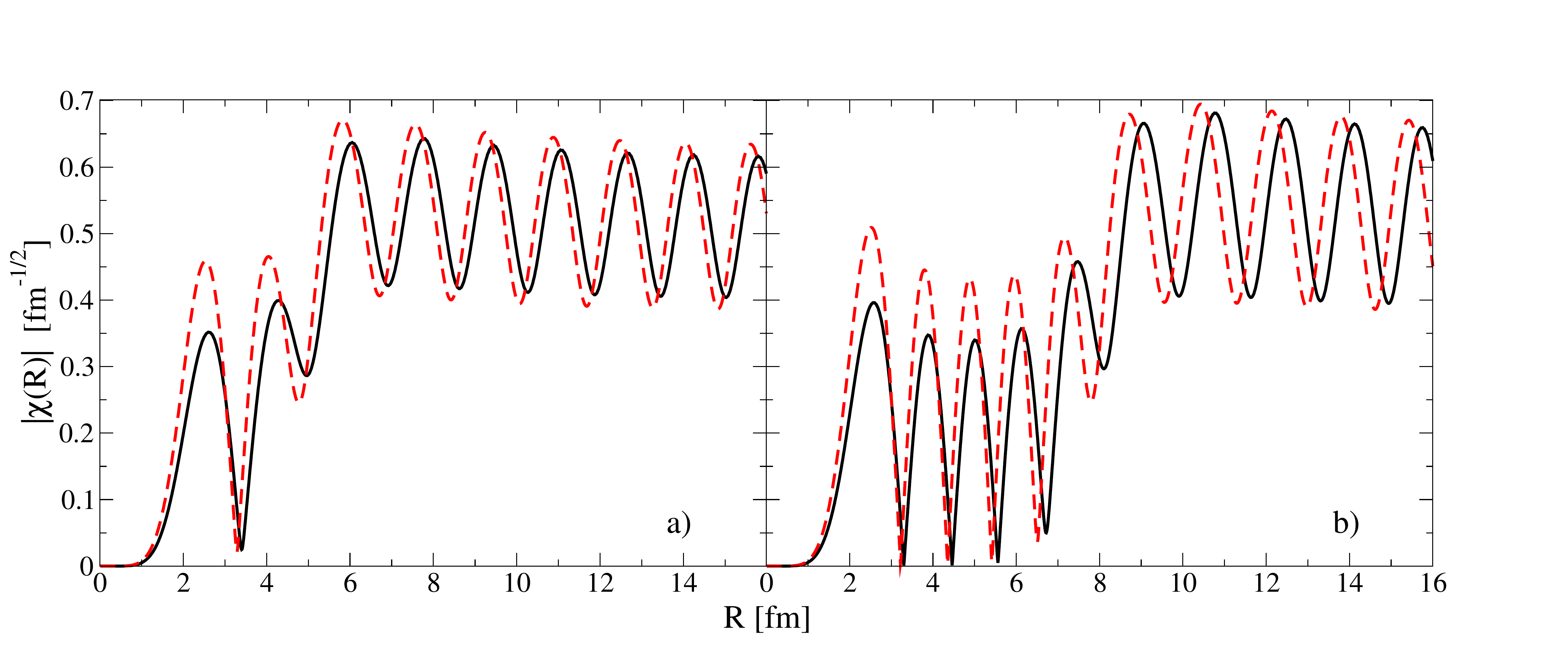}
\end{center}
\caption{Absolute value of the $d+A$ scattering wave function using the ADWA theory when nonlocal and local potentials are used. (a) $d+^{48}$Ca and (b) $d+^{208}$Pb. Both for the $L=6$ and $J=5$ partial wave at $E_d=50$ MeV in the laboratory frame. Figure reprinted from \cite{Titus_prc2015} with permission. }
\label{fig:dWF_Comparison_Leq6_Jeq5}
\end{figure}

Turning our focus to the deuteron scattering wave function, Figs. \ref{fig:dWF_Comparison_Leq1_Jeq0} and \ref{fig:dWF_Comparison_Leq6_Jeq5} show the absolute values of the $d+A$ scattering wave function when using the ADWA with nonlocal and local potentials. The solid line corresponds to the scattering wave function resulting from using the nonlocal Perey-Buck potential in Eq.(\ref{NL-Adiabatic}), while the dashed line is the scattering wave function that results from using the local adiabatic potential, Eq.(\ref{LocalAdPot}), where the necessary are used for the nucleon optical potentials. Panel (a) is for $d+^{48}$Ca while panel (b) is for $d+^{208}$Pb. 

Here, it is important to note that the individual $n+A$ and $p+A$ local optical potentials are phase equivalent to the nonlocal Perey-Buck, but the nonlocal and local adiabatic potentials are not phase equivalent. The adiabatic potential is only useful for calculating the deuteron scattering wave function within the range of the $V_{np}$ interaction, and is not applicable for calculating deuteron elastic scattering. It is for this reason that we chose for the input optical potentials to be phase equivalent, and not the full adiabatic potential. 


When compared to the source term that drives this wave function in Fig. \ref{fig:Source_Comparison_Leq1_Jeq0}, we see that both the wave function and the source term are reduced relative to the local counterpart. This is the same feature that we saw when studying the proton scattering state in $(p,d)$ reactions. The reduction of the wave function in the interior is a common feature of using nonlocal potentials, and can be understood physically in terms of the Pauli exclusion principle.

When studying proton scattering states, nonlocality only had the effect of reducing the amplitude of the scattering wave function. However, differing from the proton scattering state, the $d+A$ scattering wave function is also shifted outward relative to the wave function resulting from local potentials (see Fig. \ref{fig:dWF_Comparison_Leq1_Jeq0}). This is analogous to the bound state case where the wave function was both reduced and shifted outward due to nonlocality. This shifting outward of the $d+A$ scattering wave function changes the amplitude of the wave function at the nuclear surface. Since the surface region is where $(d,p)$ cross sections are most sensitive, the shifting outward can have a significant effect on the cross section. In fact, as we will see shortly, nonlocality in the deuteron scattering state increases the transfer cross section in most cases that were studied. 

The absolute values of the $d+A$ scattering wave functions for the $L=6$ and $J=5$ partial wave are shown in Fig. \ref{fig:dWF_Comparison_Leq6_Jeq5} for $d+^{48}$Ca and $d+^{208}$Pb at $E_d=50$ MeV. We see similar features as we did for the $L=1$ and $J=0$ case: the wave function is both reduced and pushed outward due to nonlocality. For the $d+^{208}$Pb case in particular, there is dramatic shift in the wave function around the nuclear surface ($\sim 7.5 - 8.5$ fm), which we will find is very important when calculating transfer cross sections.


\subsection{$(d,p)$ Transfer Cross Sections}

For the calculation of $(d,p)$ transfer cross sections, the nonlocal Perey-Buck potential was used for the neutron and proton optical potentials in the entrance and exit channels. The separate effects of nonlocality in the proton scattering state and the neutron bound state have already been studied. The results of such a study has been published in our previous papers \cite{Titus_prc2014,Ross_prc2015}, and is not discussed here. In addition, since we have already determined that the PCF is insufficient to take nonlocality into account, we did not investigate the PCF in this study and focused instead on the effects of explicitly including nonlocality in the entrance and exit channels in $(d,p)$ reactions.  

\begin{figure}[h!]
\begin{center}
\includegraphics[scale=0.34]{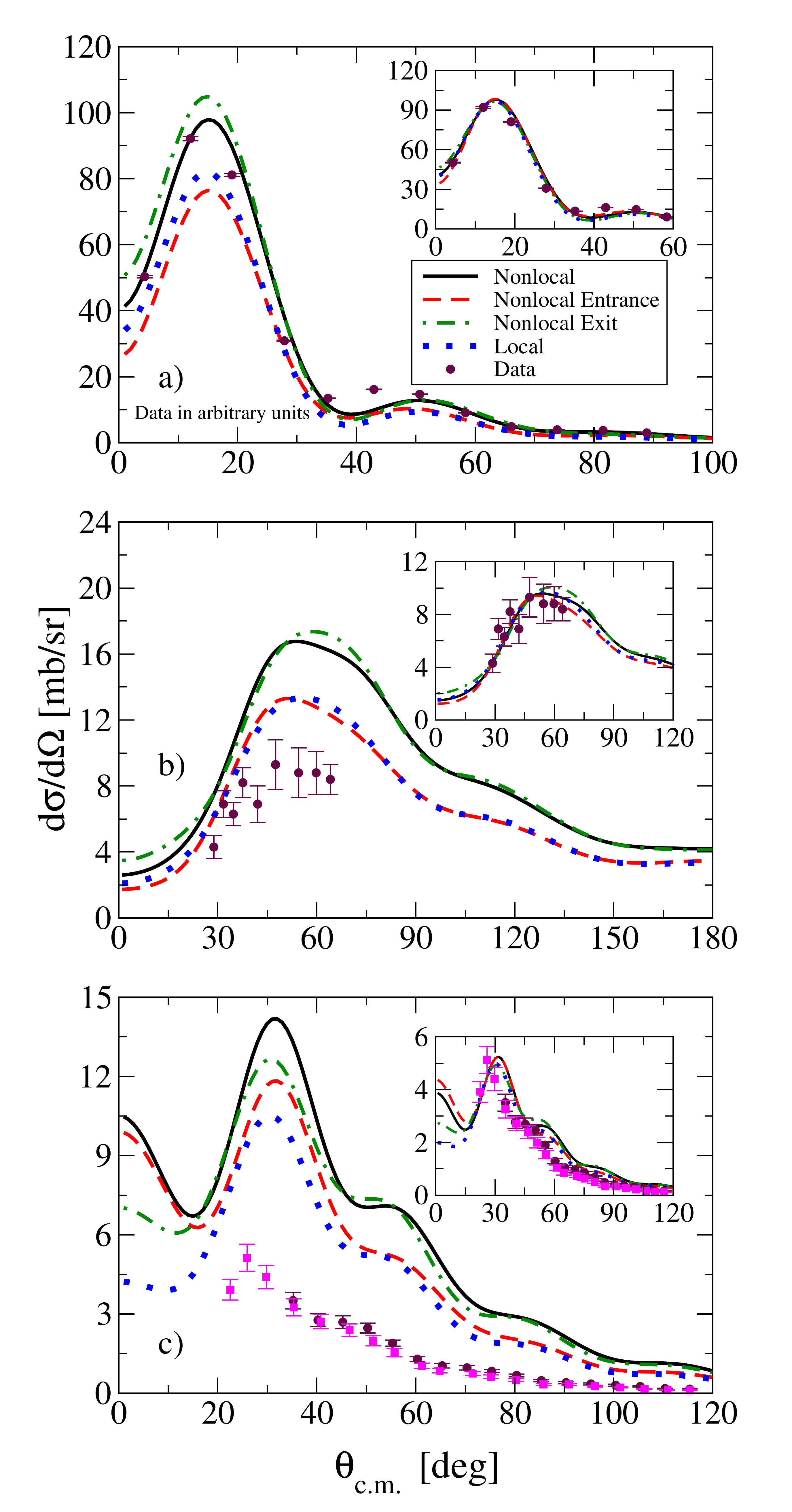}
\end{center}
\caption{Angular distributions for $(d,p)$ transfer cross sections. The insets are the theoretical distributions normalized to the peak of the data distribution. (a) $^{48}$Ca$(d,p)^{49}$Ca at $E_d=10$ MeV with data \cite{Brown_npa1970} at $E_d=10$ MeV in arbitrary units. (b) $^{132}$Sn$(d,p)^{133}$Sn at $E_d=10$ MeV with data \cite{Jones_prc2011} at $E_d=9.4$ MeV. (c) $^{208}$Pb$(d,p)^{209}$Pb at $20$ MeV with data \cite{Hirota_npa1998} (Circles) and \cite{Seichert_aipcs1981} (Squares) at $E_d=22$ MeV. Figure reprinted from \cite{Titus_prc2015} with permission. }
\label{fig:TransferCS_NL_ADWA_LowE}
\end{figure}

\begin{figure}[h!]
\begin{center}
\includegraphics[scale=0.40]{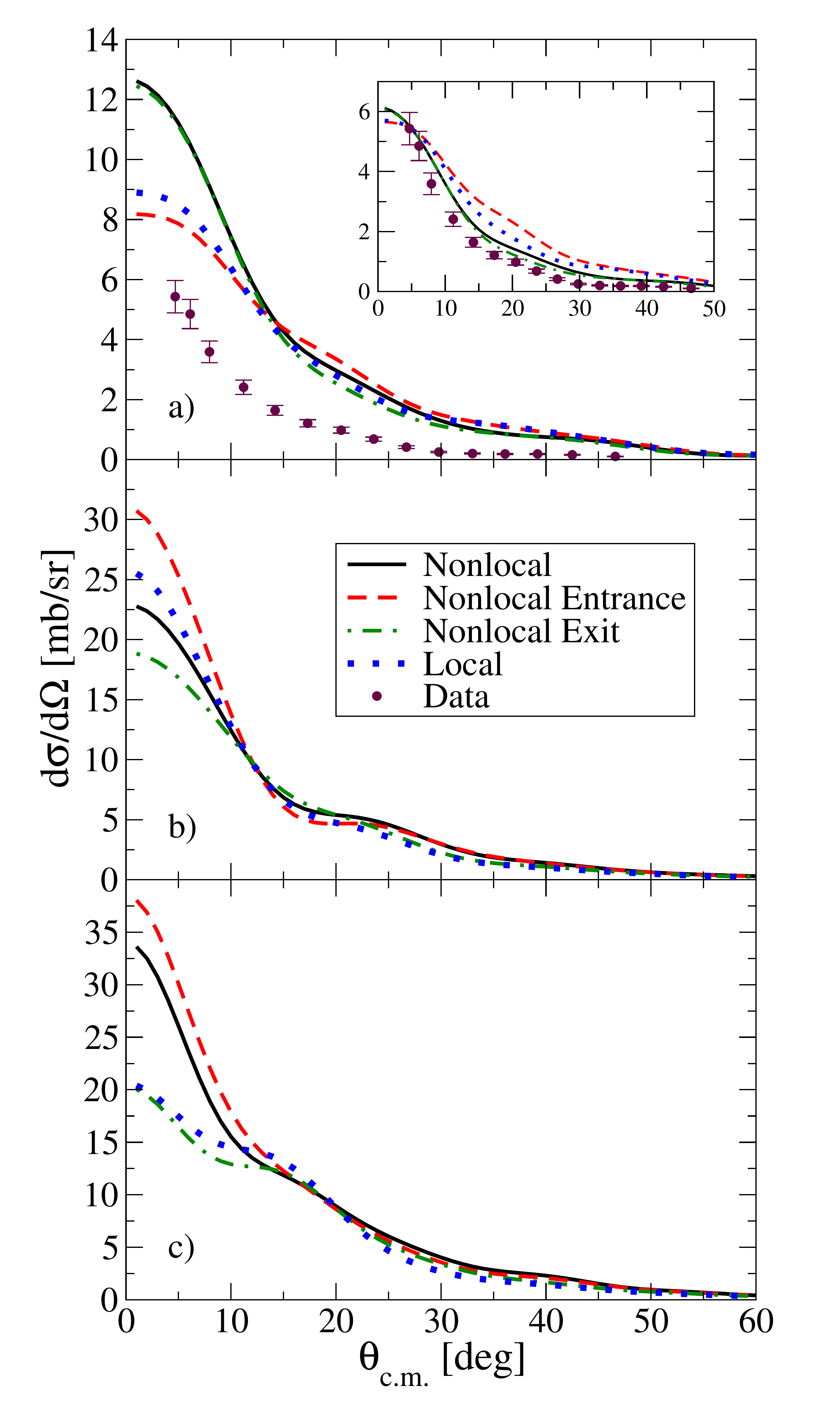}
\end{center}
\caption{Angular distributions for $(d,p)$ transfer cross sections. The inset is the theoretical distributions normalized to the peak of the data distribution. (a) $^{48}$Ca$(d,p)^{49}$Ca at $E_d=50$ MeV with data \cite{Uozumi_npa1994} at $E_d=56$ MeV. (b) $^{132}$Sn$(d,p)^{133}$Sn at $E_d=50$ MeV. (c) $^{208}$Pb$(d,p)^{209}$Pb at $50$ MeV. Figure reprinted from \cite{Titus_prc2015} with permission. }
\label{fig:TransferCS_NL_ADWA_HighE}
\end{figure}

In our analysis, we computed angular distributions for a wide variety of cases from $^{16}$O to $^{208}$Pb. Some illustrative examples are shown in Fig. \ref{fig:TransferCS_NL_ADWA_LowE} and \ref{fig:TransferCS_NL_ADWA_HighE}. Extensive tables for all cases are shown in Tables \ref{Tab:Percent_Difference_10}, \ref{Tab:Percent_Difference_20}, and \ref{Tab:Percent_Difference_50}. In the tables we show the percent difference between cross sections produced by nonlocal and local interactions, at the peak of the angular distribution, relative to a purely local calculation. In the first column we include nonlocality in all nucleon-target interactions. In the second (third) column we include nonlocality in the entrance (exit) channel only. 

In Fig. \ref{fig:TransferCS_NL_ADWA_LowE} and \ref{fig:TransferCS_NL_ADWA_HighE} we include the results when nonlocality is included consistently (solid line), only in the deuteron channel (dashed line), only in the proton channel (dot-dashed line), and where only a LPE potential is used (dotted line). In Fig. \ref{fig:TransferCS_NL_ADWA_LowE} we present $(d,p)$ calculations for deuterons impinging on: (a) $^{48}$Ca at $E_d=10$ MeV, (b) $^{132}$Sn at $E_d=10$ MeV, and (c) $^{208}$Pb at $E_d=20$ MeV. The same cases are presented in Fig. \ref{fig:TransferCS_NL_ADWA_HighE}, but for $E_d=50$ MeV. When available, we also present data points. The data in Fig. \ref{fig:TransferCS_NL_ADWA_LowE}a is published in arbitrary units. Therefore, this data set was normalized to the peak of the theoretical distribution that is generated when nonlocality is fully included.

\begin{table}[h]
\centering
\begin{tabular}{|c|r|r|r|r|r|}
\hline
                &              &              & Nonlocal      & Nonlocal    &       \\
                &Final         & Nonlocal     & Entrance      & Exit        & Angle \\
$E_{lab}$        &Bound         & Relative     & Relative      & Relative    & of    \\
$10$ MeV        &State         & to Local     & to Local      & to Local    & Peak  \\
\hline
$^{16}$O$(d,p)$    & $1d_{5/2}$  & $27.2\%$    & $-3.0\%$     & $32.7\%$      & $26 ^\circ$  \\ 
$^{16}$O$(d,p)$    & $2s_{1/2}$  & $15.5\%$    & $0.2\%$      & $13.5\%$      & $0^\circ$    \\ 
$^{40}$Ca$(d,p)$   & $1f_{7/2}$  & $48.5\%$    & $11.4\%$     & $46.5\%$      & $39^\circ$   \\ 
$^{48}$Ca$(d,p)$   & $2p_{3/2}$  & $19.4\%$    & $-6.8\%$     & $27.8\%$      & $15^\circ$   \\ 
$^{126}$Sn$(d,p)$  & $1h_{11/2}$ & $36.9\%$    & $8.7\%$      & $26.9$        & $72^\circ$    \\  
$^{132}$Sn$(d,p)$  & $2f_{7/2}$  & $25.7\%$    & $-0.2\%$     & $30.1\%$      & $55^\circ$    \\ 
$^{208}$Pb$(d,p)$  & $2g_{9/2}$  & $52.5\%$    & $2.0\%$      & $47.3\%$      & $180^\circ$   \\ 
\hline
\end{tabular}
\caption{Percent difference of the $(d,p)$ transfer cross sections at the first peak when using nonlocal potentials in entrance and exit channels (1st column), nonlocal potentials in entrance channel only (2nd column), and nonlocal potentials in exit channel only (3rd column), relative to the local calculation with the LPE potentials, for a number of reactions occurring at 
$10$ MeV.}
\label{Tab:Percent_Difference_10}
\end{table}

\begin{table}[h]
\centering
\begin{tabular}{|c|r|r|r|r|r|}
\hline
                &      &                   & Nonlocal          & Nonlocal    &         \\
                &Final & Nonlocal          & Entrance          & Exit        & Angle   \\
$E_{lab}$        &Bound & Relative          & Relative          & Relative    & of   \\
$20$ MeV        &State & to Local          & to Local          & to Local    & Peak \\
\hline
$^{16}$O$(d,p)$   & $1d_{5/2}$  & $24.9\%$  & $2.6\%$ & $25.7\%$  & $0^\circ$ \\ 
$^{16}$O$(d,p)$   & $2s_{1/2}$  & $7.1\%$   & $-0.7\%$ & $6.0\%$   & $0^\circ$  \\ 
$^{40}$Ca$(d,p)$  & $1f_{7/2}$  & $43.3\%$  & $11.0\%$ & $34.1\%$          & $26^\circ$ \\ 
$^{48}$Ca$(d,p)$  & $2p_{3/2}$  & $14.9\%$  & $7.1\%$ & $12.2\%$           & $8^\circ$ \\ 
$^{126}$Sn$(d,p)$ & $1h_{11/2}$  & $33.6\%$ & $7.7\%$ & $26.4$           & $35^\circ$ \\  
$^{132}$Sn$(d,p)$ & $2f_{7/2}$  &$3.2\%$    & $2.5\%$ & $4.2\%$            & $16^\circ$ \\ 
$^{208}$Pb$(d,p)$ & $2g_{9/2}$  &$35.0\%$   & $12.6\%$ & $20.5\%$         & $32^\circ$ \\ 
\hline
\end{tabular}
\caption{Percent difference of the $(d,p)$ transfer cross sections at the first peak when using nonlocal potentials in entrance and exit channels (1st column), nonlocal potentials in entrance channel only (2nd column), and nonlocal potentials in exit channel only (3rd column), relative to the local calculation with the LPE potentials, for a number of reactions occurring at 
$20$ MeV.}
\label{Tab:Percent_Difference_20}
\end{table}

\begin{table}[h]
\centering
\begin{tabular}{|c|r|r|r|r|r|}
\hline
                 &      &                   & Nonlocal           & Nonlocal    &         \\       
                 &Final & Nonlocal          & Entrance           & Exit        &  Angle       \\
$E_{lab}$         &Bound & Relative          & Relative           & Relative    &  of  \\
$50$ MeV         &State & to Local          & to Local           & to Local    &  Peak \\
\hline
$^{16}$O$(d,p)$   & $1d_{5/2}$  & $22.3\%$  & $3.1\%$ & $15.8\%$  & $10^\circ$  \\ 
$^{16}$O$(d,p)$   & $2s_{1/2}$  & $20.7\%$  & $0.4\%$ & $21.2\%$  & $0^\circ$  \\ 
$^{40}$Ca$(d,p)$  & $1f_{7/2}$  & $4.8\%$  & $4.4\%$ & $0.2\%$             & $0^\circ$  \\ 
$^{48}$Ca$(d,p)$  & $2p_{3/2}$  & $41.9\%$  & $-8.1\%$ & $39.9\%$          & $0^\circ$  \\ 
$^{126}$Sn$(d,p)$ & $1h_{11/2}$ & $6.9\%$  & $6.7\%$ & $-2.5$            & $13^\circ$  \\  
$^{132}$Sn$(d,p)$ & $2f_{7/2}$  & $-10.9\%$  & $20.4\%$ & $-26.2\%$       & $0^\circ$  \\ 
$^{208}$Pb$(d,p)$ & $2g_{9/2}$  & $64.8\%$  & $86.5\%$ & $-1.7\%$         & $0^\circ$  \\ 
\hline
\end{tabular}
\caption{Percent difference of the $(d,p)$ transfer cross sections at the first peak when using nonlocal potentials in entrance and exit channels (1st column), nonlocal potentials in entrance channel only (2nd column), and nonlocal potentials in exit channel only (3rd column), relative to the local calculation with the LPE potentials, for a number of reactions occurring at $50$ MeV. Figure reprinted from \cite{Titus_prc2015} with permission.}
\label{Tab:Percent_Difference_50}
\end{table}

At low energies, nonlocality in the exit channel provides a significant enhancement of the cross section for all cases, which is due to the neutron bound state. As mentioned before, the ANC of the bound state resulting from nonlocal potentials is larger than that from local potentials. Since low energy transfer reactions are primarily sensitive to the asymptotic properties of the wave functions, this results in an increase of the cross section.

The nonlocality in the proton scattering state is not felt significantly at low energies, so the reduction of the cross section due to the reduced amplitude of the proton scattering state is small. This is consistent with the results published in our previous papers, \cite{Titus_prc2014,Ross_prc2015}. At higher energies, the nonlocality of the proton scattering state becomes more significant, and there is a competition between the effects of nonlocality in the neutron bound state to enhance the cross section, and the effects of nonlocality in the proton scattering state to reduce the cross section. Nonlocality in the proton scattering state had a larger effect for the heavier nuclei due to a larger surface region being probed. This is seen in Fig. \ref{fig:TransferCS_NL_ADWA_HighE}. Comparing the dot-dashed lined with the dotted line, we see that there is an enhancement of the cross section for $^{48}$Ca, but a reduction for $^{132}$Sn and $^{208}$Pb. The net effect of nonlocality in the exit channel depends on a complex interplay of the properties of the bound state (i.e. number of nodes, binding energy, and orbital angular momentum), as well as the magnitude of the real and imaginary parts of the scattering wave function near the nuclear surface.

Depending on the case, the shifting outward of the deuteron wave functions seen in Figs. \ref{fig:dWF_Comparison_Leq1_Jeq0} and \ref{fig:dWF_Comparison_Leq6_Jeq5} did not always have the same effect on the transfer cross sections. Comparing the dashed and dotted lines in Figs. \ref{fig:TransferCS_NL_ADWA_LowE} and \ref{fig:TransferCS_NL_ADWA_HighE}, we see that for $^{48}$Ca, nonlocality in the deuteron scattering state has a similar effect as for the proton scattering state in that it reduces the cross section. As the size of the target increases, the outward shift of the wave function becomes more important. This is seen in the comparison of the $d+^{208}$Pb and the $d+^{48}$Ca wave functions in Figs. \ref{fig:dWF_Comparison_Leq1_Jeq0} and \ref{fig:dWF_Comparison_Leq6_Jeq5}. The $d+^{208}$Pb wave function is shifted outward more than the $d+^{48}$Ca wave function, which changes the amplitude at the nuclear surface, and has a significant impact on the transfer cross section. As seen in Table \ref{Tab:Percent_Difference_50}, nonlocality in the deuteron scattering state for $^{208}$Pb$(d,p)^{209}$Pb has the most significant effect of all the cases studies.

The insets in Fig. \ref{fig:TransferCS_NL_ADWA_LowE} show that when the theoretical cross sections are normalized to the data at the peak of the distribution, the low energy data cannot distinguish between the various models since the shape of the theoretical distributions are similar. However, for $^{48}$Ca$(d,p)^{49}$Ca at $E_d=50$ MeV, nonlocality significantly improves the description of the data. In all cases, if one were to extract a spectroscopic factor from the data, the results including nonlocality would differ considerably from those when only local interactions are used.


\subsection{Comparing Distorted Wave Born Approximation and the Adiabatic Distorted Wave Approximation}
\label{DWBA_vs_ADWA}

The DWBA is still the work-horse used in the analysis of most transfer cross sections. The DWBA is based on a series expansion described in Sec. \ref{Sec:BornSeries}. This expansion is usually truncated to first-order so that deuteron breakup is only included implicitly through the imaginary part of the deuteron optical potential. This is unlike the ADWA which is based on a three-body model, includes breakup explicitly, and relies on nucleon optical potentials. Here we show that the differences in the DWBA and ADWA formalism can lead to very different predictions for the $(d,p)$ cross sections.

\begin{figure}[h!]
\begin{center}
\includegraphics[scale=0.40]{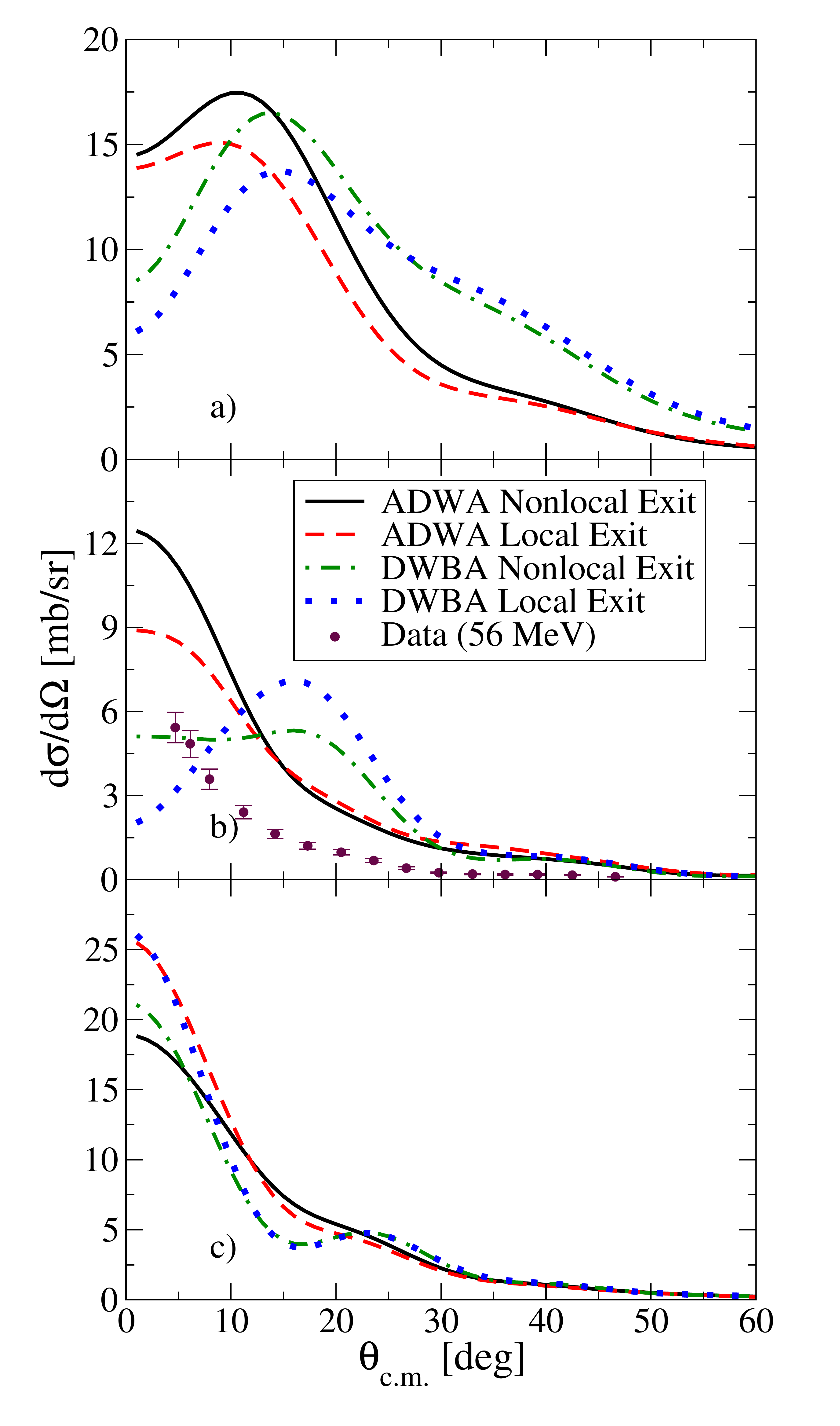}
\end{center}
\caption{Comparison of $(d,p)$ transfer cross sections when using the DWBA as compared to the ADWA. (a) $^{16}$O$(d,p)^{17}$O, (b) $^{48}$Ca$(d,p)^{49}$Ca with data from \cite{Uozumi_npa1994}. (c) $^{132}$Sn$(d,p)^{133}$Sn. All distributions at $E_d=50$ MeV. Figure reprinted from \cite{Titus_prc2015} with permission.}
\label{fig:ADWA_vs_DWBA}
\end{figure}

In Fig. \ref{fig:ADWA_vs_DWBA} the angular distributions for three different $(d,p)$ reactions obtained using the DWBA are compared to those obtained with the ADWA. There was no obvious way to compare the effect of nonlocality in the entrance channel since DWBA and ADWA treat the deuteron channel very differently, and a nonlocal global deuteron optical potential does not exist. Both local and nonlocal potentials were used in the exit channel. For a useful comparison, the same nonlocal and local potentials are used in the exit channel. For the ADWA, the LPE potentials obtained from fits to the distribution generated with the Perey-Buck potential are used, while for the DWBA we used the deuteron optical potential of Daehnick \cite{Daehnick_prc1980}.

We first focus on the local results, and compare in Fig. \ref{fig:ADWA_vs_DWBA} the DWBA (dotted line) with the ADWA (dashed line). The shapes are significantly different, as well as the magnitude of the cross section at the first peak. Including nonlocality in the exit channel does not resolve this discrepancy. We see that introducing nonlocality in the exit channel has the similar effect of increasing the cross section for both the DWBA and ADWA calculations. We also show the in Fig. \ref{fig:ADWA_vs_DWBA}b. It is clear that for this case, the DWBA is not able to describe the angular distribution from experiment. This is one example that demonstrates the need to explicitly include deuteron breakup into the calculation.


\subsection{Energy Shift Method}

Since many reaction problems are solved in coordinate based theories, local interactions have been preferred due to the simplicity of solving the equations. For this reason, Timofeyuk and Johnson, \cite{Timofeyuk_prc2013, Timofeyuk_prl2013}, developed a method to effectively include nonlocality in the deuteron scattering state within the formalism of the ADWA. Their method relies on local potentials so that the nonlocal equation does not need to be solved. Assuming the Perey-Buck form for the nonlocal potential, and through expansions, they find that by shifting the energy at which the local potentials are evaluated by $\sim 40$ MeV from the standard $E_d/2$ value, one can capture the effects of nonlocality. Since we are now able to include nonlocality explicitly, this method can be tested. 

\begin{figure}[h!]
\begin{center}
\includegraphics[scale=0.37]{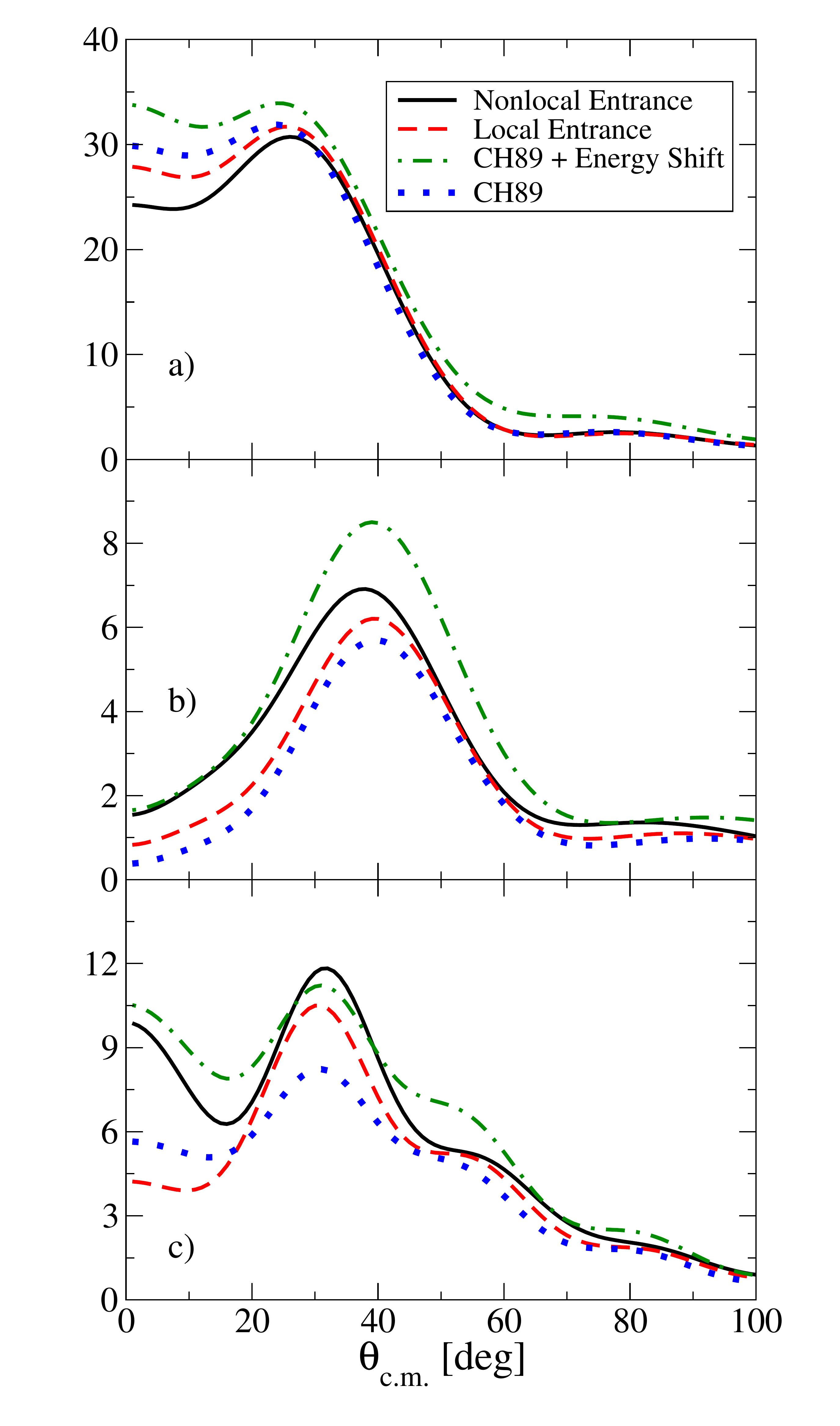}
\end{center}
\caption{Comparison of $(d,p)$ angular distributions when using the energy shift method of \cite{Timofeyuk_prc2013, Timofeyuk_prl2013}. (a) $^{16}$O$(d,p)^{17}$O at $E_d=10$ MeV (b) $^{40}$Ca$(d,p)^{41}$Ca at $E_d=10$ MeV (c) $^{208}$Pb$(d,p)^{209}$Pb at $E_d=20$ MeV. The solid line is when full nonlocality was included in the entrance channel, dashed line is when the LPE potential was used, dot-dashed line when the CH89 potential \cite{Varner_pr1991} was used with the additional energy shift quantified in \cite{Timofeyuk_prc2013}, and the dotted line when the CH89 potential was used at the standard $E_d/2$ value. Figure reprinted from \cite{Titus_prc2015} with permission. }
\label{fig:Energy_Shift}
\end{figure}

As we are only concerned with nonlocality in the deuteron channel, we fixed the potentials in the proton channel so we can make a meaningful comparison. In the exit channel, the LPE potentials found from fits to Perey-Buck proton elastic scattering distributions were used, along with the local binding potential used to reproduce the experimental binding energy. In the entrance channel, we used the nonlocal Perey-Buck potential, and the corresponding LPE potentials. To use the method of \cite{Timofeyuk_prc2013, Timofeyuk_prl2013}, we needed an energy dependent local optical potential. For this we used the CH89 potential \cite{Varner_pr1991} evaluated at the standard $E_d/2$ value, and with the additional energy shift that was quantified in \cite{Timofeyuk_prc2013}. 

The results of this study are shown in Fig. \ref{fig:Energy_Shift}. We show angular distributions for $(d,p)$ reactions on (a) $^{16}$O at $E_d=10$ MeV, (b) $^{40}$Ca at $E_d=10$ MeV, and (c) $^{208}$Pb at $E_d=20$ MeV. The solid line is the distribution form nonlocality explicitly included in the deuteron channel, the dashed line are the results of the local calculations with the LPE potentials for the deuteron scattering state, the dot-dashed line used the method of \cite{Timofeyuk_prc2013, Timofeyuk_prl2013}, and the dotted line is when the local potentials are evaluated at the standard $E_d/2$ value. 
 
Our results show that the energy shift method always increases the cross section. However, the explicit inclusion of nonlocality in the deuteron scattering state can sometimes decrease the cross section, as is seen in Fig. \ref{fig:Energy_Shift}a. Often times, the energy shift moves the transfer distribution towards larger angles, an effect also seen in the full nonlocal calculations. In some cases, the energy shift method over shoots the full nonlocal calculation, as in Fig. \ref{fig:Energy_Shift}b. In Fig. \ref{fig:Energy_Shift}c, we see an example where the energy shift does a very good job at reproducing the nonlocal effects. In general, we found that for most cases, the energy shift captured the qualitative effects of nonlocality, but was unable to provide an accurate account of the nonlocal effects.


\subsection{Summary}

In this work we studied the effects of nonlocality on $(d,p)$ transfer reactions. An extension of the ADWA theory was developed to include nonlocality in the deuteron scattering state using the Perey-Buck nonlocal nucleon optical potential \cite{Perey_np1962}. In the exit channel the Perey-Buck potential was used to describe the proton scattering state, and its real part was adjusted for the neutron bound state. For the scattering, a local phase equivalent (LPE) potential was obtained by fitting the elastic scattering generated from the corresponding nonlocal potential. Both the local and nonlocal bound states reproduced the same experimental binding energies. 

For the $(d,p)$ reactions studied, we found that the inclusion of nonlocality in both the entrance and exit channels increased the transfer cross section by $\sim 40\%$. In most cases, nonlocality in the deuteron scattering state caused a moderate increase in the transfer cross section. However, for heavy targets at high energies, this increase was large. Nonlocality in the exit channel caused, almost exclusively, an increase in the transfer cross section, except for heavy targets at high energies for which the cross sections were reduced. We also compared our ADWA result with those from DWBA and found the effects of nonlocality in the final state to be consistent in both formulations, even if quantitatively different. We also compared our ADWA results with the energy shift method introduced by Timofeyuk and Johnson \cite{Timofeyuk_prc2013,Timofeyuk_prl2013} and found that method to be qualitatively consistent with our results. 

The conclusion of the present study confirm those of \cite{Titus_prc2014,Ross_prc2015}. There are important differences in the transfer cross sections when including nonlocality explicitly as compared to when using LPE potentials. This highlights the necessity of explicitly including nonlocality to describe transfer reactions. Since the inclusion of nonlocality normally increases the cross section, a re-analysis of transfer reaction data will likely reduce currenlty accepted spectroscopic factors, such as those reported in \cite{Tsang_prl2005}.


\chapter{Conclusions and Outlook }
\label{Conclusions}

\section{Conclusions}

In this thesis we studied the effect of nonlocality of the optical potential in transfer reactions. For this purpose we developed a method for solving the integro-differential equations and extended the adiabatic distorted wave approximation for transfer $(d,p)$ reactions to include nonlocal interactions of general form. We performed several systematic studies, including a range of energies, targets, and interactions.

For the $(p,d)$ reaction study using the Perey-Buck nonlocal potential of Sec. \ref{PB_Transfer}, we considered a range of nuclei, and proton energies of $E_p=20$ and $50$ MeV \cite{Titus_prc2014}. We calculated the transfer matrix element in the distorted wave Born approximation (DWBA), and used a local optical potential to describe the deuteron scattering state. We found that the explicit inclusion of nonlocality increased the transfer cross section at the first peak by $15-35\%$, relative to when local potentials were use. We found that in all cases, the Perey correction factor traditionally used does not provide a quantitative description. Our results suggest that such a correction factor to account for nonlocality should not be used.

In Sec. \ref{DOM_Transfer} we compared the DOM potential and the Perey-Buck potential to study $(p,d)$ reactions on $^{40}$Ca at $E_p=20$, $35$, and $50$ MeV \cite{Ross_prc2015}. We included nonlocality in the entrance channel, then computed transfer cross sections in the DWBA, ignoring nonlocality in the deuteron channel. We generated two local phase equivalent (LPE) potentials, one for the DOM and one for the Perey-Buck potential. Both the DOM and the Perey-Buck potential produced very large increases in the magnitude of the transfer cross section, $\approx 15-50\%$ for the DOM potential, and $\approx 30-50\%$ for the Perey-Buck potential. Like in the first study of $(p,d)$ reactions, when nonlocality was included only in the bound state, large increases in the magnitude of the transfer cross section were seen. Typically, nonlocality in the proton scattering state acts in the opposite direction, reducing the transfer cross section. 

In the last study of Sec. \ref{NL_ADWA} involving $(d,p)$ reactions, nonlocality was included in the deuteron channel within the adiabatic distorted wave approximation (ADWA), which unlike the DWBA, includes deuteron breakup explicitly \cite{Titus_prc2015}. The formalism for the local ADWA theory had to be extended to include nonlocal potentials, and was done in Sec. \ref{Sec:Nonlocal_ADWA}. We found that the inclusion of nonlocality increased the transfer cross section by $\sim 40\%$. In most cases, nonlocality in the deuteron scattering state caused a modest increase in the transfer cross section. However, for heavy targets at high energy, the increase due to nonlocality in the deuteron channel was very large. This is in contrast to when nonlocality is added to the proton scattering state, which often times reduced the transfer cross section. The reason for the difference is that nonlocality in the deuteron scattering state had both the effect of reducing the magnitude of the scattering wave function within the nuclear interior, but also shifting the wave function outwards towards the periphery. 

All these three studies \cite{Titus_prc2014,Ross_prc2015,Titus_prc2015} demonstrate the important differences in the transfer cross sections when including nonlocality explicitly, as compared to when using local phase equivalent potentials. This emphasizes the necessity of including nonlocality to describe transfer reactions if accurate structure information is to be extracted. 

Most often, transfer reactions are performed to extract a spectroscopic factor. The analysis is done using local potentials. As explicit inclusion of nonlocality increased the predicted cross sections, one would expect lower spectroscopic factors to result from the analysis if nonlocal potentials are used. This may well contribute to solving the discrepancy between spectroscopic factors extracted from knockout and transfer \cite{Kramer_npa2001}, but that study has yet to be performed. 



\section{Outlook}

Going forward, nonlocality must be carefully taken into account in any advanced reaction theory. It will become increasingly important to construct a modern nonlocal global optical potential. Work along these lines has already been done by Tian, Pang, and Ma (TPM) with the recent publication of their nonlocal potential, \cite{Tian_ijmpe2015}. However, this potential is still energy independent, and is based on the simple Perey-Buck form. 

On physical grounds, the optical potential must be energy dependent due to channel coupling effects. Preliminary indications from an unpublished study by Bacq, Lovell, Titus, and Nunes \cite{Bacq_report2015} show that there is indeed an energy dependence in nonlocal potentials of the Perey-Buck form. Work on specifying the precise energy dependence is ongoing. It would be advantageous to construct an energy dependent nonlocal global optical potential through $\chi^2$ minimization of a large quantity of elastic scattering angular distributions, and perhaps polarization observables and other data as well.

The Perey-Buck form for the potential comprises a single nonlocality parameter, $\beta$. This simple form has been useful for many decades since it allowed for simple implementation. It is unlikely that a single nonlocality parameter is able to represent the complex nature of nonlocality in the realistic many-body problem. The DOM potential, for example, has a different $\beta$ for each term of the potential. A more sophisticated form for the nonlocal potential should be considered. 

There are several methods available to construct a microscopically based optical potential. Some of these methods were discussed in Chapter \ref{Intro}, and should be pursued now that it is known that nonlocality is very important and must be included explicitly. Since previous microscopic calculations of nonlocal potentials have shown that their form does not resemble the simple Gaussian nonlocality of the Perey-Buck potential, it is important to better understand its analytic properties.

With an improved nonlocal global optical potential, existing transfer data can be re-analyzed. The large discrepancies between spectroscopic factors extracted from the nonlocal and local calculations in this study demonstrates that the structure information of most nuclei are likely to be altered when the data is analyzed with nonlocal potentials.

While this thesis focused on $(d,p)$ and $(p,d)$ reactions, the role of nonlocality should be investigated in other reactions as well. We are currently investigating the role of nonlocality in $(d,n)$ reactions \cite{Ross_ip2015}. Surprisingly, the effects of nonlocality in $(d,n)$ reactions are even more significant than in $(d,p)$ reactions. 

Along with transfer, there are many other reactions that are performed to extract single particle structure of nuclei. Nuclear knockout reactions ($A(a,b\gamma)X$) are an alternative method to extract a spectroscopic factor. Such a reaction also requires an optical potential between the colliding nuclei. Understanding the effect of nonlocality in this case is also important. Inelastic scattering provides the transition strength between the ground state and a bound excited state in a nucleus. Since these are obtained by comparing experimental data to theoretical distributions resulting from a DWBA or coupled channel analysis, one may again expect that inclusion of nonlocality in the description of the process will improve the reliability of the extracted transition strength. Three-body models exist to calculate transfer, such as the continuum discretized coupled channel method \cite{Austern_pr1987}. In this case, coupled channel integro-differential equations would need to be solved, and new numerical methods may be needed to accomplish this task. Finally, charge-exchange reactions, such as $(p,n)$, probe the spin and isospin properties of nuclei. Once again, information from experiments are sometimes extracted from a DWBA analysis, and including nonlocality in the optical potential will improve the reliability of theoretical predictions, consequently the reliability of the extracted Gamow-Teller strengths. 

One aspect that was not addressed in this thesis concerns error quantification. Lovell and Nunes \cite{Lovell_jpg2015} are currently addressing error quantification in direct reaction theories when using local potentials. An extension of that work will be necessary as the use of nonlocal nonlocal potentials becomes more widespread.
Much work has highlighted the importance of nonlocality in the reaction dynamics. How then can we constrain nonlocality from data? The Perey-Buck potential, for example, has a Gaussian form with a nonlocality range of $\beta=0.85$ fm. Elastic scattering is not sensitive to short-range properties, so constraining $\beta$ in this way may not be the best method. Reactions sensitive to short range correlations may offer a better avenue.

The results of this thesis, along with the advent of microscopic theories to construct nonlocal potentials, improved phenomenological nonlocal global optical potentials, and ever increasing computer power, has the potential to elevate reaction theory to a new level. While we have focused on $(p,d)$ and $(d,p)$ reactions, with our increased understanding of nonlocality, it has become necessary to update the other formalisms and codes commonly used in nuclear physics to properly include nonlocality in reaction theory.





\appendices




\chapter{ Solving the Nonlocal Equation}
\label{SolvingEquation}

In order to assess the validity of the local approximation we need to solve Eq.(\ref{NLeqn}) exactly. Several methods exist for solving the scattering state, such as \cite{Kim_prc1990, Rawitscher_npa2012}. Our approach follows Perey and Buck \cite{Perey_np1962} where Eq.(\ref{NLeqn}) is solved by iteration. For simplicity, we will drop the local part of the nonlocal potential, $U_o(r)$, in our discussion, although it is included in our calculations.

To solve the partial wave equation of Eq.(\ref{NLeqn}) numerically, we need to find the kernel function $g_{L}(R,R')$. In order to do this, we first need to do a partial wave expansion of the nonlocal potential,

\begin{eqnarray}
U^{NL}_{PB}(\textbf{R},\textbf{R}')=\sum_L\frac{2L+1}{4\pi} \frac{g_L(R,R')}{RR'}P_L(\cos \theta),
\end{eqnarray}

\noindent where we defined $\theta$ as the angle between \textbf{$R$} and \textbf{$R'$}. Inserting the Perey-Buck form for the nonlocal potential, multiplying both sides by $P_{L}(\cos \theta)$, integrating over all angles, using the orthogonality of the Legendre polynomials, and solving for $g_L(R,R')$, we find that

\begin{eqnarray}
g_{L}(R,R')=2{\pi}RR'U\left(\frac{1}{2}(R+R')\right)\int_{-1}^{1}\frac{\textrm{exp}\left(-\left|\frac{\textbf{R}-\textbf{R}'}{\beta}\right|^2\right)}{\pi^{\frac{3}{2}}\beta^3}P_{L}(\cos\theta)d(\cos\theta).
\end{eqnarray}

\noindent For a moment, consider only the integral on the right:

\begin{eqnarray}
\int_{-1}^{1}\frac{\exp \left(-\left|\frac{\textbf{R}-\textbf{R}'}{\beta}\right|^2\right)}{\pi^{\frac{3}{2}}\beta^3}P_{L}(\cos\theta)d(\cos\theta)&=&\frac{\exp\left(-\frac{R^2+R'^2}{\beta^2}\right)}{\pi^{\frac{3}{2}}\beta^3}\int^1_{-1}e^{i(-izcos\theta)}P_{L}(\cos \theta)d(\cos \theta). \nonumber \\
\end{eqnarray}

\noindent Using the integral representation for the spherical Bessel functions, 

\begin{eqnarray}
j_L(x)=\frac{1}{2i^L}\int_{-1}^{1} e^{ixu}P_L(u)du,
\end{eqnarray}

\noindent with $u=\cos\theta$, we find that

\begin{eqnarray}
g_{L}(R,R')&=&\frac{2i^{L}z}{\pi^{\frac{1}{2}}\beta}j_{L}(-iz)\textrm{exp}\left(-\frac{R^2+R'^2}{\beta^2}\right)U\left(\frac{1}{2}(R+R')\right) \nonumber \\
&=&h_L(R,R')U\left(\frac{1}{2}(R+R')\right).
\end{eqnarray}

Calculating $h_L(R,R')$ numerically is difficult for $z \gg 1$ due to large cancellations between the terms, so we need to approximate this function when doing numerical calculations for large values of the argument. Making use of the asymptotic expression for the spherical Bessel function, and neglecting terms proportional to $\exp\left[-\left(\frac{R+R'}{\beta} \right)^2\right]$, we find that $h_L(R,R')$ for $z \gg 1$ can be approximated as

\begin{eqnarray}
h_{L}(R,R')&\approx&\frac{1}{\pi^{\frac{1}{2}}\beta}e^{-\left(\frac{R-R'}{\beta} \right)^2} \quad \textrm{for}\left|z \right| \gg 1.
\end{eqnarray}

Scattering solutions are considered first, where the subscript $n$ denotes the $n$th order approximation of the correct solution. The iteration scheme starts with an initialization,

\begin{equation}\label{NLeqn2}
\frac{\hbar^2}{2\mu}\left[\frac{d^2}{dr^2}-\frac{L(L+1)}{R^2}\right]\chi_{n=0}(R)+[E-U_{\textrm{init}}(R)]\chi_{n=0}(R)=0,
\end{equation}

\noindent where $U_{\textrm{init}}(R)$ is some suitable local potential used to get the iteration process started. Knowing $\chi_o(R)$ one then proceeds with solving:

\begin{eqnarray}\label{NLeqn3}
&\phantom{=}&\frac{\hbar^2}{2\mu}\left[\frac{d^2}{dr^2}-\frac{L(L+1)}{R^2} \right]\chi_n(R)+[E-U_{\textrm{init}}(R)]\chi_n(R) \nonumber \\
&=&\int g_L(R,R')\chi_{n-1}(R') dR' -U_{\textrm{init}}(R)\chi_{n-1}(R),
\end{eqnarray}

\noindent with as many iterations necessary for convergence. The number of iterations depends mostly on the 
partial wave being solved for (lower partial waves require more iterations) and the quality of $U_{\textrm{init}}(R)$. It was 
rare for any partial wave to require more than 20 iterations to converge, even with a very poor choice for $U_{\textrm{init}}(R)$. 
If the LPE potential is used as $U_{\textrm{init}}(R)$, then any partial wave converges with less than 10 iterations.

For the bound state problem, the method is somewhat different. A variety of methods exist in the literature, some developed specifically to
 handle non-analytic forms (e.g. \cite{Michel_epja2009}). Our approach may not be the most efficient, but it is straightforward, general, and easy to implement. 
To solve the bound state problem with a nonlocal potential we begin by solving Eq.(\ref{NLeqn2}). Since we are using the wave function from the 
previous iteration to calculate the nonlocal integral, we need to keep track of the different normalizations of the inward 
and outward wave functions that results from the choice for the initial conditions for each wave function. Thus, the 
equations we iterate are:

\begin{eqnarray}\label{Inward}
&\phantom{=}&\frac{\hbar^2}{2\mu}\left[\frac{d^2}{dr^2}-\frac{\ell(\ell+1)}{r^2} \right]\phi_n^{In}(r)+[E-U_{\textrm{init}}(r)]\phi_n^{In}(r) \nonumber \\
&=&\int_0^{R_{Max}} g_\ell(r,r')\phi_{n-1}^{In}(r') dr' -U_{\textrm{init}}(r)\phi_{n-1}^{In}(r),
\end{eqnarray}

\noindent and

\begin{eqnarray}\label{Outward}
&\phantom{=}&\frac{\hbar^2}{2\mu}\left[\frac{d^2}{dr^2}-\frac{\ell(\ell+1)}{r^2} \right]\phi_n^{Out}(r)+[E-U_{\textrm{init}}(r)]\phi_n^{Out}(r) \nonumber \\
&=&\int_0^{R_{Max}} g_\ell(r,r')\phi_{n-1}^{Out}(r') dr' -U_{\textrm{init}}(r)\phi_{n-1}^{Out}(r),
\end{eqnarray}

\noindent where $R_{Max}$ is some maximum radius chosen greater than the range of the nuclear interaction. Note that $\phi^{In}(r)$ is 
the wave function for integrating from the edge of the box inward and has a normalization set by  the Whittaker function 
as the initial condition, while $\phi^{Out}(r)$ is the wave function for integrating from the origin outward and has the 
normalization set  using the standard $r^{L+1}$ initial condition near the origin. 

Even though $\phi^{Out}$ and $\phi^{In}$ differ by only a constant, these two equations (Eq.(\ref{Inward}) and Eq.(\ref{Outward})) are necessary because the value 
of the normalization constant is only known after convergence. For a given iteration, $\phi^{Out}$ and $\phi^{In}$ converge 
when their logarithmic derivatives agree at the matching point. To keep the proper normalization throughout the entire 
range $[0,R_{Max}]$, we need to retain two versions of the converged wave function 
for each iteration:

\begin{equation}
 \phi_n^{\textrm{In}}(r) = \left\{
  \begin{array}{l l}
    C_n^{\textrm{Out}}\phi_n^{\textrm{Out}}(r) & \quad \textrm{for $0 \le r < R_{\textrm{Match}}$}\\
    \phi_n^{\textrm{In}}(r) & \quad \textrm{for $R_{\textrm{Match}} \le r \le R_{\textrm{Max}}$}
  \end{array} \right.
\end{equation}

\begin{equation}
 \phi_n^{\textrm{Out}}(r) = \left\{
  \begin{array}{l l}
    \phi_n^{\textrm{Out}}(r) & \quad \textrm{for $0 \le r < R_{\textrm{Match}}$}\\
    C_n^{\textrm{In}}\phi_n^{\textrm{In}}(r) & \quad \textrm{for $R_{\textrm{Match}} \le r \le R_{\textrm{Max}}$}
  \end{array} \right.
\end{equation}

\noindent where 

\begin{equation}
C^{\textrm{In(Out)}}=\frac{\phi^{\textrm{Out(In)}}(R_{\textrm{Match}})}{\phi^{\textrm{In(Out)}}(R_{\textrm{Match}})}
\end{equation}

The full iteration scheme is converged when the binding energy obtained from the previous iteration agrees with the 
binding energy from the current iteration within a desired level of accuracy. Again, although this may not be the most efficient method, it is general 
(whatever the form of nonlocality) and is very stable, providing a good option for future studies beyond the Perey-Buck potentials.


\chapter{ Deriving the Perey Correction Factor}
\label{Correction_Factor}

Here we provide details on the derivation of the PCF, Eq. (\ref{CorrectionFactor}). We also include the derivation of the transformation formulas, as well as the correct radial version of the transformation formulas which could be used to transform the nonlocal radius and diffuseness to their local counterpart.  

We start from Eq. (\ref{NL-Sch-Eq}). Let us define a function $F(\textbf{R})$ that connects the local 
wave function $\Psi^{Loc}(\textbf{R})$, resulting from the potential $U^{LE}(\textbf{R})$, with the wave function resulting
from a nonlocal potential, $\Psi^{NL}(\textbf{R})$:

\begin{equation}{\label{locNLequivalence}}
\Psi^{NL}(\textbf{R}) \equiv F(\textbf{R})\Psi^{Loc}(\textbf{R}).
\end{equation}

Since the local and nonlocal equations describe the same elastic scattering, the wave functions should be identical 
outside the nuclear interior. Thus, $F(\textbf{R}) \rightarrow 1$ as $R \rightarrow \infty$. By inserting 
Eq.(\ref{locNLequivalence}) into the nonlocal equation Eq.(\ref{NL-Sch-Eq}) we can reduce the result to the following 
local equivalent equation

\begin{equation}
-\frac{\hbar^2}{2\mu}\nabla^2\Psi^{Loc}(\textbf{R})+U^{LE}(\textbf{R})\Psi^{Loc}(\textbf{R})=E\Psi^{Loc}(\textbf{R}),
\end{equation}

\noindent where the local equivalent potential is given by:

\begin{eqnarray}\label{eqn:ULE}
U^{LE}(\textbf{R})&=&\frac{-\frac{\hbar^2}{\mu}\nabla F(\textbf{R})\cdot \nabla \Psi^{Loc}(\textbf{R})-\frac{\hbar^2}{2\mu}(\nabla^2F(\textbf{R}))\Psi^{Loc}(\textbf{R})}{F(\textbf{R})\Psi^{Loc}(\textbf{R})} \nonumber \\
&\phantom{=}& + \ \frac{\int U^{NL}(\textbf{R},\textbf{R}')F(\textbf{R}')\Psi^{Loc}(\textbf{R}')d\textbf{R}'}{{F(\textbf{R})\Psi^{Loc}(\textbf{R})}}+U_o(\textbf{R}). \nonumber \\
\end{eqnarray}

\noindent We next consider the numerator of the second term of Eq.(\ref{eqn:ULE}) and introduce the explicit nonlocal potential form of Eq.(\ref{PB-Form}). Using the definition $\textbf{s}=\textbf{R}-\textbf{R}'$,  expanding  in powers of $s$ up to first order, the integral becomes

\begin{eqnarray}
&\phantom{=}&\int U^{NL}_{WS}\left(\left|\textbf{R}-\frac{1}{2}\textbf{s}\right|\right)H(s)F(\textbf{R}-\textbf{s})\Psi^{Loc}(\textbf{R}-\textbf{s})d\textbf{s}  \nonumber \\
&\phantom{=}&\approx U_{WS}^{NL}(R)F(R)\int H(s)\Psi^{Loc}(\textbf{R}-\textbf{s})d\textbf{s}-\frac{1}{2}F(R)\nabla U_{WS}^{NL}(\textbf{R})\cdot \int \textbf{s}H(s)\Psi^{Loc}(\textbf{R}-\textbf{s})d\textbf{s} \nonumber \\
&\phantom{=}& \quad - \ U_{WS}^{NL}(R)\nabla F(\textbf{R})\cdot \int \textbf{S}H(s)\Psi^{Loc}(\textbf{R}-\textbf{s})d\textbf{s},
\end{eqnarray}

\noindent where

\begin{equation}
H(s)=\frac{\exp\left(-\frac{s^2}{\beta^2} \right)}{\pi^{3/2}\beta^3}.
\end{equation}

\noindent Therefore, the local equivalent potential becomes:

\begin{eqnarray}
U^{LE}(\textbf{R}) &\approx& \frac{1}{F(\textbf{R})\Psi^{Loc}(\textbf{R})}\left[ -\frac{\hbar^2}{\mu}(\nabla F \cdot \nabla \Psi^{Loc}) \right. \nonumber \\
&\phantom{=}& + \ \left.  U^{NL}_{WS}(R)F(R)\int H(s)\Psi^{Loc}(\textbf{R}-\textbf{s})d\textbf{s} \right. \nonumber \\
&\phantom{=}& - \ \left.  \frac{1}{2}F(R)\nabla U^{NL}_{WS} \cdot \int \textbf{s}H(s)\Psi^{Loc}(\textbf{R}-\textbf{s})d\textbf{s} \right. \nonumber \\
&\phantom{=}& - \ \left. U^{NL}_{WS}(R)\nabla F \cdot \int \textbf{s}H(s)\Psi^{Loc}(\textbf{R}-\textbf{s})d\textbf{s}\right] \nonumber \\
&\phantom{=}& - \ \frac{\hbar^2}{2\mu}\frac{\nabla^2 F(\textbf{R})}{F(\textbf{R})}+U_o(\textbf{R}).
\end{eqnarray}

Consider the four terms in the brackets. All of these terms are divided by $\Psi^{Loc}$, which has nodes. 
The first, third, and fourth terms depend on dot products and gradients of $\Psi^{Loc}$. These terms are unlikely 
to individually equal zero when $\Psi^{Loc}$ in the denominator equals zero. Thus, we require that these terms sum 
to zero so that $U^{LE}(\textbf{R})$ remains finite. As pointed out in \cite{Austern_Book}, this is not an approximation, but merely a condition for the method to work. 
Applying this condition gives us two equations:

\begin{eqnarray}\label{U_eqn}
U^{LE}(\textbf{R})&=&U^{NL}_{WS}(R)\left[\frac{\int H(s)\Psi^{Loc}(\textbf{R}-\textbf{s})d\textbf{s}}{\Psi^{Loc}(\textbf{R})} \right] + U_o({\bf R})-\frac{\hbar^2}{2\mu}\frac{\nabla^2F(\textbf{R})}{F(\textbf{R})}
\end{eqnarray}

\begin{eqnarray}\label{0_eqn}
0&=&\frac{\hbar^2}{\mu}(\nabla F \cdot \nabla \Psi^{Loc})+\left[\frac{1}{2}F(R)\nabla U^{NL}_{WS}+U^{NL}_{WS}(R) \nabla F \right] \cdot \int \textbf{s}H(s)\Psi^{Loc}(\textbf{R}-\textbf{s})d\textbf{s}. \nonumber \\
\end{eqnarray}

\noindent Instead of using the local WKB approximations as Austern did \cite{Austern_Book}, we use the operator form of the Taylor expansion to factorize the wave function:

\begin{equation}\label{expansion}
\Psi^{Loc}(\textbf{R}-\textbf{s})=e^{-i\textbf{s}\cdot\textbf{k}}\Psi^{Loc}(\textbf{R}),
\end{equation}

\noindent with $\textbf{k}=-i\nabla$. This simplifies the integrals in Eq.(\ref{U_eqn}) and Eq.(\ref{0_eqn}). 
Consider first the integral in Eq.(\ref{U_eqn})

\begin{eqnarray}\label{integral}
\int H(s)\Psi^{Loc}(\textbf{R}-\textbf{s})d\textbf{s}&=&\left[\int e^{-i\textbf{s}\cdot\textbf{k}}H(s)d\textbf{s}\right]\Psi^{Loc}(\textbf{R}) \nonumber \\
&=& \exp\left[\frac{-k^2\beta^2}{4} \right]\Psi^{Loc}(\textbf{R}).
\end{eqnarray}

\noindent Therefore, assuming the potentials are scalar functions of $R$, and replacing Eq.(\ref{integral}) into Eq.(\ref{U_eqn}) we obtain;

\begin{eqnarray}\label{U_LE}
U^{LE}(R)&=&U^{NL}_{WS}(R)\exp\left[{-\frac{\mu\beta^2}{2\hbar^2}\left(E-U^{LE}(R) \right)}\right] + U_o(R)-\frac{\hbar^2}{2\mu}\frac{\nabla^2F(R)}{F(R)},
\end{eqnarray}

\noindent where we used $k^2=-\nabla^2$ in the exponent to first order, and the Schr\"odinger's equation. Making the 
replacement $U^{LE}(R)=U^{Loc}_{WS}(R)+U_o(R)$, gives us the radial transformation formula

\begin{eqnarray}\label{radial}
U_{WS}^{NL}(R)&=&\left(U_{WS}^{Loc}(R)+\frac{\hbar^2}{2\mu}\frac{\nabla^2 F(R)}{F(R)} \right)\exp\left[\frac{\mu\beta^2}{2\hbar^2}\left(E-U_{WS}^{Loc}(R)-U_o(R) \right) \right]. \nonumber
\end{eqnarray}

\noindent The $\nabla^2F$ term is significant around the surface, but near the origin this term is negligible. 
Therefore, if we neglect this term, then we must remove the radial arguments, and consider this formula only near the origin. 
Therefore, for $R\approx0$

\begin{equation}\label{NLtoLoc3}
U_{WS}^{NL}(0)\approx U_{WS}^{Loc}(0)\exp\left[\frac{\mu\beta^2}{2\hbar^2}\left(E-U_{WS}^{Loc}(0)-U_o(0) \right) \right].
\end{equation}

The $U_{WS}(R)$ functions are of a Woods-Saxon form, and have real and imaginary parts

\begin{eqnarray}
U_{WS}(R)&=&U_R(R)+iU_I(R)  \\
&=&\frac{-V_v}{1+\exp\left(\frac{R-rA^{1/3}}{a} \right)}+4i\frac{-W_d\exp\left(\frac{R-rA^{1/3}}{a} \right)}{\left(1+\exp\left(\frac{R-rA^{1/3}}{a} \right)\right)^2}. \nonumber
\end{eqnarray}

\noindent Inserting this into Eq.(\ref{NLtoLoc3}) we obtain;

\begin{eqnarray}\label{Trans}
&\phantom{=}&U_R^{NL}(R)+iU_I^{NL}(R)=(U_R^{Loc}(R)+iU_I^{Loc}(R)) \nonumber \\
&\times& \exp\left[\frac{\mu\beta^2}{2\hbar^2}\left(E-U_o(R)-U_R^{Loc}(r)-iU_I^{Loc}(R) \right) \right]. \nonumber \\
\end{eqnarray}

\noindent Near the origin, $U_{I}^{Loc}\approx 0$ so this term can be neglected in the exponent, and $U_R\approx-V_v$. 
While the spin-orbit term diverges at the origin, it rapidly goes to zero away from the origin, so we assume the spin-orbit 
contribution is negligible. Thus, $U_o=V_c$, where $V_c$ is the Coulomb potential at the origin for a uniform sphere of charge. 
Taking the real part of the above equation and making these substitutions gives

\begin{equation}
V^{NL}_v=V^{Loc}_v\exp\left[\frac{\mu\beta^2}{2\hbar^2}\left(E-V_c+V^{Loc}_v \right) \right].
\end{equation}

\noindent For the imaginary part, we have:

\begin{equation}
U_I^{NL}(R)=U_I^{Loc}(R)\exp\left[\frac{\mu\beta^2}{2\hbar^2}\left(E-V_c+V^{Loc}_v \right) \right].
\end{equation}

\noindent While $U_I(R)\approx 0$ near the origin, the local and nonlocal terms have the same form factor, so the form factors 
exactly cancel as long as the radius and diffuseness are identical. Therefore, the imaginary part of Eq.(\ref{Trans}) gives

\begin{equation}
W^{NL}_d=W^{Loc}_d\exp\left[\frac{\mu\beta^2}{2\hbar^2}\left(E-V_c+V^{Loc}_v \right) \right],
\end{equation}

\noindent It is important to note that these equations are only valid for 
transforming the depths of the potentials, thus Eq.(\ref{NLtoLoc3}) should not be used while retaining the radial dependence. Indeed, Eq.(A13) is not valid for all R.

Now consider the integral in Eq.(\ref{0_eqn}). Using Eq.(\ref{expansion}) to expand the wave function, and evaluating the 
dot product we get

\begin{eqnarray}
0&=&\frac{\hbar^2}{\mu}(\nabla F \cdot \nabla \Psi^{Loc})+\left[\frac{1}{2}F(R)\nabla U^{NL}_{WS}+U^{NL}_{WS}(R) \nabla F \right] \nonumber \\
&\phantom{=}&\times \left[\int s\cos(\theta)H(s)e^{-i\textbf{s}\cdot\textbf{k}}d\textbf{s} \right]\Psi^{Loc}(\textbf{R}).
\end{eqnarray}

\noindent Doing the integral, we find that this becomes

\begin{eqnarray}\label{F-eqn}
0&=&\frac{\hbar^2}{\mu}\nabla F-\left[\frac{1}{2}F(R)(\nabla U^{NL}_{WS})+U^{NL}_{WS}(R)(\nabla F)\right] \nonumber \\
&\phantom{=}&\times\frac{\beta^2}{2}\exp\left[-\frac{\mu\beta^2}{2\hbar^2}\left(E-U^{LE}(R) \right) \right].
\end{eqnarray}

\noindent If we assume that the local momentum approximation is valid, this equation can be solved exactly and has the solution

\begin{equation}\label{CorrectionFactor2}
F(R)=\left[1-\frac{\mu \beta^2}{2 \hbar^2}U^{NL}_{WS}(R)\exp\left({-\frac{\mu\beta^2}{2\hbar^2}\left(E-U^{LE}(R) \right)}\right) \right]^{-\frac{1}{2}}.
\end{equation}

\noindent If the local momentum approximation is not valid, then insertion of Eq.(\ref{CorrectionFactor2}) into the $rhs$ of Eq.(\ref{F-eqn}) 
will deviate from zero by a term related to the derivative of $U^{LE}(R)$. This additional term will be significant at the surface, and thus one 
can expect discrepancies in applying Eq.(\ref{CorrectionFactor2}) in this region. 

Comparing Eq.(\ref{CorrectionFactor2}) with Eq.(\ref{U_LE}) we see that

\begin{equation}\label{FullCorrectionFactor}
F(R)=\left[1-\frac{\mu \beta^2}{2\hbar^2}\left(U^{LE}(R)-U_o(R)+\frac{\hbar^2}{2\mu}\frac{\nabla^2F(R)}{F(R)} \right) \right]^{-\frac{1}{2}}.
\end{equation}

\noindent Neglecting the term containing $\nabla^2F$ gives us Eq.(\ref{CorrectionFactor}), which is the correction factor of Austern \cite{Austern_Book}. The contribution of $\nabla^2F/F$ is only important at the surface, 
and again it is precisely for these radii that discrepancies can be expected in applying Eq.(\ref{CorrectionFactor}).


\chapter{ Nonlocal Adiabatic Potential}
\label{Nonlocal_Adiabatic}

\noindent Here we will derive the nonlocal adiabatic potential. We begin with a three-body Schr\"odinger Equation:

\begin{equation}
\left[\hat{T}_R + T_r +V_{np}(r)+\hat{U}_{nA} + \hat{U}_{pA}-E\right]\Psi(\textbf{r},\textbf{R})=0.
\end{equation}

\noindent We expand the wave function using Weinberg states,

\begin{eqnarray}
\Psi(\textbf{r},\textbf{R})=\sum_{i=0}^{\infty}\Phi_i(\textbf{r})X_i(\textbf{R}),
\end{eqnarray}

\noindent and keep only the first Weinberg State, 

\begin{eqnarray}
\Psi(\textbf{r},\textbf{R})\approx \Phi_o(\textbf{r})X_o(\textbf{R})=\Phi(\textbf{r})X(\textbf{R}).
\end{eqnarray}

\noindent Noting that

\begin{equation}
(\hat{T}_r+V_{np})\Phi(\textbf{r})=-\epsilon_d \Phi(\textbf{r}),
\end{equation}

\noindent with $E_d=E+\epsilon_d$, the Schr\"odinger equation becomes

\begin{equation}
\left[\hat{T}_R -E_d\right]\Phi(\textbf{r})X(\textbf{R})=-\left[\hat{U}_{nA} + \hat{U}_{pA}\right]\Phi(\textbf{r})X(\textbf{R}).
\end{equation}

Now we evaluate the nucleon nonlocal operator, $\hat{U}_{NA}\Phi(\textbf{r})X(\textbf{R})$. For a moment, consider just the neutron potential (with $\textbf{R}_{p,n}=\textbf{R}\pm \frac{\textbf{r}}{2}$ where the ``+'' sign is for the proton and the ``-'' sign is for the neutron)

\begin{eqnarray}\label{Argument}
\hat{U}_{nA}\Psi(\textbf{r},\textbf{R})&=&\hat{U}_{nA}\Psi(\textbf{R}_n,\textbf{R}_p) \nonumber \\
&=&\int U(\textbf{R}_n,\textbf{R}'_n)\Psi(\textbf{R}'_n, \textbf{R}_p)\delta(\textbf{R}'_p-\textbf{R}_p)d\textbf{R}'_pd\textbf{R}'_n  \\
&=&\mathcal{J}\int U\left(\textbf{R}-\frac{\textbf{r}}{2}, \textbf{R}'-\frac{\textbf{r}'}{2}\right)\Psi(\textbf{r}',\textbf{R}')\delta(\textbf{R}'_p-\textbf{R}_p)d\textbf{r}'d\textbf{R}' \nonumber \\
&=&\mathcal{J}\int U\left(\textbf{R}-\frac{\textbf{r}}{2}, \textbf{R}'-\frac{\textbf{r}'}{2}\right)\Psi(\textbf{r}',\textbf{R}')\delta \left(\textbf{R}'+\frac{\textbf{r}'}{2}-(\textbf{R}+\frac{\textbf{r}}{2}) \right)d\textbf{r}'d\textbf{R}' \nonumber \\
&=&8\mathcal{J}\int U\left(\textbf{R}-\frac{\textbf{r}}{2}, \textbf{R}'-\frac{\textbf{r}'}{2}\right)\Psi(\textbf{r}',\textbf{R}')\delta \left(\textbf{r}'-(\textbf{r}-2(\textbf{R}'-\textbf{R})) \right)d\textbf{r}'d\textbf{R}' \nonumber
\end{eqnarray}

\noindent The Jacobian for the coordinate transformation is

\begin{eqnarray}
\mathcal{J}=
\begin{vmatrix} 
\frac{\partial \textbf{R}'_n}{\partial\textbf{R}'} & \frac{\partial \textbf{R}'_n}{\partial \textbf{r}'} \\
\frac{\partial \textbf{R}'_p}{\partial\textbf{R}'} & \frac{\partial \textbf{R}'_p}{\partial \textbf{r}'}
\end{vmatrix}=
\begin{vmatrix}
1 & -\frac{1}{2} \\
1 & \frac{1}{2} \\
\end{vmatrix}=1,
\end{eqnarray}

\noindent which gives us

\begin{eqnarray}
\hat{U}_{nA}\Psi(\textbf{r},\textbf{R})&=&8\int U_{nA}\left(\textbf{R}-\frac{\textbf{r}}{2},\textbf{R}'- \frac{\textbf{r}-2(\textbf{R}'-\textbf{R})}{2}\right)\Psi\left(\textbf{r}-2(\textbf{R}'-\textbf{R}),\textbf{R}'\right)d\textbf{R}' \nonumber \\
&=&8\int U_{nA}\left(\textbf{R}-\frac{\textbf{r}}{2},2\textbf{R}'-\textbf{R}-\frac{\textbf{r}}{2}\right)\Psi(\textbf{r}-2(\textbf{R}'-\textbf{R}),\textbf{R}')d\textbf{R}'.
\end{eqnarray}
 
\noindent For the nucleon nonlocal operator, we have

\begin{eqnarray}
\hat{U}_{NA}\Phi(\textbf{r})X(\textbf{R})&=&8\int U_{NA}\left(\textbf{R}\pm\frac{\textbf{r}}{2},2\textbf{R}'-\textbf{R}\pm\frac{\textbf{r}}{2}\right)\Phi(\textbf{r}\pm 2(\textbf{R}'-\textbf{R}))X(\textbf{R}')d\textbf{R}'. \nonumber \\
\end{eqnarray}

\noindent Consider the argument of $U_{NA}$. Adding and subtracting by $\textbf{R}$ in the second argument we get

\begin{eqnarray}
U_{NA}\left(\textbf{R}\pm\frac{\textbf{r}}{2},2\textbf{R}'-\textbf{R}\pm\frac{\textbf{r}}{2}\right)&=&U_{NA}\left(\textbf{R}_{p,n},2\textbf{R}'-\textbf{R}-\textbf{R}+\textbf{R}\pm\frac{\textbf{r}}{2}\right) \nonumber \\
&=&U_{NA}\left(\textbf{R}_{p,n},2(\textbf{R}'-\textbf{R})+\textbf{R}_{p,n} \right) \nonumber \\
&=&U_{NA}\left(\textbf{R}_{p,n},\textbf{R}_{p,n}+2\textbf{s} \right),
\end{eqnarray}

\noindent where we made the definition, $\textbf{s}=\textbf{R}'-\textbf{R}$. Since $d\textbf{R}'\rightarrow d\textbf{s}$, we see that $\hat{U}_{NA}\Phi(\textbf{r})X(\textbf{R})$ becomes

\begin{eqnarray}
\hat{U}_{NA}\Phi(\textbf{r})X(\textbf{R})&=&8\int U_{NA}\left(\textbf{R}_{p,n},\textbf{R}_{p,n}+2\textbf{s} \right)\Phi(\textbf{r} \pm 2\textbf{s})X(\textbf{R}+\textbf{s})d\textbf{s}.
\end{eqnarray}

The general expansion of the full wave function for a given partial wave with total angular momentum $J_T$ and projection $M_T$, is given by

\begin{eqnarray}
\Psi(\textbf{r},\textbf{R})\approx \Phi(\textbf{r})X(\textbf{R})&=&\sum_{L' J'_p}\left\{ \left\{ \Phi(\textbf{r}) \otimes \tilde{Y}_{L'}(\hat{R})\right\}_{J'_p} \otimes \Xi_{I_t}(\xi_t) \right\}_{J_T M_T}\chi_{L' J'_p}^{J_T M_T}(R) \nonumber \\
&=& \sum_{\ell' L' J'_p}\left\{ \left\{ \left\{\Xi_{I_d}(\xi_{np})\otimes \tilde{Y}_{\ell'}(\hat{r})\right\}_{J_d} \otimes \tilde{Y}_{L'}(\hat{R})\right\}_{J'_p} \otimes \Xi_{I_t}(\xi_t) \right\}_{J_T M_T} \nonumber \\
&\phantom{=}& \times \ \phi_{\ell'}(r)\frac{\chi_{L' J'_p}^{J_T M_T}(R)}{R},
\end{eqnarray}

\noindent where

\begin{eqnarray}
\Xi_{I_d}(\xi_{np})=\left\{\Xi_{I_p}(\xi_p)\otimes \Xi_{I_n}(\xi_n) \right\}_{I_d}.
\end{eqnarray}

\noindent In these equations, $\Xi_{I_p}(\xi_p)$, $\Xi_{I_n}(\xi_n)$ and $\Xi_{I_t}(\xi_t)$ are the spin functions for the proton, neutron, and target, respectively. $I_p=I_n=1/2$ are the spin of the proton and neutron respectively, and $I_t$ is the spin of the target. $\tilde{Y}_{L}(\hat{R})$ is the spherical harmonic for the orbital motion between the projectile and target, while $\tilde{Y}_{\ell}(\hat{r})$ is for the internal orbital angular momentum of the deuteron. We are using the phase convention where there is a built in factor of $i^L$ so that $\tilde{Y}_L(\hat{R})=i^LY_{L}(\hat{R})$ where $Y_{L}(\hat{R})$ is defined on \cite{Varshalovich_Book}, p.133, Eq.(1). The spin of the deuteron is given by $I_d=1$ and the total angular momentum of the deuteron is $J_d=1$. The total angular momentum of the deuteron is coupled to the orbital angular momentum between the deuteron and target to give a total angular momentum of the projectile, $J_p$. The total angular momentum of the projectile is coupled to the spin of the target to give the total angular momentum of the system, $J_T$ with projection $M_T$. 

The Schr\"odinger equation is now

\begin{eqnarray}\label{eq:Sch-Eq-NL-ADWA}
&\phantom{=}& \left[\hat{T}_R -E_d\right]\Phi(\textbf{r})X(\textbf{R})=-\hat{U}_{NA}\Phi(\textbf{r})X(\textbf{R}) \nonumber \\
&=&-8\int U_{NA}\left(\textbf{R}_{p,n},\textbf{R}_{p,n}+2\textbf{s} \right)\Phi(\textbf{r} \pm 2\textbf{s})X(\textbf{R}+\textbf{s})d\textbf{s} \nonumber \\
&=&-\sum_{\ell' L' J'_p}8\int U_{NA}\left(\textbf{R}_{p,n},\textbf{R}_{p,n}+2\textbf{s}  \right) \nonumber \\
&\phantom{=}& \times \ \left\{ \left\{ \left\{\Xi_{I_d}(\xi_{np})\otimes \tilde{Y}_{\ell'}(\widehat{r \pm 2s})\right\}_{J_d} \otimes \tilde{Y}_{L'}(\widehat{R+s})\right\}_{J'_p} \otimes \Xi_{I_t}(\xi_t) \right\}_{J_T M_T} \nonumber \\
&\phantom{=}& \times \ \phi_{\ell' }(|\textbf{r}\pm 2\textbf{s}|)\frac{\chi_{L' J'_p}^{J_T M_T}\left(\left|\textbf{R}+\textbf{s} \right|\right)}{\left|\textbf{R}+\textbf{s} \right|} d\textbf{s}.
\end{eqnarray}

\noindent We would like to do a partial wave decomposition to get an equation for each $LJ$ combination of the scattering wave function. To do this, multiply both sides of Eq.(\ref{eq:Sch-Eq-NL-ADWA}) by

\begin{eqnarray}\label{eq:partial-wave-decomp}
 \sum_{\ell}\left\{ \left\{ \left\{\Xi_{I_d}(\xi_{np})\otimes \tilde{Y}_{\ell}(\hat{r})\right\}_{J_d} \otimes \tilde{Y}_{L}(\hat{R})\right\}_{J_p} \otimes \Xi_{I_t}(\xi_t) \right\}_{J_T M_T}^*\phi_{\ell}(r)V_{np}(r),
\end{eqnarray}

\noindent and integrate over $d\textbf{r}$, $d\Omega_R$, $d\xi_{np}$, and $d\xi_t$, where $d\xi_{np}=d\xi_nd\xi_p$. Consider first just the $lhs$ of Eq.(\ref{eq:Sch-Eq-NL-ADWA}) after multiplication of Eq.(\ref{eq:partial-wave-decomp}),

\begin{eqnarray}\label{lhs}
&\phantom{=}&\int  \sum_{\ell}\left\{ \left\{ \left\{\Xi_{I_d}(\xi_{np})\otimes \tilde{Y}_{\ell}(\hat{r})\right\}_{J_d} \otimes \tilde{Y}_{L}(\hat{R})\right\}_{J_p} \otimes \Xi_{I_t}(\xi_t) \right\}_{J_T M_T}^* \phi_{\ell}(r)V_{np}(r) \nonumber \\
&\phantom{=}& \times \ \left[\hat{T}_R -E_d\right]\sum_{\ell' L' J'_p} \left\{ \left\{ \left\{\Xi_{I_d}(\xi_{np})\otimes \tilde{Y}_{\ell'}(\hat{r})\right\}_{J_d} \otimes \tilde{Y}_{L'}(\hat{R})\right\}_{J'_p} \otimes \Xi_{I_t}(\xi_t) \right\}_{J_T M_T} \nonumber \\
&\phantom{=}& \times \ \phi_{\ell'}(r)\frac{\chi_{L' J'_p}^{J_T M_T}(R)}{R}d\textbf{r}d\Omega_R \nonumber \\
&=&-\sum_{\ell}\sum_{\ell' L' J'_p}\frac{1}{R}\left[\frac{\hbar^2}{2\mu}\left(\frac{\partial^2}{\partial R^2}-\frac{L'(L'+1)}{R^2}\right) +E_d\right]\chi_{L' J'_p}^{J_T M_T}(R)\int \phi_{\ell}(r)V_{np}(r)\phi_{\ell'}(r)r^2 dr \nonumber \\
&\phantom{=}& \times \ \int \left\{ \left\{ \left\{\Xi_{I_d}(\xi_{np})\otimes \tilde{Y}_{\ell}(\hat{r})\right\}_{J_d} \otimes \tilde{Y}_{L}(\hat{R})\right\}_{J_p} \otimes \Xi_{I_t}(\xi_t) \right\}_{J_T M_T}^* \\
&\phantom{=}& \times \ \left\{ \left\{ \left\{\Xi_{I_d}(\xi_{np})\otimes \tilde{Y}_{\ell'}(\hat{r})\right\}_{J_d} \otimes \tilde{Y}_{L'}(\hat{R})\right\}_{J'_p} \otimes \Xi_{I_t}(\xi_t) \right\}_{J_T M_T}d\Omega_r d\Omega_R d\xi_{np} d\xi_t. \nonumber
\end{eqnarray}

\noindent Next we consider the integral in the last two lines of Eq.(\ref{lhs})

\begin{eqnarray}
&\phantom{=}&\int \left\{ \left\{ \left\{\Xi_{I_d}(\xi_{np})\otimes \tilde{Y}_{\ell}(\hat{r})\right\}_{J_d} \otimes \tilde{Y}_{L}(\hat{R})\right\}_{J_p} \otimes \Xi_{I_t}(\xi_t) \right\}_{J_T M_T}^*  \nonumber \\
&\phantom{=}& \times \ \left\{ \left\{ \left\{\Xi_{I_d}(\xi_{np})\otimes \tilde{Y}_{\ell'}(\hat{r})\right\}_{J_d} \otimes \tilde{Y}_{L'}(\hat{R})\right\}_{J'_p} \otimes \Xi_{I_t}(\xi_t) \right\}_{J_T M_T}d\Omega_r d\Omega_R d\xi_{np} d\xi_t \nonumber \\
&=&\sum_{M_p M'_p}\int  \left\{ \left\{\Xi_{I_d}(\xi_{np})\otimes \tilde{Y}_{\ell}(\hat{r})\right\}_{J_d} \otimes \tilde{Y}_{L}(\hat{R})\right\}_{J_p M_p}^*  \nonumber \\
&\phantom{=}& \times \ \left\{ \left\{\Xi_{I_d}(\xi_{np})\otimes \tilde{Y}_{\ell'}(\hat{r})\right\}_{J_d} \otimes \tilde{Y}_{L'}(\hat{R})\right\}_{J'_p M'_p} d\Omega_r d\Omega_R d\xi_{np} \nonumber \\
&\phantom{=}& \times \ \sum_{\mu'_t \mu_t}C_{J_p M_p I_t \mu_t}^{J_T M_T}C_{J'_p M'_p I_t \mu'_t}^{J_T M_T} \int \Xi^*_{I_t \mu_t}(\xi_t)\Xi_{I_t \mu'_t}(\xi_t)d\xi_t.
\end{eqnarray}

\noindent Evaluating the integral, $\int \Xi^*_{I_t \mu_t}(\xi_t)\Xi_{I_t \mu'_t}(\xi_t)d\xi_t=\delta_{\mu_t \mu'_t}$, summing over $\mu'_t$, and evaluating the complex conjugate using \cite{Varshalovich_Book}, p.62, Eq.(6), gives

\begin{eqnarray}\label{eq:integral-term}
&=&\sum_{M_p M'_p}\sum_{\mu_t}C_{J_p M_p I_t \mu_t}^{J_T M_T}C_{J'_p M'_p I_t \mu_t}^{J_T M_T} \int  \left\{ \left\{\Xi_{I_d}(\xi_{np})\otimes \tilde{Y}_{\ell}(\hat{r})\right\}_{J_d} \otimes \tilde{Y}_{L}(\hat{R})\right\}_{J_p M_p}^*  \nonumber \\
&\phantom{=}& \times \ \left\{ \left\{\Xi_{I_d}(\xi_{np})\otimes \tilde{Y}_{\ell'}(\hat{r})\right\}_{J_d} \otimes \tilde{Y}_{L'}(\hat{R})\right\}_{J'_p M'_p} d\Omega_r d\Omega_R d\xi_{np} \nonumber \\
&=&\sum_{M_p M'_p}\sum_{\mu_t}C_{J_p M_p I_t \mu_t}^{J_T M_T}C_{J'_p M'_p I_t \mu_t}^{J_T M_T} \nonumber \\
&\phantom{=}& \times \ \int (-)^{J_p-M_p} \left\{ \left\{\Xi_{I_d}(\xi_{np})\otimes \tilde{Y}_{\ell}(\hat{r})\right\}_{J_d} \otimes \tilde{Y}_{L}(\hat{R})\right\}_{J_p, -M_p}  \nonumber \\
&\phantom{=}& \times \ \left\{ \left\{\Xi_{I_d}(\xi_{np})\otimes \tilde{Y}_{\ell'}(\hat{r})\right\}_{J_d} \otimes \tilde{Y}_{L'}(\hat{R})\right\}_{J'_p M'_p} d\Omega_r d\Omega_R d\xi_{np}.
\end{eqnarray}

\noindent Coupling the tensors up to zero angular momentum provides further simplifications:

\begin{eqnarray}
&\phantom{=}&\left\{ \left\{\Xi_{I_d}(\xi_{np})\otimes \tilde{Y}_{\ell}(\hat{r})\right\}_{J_d} \otimes \tilde{Y}_{L}(\hat{R})\right\}_{J_p, -M_p}  \left\{ \left\{\Xi_{I_d}(\xi_{np})\otimes \tilde{Y}_{\ell'}(\hat{r})\right\}_{J_d} \otimes \tilde{Y}_{L'}(\hat{R})\right\}_{J'_p M'_p} \nonumber \\
&=&\sum_{S M_S}C_{J_p, -M_p J'_p M'_p}^{S M_S}\left\{\left\{ \left\{\Xi_{I_d}(\xi_{np})\otimes \tilde{Y}_{\ell}(\hat{r})\right\}_{J_d}\otimes \tilde{Y}_{L}(\hat{R})\right\}_{J_p} \right. \nonumber \\
&\phantom{=}& \left. \ \otimes  \left\{ \left\{\Xi_{I_d}(\xi_{np})\otimes \tilde{Y}_{\ell'}(\hat{r})\right\}_{J_d} \otimes \tilde{Y}_{L'}(\hat{R})\right\}_{J'_p} \right\}_{S M_S} \nonumber \\
&\rightarrow& C_{J_p, -M_p J'_p M'_p}^{0 0}\left\{\left\{ \left\{\Xi_{I_d}(\xi_{np})\otimes \tilde{Y}_{\ell}(\hat{r})\right\}_{J_d} \otimes \tilde{Y}_{L}(\hat{R})\right\}_{J_p} \right. \nonumber \\
&\phantom{=}& \left. \ \otimes  \left\{ \left\{\Xi_{I_d}(\xi_{np})\otimes \tilde{Y}_{\ell'}(\hat{r})\right\}_{J_d} \otimes \tilde{Y}_{L'}(\hat{R})\right\}_{J'_p} \right\}_{0 0}. 
\end{eqnarray}

\noindent Next we consider just an $\ell=0$ deuteron. Therefore, $J_d=I_d$, and we get,

\begin{eqnarray}
\frac{1}{4\pi}C_{J_p, -M_p J'_p M'_p}^{0 0}\left\{ \left\{\Xi_{I_d}(\xi_{np}) \otimes \tilde{Y}_{L}(\hat{R})\right\}_{J_p} \otimes   \left\{\Xi_{I_d}(\xi_{np}) \otimes \tilde{Y}_{L'}(\hat{R})\right\}_{J'_p} \right\}_{0 0}.
\end{eqnarray}

\noindent Putting this into Eq.(\ref{eq:integral-term}), we get

\begin{eqnarray}
&\phantom{=}&-\sum_{L' J'_p}\frac{1}{R}\left[\frac{\hbar^2}{2\mu}\left(\frac{\partial^2}{\partial R^2}-\frac{L'(L'+1)}{R^2}\right) +E_d\right]\chi_{L' J'_p}^{J_T M_T}(R)\int \phi_{\ell}(r)V_{np}(r)\phi_{\ell'}(r)r^2 dr \nonumber \\
&\phantom{=}& \times \ \sum_{M_p M'_p}\sum_{\mu_t}C_{J_p M_p I_t \mu_t}^{J_T M_T}C_{J'_p M'_p I_t \mu_t}^{J_T M_T} \frac{1}{4\pi}C_{J_p, -M_p J'_p M'_p}^{0 0}(-)^{J_p-M_p} \\
&\phantom{=}& \times \ \int \left\{ \left\{\Xi_{I_d}(\xi_{np}) \otimes \tilde{Y}_{L}(\hat{R})\right\}_{J_p} \otimes   \left\{\Xi_{I_d}(\xi_{np}) \otimes \tilde{Y}_{L'}(\hat{R})\right\}_{J'_p} \right\}_{0 0}  d\Omega_r d\Omega_R d\xi_{np}. \nonumber
\end{eqnarray}

\noindent The integral over $d\Omega_r$ gives $4\pi$ and cancels the $1/4\pi$ already there, which leaves us with

\begin{eqnarray}\label{eq:sch-eq22}
&\phantom{=}&-\sum_{L' J'_p}\frac{1}{R}\left[\frac{\hbar^2}{2\mu}\left(\frac{\partial^2}{\partial R^2}-\frac{L'(L'+1)}{R^2}\right) +E_d\right]\chi_{L' J'_p}^{J_T M_T}(R)\int \phi_{\ell}(r)V_{np}(r)\phi_{\ell'}(r)r^2 dr \nonumber \\
&\phantom{=}& \times \ \sum_{M_p M'_p}\sum_{\mu_t}C_{J_p M_p I_t \mu_t}^{J_T M_T}C_{J'_p M'_p I_t \mu_t}^{J_T M_T} C_{J_p, -M_p J'_p M'_p}^{0 0}(-)^{J_p-M_p} \nonumber \\
&\phantom{=}& \times \ \int \left\{ \left\{\Xi_{I_d}(\xi_{np}) \otimes \tilde{Y}_{L}(\hat{R})\right\}_{J_p} \otimes   \left\{\Xi_{I_d}(\xi_{np}) \otimes \tilde{Y}_{L'}(\hat{R})\right\}_{J'_p} \right\}_{0 0} d\Omega_R d\xi_{np}
\end{eqnarray}

\noindent We can use, \cite{Varshalovich_Book}, p.248, Eq.(1)

\begin{eqnarray}\label{eq:cgc23}
C_{J_p, -M_p J_p M_p}^{0 0}&=&(-)^{J_p+M_p}\frac{\delta_{J_p J'_p}\delta_{M_p,M'_p}}{\hat{J}_p}.
\end{eqnarray}

\noindent Inserting Eq.(\ref{eq:cgc23}) into Eq.(\ref{eq:sch-eq22}), and summing over $J'_p$ and $M'_p$ (using \cite{Varshalovich_Book}, p.236, Eq.(8)) we get, 

\begin{eqnarray}\label{eq:sch-eq24}
&\phantom{=}&-\sum_{L'}\frac{1}{R}\left[\frac{\hbar^2}{2\mu}\left(\frac{\partial^2}{\partial R^2}-\frac{L'(L'+1)}{R^2}\right) +E_d\right]\chi_{L' J_p}^{J_T M_T}(R)\int \phi_{\ell}(r)V_{np}(r)\phi_{\ell'}(r)r^2 dr \nonumber \\
&\phantom{=}& \times \ \sum_{M_p}\sum_{\mu_t}C_{J_p M_p I_t \mu_t}^{J_T M_T}C_{J_p M_p I_t \mu_t}^{J_T M_T} (-)^{2J_p}\frac{1}{\hat{J}_p} \\
&\phantom{=}& \times \ \int \left\{ \left\{\Xi_{I_d}(\xi_{np}) \otimes \tilde{Y}_{L}(\hat{R})\right\}_{J_p} \otimes   \left\{\Xi_{I_d}(\xi_{np}) \otimes \tilde{Y}_{L'}(\hat{R})\right\}_{J_p} \right\}_{0 0} d\Omega_R d\xi_{np} \nonumber \\
&=&-\sum_{L'}\frac{1}{R}\left[\frac{\hbar^2}{2\mu}\left(\frac{\partial^2}{\partial R^2}-\frac{L'(L'+1)}{R^2}\right) +E_d\right]\chi_{L' J_p}^{J_T M_T}(R)\int \phi_{\ell}(r)V_{np}(r)\phi_{\ell'}(r)r^2 dr \nonumber \\
&\phantom{=}& \times \  (-)^{2J_p}\frac{1}{\hat{J}_p}\int \left\{ \left\{\Xi_{I_d}(\xi_{np}) \otimes \tilde{Y}_{L}(\hat{R})\right\}_{J_p} \otimes   \left\{\Xi_{I_d}(\xi_{np}) \otimes \tilde{Y}_{L'}(\hat{R})\right\}_{J_p} \right\}_{0 0} d\Omega_R d\xi_{np}. \nonumber
\end{eqnarray}

\noindent Coupling the spin functions together and the spherical harmonics together, each up to zero angular momentum, using \cite{Varshalovich_Book}, p.70, Eq. (11), and  p.358 Eq.(4),

\begin{eqnarray}\label{eq:tensor25}
&\phantom{=}&\left\{ \left\{\Xi_{I_d}(\xi_{np}) \otimes \tilde{Y}_{L}(\hat{R})\right\}_{J_p} \otimes   \left\{\Xi_{I_d}(\xi_{np}) \otimes \tilde{Y}_{L'}(\hat{R})\right\}_{J_p} \right\}_{0 0} \nonumber \\
&=& \hat{J}_p^2
\begin{Bmatrix}
I_d & L & J_p \\
I_d & L' & J_p \\
0 & 0 & 0
\end{Bmatrix}
\left\{ \left\{\Xi_{I_d}(\xi_{np}) \otimes \Xi_{I_d}(\xi_{np}) \right\}_{0} \otimes   \left\{\tilde{Y}_{L}(\hat{R}) \otimes \tilde{Y}_{L'}(\hat{R})\right\}_{0} \right\}_{0 0} \nonumber \\
&=&\frac{\hat{J}_p\delta_{LL'}}{\hat{I_d}\hat{L}}
\left\{ \left\{\Xi_{I_d}(\xi_{np}) \otimes \Xi_{I_d}(\xi_{np}) \right\}_{0} \otimes   \left\{\tilde{Y}_{L}(\hat{R}) \otimes \tilde{Y}_{L'}(\hat{R})\right\}_{0} \right\}_{0 0}.
\end{eqnarray}

\noindent Replacing Eq.(\ref{eq:tensor25}) into Eq.(\ref{eq:sch-eq24}), and summing over $L'$, we obtain:

\begin{eqnarray}
&\phantom{=}&-\frac{1}{R}\left[\frac{\hbar^2}{2\mu}\left(\frac{\partial^2}{\partial R^2}-\frac{L'(L'+1)}{R^2}\right) +E_d\right]\chi_{L' J_p}^{J_T M_T}(R)\int \phi_{\ell}(r)V_{np}(r)\phi_{\ell'}(r)r^2 dr \nonumber \\
&\phantom{=}& \times \  (-)^{2J_p}\left(\frac{1}{\hat{I_d}} \int   \left\{\Xi_{I_d}(\xi_{np}) \otimes \Xi_{I_d}(\xi_{np}) \right\}_{0 0} d\xi_{np}\right)\left(\frac{1}{\hat{L}}  \int  \left\{\tilde{Y}_{L}(\hat{R}) \otimes \tilde{Y}_{L}(\hat{R})\right\}_{0 0}  d\Omega_R \right) \nonumber \\
\end{eqnarray}

\noindent The integral over the two Weinberg states multiplied by $V_{np}$ gives $-1$ by the normalization condition of Eq.(\ref{eq:Weinberg-Normalization}). Also, since $I_d=1$, $J_p$ is an integer, so $(-)^{2J_p}=1$. Thus, we have

\begin{eqnarray}\label{eq:sch-eq27}
&\phantom{=}&\frac{1}{R}\left[\frac{\hbar^2}{2\mu}\left(\frac{\partial^2}{\partial R^2}-\frac{L'(L'+1)}{R^2}\right) +E_d\right]\chi_{L' J_p}^{J_T M_T}(R) \\
&\phantom{=}& \times \  \left(\frac{1}{\hat{I_d}} \int   \left\{\Xi_{I_d}(\xi_{np}) \otimes \Xi_{I_d}(\xi_{np}) \right\}_{0 0} d\xi_{np}\right)\left(\frac{1}{\hat{L}}  \int  \left\{\tilde{Y}_{L}(\hat{R}) \otimes \tilde{Y}_{L}(\hat{R})\right\}_{0 0}  d\Omega_R \right). \nonumber
\end{eqnarray}

\noindent The integral over the spin functions can be worked out with Eq.(11), p.70,  and Eq.(4), p.358 of \cite{Varshalovich_Book}:

\begin{eqnarray}
&\phantom{=}&\int \left\{\Xi_{I_d}(\xi_{np}) \otimes \Xi_{I_d}(\xi_{np}) \right\}_{0 0} d\xi_{np} \nonumber \\
&=&\int \left\{ \left\{\Xi_{1/2}(\xi_n)\otimes \Xi_{1/2}(\xi_p) \right\}_{1} \otimes \left\{\Xi_{1/2}(\xi_n)\otimes \Xi_{1/2}(\xi_p) \right\}_{1}   \right\}_{00}d\xi_n d\xi_p \nonumber \\
&=&\int \hat{1}^2
\begin{Bmatrix}
\frac{1}{2} & \frac{1}{2} & 1 \\
\frac{1}{2} & \frac{1}{2} & 1 \\
0 & 0 & 0 
\end{Bmatrix}
\left\{ \left\{\Xi_{1/2}(\xi_n)\otimes \Xi_{1/2}(\xi_n) \right\}_{0} \otimes \left\{\Xi_{1/2}(\xi_p)\otimes \Xi_{1/2}(\xi_p) \right\}_{0}   \right\}_{00}d\xi_n d\xi_p \nonumber \\
&=&\int \frac{\hat{1}^2}{\hat{\tfrac{1}{2}}^2\hat{1}}\int \left\{\Xi_{1/2}(\xi_n)\otimes \Xi_{1/2}(\xi_n) \right\}_{00} d\xi_n \int \left\{\Xi_{1/2}(\xi_p)\otimes \Xi_{1/2}(\xi_p) \right\}_{00} d\xi_p.
\end{eqnarray}

\noindent Doing the integral over the neutron spin functions by using Eq.(13), p.132, and Eq.(1), p.248, of \cite{Varshalovich_Book} we arrrive at:

\begin{eqnarray}
&\phantom{=}& \int \left\{\Xi_{1/2}(\xi_n)\otimes \Xi_{1/2}(\xi_n) \right\}_{00} d\xi_n \nonumber \\
&=& \sum_{\mu_n}\int C_{1/2, -\mu_n, 1/2, \mu_n}^{00}\Xi_{1/2,-\mu_n}(\xi_n)\Xi_{1/2,\mu_n}(\xi_n)d\xi_n \nonumber \\
&=& \sum_{\mu_n} C_{1/2, -\mu_n, 1/2, \mu_n}^{00}(-)^{-1/2-\mu_n}\int \Xi^*_{1/2 \mu_n}(\xi_n)\Xi_{1/2,\mu_n}(\xi_n)d\xi_n \nonumber \\
&=& \sum_{\mu_n} \frac{(-)^{1/2+\mu_n}}{\widehat{\left(\frac{1}{2}\right)}}(-)^{-1/2-\mu_n}\int \Xi^*_{1/2, \mu_n}(\xi_n)\Xi_{1/2, \mu_n}(\xi_n)d\xi_n \nonumber \\
&=& \sqrt{2},
\end{eqnarray} 

\noindent A similar procedure is followed for the integral over the proton spin functions. Thus, with $I_d=1$

\begin{eqnarray}
\frac{1}{\hat{I_d}} \int   \left\{\Xi_{I_d}(\xi_{np}) \otimes \Xi_{I_d}(\xi_{np}) \right\}_{0 0} d\xi_{np}=1.
\end{eqnarray}

\noindent The final integral in Eq.(\ref{eq:sch-eq27}) can be worked out using Eq.(6), p.62, and Eq.(1), p.248, of \cite{Varshalovich_Book}

\begin{eqnarray}\label{eq:integral31}
\frac{1}{\hat{L}}\int  \left\{\tilde{Y}_{L}(\hat{R}) \otimes \tilde{Y}_{L}(\hat{R})\right\}_{0 0}  d\Omega_R &=& \frac{1}{\hat{L}}\sum_{M} C_{L,-M L M}^{00}\int \tilde{Y}_{L, -M}(\hat{R})\tilde{Y}_{LM}(\hat{R})d\Omega_{R} \nonumber \\
&=&\frac{1}{\hat{L}}\sum_{M} C_{L,-M L M}^{00}(-)^{-L+M}\int \tilde{Y}_{L M}^*(\hat{R})\tilde{Y}_{LM}(\hat{R})d\Omega_{R} \nonumber \\
&=&\frac{1}{\hat{L}}\sum_{M}\frac{(-)^{L+M}}{\hat{L}}(-)^{-L+M}=1.
\end{eqnarray}

\noindent Therefore, introducing Eq.(\ref{eq:integral31}) into Eq.(\ref{eq:sch-eq27}) and joining the $rhs$ of Eq.(\ref{lhs}) we obtain:

\begin{eqnarray}\label{eq:sch-eq32}
&\phantom{=}&\frac{1}{R}\left[\frac{\hbar^2}{2\mu}\left(\frac{\partial^2}{\partial R^2}-\frac{L(L+1)}{R^2}\right) +E_d\right]\chi_{L J_p}^{J_T M_T}(R) \nonumber \\
&=&-\sum_{\ell' L' J'_p} \sum_{\ell}8\int \left\{ \left\{ \left\{\Xi_{I_d}(\xi_{np})\otimes \tilde{Y}_{\ell}(\hat{r})\right\}_{j_p} \otimes \tilde{Y}_{L}(\hat{R})\right\}_{J_p} \otimes \Xi_{I_t}(\xi_t) \right\}_{J_T M_T}^* \nonumber \\
&\phantom{=}& \times \ U_{NA}\left(\textbf{R}_{p,n},\textbf{R}_{p,n}+2\textbf{s}  \right) \nonumber \\
&\phantom{=}& \times \ \left\{ \left\{ \left\{\Xi_{I_d}(\xi_{np})\otimes \tilde{Y}_{\ell'}(\widehat{r \pm 2s})\right\}_{j_p} \otimes \tilde{Y}_{L'}(\widehat{R+s})\right\}_{J'_p} \otimes \Xi_{I_t}(\xi_t) \right\}_{J_T M_T} \nonumber \\
&\phantom{=}& \times \ \phi_\ell(r)V_{np}(r)\phi_{\ell' }(|\textbf{r}\pm 2\textbf{s}|)\frac{\chi_{L' J'_p}^{J_T M_T}\left(\left|\textbf{R}+\textbf{s} \right|\right)}{\left|\textbf{R}+\textbf{s} \right|} d\textbf{s} d\textbf{r} d\Omega_R d\xi_t d\xi_{np}
\end{eqnarray}

\noindent We now concentrate on the tensor couplings in the $rhs$ of Eq.(\ref{eq:sch-eq32}). First, we introduce $\ell=0$ for the deuteron, and integrate over $d\xi_t$:

\begin{eqnarray}\label{eq:tensor33}
&\phantom{=}&\left\{ \left\{ \left\{\Xi_{I_d}(\xi_{np})\otimes \tilde{Y}_{0}(\hat{r})\right\}_{j_p} \otimes \tilde{Y}_{L}(\hat{R})\right\}_{J_p} \otimes \Xi_{I_t}(\xi_t) \right\}_{J_T M_T}^* \nonumber \\
&\phantom{=}& \times \ \left\{ \left\{ \left\{\Xi_{I_d}(\xi_{np})\otimes \tilde{Y}_{0}(\widehat{r \pm 2s})\right\}_{j_p} \otimes \tilde{Y}_{L'}(\widehat{R+s})\right\}_{J'_p} \otimes \Xi_{I_t}(\xi_t) \right\}_{J_T M_T} \nonumber \\
&=&\frac{1}{4\pi}\left\{ \left\{\Xi_{I_d}(\xi_{np}) \otimes \tilde{Y}_{L}(\hat{R})\right\}_{J_p} \otimes \Xi_{I_t}(\xi_t) \right\}_{J_T M_T}^* \nonumber \\
&\phantom{=}& \times \ \left\{ \left\{\Xi_{I_d}(\xi_{np}) \otimes \tilde{Y}_{L'}(\widehat{R+s})\right\}_{J'_p} \otimes \Xi_{I_t}(\xi_t) \right\}_{J_T M_T} \nonumber \\
&=&\frac{1}{4\pi}\sum_{M_p M'_p}\sum_{\mu_t \mu'_t}C_{J_p M_p I_t \mu_t}^{J_T M_T}C_{J'_p M'_p I_t \mu'_t}^{J_T M_T} \left\{\Xi_{I_d}(\xi_{np}) \otimes \tilde{Y}_{L}(\hat{R})\right\}_{J_p M_p}^*    \nonumber \\
&\phantom{=}& \times \ \left\{\Xi_{I_d}(\xi_{np}) \otimes \tilde{Y}_{L'}(\widehat{R+s})\right\}_{J'_p M'_p}\int  \Xi_{I_t \mu_t}^*(\xi_t) \Xi_{I_t \mu'_t}(\xi_t) d\xi_t \nonumber \\
&=&\frac{1}{4\pi}\sum_{M_p M'_p}\sum_{\mu_t}C_{J_p M_p I_t \mu_t}^{J_T M_T}C_{J'_p M'_p I_t \mu_t}^{J_T M_T} \left\{\Xi_{I_d}(\xi_{np}) \otimes \tilde{Y}_{L}(\hat{R})\right\}_{J_p M_p}^*   \nonumber \\
&\phantom{=}& \times \ \left\{\Xi_{I_d}(\xi_{np}) \otimes \tilde{Y}_{L'}(\widehat{R+s})\right\}_{J'_p M'_p}. \nonumber \\
\end{eqnarray}

\noindent Inserting Eq.(\ref{eq:tensor33}) into Eq.(\ref{eq:sch-eq32}) we arrive at:

\begin{eqnarray}\label{eq:sch-eq34}
&\phantom{=}&\frac{1}{R}\left[\frac{\hbar^2}{2\mu}\left(\frac{\partial^2}{\partial R^2}-\frac{L(L+1)}{R^2}\right) +E_d\right]\chi_{L J_p}^{J_T M_T}(R) \nonumber \\
&=&-\frac{1}{4\pi}\sum_{L' J'_p} 8\int\phi_{0}(r)V_{np}(r) U_{NA}\left(\textbf{R}_{p,n},\textbf{R}_{p,n}+2\textbf{s}  \right) \nonumber \\
&\phantom{=}& \times \sum_{M_p M'_p}\sum_{\mu_t}C_{J_p M_p I_t \mu_t}^{J_T M_T}C_{J'_p M'_p I_t \mu_t}^{J_T M_T} \left\{\Xi_{I_d}(\xi_{np}) \otimes \tilde{Y}_{L}(\hat{R})\right\}_{J_p M_p}^*    \\
&\phantom{=}& \times \ \left\{\Xi_{I_d}(\xi_{np}) \otimes \tilde{Y}_{L'}(\widehat{R+s})\right\}_{J'_p M'_p}\phi_{0}(|\textbf{r}\pm 2\textbf{s}|)\frac{\chi_{L' J'_p}^{J_T M_T}\left(\left|\textbf{R}+\textbf{s} \right|\right)}{\left|\textbf{R}+\textbf{s} \right|} d\textbf{s} d\textbf{r} d\Omega_R  d\xi_{np}. \nonumber
\end{eqnarray}

\noindent Next we couple the two tensors together up to zero angular momentum using,  Eq.(6), p.62, and Eq.(1), p.248, of \cite{Varshalovich_Book}:

\begin{eqnarray}\label{eq:tensor35}
&\phantom{=}&\left\{\Xi_{I_d}(\xi_{np}) \otimes \tilde{Y}_{L}(\hat{R})\right\}_{J_p M_p}^*   \left\{\Xi_{I_d}(\xi_{np}) \otimes \tilde{Y}_{L'}(\widehat{R+s})\right\}_{J'_p M'_p} \nonumber \\
&=&(-)^{J_p-M_p}\left\{\Xi_{I_d}(\xi_{np}) \otimes \tilde{Y}_{L}(\hat{R})\right\}_{J_p, -M_p}   \left\{\Xi_{I_d}(\xi_{np}) \otimes \tilde{Y}_{L'}(\widehat{R+s})\right\}_{J'_p M'_p} \nonumber \\
&=&(-)^{J_p-M_p}\sum_{S M_S}C_{J_p, -M_p J'_p M_p}^{S M_S} \nonumber \\
&\phantom{=}& \times \ \left\{\left\{\Xi_{I_d}(\xi_{np}) \otimes \tilde{Y}_{L}(\hat{R})\right\}_{J_p} \otimes   \left\{\Xi_{I_d}(\xi_{np}) \otimes \tilde{Y}_{L'}(\widehat{R+s})\right\}_{J'_p}\right\}_{00} \nonumber \\
&\rightarrow&(-)^{J_p-M_p} C_{J_p, -M_p J'_p M_p}^{00} \nonumber \\
&\phantom{=}& \times \ \left\{\left\{\Xi_{I_d}(\xi_{np}) \otimes \tilde{Y}_{L}(\hat{R})\right\}_{J_p} \otimes   \left\{\Xi_{I_d}(\xi_{np}) \otimes \tilde{Y}_{L'}(\widehat{R+s})\right\}_{J'_p}\right\}_{00} \nonumber \\
&=&(-)^{J_p-M_p}(-)^{J_p+M_p} \frac{\delta_{J_p J'_p}\delta_{M_p M'_p}}{\hat{J}_p} \nonumber \\
&\phantom{=}& \times \ \left\{\left\{\Xi_{I_d}(\xi_{np}) \otimes \tilde{Y}_{L}(\hat{R})\right\}_{J_p} \otimes   \left\{\Xi_{I_d}(\xi_{np}) \otimes \tilde{Y}_{L'}(\widehat{R+s})\right\}_{J'_p}\right\}_{00}  \\
&=&(-)^{2J_p}\frac{\delta_{J_p J'_p}\delta_{M_p M'_p}}{\hat{J}_p}\left\{\left\{\Xi_{I_d}(\xi_{np}) \otimes \tilde{Y}_{L}(\hat{R})\right\}_{J_p} \otimes   \left\{\Xi_{I_d}(\xi_{np}) \otimes \tilde{Y}_{L'}(\widehat{R+s})\right\}_{J'_p}\right\}_{00}. \nonumber
\end{eqnarray}

\noindent We replace Eq.(\ref{eq:tensor35}) in Eq.(\ref{eq:sch-eq34}) and sum over $J'_p$ and $M'_p$ to obtain:

\begin{eqnarray}
&\phantom{=}&\frac{1}{R}\left[\frac{\hbar^2}{2\mu}\left(\frac{\partial^2}{\partial R^2}-\frac{L(L+1)}{R^2}\right) +E_d\right]\chi_{L J_p}^{J_T M_T}(R) \nonumber \\
&=&-\frac{1}{4\pi}\sum_{L'} 8\int\phi_{0}(r)V_{np}(r) U_{NA}\left(\textbf{R}_{p,n},\textbf{R}_{p,n}+2\textbf{s}  \right) \nonumber \\
&\phantom{=}& \times \ \sum_{M_p}\sum_{\mu_t}C_{J_p M_p I_t \mu_t}^{J_T M_T}C_{J_p M_p I_t \mu_t}^{J_T M_T} (-)^{2J_p}\frac{1}{\hat{J}_p} \nonumber \\
&\phantom{=}& \times \ \left\{\left\{\Xi_{I_d}(\xi_{np}) \otimes \tilde{Y}_{L}(\hat{R})\right\}_{J_p} \otimes   \left\{\Xi_{I_d}(\xi_{np}) \otimes \tilde{Y}_{L'}(\widehat{R+s})\right\}_{J_p}\right\}_{00} \nonumber \\
&\phantom{=}& \times \ \phi_{0}(|\textbf{r}\pm 2\textbf{s}|)\frac{\chi_{L' J_p}^{J_T M_T}\left(\left|\textbf{R}+\textbf{s} \right|\right)}{\left|\textbf{R}+\textbf{s} \right|} d\textbf{s} d\textbf{r} d\Omega_R  d\xi_{np}.
\end{eqnarray}

\noindent Next we sum over $M_p$ and $\mu_t$ using, Eq.(8), p.236, of \cite{Varshalovich_Book},

\begin{eqnarray}
&\phantom{=}&\frac{1}{R}\left[\frac{\hbar^2}{2\mu}\left(\frac{\partial^2}{\partial R^2}-\frac{L(L+1)}{R^2}\right) +E_d\right]\chi_{L J_p}^{J_T M_T}(R) \nonumber \\
&=&-\frac{1}{4\pi}\sum_{L'}8\int\phi_{0}(r)V_{np}(r) U_{NA}\left(\textbf{R}_{p,n},\textbf{R}_{p,n}+2\textbf{s}  \right) \nonumber \\
&\phantom{=}& \times  (-)^{2J_p}\frac{1}{\hat{J}_p}\left\{\left\{\Xi_{I_d}(\xi_{np}) \otimes \tilde{Y}_{L}(\hat{R})\right\}_{J_p} \otimes   \left\{\Xi_{I_d}(\xi_{np}) \otimes \tilde{Y}_{L'}(\widehat{R+s})\right\}_{J_p}\right\}_{00} \nonumber \\
&\phantom{=}& \times \ \phi_{0}(|\textbf{r}\pm 2\textbf{s}|)\frac{\chi_{L' J_p}^{J_T M_T}\left(\left|\textbf{R}+\textbf{s} \right|\right)}{\left|\textbf{R}+\textbf{s} \right|} d\textbf{s} d\textbf{r} d\Omega_R  d\xi_{np}.
\end{eqnarray}

\noindent We now use Eq.(11), p.70, and Eq.(4), p.358, of \cite{Varshalovich_Book} to couple the spin functions and the spherical harmonics to zero angular momentum,  

\begin{eqnarray}
&\phantom{=}&\left\{ \left\{\Xi_{I_d}(\xi_{np}) \otimes \tilde{Y}_{L}(\hat{R})\right\}_{J_p} \otimes   \left\{\Xi_{I_d}(\xi_{np}) \otimes \tilde{Y}_{L'}(\widehat{R+s})\right\}_{J_p}\right\}_{0 0} \nonumber \\
&=& \hat{J}_p^2
\begin{Bmatrix}
I_d & L & J_p \\
I_d & L' & J_p \\
0 & 0 & 0
\end{Bmatrix}
\left\{ \left\{\Xi_{I_d}(\xi_{np}) \otimes \Xi_{I_d}(\xi_{np}) \right\}_{0} \otimes  \left\{\tilde{Y}_{L}(\hat{R}) \otimes \tilde{Y}_{L'}(\widehat{R+s})\right\}_{0} \right\}_{0 0} \nonumber \\
&=&\delta_{LL'}\frac{\hat{J}_p}{\hat{I_d}\hat{L}}
\left\{ \left\{\Xi_{I_d}(\xi_{np}) \otimes \Xi_{I_d}(\xi_{np}) \right\}_{0} \otimes  \left\{\tilde{Y}_{L}(\hat{R}) \otimes \tilde{Y}_{L'}(\widehat{R+s})\right\}_{0} \right\}_{0 0}.
\end{eqnarray}

\noindent Using $(-)^{2J_p}=1$ and summing over $L'$,

\begin{eqnarray}
&\phantom{=}&\frac{1}{R}\left[\frac{\hbar^2}{2\mu}\left(\frac{\partial^2}{\partial R^2}-\frac{L(L+1)}{R^2}\right) +E_d\right]\chi_{L J_p}^{J_T M_T}(R) \nonumber \\
&=&-\frac{1}{4\pi}\sum_{L'} 8\int\phi_{0}(r)V_{np}(r) U_{NA}\left(\textbf{R}_{p,n},\textbf{R}_{p,n}+2\textbf{s}  \right) \nonumber \\
&\phantom{=}& \times  (-)^{2J_p}\frac{1}{\hat{I_d}\hat{L}}
 \left\{\Xi_{I_d}(\xi_{np}) \otimes \Xi_{I_d}(\xi_{np}) \right\}_{00}  \left\{\tilde{Y}_{L}(\hat{R}) \otimes \tilde{Y}_{L}(\widehat{R+s})\right\}_{00}  \nonumber \\
&\phantom{=}& \times \ \phi_{0}(|\textbf{r}\pm 2\textbf{s}|)\frac{\chi_{L J_p}^{J_T M_T}\left(\left|\textbf{R}+\textbf{s} \right|\right)}{\left|\textbf{R}+\textbf{s} \right|} d\textbf{s} d\textbf{r} d\Omega_R  d\xi_{np} \nonumber \\
&=&-\frac{1}{4\pi}\sum_{L'} 8\int\phi_{0}(r)V_{np}(r) U_{NA}\left(\textbf{R}_{p,n},\textbf{R}_{p,n}+2\textbf{s}  \right)  \\
&\phantom{=}& \times  \frac{1}{\hat{L}}\left\{\tilde{Y}_{L}(\hat{R}) \otimes \tilde{Y}_{L}(\widehat{R+s})\right\}_{00} \phi_{0}(|\textbf{r}\pm 2\textbf{s}|)\frac{\chi_{L J_p}^{J_T M_T}\left(\left|\textbf{R}+\textbf{s} \right|\right)}{\left|\textbf{R}+\textbf{s} \right|} d\textbf{s} d\textbf{r} d\Omega_R. \nonumber
\end{eqnarray}

\noindent Bringing the $1/R$ term from the $lhs$ over to the $rhs$ gives us:

\begin{eqnarray}\label{eq:sch-eq40}
&\phantom{=}&\left[\frac{\hbar^2}{2\mu}\left(\frac{\partial^2}{\partial R^2}-\frac{L(L+1)}{R^2}\right) +E_d\right]\chi_{L J_p}^{J_T M_T}(R) \nonumber \\
&=&-\frac{R}{\hat{L}}\frac{8}{4\pi}\int \phi_{0}(r)V_{np}(r) U_{NA}\left(\textbf{R}_{p,n},\textbf{R}_{p,n}+2\textbf{s}  \right) \left\{\tilde{Y}_{L}(\hat{R}) \otimes \tilde{Y}_{L}(\widehat{R+s})\right\}_{00}  \nonumber \\
&\phantom{=}& \times \ \phi_{0}(|\textbf{r}\pm 2\textbf{s}|)\frac{\chi_{L J_p}^{J_T M_T}\left(\left|\textbf{R}+\textbf{s} \right|\right)}{\left|\textbf{R}+\textbf{s} \right|} d\textbf{s} d\textbf{r} d\Omega_R  \nonumber \\
&=&-\frac{R}{\hat{L}}\frac{8}{4\pi}\sum_{M}C_{L, -M L M}^{00}\int \phi_{0}(r)V_{np}(r) U_{NA}\left(\textbf{R}_{p,n},\textbf{R}_{p,n}+2\textbf{s}  \right) \tilde{Y}_{L,-M}(\hat{R})\tilde{Y}_{LM}(\widehat{R+s})  \nonumber \\
&\phantom{=}& \times \ \phi_{0 }(|\textbf{r}\pm 2\textbf{s}|)\frac{\chi_{L J_p}^{J_T M_T}\left(\left|\textbf{R}+\textbf{s} \right|\right)}{\left|\textbf{R}+\textbf{s} \right|} d\textbf{s} d\textbf{r} d\Omega_R \nonumber \\
&=&-\frac{R}{\hat{L}}\frac{8}{4\pi}\sum_{M}\frac{(-)^{L-M}}{\hat{L}}\int \phi_{0}(r)V_{np}(r) U_{NA}\left(\textbf{R}_{p,n},\textbf{R}_{p,n}+2\textbf{s}  \right) \tilde{Y}_{L,-M}(\hat{R})\tilde{Y}_{LM}(\widehat{R+s})  \nonumber \\
&\phantom{=}& \times \ \phi_{0}(|\textbf{r}\pm 2\textbf{s}|)\frac{\chi_{L J_p}^{J_T M_T}\left(\left|\textbf{R}+\textbf{s} \right|\right)}{\left|\textbf{R}+\textbf{s} \right|} d\textbf{s} d\textbf{r} d\Omega_R.
\end{eqnarray}

\noindent Since the integrand is coupled to zero angular momentum, it is spherically symmetric, which means that it is invariant under rotations of the three vectors $\textbf{R}$, $\textbf{r}$, and $\textbf{s}$. Thus, we can evaluate it in any configuration we want. By placing the $\textbf{R}$ in the $\hat{z}$-direction, $M=0$, and $\tilde{Y}_{L0}(\hat{z})=\frac{i^L\hat{L}}{\sqrt{4\pi}}$. We will place $\textbf{r}$ in the $xz$-plane so that the $\phi_r$-dependence is removed. Integration over $d\Omega_R$ yields a factor of $4\pi$ for all other choices for the direction of $\textbf{R}$. Since we are fixing $\textbf{r}$ to be in the $xz$-plane, we get a factor of $2\pi$ from each vector to take care of rotations around the z-axis. Thus, we need to multiply the integral by $(4\pi)*(2\pi)=8\pi^2$. There is no additional symmetry to fix $\textbf{s}$. Finally, introducing these symmetries in the integral of Eq.(\ref{eq:sch-eq40}) and the phase $i^L$ of the spherical harmonics, we arrive at the simplified expression we have used in our implementation:

\begin{eqnarray}
&\phantom{=}&\left[\frac{\hbar^2}{2\mu}\left(\frac{\partial^2}{\partial R^2}-\frac{L(L+1)}{R^2}\right) +E_d\right]\chi_{L J_p}^{J_T M_T}(R) \nonumber \\
&=&-\frac{R}{\hat{L}}\frac{8}{4\pi}\frac{\hat{L}}{\sqrt{4\pi}}(-)^Li^{2L}\frac{1}{\hat{L}}8\pi^2\int \phi_{0}(r)V_{np}(r) U_{NA}\left(\textbf{R}_{p,n},\textbf{R}_{p,n}+2\textbf{s}  \right) Y_{L0}(\widehat{R+s})  \nonumber \\
&\phantom{=}& \times \ \phi_{0}(|\textbf{r}\pm 2\textbf{s}|)\frac{\chi_{L J_p}^{J_T M_T}\left(\left|\textbf{R}+\textbf{s} \right|\right)}{\left|\textbf{R}+\textbf{s} \right|} d\textbf{s}  r^2 dr \sin\theta_{r} d\theta_{r} \nonumber \\
&=&-\frac{8R \sqrt{\pi}}{\hat{L}}\int \phi_{0}(r)V_{np}(r) U_{NA}\left(\textbf{R}_{p,n},\textbf{R}_{p,n}+2\textbf{s}  \right) Y_{L0}(\widehat{R+s})  \nonumber \\
&\phantom{=}& \times \ \phi_{0}(|\textbf{r}\pm 2\textbf{s}|)\frac{\chi_{L J_p}^{J_T M_T}\left(\left|\textbf{R}+\textbf{s} \right|\right)}{\left|\textbf{R}+\textbf{s} \right|} d\textbf{s} r^2 dr \sin\theta_{r}d\theta_{r}
\end{eqnarray}

\noindent where $\textbf{R}_{p,n}$, $\textbf{R}$, and $\textbf{s}$ are to be evaluated in the configuration described before. $U_{NA}$ is the nucleon optical potential for either the proton or neutron. Making the replacement $U_{NA}\rightarrow U_{nA} + U_{pA}$ gives us the nonlocal adiabatic potential, and the resulting partial wave equation for the deuteron scattering state when using nonlocal potentials within the ADWA. 


\chapter{ Deriving the T-Matrix}
\label{tmatrix}

Here we will derive the explicit form for the T-matrix for $(d,p)$ transfer reactions. This equation was given in the post form in Eq.(\ref{Tmatrix}), and is repeated here without the remnant term 

\begin{eqnarray}\label{eqD:tmatrix}
T_{\mu_A M_d \mu_p M_B}(\textbf{k}_f,\textbf{k}_i)=\langle \Psi_{\textbf{k}_f}^{\mu_p M_B}|V_{np}|\Psi_{\textbf{k}_i}^{\mu_A M_d}\rangle.
\end{eqnarray}

\noindent We need to define the explicit partial wave for all wave functions in Eq.(\ref{eqD:tmatrix}). We begin by defining the wave function for relative motion between $d+A$, which is given by:

\begin{eqnarray}\label{eqD:wf2}
\Psi_{\ell_i j_i}&=&\sum_{L_i}\sum_{J_{P_i} M_{P_i}}\left\{\left\{\Xi_{I_p}(\xi_p) \otimes \Phi_{j_i}(\textbf{r}_{np},\xi_n)\right\}_{J_d} \otimes \tilde{Y}_{L_i}(\hat{R}_{dA})\right\}_{J_{P_i} M_{P_i}} \nonumber \\
&\phantom{=}& \times \ \Xi_{I_A \mu_A}(\xi_A) \frac{\chi_{L_i J_{P_i}}(R_{dA})}{R_{dA}}  \nonumber \\
&=&\sum_{L_i}\sum_{J_{P_i} M_{P_i}}\left\{\left\{\Xi_{I_p}(\xi_p) \otimes \left\{\tilde{Y}_{\ell_i}(\hat{r}_{np}) \otimes \Xi_{I_n}(\xi_n)\right\}_{j_i} \right\}_{J_d}\otimes \tilde{Y}_{L_i}(\hat{R}_{dA})\right\}_{J_{P_i} M_{P_i}} \nonumber \\
&\phantom{=}& \times \ \Xi_{I_A \mu_A}(\xi_A)\phi_{j_i}(r_{np})\frac{\chi_{L_i J_{p_i}}(R_{dA})}{R_{dA}}.
\end{eqnarray}

\noindent As in Appendix \ref{Nonlocal_Adiabatic}, $\Xi_{I_p}(\xi_p)$, $\Xi_{I_n}(\xi_n)$ and $\Xi_{I_A}(\xi_A)$ are the spin functions for the proton, neutron, and target, respectively, each with projections $\mu_p$, $\mu_n$, and $\mu_A$. $\tilde{Y}_{\ell_i}$ is the spherical harmonics for the relative motion between the neutron and proton in the deuteron, and $\tilde{Y}_{L_i}$ is the spherical harmonic for the relative motion between the deuteron and the target. As in Appendix \ref{Nonlocal_Adiabatic}, we are defining our tensors with the built in factor of $i^L$ so that $\tilde{Y}_{\ell_i}=i^{\ell_i}Y_{\ell_i}$ with $Y_{\ell_i}$ defined on p.133, Eq.(1), of \cite{Varshalovich_Book}. As a reminder, $\phi_{j_i}(r_{np})$ is the radial wave function for the bound state, and $j_i$ results from coupling the orbital motion of the deuteron bound state with the spin of the neutron. $\chi_{L_i J_{p_i}}(R_{dA})$ is the radial wave function for the deuteron scattering state, and $J_{p_i}$ results from coupling the spin of the deuteron, $J_d=1$ to the orbital motion between the deuteron and the target.

The incoming distored wave should depend only on the projections of the projectile and target, and reduce to a plane wave in the limit of zero potential. Therefore, we multiply Eq.(\ref{eqD:wf2}) by the incoming coefficient $\frac{4\pi}{k_i}i^{L_i}e^{i\sigma_{L_i}}\sum_{M'_i}\tilde{Y}^*_{L_i M'_i}(\hat{k}_i)C_{J_d M_d L_i M'_i}^{J_{P_i} M_{P_i}}$ giving us:

\begin{eqnarray}\label{eqD:wf3}
\Psi_{\ell_i j_i}^{M_d \mu_A}&=&\frac{4\pi}{k_i}\sum_{L_i J_{P_i}}i^{L_i}e^{i\sigma_{L_i}}\Xi_{I_A \mu_A}(\xi_A)\phi_{j_i}(r_{np})\frac{\chi_{L_i J_{p_i}}(R_{dA})}{R_{dA}}\sum_{M'_i M_{P_i}}\tilde{Y}^*_{L_i M'_i}(\hat{k}_i)C_{J_d M_d L_i M'_i}^{J_{P_i} M_{P_i}} \nonumber \\
&\phantom{=}& \times \ \left\{\left\{\Xi_{I_p}(\xi_p) \otimes \left\{\tilde{Y}_{\ell_i}(\hat{r}_{np}) \otimes \Xi_{I_n}(\xi_n)\right\}_{j_i} \right\}_{J_d}\otimes \tilde{Y}_{L_i}(\hat{R}_{dA})\right\}_{J_{P_i} M_{P_i}}.
\end{eqnarray}

\noindent Now, we make the following substitutions using Eq.(13), p.132, and Eq.(10), p.245, of \cite{Varshalovich_Book}

\begin{eqnarray}
\tilde{Y}^*_{L_i M'_i}(\hat{k}_i)&=&(-)^{L_i+ M'_i}\tilde{Y}_{L_i, -M'_i}(\hat{k}_i) \nonumber \\
C_{J_d M_d L_i M'_i}^{J_{P_i} M_{P_i}}&=&(-)^{L_i+M'_i}\frac{\hat{J}_{P_i}}{\hat{J}_d}C_{L_i,-M'_i J_{P_i} M_{P_i}}^{J_d M_d},
\end{eqnarray}

\noindent and insert $(-)^{2(L_i+M'_i)}=1$, Eq.(\ref{eqD:wf3}) becomes:

\begin{eqnarray}\label{eqD:wf5}
|\Psi_{i}^{M_d \mu_A}\rangle &=&\frac{4\pi}{k_i}\sum_{L_i J_{P_i}}i^{L_i}e^{i\sigma_{L_i}}\Xi_{I_A \mu_A}(\xi_A)\phi_{j_i}(r_{np})\frac{\chi_{L_i J_{p_i}}(R_{dA})}{R_{dA}}\frac{\hat{J}_{P_i}}{\hat{J}_d} \nonumber \\
&\phantom{=}& \times \ \sum_{M'_i M_{P_i}}C_{L_i,-M'_i J_{P_i} M_{P_i}}^{J_d M_d}\tilde{Y}_{L_i, -M'_i}(\hat{k}_i) \nonumber \\
&\phantom{=}& \times \ \left\{\left\{\Xi_{I_p}(\xi_p) \otimes \left\{\tilde{Y}_{\ell_i}(\hat{r}_{np}) \otimes \Xi_{I_n}(\xi_n)\right\}_{j_i} \right\}_{J_d}\otimes \tilde{Y}_{L_i}(\hat{R}_{dA})\right\}_{J_{P_i} M_{P_i}} \nonumber \\
&=&\frac{4\pi}{k_i}\sum_{L_i J_{P_i}}i^{L_i}e^{i\sigma_{L_i}}\Xi_{I_A \mu_A}(\xi_A)\phi_{j_i}(r_{np})\frac{\chi_{L_i J_{p_i}}(R_{dA})}{R_{dA}}\frac{\hat{J}_{P_i}}{\hat{J}_d} \nonumber \\
&\phantom{=}& \times \  \left\{\tilde{Y}_{L_i}(\hat{k}_i)\otimes \left\{\left\{\Xi_{I_p}(\xi_p) \otimes \left\{\tilde{Y}_{\ell_i}(\hat{r}_{np}) \otimes \Xi_{I_n}(\xi_n)\right\}_{j_i} \right\}_{J_d} \right. \right. \nonumber \\
&\phantom{=}& \left. \left. \ \otimes \tilde{Y}_{L_i}(\hat{R}_{dA})\right\}_{J_{P_i}}\right\}_{J_d M_d} \nonumber \\
&=&\Xi_{I_A \mu_A}(\xi_A)\phi_{j_i}(r_{np})\chi_i^{(+)}(\textbf{k}_i,\textbf{r}_{np},\textbf{R}_{dA},\xi_p,\xi_n).
\end{eqnarray}

\noindent The partial wave for the incoming distorted wave is written as:

\begin{eqnarray}
&\phantom{=}& \chi_i^{(+)}(\textbf{k}_i,\textbf{r}_{np},\textbf{R}_{dA},\xi_p,\xi_n)=\frac{4\pi}{k_i}\sum_{L_i J_{P_i}}i^{L_i}e^{i\sigma_{L_i}}\frac{\hat{J}_{P_i}}{\hat{J}_d}\frac{\chi_{L_i J_{p_i}}(R_{dA})}{R_{dA}} \\
&\phantom{=}& \times \ \left\{\tilde{Y}_{L_i}(\hat{k}_i)\otimes \left\{\left\{\Xi_{I_p}(\xi_p) \otimes \left\{\tilde{Y}_{\ell_i}(\hat{r}_{np}) \otimes \Xi_{I_n}(\xi_n)\right\}_{j_i} \right\}_{J_d}\otimes \tilde{Y}_{L_i}(\hat{R}_{dA})\right\}_{J_{P_i}}\right\}_{J_d M_d}. \nonumber
\end{eqnarray}

\noindent The wave function for relative motion between $p+B$ is given by

\begin{eqnarray}
\Psi_{\ell_f j_f}&=&\left\{\Xi_{I_A}(\xi_A)\otimes\Phi_{j_f}(\textbf{r}_{nA},\xi_n) \right\}_{J_B M_B} \nonumber \\
&\phantom{=}& \times \ \sum_{L_f} \sum_{J_{P_f} M_{P_f}} \left\{\Xi_{I_p}(\xi_p) \otimes \tilde{Y}_{L_f}(\hat{R}_{pB}) \right\}_{J_{P_f} M_{P_f}}\frac{\chi_{L_f J_{P_f}}(R_{pB})}{R_{pB}} \nonumber \\
&=&\left\{\Xi_{I_A}(\xi_A)\otimes \left\{ \tilde{Y}_{\ell_f}(\hat{r}_{nA})\otimes \Xi_{I_n}(\xi_n) \right\}_{j_f}\right\}_{J_B M_B}\phi_{j_f}(r_{nA}) \nonumber \\
&\phantom{=}& \times \ \sum_{L_f} \sum_{J_{P_f} M_{P_f}}\left\{\Xi_{I_p}(\xi_p) \otimes \tilde{Y}_{L_f}(\hat{R}_{pB}) \right\}_{J_{P_f} M_{P_f}}\frac{\chi_{L_f J_{P_f}}(R_{pB})}{R_{pB}}. \nonumber \\
\end{eqnarray}

\noindent In this equation $\ell_f$ is the orbital angular momentum between the target and the bound neutron, $j_f$ is the quantum number resulting from coupling $\ell_f$ to the spin of the target, $I_A$. The total angular momentum of the target is given by $J_B$ and results from coupling $j_f$ to $I_A$. The orbital angular momentum between the proton and the target is given by $L_f$, and the total angular momentum of the projectile, $J_{p_f}$ results from coupling $L_f$ to the spin of the proton, $I_p$. 

As we did for the entrance channel, we need to multiply the exit channel wave function by the outgoing coefficient: $\frac{4\pi}{k_f}i^{L_f}e^{i\sigma_{L_f}}\sum_{M'_f}\tilde{Y}^*_{L_f M'_f}(\hat{k}_f)C_{I_p \mu_p L_f M'_f}^{J_{P_f} M_{P_f}}$ so that the remaining quantum numbers are for the projections of the projectile and target in the exit channel:

\begin{eqnarray}
|\Psi_{\ell_f j_f}^{\mu_p M_B}\rangle &=&\frac{4\pi}{k_f}\left\{\Xi_{I_A}(\xi_A)\otimes \left\{ \tilde{Y}_{\ell_f}(\hat{r}_{nA})\otimes \Xi_{I_n}(\xi_n) \right\}_{j_f}\right\}_{J_B M_B} \nonumber \\
&\phantom{=}& \times \ \phi_{j_f}(r_{nA})\sum_{L_f J_{P_f}}i^{L_f}e^{i\sigma_{L_f}}\frac{\chi_{L_f J_{P_f}}(R_{pB})}{R_{pB}}  \nonumber \\
&\phantom{=}& \times \ \sum_{M'_f M_{P_f}}Y^*_{L_f M'_f}(\hat{k}_f)C_{I_p \mu_p L_f M'_f}^{J_{P_f}M_{P_f}}\left\{\Xi_{I_p}(\xi_p) \otimes \tilde{Y}_{L_f}(\hat{R}_{pB}) \right\}_{J_{P_f} M_{P_f}}. \nonumber \\
\end{eqnarray}

\noindent We follow the same steps as before, using Eq.(13), p.132, and Eq.(10), p.245, of \cite{Varshalovich_Book}

\begin{eqnarray}
Y^*_{L_f M'_f}(\hat{k}_f)&=&(-)^{L_f+ M'_f}\tilde{Y}_{L_f, -M'_f}(\hat{k}_f) \nonumber \\
C_{I_p \mu_p L_f M'_f}^{J_{P_f}M_{P_f}}&=&(-)^{L_f+M'_f}\frac{\hat{J}_{P_f}}{\hat{I}_p}C_{L_f,-M'_f J_{P_f} M_{P_f}}^{I_p \mu_p},
\end{eqnarray}

\noindent and use $(-)^{2(L_f+M'_f)}=1$. This results in: 

\begin{eqnarray}
|\Psi_{f}^{\mu_p M_B}\rangle &=&\frac{4\pi}{k_f}\left\{\Xi_{I_A}(\xi_A)\otimes \left\{ \tilde{Y}_{\ell_f}(\hat{r}_{nA})\otimes \Xi_{I_n}(\xi_n) \right\}_{j_f}\right\}_{J_B M_B}\phi_{j_f}(r_{nA})\sum_{L_f J_{P_f}}i^{L_f}e^{i\sigma_{L_f}} \nonumber \\
&\phantom{=}& \times \ \frac{\chi_{L_f J_{P_f}}(R_{pB})}{R_{pB}}\frac{\hat{J}_{P_f}}{\hat{I}_p}\sum_{M'_f M_{P_f}}C_{L_f,-M'_f J_{P_f} M_{P_f}}^{I_p \mu_p}\tilde{Y}_{L_f, -M'_f}(\hat{k}_f) \nonumber \\
&\phantom{=}& \times \ \left\{\Xi_{I_p}(\xi_p) \otimes \tilde{Y}_{L_f}(\hat{R}_{pB}) \right\}_{J_{P_f} M_{P_f}} \nonumber \\
&=&\frac{4\pi}{k_f}\left\{\Xi_{I_A}(\xi_A)\otimes \left\{ \tilde{Y}_{\ell_f}(\hat{r}_{nA})\otimes \Xi_{I_n}(\xi_n) \right\}_{j_f}\right\}_{J_B M_B}\phi_{j_f}(r_{nA})\sum_{L_f J_{P_f}}i^{L_f}e^{i\sigma_{L_f}} \nonumber \\
&\phantom{=}& \times \  \frac{\chi_{L_f J_{P_f}}(R_{pB})}{R_{pB}}\frac{\hat{J}_{P_f}}{\hat{I}_p}\left\{\tilde{Y}_{L_f}(\hat{k}_f)\otimes \left\{\Xi_{I_p}(\xi_p) \otimes \tilde{Y}_{L_f}(\hat{R}_{pB}) \right\}_{J_{P_f}} \right\}_{I_p \mu_p} \nonumber \\
&=&\left\{\Xi_{I_A}(\xi_A)\otimes \left\{ \tilde{Y}_{\ell_f}(\hat{r}_{nA})\otimes \Xi_{I_n}(\xi_n) \right\}_{j_f}\right\}_{J_B M_B}\phi_{j_f}(r_{nA})\chi_f^{(+)}(\textbf{k}_f,\textbf{R}_{pB},\xi_p). \nonumber \\
\end{eqnarray}

\noindent In the T-Matrix, Eq.(\ref{eqD:tmatrix}) the exit channel appears as a bra:

\begin{eqnarray}\label{eqD:wf11}
\langle \Psi_{f}^{\mu_p M_B}|&=&\left\{\Xi_{I_A}(\xi_A)\otimes \left\{ \tilde{Y}_{\ell_f}(\hat{r}_{nA})\otimes \Xi_{I_n}(\xi_n) \right\}_{j_f}\right\}^*_{J_B M_B}\phi_{j_f}(r_{nA})\chi_f^{(-)*}(\textbf{k}_f,\textbf{R}_{pB}), \nonumber \\
\end{eqnarray}

\noindent where the outgoing distorted wave $\chi^{(-)}(\textbf{k},\textbf{R})$ is the time reverse of $\chi^{(+)}$, so that $\chi^{(-)}(\textbf{k},\textbf{R})=\chi^{(+)}(-\textbf{k},\textbf{R})^*$. Therefore, to make this more explicit we use Eq.(2), p.141, of \cite{Varshalovich_Book}

\begin{eqnarray}
\langle \chi^{(-)}(\textbf{k},\textbf{R})|&=&\chi^{(-)*}(\textbf{k},\textbf{R}) \nonumber \\
&=&\left(\chi^{(+)}(-\textbf{k},\textbf{R})^*\right)^* \nonumber \\
&=&\chi^{(+)}(-\textbf{k},\textbf{R}) \nonumber \\
&=&(-)^{L}\chi^{(+)}(\textbf{k},\textbf{R}),
\end{eqnarray}

\noindent where $\textbf{k}\rightarrow -\textbf{k}$ gives a factor of $(-)^{L}$ from the spherical harmonics, as seen in Eq.(2), p.141, of \cite{Varshalovich_Book}, and the two complex conjugations cancel. 

The incoming and outgoing distorted waves are given by 

\begin{eqnarray}
&\phantom{=}& \chi_i^{(+)}(\textbf{k}_i,\textbf{r}_{np},\textbf{R}_{dA},\xi_p,\xi_n)=\frac{4\pi}{k_i\hat{J}_d}\sum_{L_i J_{P_i}}i^{L_i}e^{i\sigma_{L_i}}\hat{J}_{P_i}\frac{\chi_{L_i J_{p_i}}(R_{dA})}{R_{dA}} \\
&\phantom{=}& \times \ \left\{\tilde{Y}_{L_i}(\hat{k}_i)\otimes \left\{\left\{\Xi_{I_p}(\xi_p) \otimes \left\{\tilde{Y}_{\ell_i}(\hat{r}_{np}) \otimes \Xi_{I_n}(\xi_n)\right\}_{j_i} \right\}_{J_d}\otimes \tilde{Y}_{L_i}(\hat{R}_{dA})\right\}_{J_{P_i}}\right\}_{J_d M_d} \nonumber 
\end{eqnarray}

\noindent and,

\begin{eqnarray}
&\phantom{=}& \chi_f^{(-)*}(\textbf{k}_f,\textbf{R}_{pB},\xi_p)=\frac{4\pi}{k_f\hat{I}_p}\sum_{L_f J_{P_f}}i^{-L_f}e^{i\sigma_{L_f}}\hat{J}_{P_f} \frac{\chi_{L_f J_{P_f}}(R_{pB})}{R_{pB}} \nonumber \\
&\phantom{=}& \times \ \left\{\tilde{Y}_{L_f}(\hat{k}_f)\otimes \left\{\Xi_{I_p}(\xi_p) \otimes \tilde{Y}_{L_f}(\hat{R}_{pB}) \right\}_{J_{P_f}} \right\}_{I_p \mu_p}.
\end{eqnarray}

\noindent In the adiabatic theory, $\chi_{L_i J_{P_i}}(R_{dA})$ satisfies the equation

\begin{eqnarray}
\left[-\frac{\hbar^2}{2\mu_i}\left(\frac{\partial^2}{\partial R_{dA}^2}-\frac{L_i(L_i+1)}{R_{dA}^2} \right)+U^{ad}+V^{SO}_{1 L_i J_{P_i}}+V_C(R_{dA})-E \right]\chi_{L_i J_{P_i}}(R_{dA})=0 \nonumber \\
\end{eqnarray}

\noindent where $U^{ad}$ is the adiabatic potential, and $V^{SO}_{1 L_i J_{P_i}}$ is the spin-orbit potential. The function $\chi_{L_f J_{P_f}}(R_{pB})$ satisfies a single channel optical model equation:

\begin{eqnarray}
\left[-\frac{\hbar^2}{2\mu_f}\left(\frac{\partial^2}{\partial R_{pB}^2}-\frac{L_f(L_f+1)}{R_{pB}^2} \right)+U^{pB}+V^{SO}_{I_p L_f J_{P_f}}+V_C(R_{pB})-E \right]\chi_{L_f J_{P_f}}(R_{pB})=0 \nonumber \\
\end{eqnarray}

\noindent with $U^{pB}$ being a nucleon optical potential. In these equations $\mu_i$ and $\mu_f$ are the reduced mass in the initial and final states, not to be confused with spin projections $\mu_A$, $\mu_p$, and $\mu_n$. Also, $U^{ad}$ and $U^{pB}$ can be either local or nonlocal. 

As mentioned in Sec. \ref{Sec:Three_Body_Tmatrix}, the scattering amplitude is related to the T-matrix by

\begin{eqnarray}
f_{\mu_A M_d \mu_p M_B}(\textbf{k}_f,\textbf{k}_i)&=&-\frac{\mu_f}{2\pi\hbar^2}\tilde{T}_{\mu_A M_d \mu_p M_B}(\textbf{k}_f,\textbf{k}_i) \nonumber \\
&=&-\frac{\mu_f}{2\pi\hbar^2}\sqrt{\frac{v_f}{v_i}}T_{\mu_A M_d \mu_p M_B}(\textbf{k}_f,\textbf{k}_i) \nonumber \\
&=&-\frac{\mu_f}{2\pi\hbar^2}\sqrt{\frac{\frac{\hbar k_f}{\mu_f}}{\frac{\hbar k_i}{\mu_i}}}\langle \Psi^{\mu_p M_B}_f |V_{np}|\Psi^{\mu_A M_d}_i\rangle.
\end{eqnarray}

\noindent The differential cross section is obtained, by averaging the mod of the scattering amplitude squared over initial states, and summing over final $m$-states:

\begin{eqnarray}
\frac{d\sigma}{d\Omega}&=&\frac{1}{\hat{J}^2_d \hat{J}^2_A}\sum_{\mu_A M_d \mu_p M_B}\left|f_{\mu_A M_d \mu_p M_B}(\textbf{k}_f,\textbf{k}_i) \right|^2 \nonumber \\
&=&=\frac{k_f}{k_i}\frac{\mu_i \mu_f}{4\pi^2\hbar^4}\frac{1}{\hat{J}^2_d \hat{J}^2_A}\sum_{\mu_A M_d M_B \mu_p}\left|\langle \Psi^{\mu_p M_B}_f |V_{np}|\Psi^{\mu_A M_d}_i\rangle \right|^2.
\end{eqnarray}

\noindent We now put Eq.(\ref{eqD:wf5}) and Eq.(\ref{eqD:wf11}) into $\langle \Psi^{\mu_p M_B}_f |V_{np}|\Psi^{\mu_A M_d}_i\rangle$:

\begin{eqnarray}
&\phantom{=}&\langle \Psi^{\mu_p M_B}_f |V_{np}|\Psi^{\mu_A M_d}_i\rangle=\int \left\{\Xi_{I_A}(\xi_A)\otimes \left\{ \tilde{Y}_{\ell_f}(\hat{r}_{nA})\otimes \Xi_{I_n}(\xi_n) \right\}_{j_f}\right\}^*_{J_B M_B}\phi_{j_f}(r_{nA}) \nonumber \\
&\phantom{=}& \times \ \frac{4\pi}{k_f\hat{I}_p}\sum_{L_f J_{P_f}}i^{-L_f}e^{i\sigma_{L_f}}\hat{J}_{P_f} \frac{\chi_{L_f J_{P_f}}(R_{pB})}{R_{pB}} \nonumber \\
&\phantom{=}& \times \ \left\{\tilde{Y}_{L_f}(\hat{k}_f)\otimes \left\{\Xi_{I_p}(\xi_p) \otimes \tilde{Y}_{L_f}(\hat{R}_{pB}) \right\}_{J_{P_f}} \right\}_{I_p \mu_p} \\
&\phantom{=}& \times \ V(r_{np})\Xi_{I_A \mu_A}(\xi_A)\phi_{j_i}(r_{np})\frac{4\pi}{k_i\hat{J}_d}\sum_{L_i J_{P_i}}i^{L_i}e^{i\sigma_{L_i}}\hat{J}_{P_i}\frac{\chi_{L_i J_{p_i}}(R_{dA})}{R_{dA}} \nonumber \\
&\phantom{=}& \times \ \left\{\tilde{Y}_{L_i}(\hat{k}_i)\otimes \left\{\left\{\Xi_{I_p}(\xi_p) \otimes \left\{\tilde{Y}_{\ell_i}(\hat{r}_{np}) \otimes \Xi_{I_n}(\xi_n)\right\}_{j_i} \right\}_{J_d}\otimes \tilde{Y}_{L_i}(\hat{R}_{dA})\right\}_{J_{P_i}}\right\}_{J_d M_d} \nonumber \\
&\phantom{=}& \times \ d\textbf{R}_{pB}d\textbf{r}_{nA}d\xi_n d\xi_p d\xi_A. \nonumber
\end{eqnarray}

Breaking the coupling between the target and the final bound state, and grouping the two spin functions for the target together, we obtain:

\begin{eqnarray}
&\phantom{=}&\langle \Psi^{\mu_p M_B}_f |V_{np}|\Psi^{\mu_A M_d}_i\rangle=\left(\sum_{\mu'_A}\int \Xi^*_{I_A \mu'_A}(\xi_A)\Xi_{I_A \mu_A}(\xi_A) d\xi_A \right) \nonumber \\
&\phantom{=}& \times \ \int \sum_{m_f}C_{I_A \mu'_A j_f m_f}^{J_B M_B}\left\{ \tilde{Y}_{\ell_f}(\hat{r}_{nA})\otimes \Xi_{I_n}(\xi_n) \right\}^*_{j_f m_f}\phi_{j_f}(r_{nA}) \nonumber \\
&\phantom{=}& \times \ \frac{4\pi}{k_f\hat{I}_p}\sum_{L_f J_{P_f}}i^{-L_f}e^{i\sigma_{L_f}}\hat{J}_{P_f} \frac{\chi_{L_f J_{P_f}}(R_{pB})}{R_{pB}} \nonumber \\
&\phantom{=}& \times \ \left\{\tilde{Y}_{L_f}(\hat{k}_f)\otimes \left\{\Xi_{I_p}(\xi_p) \otimes \tilde{Y}_{L_f}(\hat{R}_{pB}) \right\}_{J_{P_f}} \right\}_{I_p \mu_p}  \\
&\phantom{=}& \times \ V(r_{np})\phi_{j_i}(r_{np})\frac{4\pi}{k_i\hat{J}_d}\sum_{L_i J_{P_i}}i^{L_i}e^{i\sigma_{L_i}}\hat{J}_{P_i}\frac{\chi_{L_i J_{p_i}}(R_{dA})}{R_{dA}} \nonumber \\
&\phantom{=}& \times \ \left\{\tilde{Y}_{L_i}(\hat{k}_i)\otimes \left\{\left\{\Xi_{I_p}(\xi_p) \otimes \left\{\tilde{Y}_{\ell_i}(\hat{r}_{np}) \otimes \Xi_{I_n}(\xi_n)\right\}_{j_i} \right\}_{J_d}\otimes \tilde{Y}_{L_i}(\hat{R}_{dA})\right\}_{J_{P_i}}\right\}_{J_d M_d} \nonumber \\
&\phantom{=}& \times \ d\textbf{R}_{pB}d\textbf{r}_{nA}d\xi_n d\xi_p. \nonumber 
\end{eqnarray}

\noindent The integral in the first line gives $\delta_{\mu'_A \mu_A}$. Performing the sum over $\mu'_A$ provides:

\begin{eqnarray}
&\phantom{=}&\langle \Psi^{\mu_p M_B}_f |V_{np}|\Psi^{\mu_A M_d}_i\rangle=\int \sum_{m_f}C_{I_A \mu_A j_f m_f}^{J_B M_B}\left\{ \tilde{Y}_{\ell_f  }(\hat{r}_{nA})\otimes \Xi_{I_n}(\xi_n) \right\}^*_{j_f m_f}\phi_{j_f}(r_{nA}) \nonumber \\
&\phantom{=}& \times \ \frac{4\pi}{k_f\hat{I}_p}\sum_{L_f J_{P_f}}i^{-L_f}e^{i\sigma_{L_f}}\hat{J}_{P_f} \frac{\chi_{L_f J_{P_f}}(R_{pB})}{R_{pB}} \nonumber \\
&\phantom{=}& \times \ \left\{\tilde{Y}_{L_f}(\hat{k}_f)\otimes \left\{\Xi_{I_p}(\xi_p) \otimes \tilde{Y}_{L_f}(\hat{R}_{pB}) \right\}_{J_{P_f}} \right\}_{I_p \mu_p} \\
&\phantom{=}& \times \ V(r_{np})\phi_{j_i}(r_{np})\frac{4\pi}{k_i\hat{J}_d}\sum_{L_i J_{P_i}}i^{L_i}e^{i\sigma_{L_i}}\hat{J}_{P_i}\frac{\chi_{L_i J_{p_i}}(R_{dA})}{R_{dA}} \nonumber \\
&\phantom{=}& \times \ \left\{\tilde{Y}_{L_i}(\hat{k}_i)\otimes \left\{\left\{\Xi_{I_p}(\xi_p) \otimes \left\{\tilde{Y}_{\ell_i}(\hat{r}_{np}) \otimes \Xi_{I_n}(\xi_n)\right\}_{j_i} \right\}_{J_d}\otimes \tilde{Y}_{L_i}(\hat{R}_{dA})\right\}_{J_{P_i}}\right\}_{J_d M_d} \nonumber \\
&\phantom{=}& \times \ d\textbf{R}_{pB}d\textbf{r}_{nA}d\xi_n d\xi_p. \nonumber
\end{eqnarray}

\noindent We now couple the following tensors together:

\begin{eqnarray}
&\phantom{=}& \left\{\tilde{Y}_{L_f}(\hat{k}_f)\otimes \left\{\Xi_{I_p}(\xi_p) \otimes \tilde{Y}_{L_f}(\hat{R}_{pB}) \right\}_{J_{P_f}} \right\}_{I_p \mu_p} \nonumber \\
&\phantom{=}& \times \ \left\{\tilde{Y}_{L_i}(\hat{k}_i)\otimes \left\{\left\{\Xi_{I_p}(\xi_p) \otimes \left\{\tilde{Y}_{\ell_i}(\hat{r}_{np}) \otimes \Xi_{I_n}(\xi_n)\right\}_{j_i} \right\}_{J_d}\otimes \tilde{Y}_{L_i}(\hat{R}_{dA})\right\}_{J_{P_i}}\right\}_{J_d M_d} \nonumber \\
&=&\sum_{Q M_Q}C_{I_p \mu_p J_d M_d}^{Q M_Q}\left\{\left\{\tilde{Y}_{L_f}(\hat{k}_f)\otimes \left\{\Xi_{I_p}(\xi_p) \otimes \tilde{Y}_{L_f}(\hat{R}_{pB}) \right\}_{J_{P_f}} \right\}_{I_p} \right. \\
&\phantom{=}& \left. \otimes \left\{\tilde{Y}_{L_i}(\hat{k}_i)\otimes \left\{\left\{\Xi_{I_p}(\xi_p) \otimes \left\{\tilde{Y}_{\ell_i}(\hat{r}_{np}) \otimes \Xi_{I_n}(\xi_n)\right\}_{j_i} \right\}_{J_d}\otimes \tilde{Y}_{L_i}(\hat{R}_{dA})\right\}_{J_{P_i}}\right\}_{J_d} \right\}_{Q M_Q}, \nonumber
\end{eqnarray}

\noindent to obtain:

\begin{eqnarray}\label{eqD:MatrixElement23}
&\phantom{=}&\langle \Psi^{\mu_p M_B}_f |V_{np}|\Psi^{\mu_A M_d}_i\rangle=\frac{(4\pi)^2}{k_ik_f\hat{J}_d\hat{I}_p}\int \sum_{m_f}\sum_{Q M_Q}C_{I_A \mu_A j_f m_f}^{J_B M_B}C_{I_p \mu_p J_d M_d}^{Q M_Q} \nonumber \\
&\phantom{=}& \times \ \left\{ \tilde{Y}_{\ell_f  }(\hat{r}_{nA})\otimes \Xi_{I_n}(\xi_n) \right\}^*_{j_f m_f}\phi_{j_f}(r_{nA})\sum_{L_i J_{P_i}}\sum_{L_f J_{P_f}} \nonumber \\
&\phantom{=}& \times \ i^{-L_f}e^{i\sigma_{L_f}}\hat{J}_{P_f} \frac{\chi_{L_f J_{P_f}}(R_{pB})}{R_{pB}}V(r_{np})\phi_{j_i}(r_{np})i^{L_i}e^{i\sigma_{L_i}}\hat{J}_{P_i}\frac{\chi_{L_i J_{p_i}}(R_{dA})}{R_{dA}} \nonumber \\
&\phantom{=}& \times \ \left\{\left\{\tilde{Y}_{L_f}(\hat{k}_f)\otimes \left\{\Xi_{I_p}(\xi_p) \otimes \tilde{Y}_{L_f}(\hat{R}_{pB}) \right\}_{J_{P_f}} \right\}_{I_p} \right. \nonumber \\
&\phantom{=}& \left. \otimes \ \left\{\tilde{Y}_{L_i}(\hat{k}_i)\otimes \left\{\left\{\Xi_{I_p}(\xi_p) \otimes \left\{\tilde{Y}_{\ell_i}(\hat{r}_{np}) \otimes \Xi_{I_n}(\xi_n)\right\}_{j_i} \right\}_{J_d}\otimes \tilde{Y}_{L_i}(\hat{R}_{dA})\right\}_{J_{P_i}}\right\}_{J_d} \right\}_{Q M_Q} \nonumber \\
&\phantom{=}& \times \ d\textbf{R}_{pB}d\textbf{r}_{nA}d\xi_n d\xi_p. \nonumber \\
\end{eqnarray}

\noindent We can rewrite Eq.(\ref{eqD:MatrixElement23}) in a more compact form:

\begin{eqnarray}
\langle \Psi^{\mu_p M_B}_f |V_{np}|\Psi^{\mu_A M_d}_i\rangle&=&\sum_{m_f}\sum_{Q M_Q}C_{I_A \mu_A j_f m_f}^{J_B M_B}C_{I_p \mu_p J_d M_d}^{Q M_Q}T_{Q M_Q m_f},
\end{eqnarray}

\noindent so that

\begin{eqnarray}\label{eqD:tmatrix25}
&\phantom{=}& \sum_{\mu_A M_d M_B \mu_p}|\langle \Psi^{\mu_p M_B}_f |V_{np}|\Psi^{\mu_A M_d}_i\rangle|^2  \nonumber \\
&=&\sum_{\mu_A M_d M_B \mu_p}\sum_{m_f Q M_Q}\sum_{m'_f Q' M'_Q}C_{I_A \mu_A j_f m_f}^{J_B M_B}C_{I_A \mu_A j_f m'_f}^{J_B M_B}C_{I_p \mu_p J_d M_d}^{Q M_Q}C_{I_p \mu_p J_d M_d}^{Q' M'_Q} \nonumber \\
&\phantom{=}& \times \ T_{Q M_Q m_f}T^*_{Q' M'_Q m'_f}.
\end{eqnarray}

\noindent Now we consider the first pair of Clebsch-Gordans and use Eq.(10), p.245, of \cite{Varshalovich_Book},

\begin{eqnarray}
C_{I_A \mu_A j_f m_f}^{J_B M_B}C_{I_A \mu_A j_f m'_f}^{J_B M_B}&=&\left((-)^{I_A-\mu_A}\frac{\hat{J}_B}{\hat{j}_f}\right)^2C_{I_A \mu_A J_B, -M_B}^{j_f, -m_f}C_{I_A \mu_A J_B,-M_B}^{j_f, -m'_f}. \nonumber \\
\end{eqnarray}

\noindent This together with Eq.(8), p.236, of \cite{Varshalovich_Book} allows us to simplify Eq.(\ref{eqD:tmatrix25}) to

\begin{eqnarray}
&\phantom{=}&\sum_{\mu_A M_d M_B \mu_p}|\langle \Psi^{\mu_p M_B}_f |V_{np}|\Psi^{\mu_A M_d}_i\rangle|^2 \nonumber \\
&=&\frac{\hat{J}_B^2}{\hat{j}_f^2}\sum_{m_f Q M_Q}\sum_{m'_f Q' M'_Q}\left(\sum_{\mu_A M_B}C_{I_A \mu_A J_B, -M_B}^{j_f, -m_f}C_{I_A \mu_A J_B,-M_B}^{j_f, -m'_f}\right) \nonumber \\
&\phantom{=}& \times \left(\sum_{\mu_p M_d} C_{I_p \mu_p J_d M_d}^{Q M_Q}C_{I_p \mu_p J_d M_d}^{Q' M'_Q}\right)T_{Q M_Q m_f}T^*_{Q' M'_Q m'_f} \nonumber \\
&=&\frac{\hat{J}_B^2}{\hat{j}_f^2}\sum_{m_f Q M_Q}\sum_{m'_f Q' M'_Q}\delta_{m_f m'_f}\delta_{Q Q'}\delta_{M_Q M'_Q}T_{Q M_Q m_f}T^*_{Q' M'_Q m'_f} \nonumber \\
&=&\frac{\hat{J}_B^2}{\hat{j}_f^2}\sum_{m_f Q M_Q}T_{Q M_Q m_f}T^*_{Q M_Q m_f},
\end{eqnarray}

\noindent and then the differential cross section is

\begin{eqnarray}\label{eqD:CrossSection28}
\frac{d\sigma}{d\Omega}&=&\frac{k_f}{k_i}\frac{\mu_i \mu_f}{4\pi^2\hbar^4}\frac{1}{\hat{J}^2_d \hat{J}^2_A}\sum_{\mu_A M_d M_B \mu_p}\left|\langle \Psi^{\mu_p M_B}_f |V_{np}|\Psi^{\mu_A M_d}_i\rangle \right|^2 \nonumber \\
&=&\frac{k_f}{k_i}\frac{\mu_i \mu_f}{4\pi^2\hbar^4}\frac{\hat{J}_B^2}{\hat{J}^2_d \hat{J}^2_A \hat{j}_f^2}\sum_{m_f Q M_Q}T_{Q M_Q m_f}T^*_{Q M_Q m_f}.
\end{eqnarray}

\noindent In essence, our task is to work out, explicitly, $T_{Q M_Q m_f}$

\begin{eqnarray}\label{eqD:tmatrix29}
&\phantom{=}& T_{Q M_Q m_f}=\frac{(4\pi)^2}{k_ik_f\hat{J}_d\hat{I}_p}\sum_{L_i J_{P_i}}\sum_{L_f J_{P_f}}i^{L_i-L_f}e^{i(\sigma_{L_i}+\sigma_{L_f})}\hat{J}_{P_i}\hat{J}_{P_f} \nonumber \\
&\phantom{=}& \times \int \frac{\phi_{j_f}(r_{nA})\chi_{L_f J_{P_f}}(R_{pB})V(r_{np})\phi_{j_i}(r_{np})\chi_{L_i J_{p_i}}(R_{dA})}{R_{pB}R_{dA}}     \nonumber \\
&\phantom{=}& \times \ \left\{\left\{\tilde{Y}_{L_f}(\hat{k}_f)\otimes \left\{\Xi_{I_p}(\xi_p) \otimes \tilde{Y}_{L_f}(\hat{R}_{pB}) \right\}_{J_{P_f}} \right\}_{I_p} \right. \nonumber \\
&\phantom{=}& \left. \otimes \left\{\tilde{Y}_{L_i}(\hat{k}_i)\otimes \left\{\left\{\Xi_{I_p}(\xi_p) \otimes \left\{\tilde{Y}_{\ell_i}(\hat{r}_{np}) \otimes \Xi_{I_n}(\xi_n)\right\}_{j_i} \right\}_{J_d}\otimes \tilde{Y}_{L_i}(\hat{R}_{dA})\right\}_{J_{P_i}}\right\}_{J_d} \right\}_{Q M_Q} \nonumber \\
&\phantom{=}& \times \ \left\{ \tilde{Y}_{\ell_f  }(\hat{r}_{nA})\otimes \Xi_{I_n \mu_n}(\xi_n) \right\}^*_{j_f m_f} d\textbf{R}_{pB}d\textbf{r}_{nA}d\xi_n d\xi_p. \nonumber \\
\end{eqnarray}

\noindent Our strategy is to couple the spherical harmonics with the argument $\hat{k}$ together so we can pull them out of the integral. We want to couple the spherical harmonics with the arguments $\hat{r}$ and $\hat{R}$ together up to zero angular momentum so we can use symmetry to reduce the dimensionality of the angular integral. Also, we want to couple the spinors with common arguments up to zero angular momentum so we can integrate them out. This is detailed in the next few pages.

We can group the $\hat{k}$ spherical harmonics together right away. Let us introduce the definitions

\begin{eqnarray}
\mathcal{A}_{J_{P_f}}&=&\left\{\Xi_{I_p}(\xi_p) \otimes \tilde{Y}_{L_f}(\hat{R}_{pB}) \right\}_{J_{P_f}} \nonumber \\
\mathcal{B}_{J_{P_i}}&=&\left\{\left\{\Xi_{I_p}(\xi_p) \otimes \left\{\tilde{Y}_{\ell_i}(\hat{r}_{np}) \otimes \Xi_{I_n}(\xi_n)\right\}_{j_i} \right\}_{J_d}\otimes \tilde{Y}_{L_i}(\hat{R}_{dA})\right\}_{J_{P_i}}
\end{eqnarray}

\noindent so the first tensor Eq.(\ref{eqD:tmatrix29}) is

\begin{eqnarray}\label{eqD:tensor31}
&\phantom{=}&\left\{\left\{\tilde{Y}_{L_f}(\hat{k}_f)\otimes \mathcal{A}_{J_{P_f}} \right\}_{I_p}\otimes\left\{\tilde{Y}_{L_i}(\hat{k}_i)\otimes \mathcal{B}_{J_{P_i}} \right\}_{J_d} \right\}_{Q M_Q}=|L_f J_{P_f} (I_p) L_i J_{P_i} (J_d) Q M_Q \rangle \nonumber \\
&=&\sum_{gh}|L_f L_i (g) J_{P_f} J_{P_i} (h) Q M_Q\rangle\langle L_f L_i (g) J_{P_f} J_{P_i} (h) Q M_Q|L_f J_{P_f} (I_p) L_i J_{P_i} (J_d) Q M_Q \rangle \nonumber \\
&=&\sum_{gh}\langle L_f L_i (g) J_{P_f} J_{P_i} (h) Q M_Q|L_f J_{P_f} (I_p) L_i J_{P_i} (J_d) Q M_Q \rangle \nonumber \\
&\phantom{=}& \times \left\{\left\{\tilde{Y}_{L_f}(\hat{k}_f)\otimes \tilde{Y}_{L_i}(\hat{k}_i) \right\}_{g} \otimes \left\{\mathcal{A}_{J_{P_f}}\otimes \mathcal{B}_{J_{P_i}} \right\}_{h}\right\}_{Q M_Q},
\end{eqnarray}

\noindent where we used the definition of the 9j in Eq.(5), p.334, of \cite{Varshalovich_Book}:

\begin{eqnarray}
\langle j_1 j_2 (j_{12}) j_3 j_4 (j_{34}) jm| j_1 j_3 (j_{13})j_2 j_4 (j_{24}) j' m' \rangle &=& \delta_{jj'}\delta_{mm'}\hat{j}_{12}\hat{j}_{13}\hat{j}_{24}\hat{j}_{34}
\begin{Bmatrix}
j_1 & j_2 & j_{12} \\
j_3 & j_4 & j_{34} \\
j_{13} & j_{24} & j
\end{Bmatrix}. \nonumber \\
\end{eqnarray}

\noindent Inserting Eq.(\ref{eqD:tensor31}) into Eq.(\ref{eqD:tmatrix29}):

\begin{eqnarray}\label{eqD:tmatrix33}
T_{Q M_Q m_f}&=&\frac{(4\pi)^2}{k_ik_f\hat{J}_d\hat{I}_p}\sum_{L_i J_{P_i}}\sum_{L_f J_{P_f}}i^{L_i-L_f}e^{i(\sigma_{L_i}+\sigma_{L_f})}\hat{J}_{P_i}\hat{J}_{P_f} \nonumber \\
&\phantom{=}& \times \ \int \frac{\phi_{j_f}(r_{nA})\chi_{L_f J_{P_f}}(R_{pB})V(r_{np})\phi_{j_i}(r_{np})\chi_{L_i J_{p_i}}(R_{dA})}{R_{pB}R_{dA}}     \nonumber \\
&\phantom{=}& \times \ \sum_{g h}\langle L_f L_i (g) J_{P_f} J_{P_i} (h) Q M_Q|L_f J_{P_f} (I_p) L_i J_{P_i} (J_d) Q M_Q \rangle \nonumber \\
&\phantom{=}& \times \ \left\{\left\{\tilde{Y}_{L_f}(\hat{k}_f)\otimes \tilde{Y}_{L_i}(\hat{k}_i) \right\}_{g} \otimes \left\{\mathcal{A}_{J_{P_f}}\otimes \mathcal{B}_{J_{P_i}} \right\}_{h}\right\}_{Q M_Q}\nonumber \\
&\phantom{=}& \times \ \left\{ \tilde{Y}_{\ell_f  }(\hat{r}_{nA})\otimes \Xi_{I_n \mu_n}(\xi_n) \right\}^*_{j_f m_f} d\textbf{R}_{pB}d\textbf{r}_{nA}d\xi_n d\xi_p.
\end{eqnarray}

\noindent Now we consider the product $\left\{\mathcal{A}\otimes \mathcal{B}\right\}$:

\begin{eqnarray}\label{eqD:tensor34}
&\phantom{=}&\left\{\mathcal{A}_{J_{P_f}}\otimes \mathcal{B}_{J_{P_i}} \right\}_{h}\nonumber \\
&=&\left\{\left\{\Xi_{I_p}(\xi_p) \otimes \tilde{Y}_{L_f}(\hat{R}_{pB}) \right\}_{J_{P_f}} \right. \nonumber \\
&\phantom{=}& \left. \ \otimes \left\{\left\{\Xi_{I_p}(\xi_p) \otimes \left\{\tilde{Y}_{\ell_i}(\hat{r}_{np}) \otimes \Xi_{I_n}(\xi_n)\right\}_{j_i} \right\}_{J_d}\otimes \tilde{Y}_{L_i}(\hat{R}_{dA})\right\}_{J_{P_i}} \right\}_{h}  \nonumber \\
&=&|I_p L_f (J_{P_f}) J_d L_i (J_{P_i}) h m_h \rangle \nonumber \\
&=&\sum_{g'h'}|I_p J_d (g') L_f L_i (h') h m_h \rangle \langle I_p J_d (g') L_f L_i (h') h m_h|I_p L_f (J_{P_f}) J_d L_i (J_{P_i}) h m_h \rangle \nonumber \\
&=&\sum_{g'h'}\langle I_p J_d (g') L_f L_i (h') h m_h|I_p L_f (J_{P_f}) J_d L_i (J_{P_i}) h m_h \rangle \nonumber \\
&\phantom{=}& \times \ \left\{\left\{\Xi_{I_p}(\xi_p) \otimes \left\{\Xi_{I_p}(\xi_p) \otimes \left\{\tilde{Y}_{\ell_i}(\hat{r}_{np}) \otimes \Xi_{I_n}(\xi_n)\right\}_{j_i} \right\}_{J_d} \right\}_{g'} \right. \nonumber \\
&\phantom{=}& \left. \ \otimes \left\{\tilde{Y}_{L_f}(\hat{R}_{pB}) \otimes \tilde{Y}_{L_i}(\hat{R}_{dA}) \right\}_{h'} \right\}_h.  
\end{eqnarray}

\noindent Inserting Eq.(\ref{eqD:tensor34}) into Eq.(\ref{eqD:tmatrix33}) we arrive at:

\begin{eqnarray}\label{eqD:tmatrix35}
&\phantom{=}&T_{Q M_Q m_f}=\frac{(4\pi)^2}{k_ik_f\hat{J}_d\hat{I}_p}\sum_{L_i J_{P_i}}\sum_{L_f J_{P_f}}i^{L_i-L_f}e^{i(\sigma_{L_i}+\sigma_{L_f})}\hat{J}_{P_i}\hat{J}_{P_f} \nonumber \\
&\phantom{=}& \times \int \frac{\phi_{j_f}(r_{nA})\chi_{L_f J_{P_f}}(R_{pB})V(r_{np})\phi_{j_i}(r_{np})\chi_{L_i J_{p_i}}(R_{dA})}{R_{pB}R_{dA}}     \nonumber \\
&\phantom{=}& \times \ \sum_{g h}\langle L_f L_i (g) J_{P_f} J_{P_i} (h) Q M_Q|L_f J_{P_f} (I_p) L_i J_{P_i} (J_d) Q M_Q \rangle \nonumber \\
&\phantom{=}& \times \ \sum_{g'h'}\langle I_p J_d (g') L_f L_i (h') h m_h|I_p L_f (J_{P_f}) J_d L_i (J_{P_i}) h m_h \rangle  \nonumber \\
&\phantom{=}& \times \ \left\{\left\{\tilde{Y}_{L_f}(\hat{k}_f)\otimes \tilde{Y}_{L_i}(\hat{k}_i) \right\}_{g} \otimes \left\{\left\{\Xi_{I_p}(\xi_p) \otimes \left\{\Xi_{I_p}(\xi_p) \otimes \left\{\tilde{Y}_{\ell_i}(\hat{r}_{np}) \otimes \Xi_{I_n}(\xi_n)\right\}_{j_i} \right\}_{J_d} \right\}_{g'} \right. \right. \nonumber \\
&\phantom{=}& \left. \left. \otimes \left\{\tilde{Y}_{L_f}(\hat{R}_{pB}) \otimes \tilde{Y}_{L_i}(\hat{R}_{dA}) \right\}_{h'} \right\}_h  \right\}_{Q M_Q}\left\{ \tilde{Y}_{\ell_f  }(\hat{r}_{nA})\otimes \Xi_{I_n \mu_n}(\xi_n) \right\}^*_{j_f m_f} \nonumber \\
&\phantom{=}& \times \ d\textbf{R}_{pB}d\textbf{r}_{nA}d\xi_n d\xi_p. 
\end{eqnarray}

\noindent The following tensor in Eq.(\ref{eqD:tmatrix35}) can be simplified using, Eq.(27), p.64, and Eq.(8), p.70, of \cite{Varshalovich_Book}

\begin{eqnarray}\label{eqD:tensor36}
&\phantom{=}&\left\{\Xi_{I_p}(\xi_p) \otimes \left\{\Xi_{I_p}(\xi_p) \otimes \left\{\tilde{Y}_{\ell_i}(\hat{r}_{np}) \otimes \Xi_{I_n}(\xi_n)\right\}_{j_i} \right\}_{J_d} \right\}_{g'} \nonumber \\
&=&(-)^{I_p+J_d-g'}\left\{\left\{\Xi_{I_p}(\xi_p) \otimes \left\{\tilde{Y}_{\ell_i}(\hat{r}_{np}) \otimes \Xi_{I_n}(\xi_n)\right\}_{j_i} \right\}_{J_d} \otimes \Xi_{I_p}(\xi_p) \right\}_{g'} \nonumber \\
&=&(-)^{I_p+J_d-g'}(-)^{J_d+I_p+g'}\sum_{q}\hat{J}_d\hat{q}
\begin{Bmatrix}
I_p & j_i & J_d \\
g' & I_p & q
\end{Bmatrix} \nonumber \\
&\phantom{=}& \times \ \left\{ \left\{\tilde{Y}_{\ell_i}(\hat{r}_{np}) \otimes \Xi_{I_n}(\xi_n)\right\}_{j_i} \otimes \left\{\Xi_{I_p}(\xi_p)\otimes\Xi_{I_p}(\xi_p) \right\}_q \right\}_{g'}.
\end{eqnarray}

\noindent Since the spin functions must be coupled to zero angular momentum, this implies that $q=0$ and $g'=j_i$. Therefore, Eq.(\ref{eqD:tensor36}) simplifies to, with $(-)^{2J_d}=1$, and using Eq.(1), p.299, of \cite{Varshalovich_Book}

\begin{eqnarray}\label{eqD:tensor37}
&\phantom{=}&(-)^{2I_p}\hat{J}_d
\begin{Bmatrix}
I_p & j_i & J_d \\
j_i & I_p & 0
\end{Bmatrix}
\left\{ \left\{\tilde{Y}_{\ell_i}(\hat{r}_{np}) \otimes \Xi_{I_n}(\xi_n)\right\}_{j_i} \otimes \left\{\Xi_{I_p}(\xi_p)\otimes\Xi_{I_p}(\xi_p) \right\}_0 \right\}_{j_i} \nonumber \\
&=&(-)^{2I_p}(-)^{I_p+j_i+J_d}\frac{\hat{J}_d}{\hat{I}_p\hat{j}_i}
\left\{ \left\{\tilde{Y}_{\ell_i}(\hat{r}_{np}) \otimes \Xi_{I_n}(\xi_n)\right\}_{j_i} \otimes \left\{\Xi_{I_p}(\xi_p)\otimes\Xi_{I_p}(\xi_p) \right\}_0 \right\}_{j_i}. \nonumber \\
\end{eqnarray}

\noindent Remembering that $g'=j_i$, inserting Eq.(\ref{eqD:tensor37}) into Eq.(\ref{eqD:tmatrix35}) we obtain:

\begin{eqnarray}\label{eqD:tmatrix38}
&\phantom{=}&T_{Q M_Q m_f}=\frac{(4\pi)^2}{\hat{I}_p^2k_ik_f}\frac{(-)^{3I_p+j_i+J_d}}{\hat{j}_i} \sum_{L_i J_{P_i}}\sum_{L_f J_{P_f}}i^{L_i-L_f}e^{i(\sigma_{L_i}+\sigma_{L_f})}\hat{J}_{P_i}\hat{J}_{P_f} \nonumber \\
&\phantom{=}& \times \ \int \frac{\phi_{j_f}(r_{nA})\chi_{L_f J_{P_f}}(R_{pB})V(r_{np})\phi_{j_i}(r_{np})\chi_{L_i J_{p_i}}(R_{dA})}{R_{pB}R_{dA}}     \nonumber \\
&\phantom{=}& \times \ \sum_{g h}\langle L_f L_i (g) J_{P_f} J_{P_i} (h) Q M_Q|L_f J_{P_f} (I_p) L_i J_{P_i} (J_d) Q M_Q \rangle \nonumber \\
&\phantom{=}& \times \ \sum_{h'}\langle I_p J_d (j_i) L_f L_i (h') h m_h|I_p L_f (J_{P_f}) J_d L_i (J_{P_i}) h m_h \rangle \\
&\phantom{=}& \times \ \left\{\left\{\tilde{Y}_{L_f}(\hat{k}_f)\otimes \tilde{Y}_{L_i}(\hat{k}_i) \right\}_{g} \otimes \left\{\left\{ \left\{\tilde{Y}_{\ell_i}(\hat{r}_{np}) \otimes \Xi_{I_n}(\xi_n)\right\}_{j_i} \otimes \left\{\Xi_{I_n}(\xi_p)\otimes\Xi_{I_n}(\xi_p) \right\}_0 \right\}_{j_i} \right. \right. \nonumber \\
&\phantom{=}& \otimes \  \left. \left. \left\{\tilde{Y}_{L_f}(\hat{R}_{pB}) \otimes \tilde{Y}_{L_i}(\hat{R}_{dA}) \right\}_{h'} \right\}_h  \right\}_{Q M_Q}\left\{ \tilde{Y}_{\ell_f  }(\hat{r}_{nA})\otimes \Xi_{I_n \mu_n}(\xi_n) \right\}^*_{j_f m_f} \nonumber \\
&\phantom{=}& \times \ d\textbf{R}_{pB}d\textbf{r}_{nA}d\xi_n d\xi_p. \nonumber 
\end{eqnarray}

\noindent We now take the last two lines of Eq.(\ref{eqD:tmatrix38}), break all the couplings between the pairs, and introduce the necessary Clebsch-Gordan coefficients:

\begin{eqnarray}\label{eqD:tensor39}
&\phantom{=}& \times \ \left\{\left\{\tilde{Y}_{L_f}(\hat{k}_f)\otimes \tilde{Y}_{L_i}(\hat{k}_i) \right\}_{g} \otimes \left\{\left\{ \left\{\tilde{Y}_{\ell_i}(\hat{r}_{np}) \otimes \Xi_{I_n}(\xi_n)\right\}_{j_i} \otimes \left\{\Xi_{I_p}(\xi_p)\otimes\Xi_{I_p}(\xi_p) \right\}_0 \right\}_{j_i} \right. \right. \nonumber \\
&\phantom{=}& \left. \left. \ \otimes \left\{\tilde{Y}_{L_f}(\hat{R}_{pB}) \otimes \tilde{Y}_{L_i}(\hat{R}_{dA}) \right\}_{h'} \right\}_h  \right\}_{Q M_Q} \nonumber \\
&=&\sum_{m_g m_h}C_{g m_g h m_h}^{Q M_Q}\left\{\tilde{Y}_{L_f}(\hat{k}_f)\otimes \tilde{Y}_{L_i}(\hat{k}_i) \right\}_{g m_g}\sum_{m_i m_{h'}}C_{j_i m_i h' m_{h'}}^{h m_h}\left\{\tilde{Y}_{\ell_i}(\hat{r}_{np}) \otimes \Xi_{I_n}(\xi_n)\right\}_{j_i m_i} \nonumber \\
&\phantom{=}& \times \ \left\{\Xi_{I_p}(\xi_p)\otimes\Xi_{I_p}(\xi_p) \right\}_{00} \left\{\tilde{Y}_{L_f}(\hat{R}_{pB}) \otimes \tilde{Y}_{L_i}(\hat{R}_{dA}) \right\}_{h' m_{h'}}.
\end{eqnarray}

\noindent Inserting Eq.(\ref{eqD:tensor39}) back into Eq.(\ref{eqD:tmatrix38}) we obtain:

\begin{eqnarray}
&\phantom{=}&T_{Q M_Q m_f}=\frac{(4\pi)^2}{\hat{I}_p^2k_ik_f}\frac{(-)^{3I_p+j_i+J_d}}{\hat{j}_i}\sum_{L_i J_{P_i}}\sum_{L_f J_{P_f}}i^{L_i-L_f}e^{i(\sigma_{L_i}+\sigma_{L_f})}\hat{J}_{P_i}\hat{J}_{P_f} \nonumber \\
&\phantom{=}& \times \ \int \frac{\phi_{j_f}(r_{nA})\chi_{L_f J_{P_f}}(R_{pB})V(r_{np})\phi_{j_i}(r_{np})\chi_{L_i J_{p_i}}(R_{dA})}{R_{pB}R_{dA}}     \nonumber \\
&\phantom{=}& \times \ \sum_{g h}\langle L_f L_i (g) J_{P_f} J_{P_i} (h) Q M_Q|L_f J_{P_f} (I_p) L_i J_{P_i} (J_d) Q M_Q \rangle \nonumber \\
&\phantom{=}& \times \ \sum_{h'}\langle I_p J_d (j_i) L_f L_i (h') h m_h|I_p L_f (J_{P_f}) J_d L_i (J_{P_i}) h m_h \rangle \nonumber \\
&\phantom{=}& \times \ \sum_{m_g m_h}C_{g m_g h m_h}^{Q M_Q}\sum_{m_i m_{h'}}C_{j_i m_i h' m_{h'}}^{h m_h}\left\{\tilde{Y}_{L_f}(\hat{k}_f)\otimes \tilde{Y}_{L_i}(\hat{k}_i) \right\}_{g m_g} \nonumber \\
&\phantom{=}& \times \ \left\{\tilde{Y}_{\ell_i}(\hat{r}_{np}) \otimes \Xi_{I_n}(\xi_n)\right\}_{j_i m_i}\left\{\Xi_{I_p}(\xi_p)\otimes\Xi_{I_p}(\xi_p) \right\}_{00} \nonumber \\
&\phantom{=}& \times \ \left\{\tilde{Y}_{L_f}(\hat{R}_{pB}) \otimes \tilde{Y}_{L_i}(\hat{R}_{dA}) \right\}_{h' m_{h'}}\left\{ \tilde{Y}_{\ell_f  }(\hat{r}_{nA})\otimes \Xi_{I_n \mu_n}(\xi_n) \right\}^*_{j_f m_f} \nonumber \\
&\phantom{=}& \times \ d\textbf{R}_{pB}d\textbf{r}_{nA}d\xi_n d\xi_p.
\end{eqnarray}

\noindent We now consider $\left\{\tilde{Y}_{L_f}(\hat{R}_{pB}) \otimes \tilde{Y}_{L_i}(\hat{R}_{dA}) \right\}_{h' m_{h'}}\left\{ \tilde{Y}_{\ell_f  }(\hat{r}_{nA})\otimes \Xi_{I_n \mu_n}(\xi_n) \right\}^*_{j_f m_f}$ and use Eq.(23), p.64, of \cite{Varshalovich_Book} to obtain:

\begin{eqnarray}\label{eqD:tensor41}
&\phantom{=}&(-)^{j_f-m_f}\left\{ \tilde{Y}_{\ell_f  }(\hat{r}_{nA})\otimes \Xi_{I_n}(\xi_n) \right\}_{j_f, -m_f}\left\{\tilde{Y}_{\ell_i}(\hat{r}_{np}) \otimes \Xi_{I_n}(\xi_n)\right\}_{j_i m_i} \nonumber \\
&=&(-)^{j_f-m_f}\sum_{K M}C_{j_f,-m_f j_i m_i}^{K M}\left\{\left\{ \tilde{Y}_{\ell_f  }(\hat{r}_{nA})\otimes \Xi_{I_n}(\xi_n) \right\}_{j_f}\otimes \left\{\tilde{Y}_{\ell_i}(\hat{r}_{np}) \otimes \Xi_{I_n}(\xi_n)\right\}_{j_i} \right\}_{K M} \nonumber \\
&=&(-)^{j_f-m_f}\sum_{K M}C_{j_f,-m_f j_i m_i}^{K M}| \ell_f I_n (j_f) \ell_i I_n (j_i) K M \rangle \nonumber \\
&=&(-)^{j_f-m_f}\sum_{K M}C_{j_f,-m_f j_i m_i}^{K M}\sum_{g'' h''} \langle \ell_f \ell_i (g'') I_n I_n (h'') K M| \ell_f I_n (j_f) \ell_i I_n (j_i) K M \rangle \nonumber \\
&\phantom{=}& \times \ |\ell_f \ell_i (g'') I_n I_n (h'') K M \rangle \nonumber \\
&=&(-)^{j_f-m_f}\sum_{K M}C_{j_f,-m_f j_i m_i}^{K M}\sum_{g'' h''} \langle \ell_f \ell_i (g'') I_n I_n (h'') K M| \ell_f I_n (j_f) \ell_i I_n (j_i) K M \rangle \nonumber \\
&\phantom{=}& \times \ \left\{ \left\{ \tilde{Y}_{\ell_f  }(\hat{r}_{nA})\otimes \tilde{Y}_{\ell_i}(\hat{r}_{np})\right\}_{g''}\otimes \left\{ \Xi_{I_n}(\xi_n)\otimes \Xi_{I_n}(\xi_n) \right\}_{h''}    \right\}_{K M}.
\end{eqnarray}

\noindent Since the spins must be coupled up to zero, we see that $h''=0$ and $g''=K$. Imposing this condition in Eq.(\ref{eqD:tensor41}):

\begin{eqnarray}
T_{Q M_Q m_f}&=&\frac{(4\pi)^2}{\hat{I}_p^2k_ik_f}\frac{(-)^{3I_p+j_i+J_d+j_f-m_f}}{\hat{j}_i}\sum_{L_i J_{P_i}}\sum_{L_f J_{P_f}}i^{L_i-L_f}e^{i(\sigma_{L_i}+\sigma_{L_f})}\hat{J}_{P_i}\hat{J}_{P_f} \nonumber \\
&\phantom{=}& \times \ \int \frac{\phi_{j_f}(r_{nA})\chi_{L_f J_{P_f}}(R_{pB})V(r_{np})\phi_{j_i}(r_{np})\chi_{L_i J_{p_i}}(R_{dA})}{R_{pB}R_{dA}}     \nonumber \\
&\phantom{=}& \times \ \sum_{g h} \langle L_f L_i (g) J_{P_f} J_{P_i} (h) Q M_Q|L_f J_{P_f} (I_p) L_i J_{P_i} (J_d) Q M_Q \rangle \nonumber \\
&\phantom{=}& \times \ \sum_{h'}\langle I_p J_d (j_i) L_f L_i (h') h m_h|I_p L_f (J_{P_f}) J_d L_i (J_{P_i}) h m_h \rangle\nonumber \\
&\phantom{=}& \times \ \sum_{m_g m_h}C_{g m_g h m_h}^{Q M_Q}\sum_{m_i m_{h'}}C_{j_i m_i h' m_{h'}}^{h m_h}\sum_{K M}C_{j_f,-m_f j_i m_i}^{K M}  \nonumber \\ 
&\phantom{=}& \times \ \langle \ell_f \ell_i (K) I_n I_n (0) K M| \ell_f I_n (j_f) \ell_i I_n (j_i) K M \rangle  \left\{ \tilde{Y}_{\ell_f  }(\hat{r}_{nA})\otimes \tilde{Y}_{\ell_i}(\hat{r}_{np})\right\}_{KM} \nonumber \\
&\phantom{=}& \times \ \left\{ \Xi_{I_n}(\xi_n)\otimes \Xi_{I_n}(\xi_n) \right\}_{00}\left\{\tilde{Y}_{L_f}(\hat{k}_f)\otimes \tilde{Y}_{L_i}(\hat{k}_i) \right\}_{g m_g}\left\{\Xi_{I_p}(\xi_p)\otimes\Xi_{I_p}(\xi_p) \right\}_{00} \nonumber \\
&\phantom{=}& \times \  \left\{\tilde{Y}_{L_f}(\hat{R}_{pB}) \otimes \tilde{Y}_{L_i}(\hat{R}_{dA}) \right\}_{h' m_{h'}} d\textbf{R}_{pB}d\textbf{r}_{nA}d\xi_n d\xi_p. 
\end{eqnarray}

\noindent Now we couple the $\hat{r}$ and $\hat{R}$ spherical harmonics up to zero using Eq.(1), p.248, of \cite{Varshalovich_Book}

\begin{eqnarray}
&\phantom{=}&\left\{\tilde{Y}_{L_f}(\hat{R}_{pB}) \otimes \tilde{Y}_{L_i}(\hat{R}_{dA}) \right\}_{h' m_{h'}}\left\{ \tilde{Y}_{\ell_f  }(\hat{r}_{nA})\otimes \tilde{Y}_{\ell_i}(\hat{r}_{np})\right\}_{K M} \nonumber \\
&=&\sum_{S M_S}C_{h' m_{h'} K M}^{S M_S}\left\{\left\{\tilde{Y}_{L_f}(\hat{R}_{pB}) \otimes \tilde{Y}_{L_i}(\hat{R}_{dA}) \right\}_{h'} \otimes \left\{ \tilde{Y}_{\ell_f  }(\hat{r}_{nA})\otimes \tilde{Y}_{\ell_i}(\hat{r}_{np})\right\}_{K} \right\}_{S M_S} \nonumber \\
&\rightarrow& C_{h' m_{h'} K M}^{00}\left\{\left\{\tilde{Y}_{L_f}(\hat{R}_{pB}) \otimes \tilde{Y}_{L_i}(\hat{R}_{dA}) \right\}_{h'} \otimes \left\{ \tilde{Y}_{\ell_f  }(\hat{r}_{nA})\otimes \tilde{Y}_{\ell_i}(\hat{r}_{np})\right\}_{K} \right\}_{00} \nonumber \\
&=& (-)^{h'-m_{h'}}\frac{\delta_{h' K}\delta_{m_{h'},-M}}{\hat{h}'}\left\{\left\{\tilde{Y}_{L_f}(\hat{R}_{pB}) \otimes \tilde{Y}_{L_i}(\hat{R}_{dA}) \right\}_{h'} \otimes \left\{ \tilde{Y}_{\ell_f  }(\hat{r}_{nA})\otimes \tilde{Y}_{\ell_i}(\hat{r}_{np})\right\}_{K} \right\}_{00} \nonumber \\
&=& (-)^{h'-m_{h'}}\frac{\delta_{h' K}\delta_{m_{h'},-M}}{\hat{h}'}\sum_{M_K}\frac{(-)^{K+M_K}}{\hat{K}}\left\{\tilde{Y}_{L_f}(\hat{R}_{pB}) \otimes \tilde{Y}_{L_i}(\hat{R}_{dA}) \right\}_{K,-M_K} \nonumber \\
&\phantom{=}& \times \ \left\{ \tilde{Y}_{\ell_f  }(\hat{r}_{nA})\otimes \tilde{Y}_{\ell_i}(\hat{r}_{np})\right\}_{K M_K},
\end{eqnarray}

\noindent to arrive at:

\begin{eqnarray}\label{eqD:tmatrix44}
T_{Q M_Q m_f}&=&\frac{(4\pi)^2}{\hat{I}_p^2k_ik_f}\frac{(-)^{3I_p+j_i+J_d+j_f-m_f}}{\hat{j}_i}\sum_{L_i J_{P_i}}\sum_{L_f J_{P_f}}i^{L_i-L_f}e^{i(\sigma_{L_i}+\sigma_{L_f})}\hat{J}_{P_i}\hat{J}_{P_f} \nonumber \\
&\phantom{=}& \times \ \int \frac{\phi_{j_f}(r_{nA})\chi_{L_f J_{P_f}}(R_{pB})V(r_{np})\phi_{j_i}(r_{np})\chi_{L_i J_{p_i}}(R_{dA})}{R_{pB}R_{dA}}     \nonumber \\
&\phantom{=}& \times \ \sum_{g h} \langle L_f L_i (g) J_{P_f} J_{P_i} (h) Q M_Q|L_f J_{P_f} (I_p) L_i J_{P_i} (J_d) Q M_Q \rangle \nonumber \\
&\phantom{=}& \times \ \sum_{h'}\langle I_p J_d (j_i) L_f L_i (h') h m_h|I_p L_f (J_{P_f}) J_d L_i (J_{P_i}) h m_h \rangle \left\{ \Xi_{I_n}(\xi_n)\otimes \Xi_{I_n}(\xi_n) \right\}_{00}  \nonumber \\
&\phantom{=}& \times \ \sum_{m_g m_h}C_{g m_g h m_h}^{Q M_Q}\sum_{m_i m_{h'}}C_{j_i m_i h' m_{h'}}^{h m_h}\sum_{K M}C_{j_f,-m_f j_i m_i}^{K M} \left\{\tilde{Y}_{L_f}(\hat{k}_f)\otimes \tilde{Y}_{L_i}(\hat{k}_i) \right\}_{g m_g} \nonumber \\ 
&\phantom{=}& \times \ \langle \ell_f \ell_i (K) I_n I_n (0) K M| \ell_f I_n (j_f) \ell_i I_n (j_i) K M \rangle   \left\{\Xi_{I_p}(\xi_p)\otimes\Xi_{I_p}(\xi_p) \right\}_{00}   \nonumber \\
&\phantom{=}& \times \ (-)^{h'-m_{h'}}\frac{\delta_{h' K}\delta_{m_{h'},-M}}{\hat{h}'}\sum_{M_K}\frac{(-)^{K+M_K}}{\hat{K}}\left\{\tilde{Y}_{L_f}(\hat{R}_{pB}) \otimes \tilde{Y}_{L_i}(\hat{R}_{dA}) \right\}_{K,-M_K} \nonumber \\
&\phantom{=}& \times \ \left\{ \tilde{Y}_{\ell_f  }(\hat{r}_{nA})\otimes \tilde{Y}_{\ell_i}(\hat{r}_{np})\right\}_{K M_K} d\textbf{R}_{pB}d\textbf{r}_{nA}d\xi_n d\xi_p. 
\end{eqnarray}

\noindent Next we sum over $h'$ and $m_{h'}$, so that $h'=K$, and $m_{h'}=-M$. Then Eq.(\ref{eqD:tmatrix44}) becomes:

\begin{eqnarray}
T_{Q M_Q m_f}&=&\frac{(4\pi)^2}{\hat{I}_p^2k_ik_f}\frac{(-)^{3I_p+j_i+J_d+j_f-m_f}}{\hat{j}_i}\sum_{L_i J_{P_i}}\sum_{L_f J_{P_f}}i^{L_i-L_f}e^{i(\sigma_{L_i}+\sigma_{L_f})}\hat{J}_{P_i}\hat{J}_{P_f} \nonumber \\
&\phantom{=}& \times \ \int \frac{\phi_{j_f}(r_{nA})\chi_{L_f J_{P_f}}(R_{pB})V(r_{np})\phi_{j_i}(r_{np})\chi_{L_i J_{p_i}}(R_{dA})}{R_{pB}R_{dA}}     \nonumber \\
&\phantom{=}& \times \ \sum_{g h} \langle L_f L_i (g) J_{P_f} J_{P_i} (h) Q M_Q|L_f J_{P_f} (I_p) L_i J_{P_i} (J_d) Q M_Q \rangle  \nonumber \\
&\phantom{=}& \times \ \langle I_p J_d (j_i) L_f L_i (K) h m_h|I_p L_f (J_{P_f}) J_d L_i (J_{P_i}) h m_h \rangle \left\{ \Xi_{I_n}(\xi_n)\otimes \Xi_{I_n}(\xi_n) \right\}_{00} \nonumber \\
&\phantom{=}& \times \ \sum_{m_g m_h}C_{g m_g h m_h}^{Q M_Q}\sum_{m_i}C_{j_i m_i K -M}^{h m_h}\sum_{K M}C_{j_f,-m_f j_i m_i}^{K M}\left\{\tilde{Y}_{L_f}(\hat{k}_f)\otimes \tilde{Y}_{L_i}(\hat{k}_i) \right\}_{g m_g}  \nonumber \\ 
&\phantom{=}& \times \ \langle \ell_f \ell_i (K) I_n I_n (0) K M| \ell_f I_n (j_f) \ell_i I_n (j_i) K M \rangle   \left\{\Xi_{I_p}(\xi_p)\otimes\Xi_{I_p}(\xi_p) \right\}_{00}   \nonumber \\
&\phantom{=}& \times \ \frac{(-)^{K+M}}{\hat{K}}\sum_{M_K}\frac{(-)^{K+M_K}}{\hat{K}}\left\{\tilde{Y}_{L_f}(\hat{R}_{pB}) \otimes \tilde{Y}_{L_i}(\hat{R}_{dA}) \right\}_{K,-M_K} \nonumber \\
&\phantom{=}& \times \ \left\{ \tilde{Y}_{\ell_f  }(\hat{r}_{nA})\otimes \tilde{Y}_{\ell_i}(\hat{r}_{np})\right\}_{K M_K} d\textbf{R}_{pB}d\textbf{r}_{nA}d\xi_n d\xi_p. 
\end{eqnarray}

\noindent The integrals over $d\xi_n$ and $d\xi_p$ give $\hat{I}_p\hat{I}_n$, so that:

\begin{eqnarray}\label{eqD:tmatrix46}
T_{Q M_Q m_f}&=& \frac{(4\pi)^2}{\hat{I}_pk_ik_f}\frac{(-)^{3I_p+j_i+J_d+j_f-m_f}}{\hat{j}_i}\hat{I}_n\sum_{L_i J_{P_i}}\sum_{L_f J_{P_f}}i^{L_i-L_f}e^{i(\sigma_{L_i}+\sigma_{L_f})}\hat{J}_{P_i}\hat{J}_{P_f} \nonumber \\
&\phantom{=}& \times \ \int \frac{\phi_{j_f}(r_{nA})\chi_{L_f J_{P_f}}(R_{pB})V(r_{np})\phi_{j_i}(r_{np})\chi_{L_i J_{p_i}}(R_{dA})}{R_{pB}R_{dA}}     \nonumber \\
&\phantom{=}& \times \ \sum_{K}\frac{1}{\hat{K}^2}\sum_{g h} \langle L_f L_i (g) J_{P_f} J_{P_i} (h) Q M_Q|L_f J_{P_f} (I_p) L_i J_{P_i} (J_d) Q M_Q \rangle \nonumber \\
&\phantom{=}& \times \ \langle I_p J_d (j_i) L_f L_i (K) h m_h|I_p L_f (J_{P_f}) J_d L_i (J_{P_i}) h m_h \rangle\nonumber \\
&\phantom{=}& \times \ \sum_{m_g m_h}C_{g m_g h m_h}^{Q M_Q}\sum_{m_i M}(-)^{M}C_{j_i m_i K -M}^{h m_h}C_{j_f,-m_f j_i m_i}^{K M} \nonumber \\
&\phantom{=}& \times \ \sum_{M_K}(-)^{M_K}\langle \ell_f \ell_i (K) I_n I_n (0) K M| \ell_f I_n (j_f) \ell_i I_n (j_i) K M \rangle  \nonumber \\
&\phantom{=}& \times \ \left\{\tilde{Y}_{L_f}(\hat{k}_f)\otimes \tilde{Y}_{L_i}(\hat{k}_i) \right\}_{g m_g}\left\{\tilde{Y}_{L_f}(\hat{R}_{pB}) \otimes \tilde{Y}_{L_i}(\hat{R}_{dA}) \right\}_{K,-M_K}  \nonumber \\
&\phantom{=}& \times \ \left\{ \tilde{Y}_{\ell_f  }(\hat{r}_{nA})\otimes \tilde{Y}_{\ell_i}(\hat{r}_{np})\right\}_{K M_K} d\textbf{R}_{pB}d\textbf{r}_{nA} \nonumber \\
\end{eqnarray}

\noindent We next consider the sum over Clebsch-Gordan coefficients, and use Eq.(11), p.245, of \cite{Varshalovich_Book}:

\begin{eqnarray}\label{eqD:cgc47}
&\phantom{=}&\sum_{m_i M}(-)^{M}C_{j_i m_i K,-M}^{h m_h}C_{j_f,-m_f j_i m_i}^{K M} \nonumber \\
&=&\sum_{m_i M}(-)^{M}C_{j_i m_i K,-M}^{h m_h}(-)^{j_f+m_f}\frac{\hat{K}}{\hat{j}_i}C_{j_f, -m_f K, -M}^{j_i,-m_i} \nonumber \\
&=&\sum_{m_i M}(-)^{M}C_{j_i m_i K,-M}^{h m_h}(-)^{j_f+m_f}\frac{\hat{K}}{\hat{j}_i}(-)^{K-M}\frac{\hat{j}_i}{\hat{j}_f}C_{j_i m_i K, -M}^{j_f m_f} \nonumber \\
&=&(-)^{j_f+m_f+K}\frac{\hat{K}}{\hat{j}_f}\sum_{m_i M}C_{j_i m_i K,-M}^{h m_h}C_{j_i m_i K, -M}^{j_f m_f} \nonumber \\
&=&(-)^{j_f+m_f+K}\frac{\hat{K}}{\hat{j}_f}\delta_{h j_f}\delta_{m_h m_f}.
\end{eqnarray}

\noindent Inserting Eq.(\ref{eqD:cgc47}) into Eq.(\ref{eqD:tmatrix46}), then summing over $h$ and $m_h$, we obtain

\begin{eqnarray}
T_{Q M_Q m_f}&=& \frac{(4\pi)^2}{\hat{I}_pk_ik_f}\frac{(-)^{3I_p+j_i+J_d+2j_f}}{\hat{j}_i\hat{j}_f}\hat{I}_n  \sum_{L_i J_{P_i}}\sum_{L_f J_{P_f}}i^{L_i-L_f}e^{i(\sigma_{L_i}+\sigma_{L_f})}\hat{J}_{P_i}\hat{J}_{P_f} \nonumber \\
&\phantom{=}& \times \ \int \frac{\phi_{j_f}(r_{nA})\chi_{L_f J_{P_f}}(R_{pB})V(r_{np})\phi_{j_i}(r_{np})\chi_{L_i J_{p_i}}(R_{dA})}{R_{pB}R_{dA}}     \nonumber \\
&\phantom{=}& \times \ \sum_{K}\frac{(-)^K}{\hat{K}}\sum_{g} \langle L_f L_i (g) J_{P_f} J_{P_i} (j_f) Q M_Q|L_f J_{P_f} (I_p) L_i J_{P_i} (J_d) Q M_Q \rangle \nonumber \\
&\phantom{=}& \times \ \langle I_p J_d (j_i) L_f L_i (K) j_f m_f|I_p L_f (J_{P_f}) J_d L_i (J_{P_i}) j_f m_f \rangle \sum_{m_g}C_{g m_g j_f m_f}^{Q M_Q} \nonumber \\
&\phantom{=}& \times \ \langle \ell_f \ell_i (K) I_n I_n (0) K M| \ell_f I_n (j_f) \ell_i I_n (j_i) K M \rangle  \left\{\tilde{Y}_{L_f}(\hat{k}_f)\otimes \tilde{Y}_{L_i}(\hat{k}_i) \right\}_{g m_g} \nonumber \\
&\phantom{=}& \times \ \sum_{M_K}(-)^{M_K}\left\{\tilde{Y}_{L_f}(\hat{R}_{pB}) \otimes \tilde{Y}_{L_i}(\hat{R}_{dA}) \right\}_{K,-M_K}  \nonumber \\
&\phantom{=}& \times \ \left\{ \tilde{Y}_{\ell_f  }(\hat{r}_{nA})\otimes \tilde{Y}_{\ell_i}(\hat{r}_{np})\right\}_{K M_K} d\textbf{R}_{pB}d\textbf{r}_{nA}. 
\end{eqnarray}

\noindent We next reorganize the sums:

\begin{eqnarray}\label{eqD:tmatrix49}
T_{Q M_Q m_f}&=& \frac{(4\pi)^2}{\hat{I}_pk_ik_f}\frac{(-)^{3I_p+j_i+J_d+2j_f}}{\hat{j}_i\hat{j}_f}\hat{I}_n\sum_{K}\frac{(-)^K}{\hat{K}} \nonumber \\
&\phantom{=}& \times \ \langle \ell_f \ell_i (K) I_n I_n (0) K M| \ell_f I_n (j_f) \ell_i I_n (j_i) K M \rangle  \nonumber \\
&\phantom{=}& \times \ \sum_{L_i J_{P_i}}\sum_{L_f J_{P_f}}i^{L_i-L_f}e^{i(\sigma_{L_i}+\sigma_{L_f})}\hat{J}_{P_i}\hat{J}_{P_f} \nonumber \\
&\phantom{=}& \times \ \langle I_p J_d (j_i) L_f L_i (K) j_f m_f|I_p L_f (J_{P_f}) J_d L_i (J_{P_i}) j_f m_f \rangle \nonumber \\
&\phantom{=}& \times \ \sum_{g} \langle L_f L_i (g) J_{P_f} J_{P_i} (j_f) Q M_Q|L_f J_{P_f} (I_p) L_i J_{P_i} (J_d) Q M_Q \rangle \nonumber \\
&\phantom{=}& \times \ \sum_{m_g}C_{g m_g j_f m_f}^{Q M_Q}    \left\{\tilde{Y}_{L_f}(\hat{k}_f)\otimes \tilde{Y}_{L_i}(\hat{k}_i) \right\}_{g m_g}  \\
&\phantom{=}& \times \ \int \frac{\phi_{j_f}(r_{nA})\chi_{L_f J_{P_f}}(R_{pB})V(r_{np})\phi_{j_i}(r_{np})\chi_{L_i J_{p_i}}(R_{dA})}{R_{pB}R_{dA}}\sum_{M_K}(-)^{M_K}   \nonumber  \\
&\phantom{=}& \times \ \left\{\tilde{Y}_{L_f}(\hat{R}_{pB}) \otimes \tilde{Y}_{L_i}(\hat{R}_{dA}) \right\}_{K,-M_K} \left\{ \tilde{Y}_{\ell_f  }(\hat{r}_{nA})\otimes \tilde{Y}_{\ell_i}(\hat{r}_{np})\right\}_{K M_K} d\textbf{R}_{pB}d\textbf{r}_{nA}. \nonumber
\end{eqnarray}

\noindent We place $\hat{k}_i$ in the $\hat{z}$-direction so that

\begin{eqnarray}
\left\{\tilde{Y}_{L_f}(\hat{k}_f)\otimes \tilde{Y}_{L_i}(\hat{k}_i) \right\}_{g m_g}&=&i^{L_f+L_i}\sum_{\tilde{M}_i \tilde{M}_f}C_{L_f \tilde{M}_f L_i \tilde{M}_i}^{g m_g}Y_{L_f \tilde{M}_f}(\hat{k}_f)Y_{L_i \tilde{M}_i}(\hat{k}_i \nonumber \\
&=&i^{L_f+L_i}\sum_{\tilde{M}_f}C_{L_f \tilde{M}_f L_i 0}^{g m_g}Y_{L_f \tilde{M}_f}(\hat{k}_f)\frac{\hat{L}_i}{\sqrt{4\pi}}\delta_{\tilde{M}_f m_g} \nonumber \\
&=&i^{L_f+L_i}C_{L_f m_g L_i 0}^{g m_g}Y_{L_f m_g}(\hat{k}_f)\frac{\hat{L}_i}{\sqrt{4\pi}}.
\end{eqnarray}

\noindent Then Eq.(\ref{eqD:tmatrix49}) becomes:

\begin{eqnarray}
T_{Q M_Q m_f}&=& \frac{(4\pi)^{3/2}}{\hat{I}_pk_ik_f}\frac{(-)^{3I_p+j_i+J_d+2j_f}}{\hat{j}_i\hat{j}_f}\hat{I}_n\sum_{K}\frac{(-)^K}{\hat{K}} \nonumber \\
&\phantom{=}& \times \ \langle \ell_f \ell_i (K) I_n I_n (0) K M| \ell_f I_n (j_f) \ell_i I_n (j_i) K M \rangle  \nonumber \\
&\phantom{=}& \times \ \sum_{L_i J_{P_i}}\sum_{L_f J_{P_f}}i^{2L_i}e^{i(\sigma_{L_i}+\sigma_{L_f})}\hat{L}_i\hat{J}_{P_i}\hat{J}_{P_f} \nonumber \\
&\phantom{=}& \times \ \langle I_p J_d (j_i) L_f L_i (K) j_f m_f|I_p L_f (J_{P_f}) J_d L_i (J_{P_i}) j_f m_f \rangle \nonumber \\
&\phantom{=}& \times \ \sum_{g} \langle L_f L_i (g) J_{P_f} J_{P_i} (j_f) Q M_Q|L_f J_{P_f} (I_p) L_i J_{P_i} (J_d) Q M_Q \rangle \nonumber \\
&\phantom{=}& \times \ \sum_{m_g}C_{g m_g j_f m_f}^{Q M_Q}C_{L_f m_g L_i 0}^{g m_g}Y_{L_f m_g}(\hat{k}_f)  \\
&\phantom{=}& \times \ \int \frac{\phi_{j_f}(r_{nA})\chi_{L_f J_{P_f}}(R_{pB})V(r_{np})\phi_{j_i}(r_{np})\chi_{L_i J_{p_i}}(R_{dA})}{R_{pB}R_{dA}} \sum_{M_K}(-)^{M_K}    \nonumber \\
&\phantom{=}& \times \ \left\{\tilde{Y}_{L_f}(\hat{R}_{pB}) \otimes \tilde{Y}_{L_i}(\hat{R}_{dA}) \right\}_{K,-M_K} \left\{ \tilde{Y}_{\ell_f  }(\hat{r}_{nA})\otimes \tilde{Y}_{\ell_i}(\hat{r}_{np})\right\}_{K M_K} d\textbf{R}_{pB}d\textbf{r}_{nA}. \nonumber
\end{eqnarray}

\noindent We can now break the remaining couplings to obtain:

\begin{eqnarray}\label{eqD:tmatrix52}
T_{Q M_Q m_f}&=& \frac{(4\pi)^{3/2}}{\hat{I}_pk_ik_f}\frac{(-)^{3I_p+j_i+J_d+2j_f}}{\hat{j}_i\hat{j}_f}\hat{I}_n\sum_{K}\frac{(-)^K}{\hat{K}} \nonumber \\
&\phantom{=}& \times \ \langle \ell_f \ell_i (K) I_n I_n (0) K M| \ell_f I_n (j_f) \ell_i I_n (j_i) K M \rangle  \nonumber \\
&\phantom{=}& \times \ \sum_{L_i J_{P_i}}\sum_{L_f J_{P_f}}i^{2L_i}e^{i(\sigma_{L_i}+\sigma_{L_f})}\hat{L}_i\hat{J}_{P_i}\hat{J}_{P_f} \nonumber \\
&\phantom{=}& \times \ \langle I_p J_d (j_i) L_f L_i (K) j_f m_f|I_p L_f (J_{P_f}) J_d L_i (J_{P_i}) j_f m_f \rangle \nonumber \\
&\phantom{=}& \times \ \sum_{g} \langle L_f L_i (g) J_{P_f} J_{P_i} (j_f) Q M_Q|L_f J_{P_f} (I_p) L_i J_{P_i} (J_d) Q M_Q \rangle \nonumber \\
&\phantom{=}& \times \ \sum_{m_g}C_{g m_g j_f m_f}^{Q M_Q}C_{L_f m_g L_i 0}^{g m_g}Y_{L_f m_g}(\hat{k}_f)  \nonumber \\
&\phantom{=}& \times \ \int \frac{\phi_{j_f}(r_{nA})\chi_{L_f J_{P_f}}(R_{pB})V(r_{np})\phi_{j_i}(r_{np})\chi_{L_i J_{p_i}}(R_{dA})}{R_{pB}R_{dA}}     \nonumber \\
&\phantom{=}& \times \ \sum_{M_K}(-)^{M_K}\sum_{M_f M_i}i^{L_f+L_i}C_{L_f M_f L_i M_i}^{K,-M_K} Y_{L_f M_f}(\hat{R}_{pB}) Y_{L_i M_i}(\hat{R}_{dA}) \nonumber \\
&\phantom{=}& \times \ \sum_{\tilde{m}_f\tilde{m}_i}i^{\ell_f+\ell_i}C_{\ell_f \tilde{m}_f \ell_i \tilde{m}_i}^{K M_K}Y_{\ell_f \tilde{m}_f}(\hat{r}_{nA})Y_{\ell_i \tilde{m}_i}(\hat{r}_{np}) d\textbf{R}_{pB}d\textbf{r}_{nA}. \nonumber \\
\end{eqnarray}

\noindent $\textbf{R}_{pB}$ is another independent variable. We place $\hat{R}_{pB}$ in the $\hat{z}$-direction. In that case, $M_f=0$, $Y_{L_f M_f}(\hat{R}_{pB})=\hat{L}_f/\sqrt{4\pi}$, and $M_i=-M_K$. Eq.(\ref{eqD:tmatrix52}) is then simplified to:

\begin{eqnarray}\label{eqD:tmatrix53}
T_{Q M_Q m_f}&=&\frac{4\pi}{\hat{I}_pk_ik_f}\frac{(-)^{3I_p+j_i+J_d+2j_f}}{\hat{j}_i\hat{j}_f}\hat{I}_n\sum_{K}\frac{(-)^K}{\hat{K}} \nonumber \\
&\phantom{=}& \times \ \langle \ell_f \ell_i (K) I_n I_n (0) K M| \ell_f I_n (j_f) \ell_i I_n (j_i) K M \rangle  \nonumber \\
&\phantom{=}& \times \ \sum_{L_i J_{P_i}}\sum_{L_f J_{P_f}}i^{3L_i+L_f+\ell_f+\ell_i}e^{i(\sigma_{L_i}+\sigma_{L_f})}\hat{L}_i\hat{L}_f\hat{J}_{P_i}\hat{J}_{P_f} \nonumber \\
&\phantom{=}& \times \ \langle I_p J_d (j_i) L_f L_i (K) j_f m_f|I_p L_f (J_{P_f}) J_d L_i (J_{P_i}) j_f m_f \rangle \nonumber \\
&\phantom{=}& \times \ \sum_{g} \langle L_f L_i (g) J_{P_f} J_{P_i} (j_f) Q M_Q|L_f J_{P_f} (I_p) L_i J_{P_i} (J_d) Q M_Q \rangle \nonumber \\
&\phantom{=}& \times \ \sum_{m_g}C_{g m_g j_f m_f}^{Q M_Q}C_{L_f m_g L_i 0}^{g m_g}Y_{L_f m_g}(\hat{k}_f)  \nonumber \\
&\phantom{=}& \times \ \int \frac{\phi_{j_f}(r_{nA})\chi_{L_f J_{P_f}}(R_{pB})V(r_{np})\phi_{j_i}(r_{np})\chi_{L_i J_{p_i}}(R_{dA})}{R_{pB}R_{dA}}     \nonumber \\
&\phantom{=}& \times \ \sum_{M_K}(-)^{M_K}C_{L_f 0 L_i, -M_K}^{K,-M_K}  Y_{L_i, -M_K}(\hat{R}_{dA}) \\
&\phantom{=}& \times \ \sum_{\tilde{m}_f\tilde{m}_i}C_{\ell_f \tilde{m}_f \ell_i \tilde{m}_i}^{K M_K}Y_{\ell_f \tilde{m}_f}(\hat{r}_{nA})Y_{\ell_i \tilde{m}_i}(\hat{r}_{np})R^2_{pB}d\Omega_{R_{pB}}dR_{pB}r^2_{nA}\sin\theta dr_{nA}d\theta d\phi. \nonumber 
\end{eqnarray}

\noindent Since we are fixing $\hat{R}_{pB}$ in the $\hat{z}$-direction, integrating over $d\Omega_{R_{pB}}$ results only in a factor of $4\pi$. We also fix the other vectors to be in the $xz$-plane, which means that the integral over $d\phi$ provides an additional factor of $2\pi$: Introducing these into Eq.(\ref{eqD:tmatrix53}) we obtain:

\begin{eqnarray}\label{eqD:tmatrix54}
T_{Q M_Q m_f}&=&\frac{32\pi^3}{\hat{I}_pk_ik_f}\frac{(-)^{3I_p+j_i+J_d+2j_f}}{\hat{j}_i\hat{j}_f}\hat{I}_n\sum_{K}\frac{(-)^K}{\hat{K}} \nonumber \\
&\phantom{=}& \times \ \langle \ell_f \ell_i (K) I_n I_n (0) K M| \ell_f I_n (j_f) \ell_i I_n (j_i) K M \rangle  \nonumber \\
&\phantom{=}& \times \ \sum_{L_i J_{P_i}}\sum_{L_f J_{P_f}}i^{3L_i+L_f+\ell_f+\ell_i}e^{i(\sigma_{L_i}+\sigma_{L_f})}\hat{L}_i\hat{L}_f\hat{J}_{P_i}\hat{J}_{P_f} \nonumber \\
&\phantom{=}& \times \ \langle I_p J_d (j_i) L_f L_i (K) j_f m_f|I_p L_f (J_{P_f}) J_d L_i (J_{P_i}) j_f m_f \rangle \nonumber \\
&\phantom{=}& \times \ \sum_{g} \langle L_f L_i (g) J_{P_f} J_{P_i} (j_f) Q M_Q|L_f J_{P_f} (I_p) L_i J_{P_i} (J_d) Q M_Q \rangle \nonumber \\
&\phantom{=}& \times \ \sum_{m_g}C_{g m_g j_f m_f}^{Q M_Q}C_{L_f m_g L_i 0}^{g m_g}Y_{L_f m_g}(\hat{k}_f)  \nonumber \\
&\phantom{=}& \times \ \int \phi_{j_f}(r_{nA})\chi_{L_f J_{P_f}}(R_{pB})V(r_{np})\phi_{j_i}(r_{np})\chi_{L_i J_{p_i}}(R_{dA}) \frac{R_{pB}r^2_{nA}}{R_{dA}}     \nonumber \\
&\phantom{=}& \times \ \sum_{M_K}(-)^{M_K}C_{L_f 0 L_i, -M_K}^{K,-M_K}  Y_{L_i, -M_K}(\hat{R}_{dA}) \nonumber \\
&\phantom{=}& \times \ \sum_{\tilde{m}_f\tilde{m}_i}C_{\ell_f \tilde{m}_f \ell_i \tilde{m}_i}^{K M_K}Y_{\ell_f \tilde{m}_f}(\hat{r}_{nA})Y_{\ell_i \tilde{m}_i}(\hat{r}_{np})\sin\theta dR_{pB}dr_{nA}d\theta.
\end{eqnarray}

\noindent Eq.(\ref{eqD:tmatrix54}) is valid for a general $\ell_i$ and $\ell_f$. Since we are interested in applying the formalism to $(d,p)$ we use $\ell_i=0$ deuteron. Therefore $Y_{\ell_i \tilde{m}_i}(\hat{r}_{np})=1/\sqrt{4\pi}$, $\tilde{m}_i=0$, $\tilde{m}_f=M_K$, and $K=\ell_f$. Introducing this simplification into Eq.(\ref{eqD:tmatrix54}) we arrive at:

\begin{eqnarray}\label{eqD:tmatrix55}
T_{Q M_Q m_f}&=&\frac{32\pi^3}{\hat{I}_p\sqrt{4\pi}k_ik_f}\frac{(-)^{3I_p+j_i+J_d+2j_f+\ell_f}}{\hat{j}_i\hat{j}_f \hat{\ell}_f}\hat{I}_n \nonumber \\
&\phantom{=}& \times \ \langle \ell_f 0 (\ell_f) I_n I_n (0) \ell_f M| \ell_f I_n (j_f) 0 I_n (j_i) \ell_f M \rangle  \nonumber \\
&\phantom{=}& \times \ \sum_{L_i J_{P_i}}\sum_{L_f J_{P_f}}i^{3L_i+L_f+\ell_f+\ell_i}e^{i(\sigma_{L_i}+\sigma_{L_f})}\hat{L}_i\hat{L}_f\hat{J}_{P_i}\hat{J}_{P_f} \nonumber \\
&\phantom{=}& \times \ \langle I_p J_d (j_i) L_f L_i (\ell_f) j_f m_f|I_p L_f (J_{P_f}) J_d L_i (J_{P_i}) j_f m_f \rangle \nonumber \\
&\phantom{=}& \times \ \sum_{g} \langle L_f L_i (g) J_{P_f} J_{P_i} (j_f) Q M_Q|L_f J_{P_f} (I_p) L_i J_{P_i} (J_d) Q M_Q \rangle \nonumber \\
&\phantom{=}& \times \ \sum_{m_g}C_{g m_g j_f m_f}^{Q M_Q}C_{L_f m_g L_i 0}^{g m_g}Y_{L_f m_g}(\hat{k}_f)\sum_{M_K}(-)^{M_K}C_{L_f 0 L_i, -M_K}^{\ell_f,-M_K} \nonumber  \\
&\phantom{=}& \times \ \int \phi_{j_f}(r_{nA})\chi_{L_f J_{P_f}}(R_{pB})V(r_{np})\phi_{j_i}(r_{np})\chi_{L_i J_{p_i}}(R_{dA}) \frac{R_{pB}r^2_{nA}}{R_{dA}} \nonumber \\
&\phantom{=}& \times \ Y_{L_i, -M_K}(\hat{R}_{dA})Y_{\ell_f M_K}(\hat{r}_{nA})\sin\theta dR_{pB}dr_{nA}d\theta.
\end{eqnarray}

\noindent We can further simplify Eq.(\ref{eqD:tmatrix55}) by using,  Eq.(5), p.334, Eq.(1), p.357 Eq. (1), and Eq.(1), p.299, of \cite{Varshalovich_Book}:

\begin{eqnarray}
&\phantom{=}&\langle \ell_f 0 (\ell_f) I_n I_n (0) \ell_f M| \ell_f I_n (j_f) 0 I_n (I_n) \ell_f M \rangle \nonumber \\
&=&\hat{\ell}_f\hat{0}\hat{j}_f\hat{I}_n
\begin{Bmatrix}
\ell_f & 0 & \ell_f \\
I_n & I_n & 0 \\
j_f & I_n & \ell_f
\end{Bmatrix} \nonumber \\
&=&\hat{\ell}_f\hat{0}\hat{j}_f\hat{I}_n
\begin{Bmatrix}
j_f & I_n & \ell_f \\
\ell_f & 0 & \ell_f \\
I_n & I_n & 0 \\
\end{Bmatrix} \nonumber \\
&=&\hat{\ell}_f\hat{0}\hat{j}_f\hat{I}_n(-)^{I_n+\ell_f+\ell_f+I_n}\frac{1}{\hat{\ell}_f\hat{I}_n}
\begin{Bmatrix}
j_f & I_n & \ell_f \\
0 & \ell_f & I_n
\end{Bmatrix} \nonumber \\
&=&\hat{\ell}_f\hat{0}\hat{j}_f\hat{I}_n(-)^{I_n+\ell_f+\ell_f+I_n}\frac{1}{\hat{\ell}_f\hat{I}_n}(-)^{j_f+I_n+\ell_f}\frac{1}{\hat{I}_n\hat{\ell}_f} \nonumber \\
&=&(-)^{3I_n+\ell_f+j_f}\frac{\hat{j}_f}{\hat{\ell}_f\hat{I}_n}.
\end{eqnarray}

\noindent Since $\ell_f$ is an integer, $(-)^{2\ell_f}=1$, and we obtain:

\begin{eqnarray}\label{eqD:tmatrix57}
T_{Q M_Q m_f}&=&\frac{32\pi^3}{\hat{I}_p\sqrt{4\pi}k_ik_f}\frac{(-)^{3I_p+3I_n+j_i+J_d+3j_f}}{\hat{j}_i \hat{\ell}^2_f} \nonumber \\
&\phantom{=}& \times \ \sum_{L_i J_{P_i}}\sum_{L_f J_{P_f}}i^{3L_i+L_f+\ell_f+\ell_i}e^{i(\sigma_{L_i}+\sigma_{L_f})}\hat{L}_i\hat{L}_f\hat{J}_{P_i}\hat{J}_{P_f} \nonumber \\
&\phantom{=}& \times \ \langle I_p J_d (j_i) L_f L_i (\ell_f) j_f m_f|I_p L_f (J_{P_f}) J_d L_i (J_{P_i}) j_f m_f \rangle \nonumber \\
&\phantom{=}& \times \ \sum_{g} \langle L_f L_i (g) J_{P_f} J_{P_i} (j_f) Q M_Q|L_f J_{P_f} (I_p) L_i J_{P_i} (J_d) Q M_Q \rangle \nonumber \\
&\phantom{=}& \times \ \sum_{m_g}C_{g m_g j_f m_f}^{Q M_Q}C_{L_f m_g L_i 0}^{g m_g}Y_{L_f m_g}(\hat{k}_f)\sum_{M_K}(-)^{M_K}C_{L_f 0 L_i, -M_K}^{\ell_f,-M_K}   \\
&\phantom{=}& \times \ \int \phi_{j_f}(r_{nA})\chi_{L_f J_{P_f}}(R_{pB})V(r_{np})\phi_{j_i}(r_{np})\chi_{L_i J_{p_i}}(R_{dA}) \frac{R_{pB}r^2_{nA}}{R_{dA}} \nonumber \\
&\phantom{=}& \times \ Y_{L_i, -M_K}(\hat{R}_{dA})Y_{\ell_f M_K}(\hat{r}_{nA})\sin\theta dR_{pB}dr_{nA}d\theta. \nonumber
\end{eqnarray}

\noindent Expand now expand the 9js, as in Eq.(5), p.334, of \cite{Varshalovich_Book}

\begin{eqnarray}\label{eqD:9j58}
\langle L_f L_i (g) J_{P_f} J_{P_i} (j_f) Q M_Q|L_f J_{P_f} (I_p) L_i J_{P_i} (J_d) Q M_Q \rangle&=&\hat{g}\hat{j}_f\hat{I_p}\hat{J}_{d}
\begin{Bmatrix}
L_f & L_i & g \\
J_{P_f} & J_{P_i} & j_f \\
I_p & J_d & Q
\end{Bmatrix} \nonumber \\
 \langle I_p J_d (j_i) L_f L_i (\ell_f) j_f m_f|I_p L_f (J_{P_f}) J_d L_i (J_{P_i}) j_f m_f \rangle&=&\hat{j}_i\hat{\ell}_f\hat{J}_{P_f}\hat{J}_{P_i}
\begin{Bmatrix}
I_p & J_d & j_i \\
L_f & L_i & \ell_f \\
J_{P_f} & J_{P_i} & j_f
\end{Bmatrix}, \nonumber \\
\end{eqnarray}

\noindent Finally, inserting Eq.(\ref{eqD:9j58}) into Eq.(\ref{eqD:tmatrix57}) gives us the form for $T_{Q M_Q m_f}$ which we implement in NLAT:

\begin{eqnarray}
T_{Q M_Q m_f}&=&\frac{32\pi^3}{\sqrt{4\pi}k_ik_f}(-)^{3I_p+3I_n+j_i+J_d+3j_f}\frac{\hat{j}_f\hat{J}_{d}}{ \hat{\ell}_f}\sum_{L_i J_{P_i}} \nonumber \\
&\phantom{=}& \times \ \sum_{L_f J_{P_f}}i^{3L_i+L_f+\ell_f+\ell_i}e^{i(\sigma_{L_i}+\sigma_{L_f})}\hat{L}_i\hat{L}_f\hat{J}^2_{P_i}\hat{J}^2_{P_f} \nonumber \\
&\phantom{=}& \times \ 
\begin{Bmatrix}
I_p & J_d & j_i \\
L_f & L_i & \ell_f \\
J_{P_f} & J_{P_i} & j_f
\end{Bmatrix}\sum_{g} \hat{g}
\begin{Bmatrix}
L_f & L_i & g \\
J_{P_f} & J_{P_i} & j_f \\
I_p & J_d & Q
\end{Bmatrix}\sum_{m_g}C_{g m_g j_f m_f}^{Q M_Q}C_{L_f m_g L_i 0}^{g m_g} \nonumber \\
&\phantom{=}& \times \ Y_{L_f m_g}(\hat{k}_f)\sum_{M_K}(-)^{M_K}C_{L_f 0 L_i, -M_K}^{\ell_f,-M_K} \nonumber \\
&\phantom{=}& \times \ \int \phi_{j_f}(r_{nA})\chi_{L_f J_{P_f}}(R_{pB})V(r_{np})\phi_{j_i}(r_{np}) \nonumber \\
&\phantom{=}& \times \ \chi_{L_i J_{p_i}}(R_{dA}) \frac{R_{pB}r^2_{nA}}{R_{dA}}Y_{L_i, -M_K}(\hat{R}_{dA})Y_{\ell_f M_K}(\hat{r}_{nA})\sin\theta dR_{pB}dr_{nA}d\theta, \nonumber \\
\end{eqnarray}

\noindent The observable is the differential cross section, which as we saw in Eq.(\ref{eqD:CrossSection28}), is given by

\begin{eqnarray}
\frac{d\sigma}{d\Omega}&=&\frac{k_f}{k_i}\frac{\mu_i \mu_f}{4\pi^2\hbar^4}\frac{\hat{J}_B^2}{\hat{J}^2_d \hat{J}^2_A \hat{j}_f^2}\sum_{m_f Q M_Q}T_{Q M_Q m_f}T^*_{Q M_Q m_f}.
\end{eqnarray}


\chapter{ Checks of the Code NLAT}
\label{Checks}

To perform the calculations in this thesis, the code ``Nonlocal Adiabatic Transfer'' (NLAT) was written to calculate $(d,p)$ transfer reactions with the inclusion of nonlocality. In order to ensure that the code works properly, multiple checks were performed, and are discussed in the following sections. When making comparisons to local calculations, we will compare with the code \textsc{FRESCO} \cite{fresco}. 

\section*{Local Elastic Scattering}

\begin{figure}[h!]
\begin{center}
\includegraphics[scale=0.35]{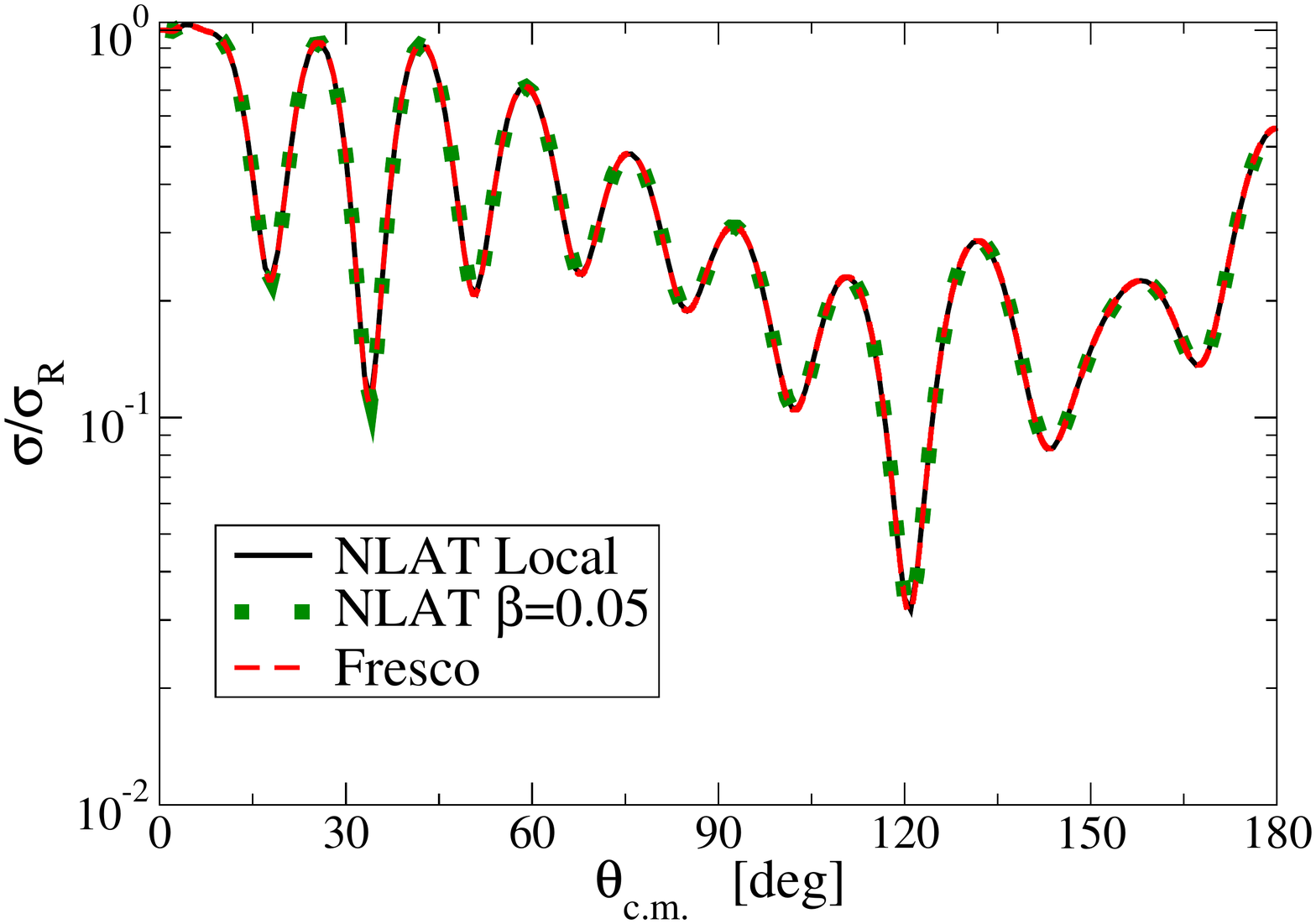}
\end{center}
\caption{Differential elastic scattering relative to Rutherfored as a function of scattering angle. $^{209}$Pb$(p,p)^{209}$Pb at $E_p=50.0$ MeV: The solid line is obtained from NLAT, the dotted line is obtained from NLAT and setting $\beta=0.05$ fm, and the dashed line is from \textsc{FRESCO}. }
\label{fig:p209Pb_50-0_Local_Elastic_Check}
\end{figure}

First, we look at the local elastic scattering distribution. In Fig. \ref{fig:p209Pb_50-0_Local_Elastic_Check} we show this check for the reaction $^{209}$Pb$(p,p)^{209}$Pb at $E_p=50.0$ MeV. The solid line is a local calculation using NLAT, the dotted line is a nonlocal calculation, but with $\beta=0.05$ fm so that it reduces to the local calculation, and the dashed line is \textsc{FRESCO}. We used $\beta=0.05$ fm rather than $\beta=0$ fm since we would have numerical problems with dividing by zero if we set $\beta$ exactly equal to zero. For these calculations, we used a step size of $0.01$ fm, a maximum radius of $30$ fm, and included partial waves up to $L=20$. These calculations are converged in that a smaller step size or more partial waves does not change the results of the calculation.

\section*{Nonlocal Elastic Scattering}

\begin{figure}[h!]
\begin{center}
\includegraphics[scale=0.35]{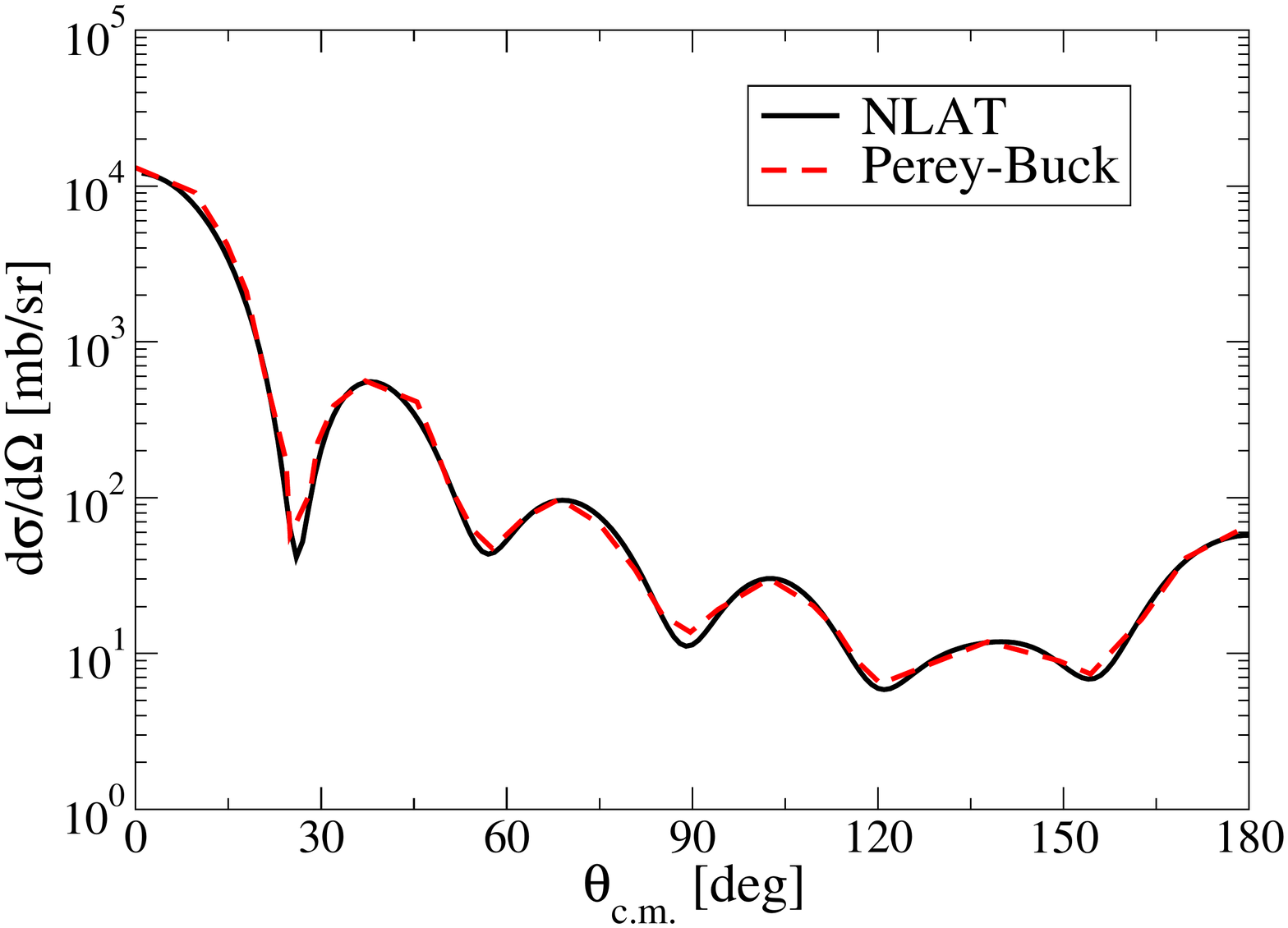}
\end{center}
\caption{Differential elastic scattering as a function of scattering angle. $^{208}$Pb$(n,n)^{208}$Pb at $E_p=14.5$ MeV: The solid line is obtained a nonlocal calculation using NLAT, and the dashed line is the nonlocal calculation published by Perey and Buck \cite{Perey_np1962}. }
\label{fig:PereyBuck_Elastic_Check}
\end{figure}

Next, we look at the nonlocal elastic scattering distribution. In Fig. \ref{fig:PereyBuck_Elastic_Check} we present $^{208}$Pb$(n,n)^{208}$Pb at $E_n=14.5$ MeV. The solid line is a nonlocal calculation with $\beta=0.85$ fm using NLAT. The dashed line is the digitized results of the same calculation from the paper of Perey and Buck \cite{Perey_np1962}. The two calculations agree quite well, indicating that NLAT calculates elastic scattering with a nonlocal potential properly. The calculations of Perey and Buck were digitized from their paper, so any discrepancies between the results shown here and theirs is a result of errors in the digitizing process.

\section*{Bound States}

\begin{figure}[h!]
\begin{center}
\includegraphics[scale=0.35]{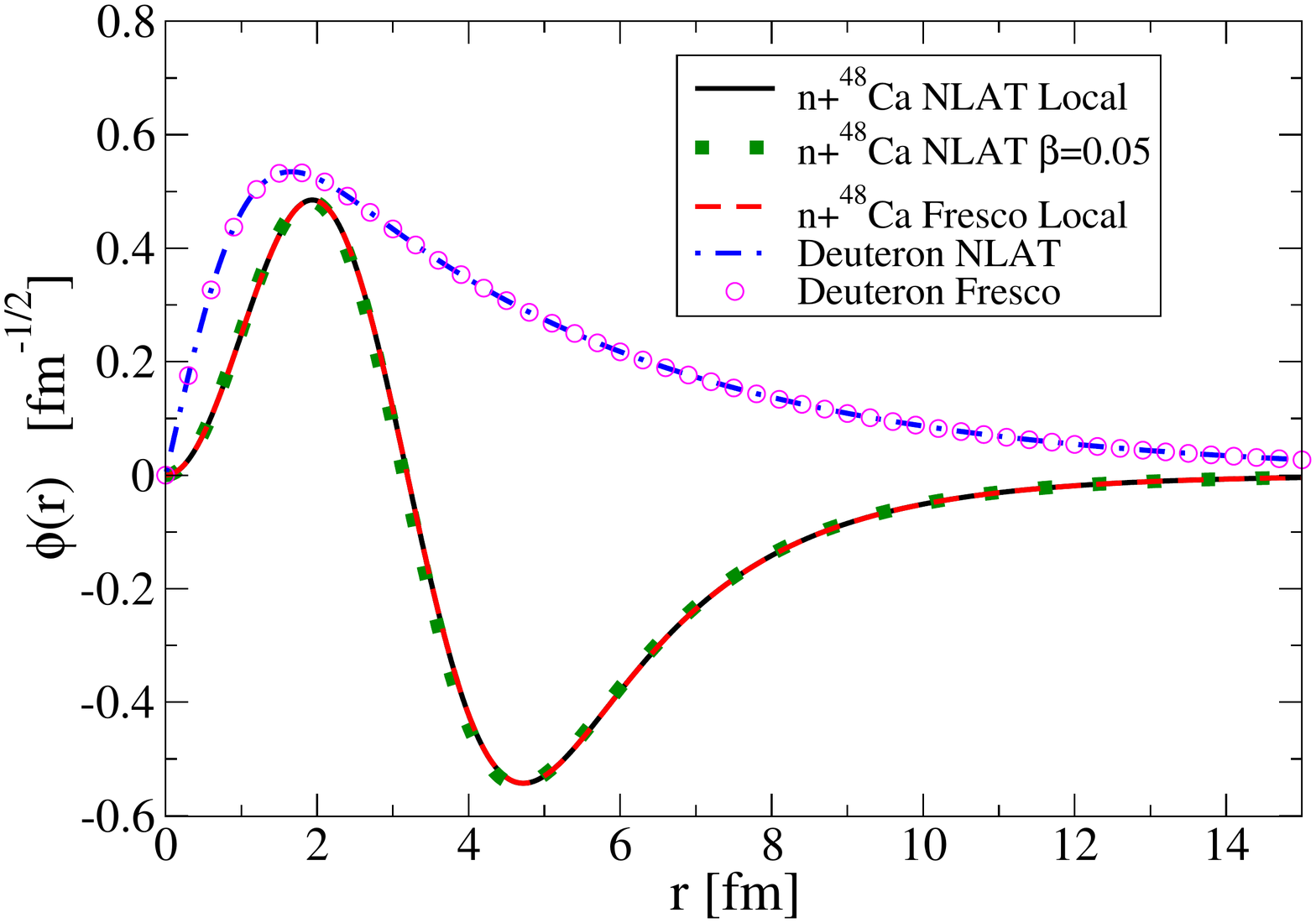}
\end{center}
\caption{$n+^{48}$Ca bound wave function, and the deuteron bound wave function. $n+^{48}$Ca: The solid line is obtained from NLAT, the dotted line is obtained from NLAT and setting $\beta=0.05$ fm, and the dashed line is from \textsc{FRESCO}. Deuteron: Dot-dashed line is deuteron bound wave function obtained from NLAT, and the open circles are obtained with \textsc{FRESCO}. }
\label{fig:n48Ca_Bound_Check}
\end{figure}

Next, we examine the bound wave functions. In Fig. \ref{fig:n48Ca_Bound_Check} we show the $n+^{48}$Ca bound wave function as well as the deuteron bound wave function. For the $n+^{48}$Ca wave functions, the solid line is obtained from a local calculation with NLAT, the dotted line is a nonlocal calculation with $\beta=0.05$ fm, and the dashed line is obtained from \textsc{FRESCO}. For the deuteron bound wave function, the dot-dashed line results from a local calculation using NLAT, and the open circles are from \textsc{FRESCO}. For all calculations we used a step size of $0.01$ fm, a matching radius of $1.5$ fm, and a maximum radius of $30$ fm. This model space produces converged wave functions.

\section*{Adiabatic Potential}
\label{Sec:Adiabatic_Potential}

\begin{figure}[h!]
\begin{center}
\includegraphics[width=0.9\textwidth]{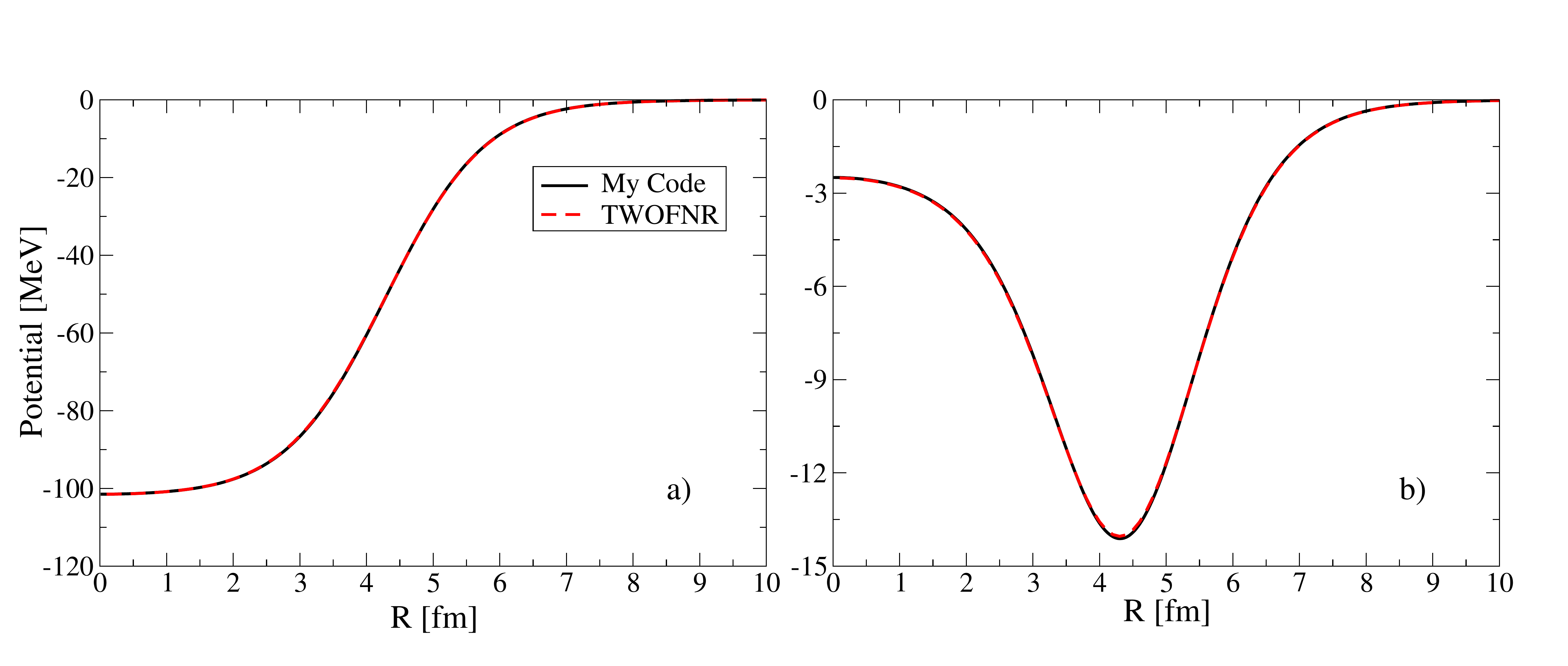}
\end{center}
\caption{The local adiabatic potential for $d+^{48}$Ca at $E_d=20$ MeV calculated with NLAT and with \textsc{TWOFNR} \cite{twofnr}. (a) Real part, (b) Imaginary part. }
\label{fig:Local_AdPot_Check}
\end{figure}

Next, we check the adiabatic potential. In Fig. \ref{fig:Local_AdPot_Check} we show the local adiabatic potential for $d+^{48}$Ca at $E_d=20$ MeV calculated with the CH89 global optical potential \cite{Varner_pr1991}. The comparison is with the code \textsc{TWOFNR} \cite{twofnr}. Panel (a) is the real part of the adiabatic potential, and (b) is the imaginary part.

\begin{figure}[h!]
\begin{center}
\includegraphics[scale=0.35]{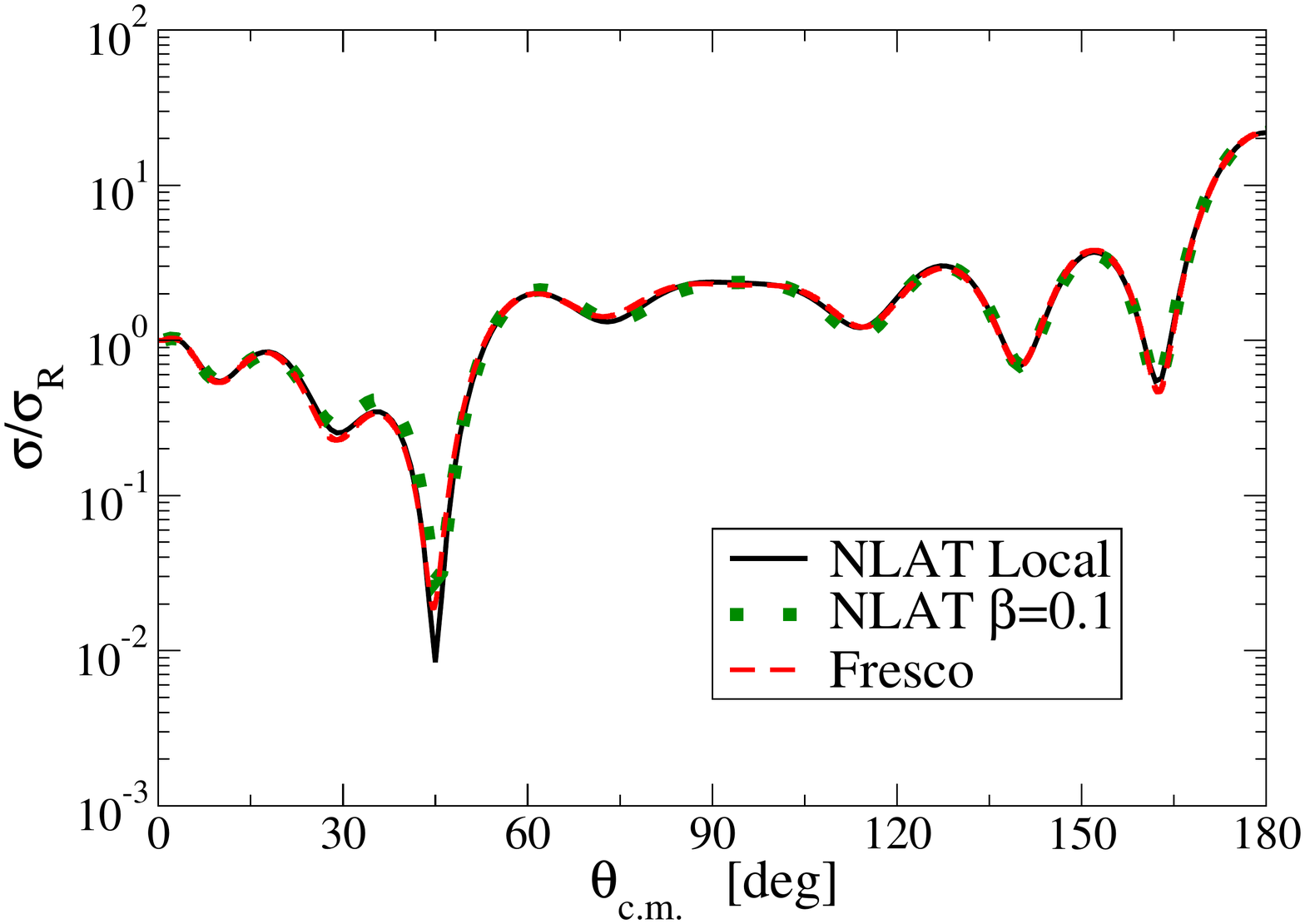}
\end{center}
\caption{$^{48}$Ca$(d,d)^{48}$Ca at $E_d=20$ MeV. The solid line is when using the local adiabatic potential, the dotted line is when doing a nonlocal calculation with $\beta=0.1$ fm in the nucleon optical potentials, and the dashed line is a calculation done in \textsc{FRESCO}. }
\label{fig:d48Ca_20-0_Elastic_Adiabatic_Check}
\end{figure}

In Fig. \ref{fig:d48Ca_20-0_Elastic_Adiabatic_Check} we show elastic scattering normalized to Rutherford for $^{48}$Ca$(d,d)^{48}$Ca at $E_d=20$ MeV when using the adiabatic potential. While the adiabatic potential is not suitable for accurately describing elastic scattering, this comparison is to show that NLAT calculates the adiabatic potential properly, and correctly does the scattering calculation. The nonlocal calculation used $\beta=0.1$ since accuracy was lost with a smaller $\beta$ due to inaccuracies in calculating the nonlocal integral. The agreement between the nonlocal and local calculations demonstrates that the nonlocal adiabatic integral is being calculated properly since it reduces to the local calculation in the limit of $\beta \rightarrow 0$, as it should. The solid line is a calculation done with NLAT using the local adiabatic potential, the dotted line is a nonlocal calculation with $\beta=0.1$ fm in the nucleon optical potentials, and the dashed line is a local calculation done with \textsc{FRESCO}. For these calculations, we used a $0.01$ fm step size, a maximum radius of $30$ fm, and partial waves up to $L=20$. 

\section*{Transfer}
\label{Sec:Transfer}

\begin{figure}[h!]
\begin{center}
\includegraphics[scale=0.35]{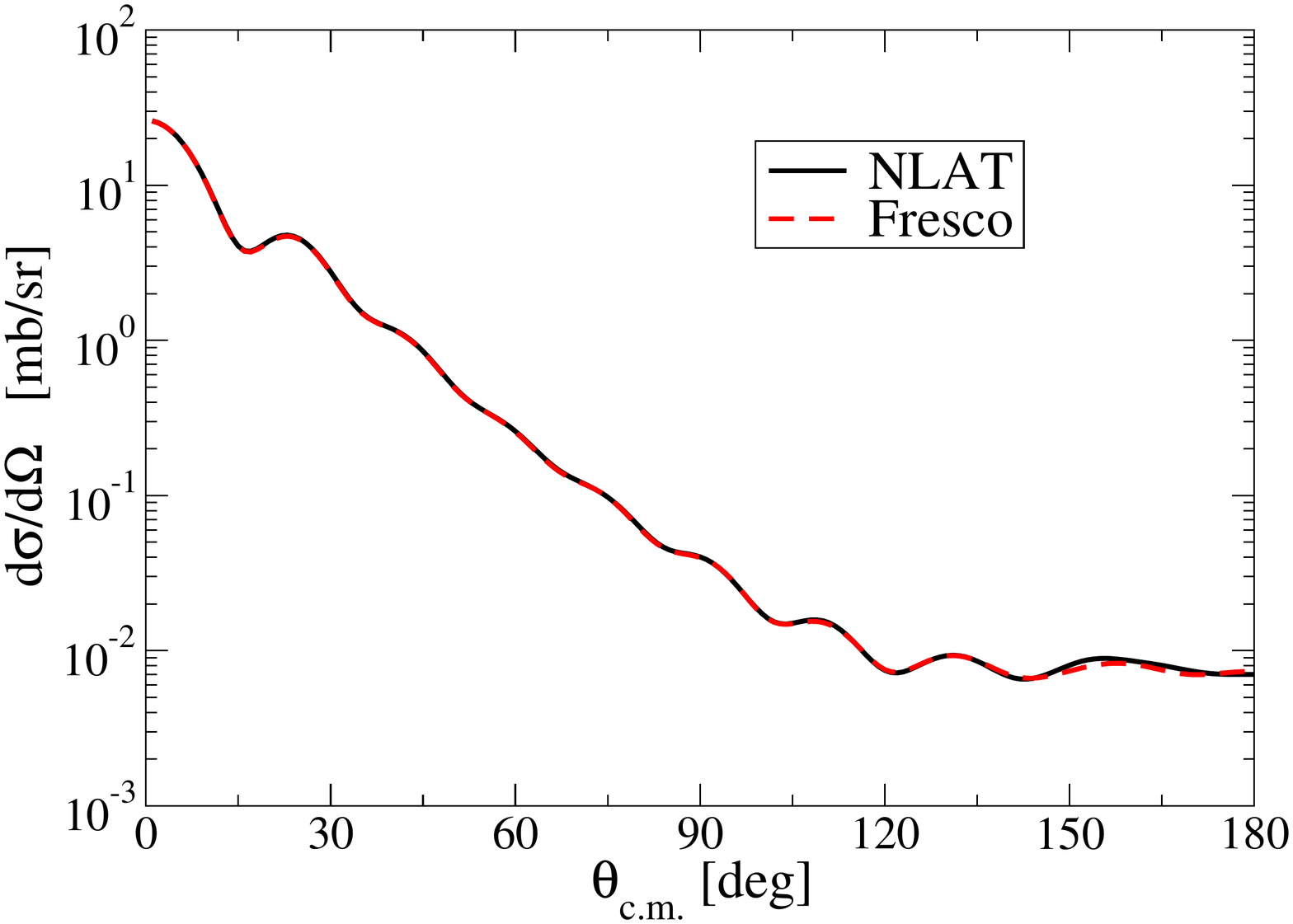}
\end{center}
\caption{$^{132}$Sn$(d,p)^{133}$Sn at $E_d=50$ MeV. Solid line is a local DWBA calculation with NLAT, the dashed line is a calculation done with \textsc{FRESCO}.}
\label{fig:dp132Sn_50-0_Check}
\end{figure}

Next, we check the T-matrix calculation. In Fig. \ref{fig:dp132Sn_50-0_Check} we show DWBA transfer cross sections for $^{132}$Sn$(d,p)^{133}$Sn at $E_d=50$ MeV. The solid line is a calculation using NLAT, and the dashed line is a calculation done using \textsc{FRESCO}. The agreement with \textsc{FRESCO} demonstrates that NLAT calculates the T-matrix for $(d,p)$ transfer reactions properly. Therefore, as long as the wave functions going into the T-matrix are correct, the correct cross section will be calculated. The previous checks have demonstrated that the nonlocal wave functions being calculated are correct, so we can trust that the transfer results when using nonlocal potentials will be correct as well. For this calculation we used a step size of $0.01$ fm, a maximum radius of $30$ fm, and partial waves up to $L=30$. 

\section*{Nonlocal Source}
\label{Nonlocal-Source}

Finally, we examine the nonlocal source. In Sec. \ref{Sec:Adiabatic_Potential} we showed that the nonlocal adiabatic source is calculated accurately for $\beta \approx 0$ fm. For larger $\beta$ values, a check with Mathematica \cite{Mathematica} was done, and the results of this comparison are shown in Table \ref{Tab:Check_With_Mathematica}. For this comparison, we used analytic expressions for the wave functions that mimicked the behavior of the numerical wave functions. For the bound wave function we used

\begin{eqnarray}
\phi(r)=\frac{2}{r+3}e^{-0.3r},
\end{eqnarray} 

\noindent for the scattering wave function we used

\begin{eqnarray}
\chi(R)=\frac{\sin(4R)}{6R}-i\frac{\sin(3R)}{5R},
\end{eqnarray}

\noindent and the $V_{np}(r)$ potential was a central Gaussian:

\begin{eqnarray}
V_{np}(r)=-72.15e^{-\left(\frac{r}{1.494} \right)^2}.
\end{eqnarray}

\begin{table}[h]
\centering
\begin{tabular}{|c|r|r|r|r|}
\hline
                  &  L   & R       &  Mathematica     & NLAT           \\
\hline
$\beta=0.45$ fm   &  0   & 0.05    &  13.70+$i$16.53  &  13.71+$i$16.54   \\ 
                  &  0   & 2.00    &  1.69+$i$0.78    &  1.69+$i$0.78     \\ 
                  &  0   & 5.00    &  0.41-$i$0.29    &  0.41-$i$0.29     \\ 
                  &  1   & 0.05    &  1.67-$i$2.30    &  1.70-$i$2.33     \\ 
                  &  1   & 2.00    &  2.86+$i$1.31    &  2.86+$i$1.30     \\  
                  &  1   & 5.00    &  0.71-$i$0.50    &  0.71-$i$0.51     \\ 
                  &  5   & 0.05    &  0.00-$i$0.0002  &  -0.015+$i$0.016  \\
                  &  5   & 2.00    &  4.07+$i$1.62    &  4.08+$i$1.62     \\  
                  &  5   & 5.00    &  1.30-$i$0.91    &  1.29-$i$0.92     \\ 
\hline
$\beta=0.85$ fm   &  0   & 0.05    &  24.00-$i$23.61    & 24.01-$i$23.62  \\ 
                  &  0   & 2.00    &   2.96+$i$1.10     &  2.95+$i$1.10   \\ 
                  &  0   & 5.00    &   0.71-$i$0.33     &  0.71-$i$0.33   \\ 
                  &  1   & 0.05    &   6.52-$i$6.61     &  6.55-$i$6.64   \\ 
                  &  1   & 2.00    &   5.08+$i$1.90     &  5.08+$i$1.89   \\ 
                  &  1   & 5.00    &   1.23-$i$0.56     &  1.23-$i$0.57   \\ 
                  &  5   & 0.05    &   0.007-$i$0.0007  &  -0.02+$i$0.01  \\ 
                  &  5   & 2.00    &   8.91+$i$3.29     &  8.92+$i$3.28   \\ 
                  &  5   & 5.00    &   2.33-$i$1.06     &  2.32-$i$1.08   \\ 

\hline
\end{tabular}
\caption{The nonlocal adiabatic integral, $rhs$ of Eq.(\ref{NL-Adiabatic}), calculated with Mathematica and NLAT using analytic expressions for the wave functions and potentials.}
\label{Tab:Check_With_Mathematica}
\end{table}

\begin{figure}[h!]
\begin{center}
\includegraphics[scale=0.35]{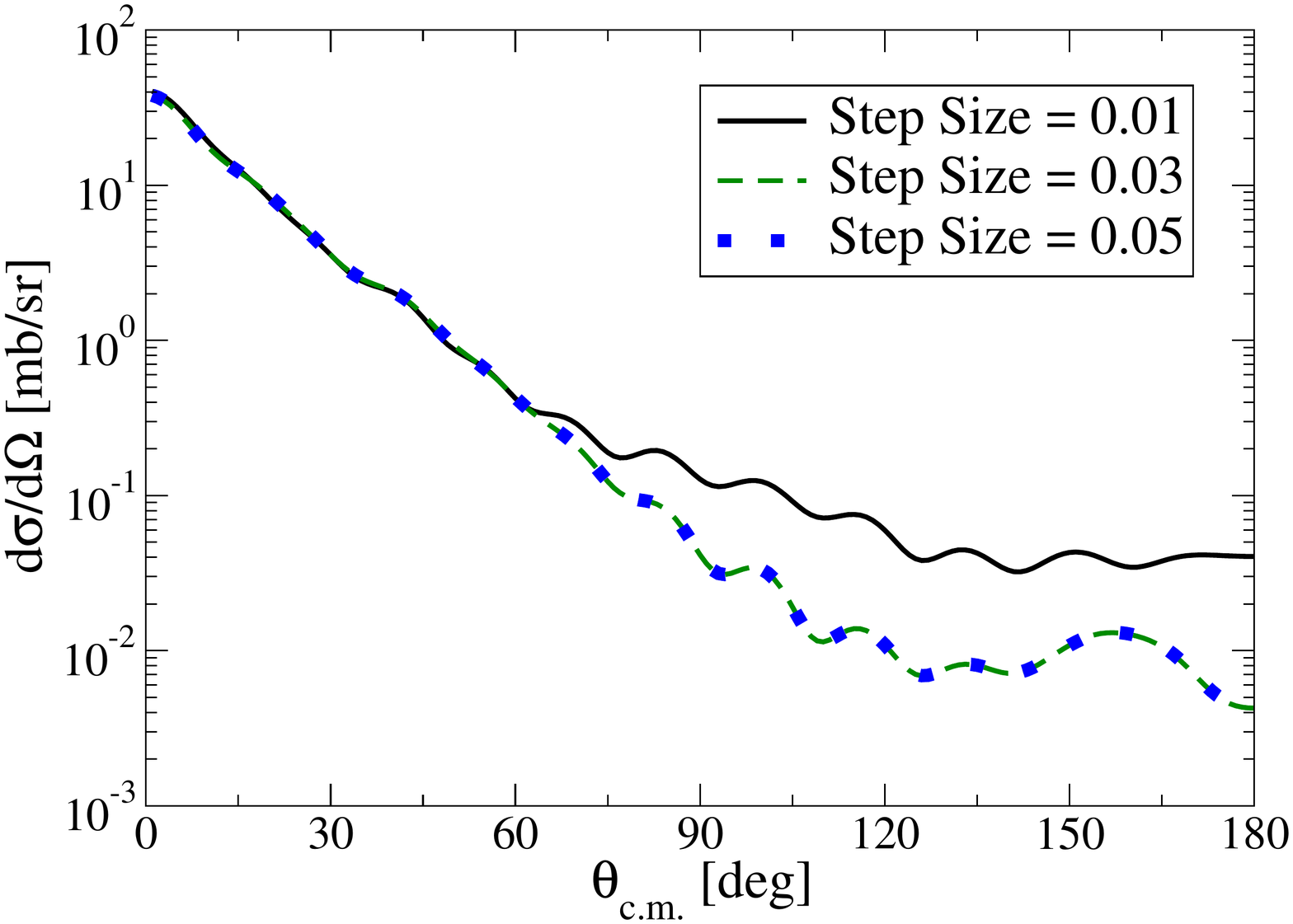}
\end{center}
\caption{Angular distributions for $^{208}$Pb$(d,p)^{209}$Pb at $E_d=50$ MeV obtained by using different step sizes to calculate the $rhs$ of Eq.(\ref{NL-Adiabatic}). The solid line uses a step size of $0.01$ fm, the dashed line a step size of $0.03$ fm, and the dotted line a step size of $0.05$ fm.}
\label{fig:StepSizeComparison}
\end{figure}

There is one additional complication, namely, in order to calculate the T-matrix accurately, we would like our $d+A$ scattering wave function to be calculated in steps of $0.01$ fm. To do this, we need to know our source term $S(R)$ (the $rhs$ of Eq.(\ref{NL-Adiabatic})) in steps of $0.01$ fm as well. However, it requires a significant amount of computer time to calculate  $S(R)$ with such a fine grid. Therefore, in practice, $S(R)$ is calculated in steps greater than $0.01$ fm, and then linear interpolation is used to construct $S(R)$ in steps of $0.01$ fm. To save computer time, we would like the step size we calculate $S(R)$ with to be as large as possible while still maintaining the desired level of accuracy. In Fig. \ref{fig:StepSizeComparison} we show $^{208}$Pb$(d,p)^{209}$Pb using various step sizes for $S(R)$. It is seen that the larger two step sizes agree, while the step size of $0.01$ fm disagrees with the other two calculations. In fact, all calculations with a step size ranging from $0.02 - 0.05$ fm agree, and only when a step size of $0.01$ fm was used did we find disagreement. This required further investigation to determine which calculation is correct.

When calculating the wave function numerically for high values of the angular momentum, $L$, there are difficulties near the origin due to the large centrifugal barrier. To remedy this problem, what is often done is the wave function is set equal to zero near the origin. This is done with a parameter we will call ``CutL''. From the origin to a distance of (StepSize)$\times$(CutL)$\times$($L$), the wave function is set equal to zero. For all the calculations done in this study, we used CutL=2. It was suspected that the discrepancy between the calculation with a step size of $0.01$ fm and the other two in Fig. \ref{fig:StepSizeComparison} was because CutL was not big enough. This can cause problems if we try to calculate the wave function below a very large centrifugal barrier, because numerical inaccuracies will propagate to the rest of the wave function as we continue to integrate outward.

\begin{figure}[h!]
\begin{center}
\includegraphics[scale=0.35]{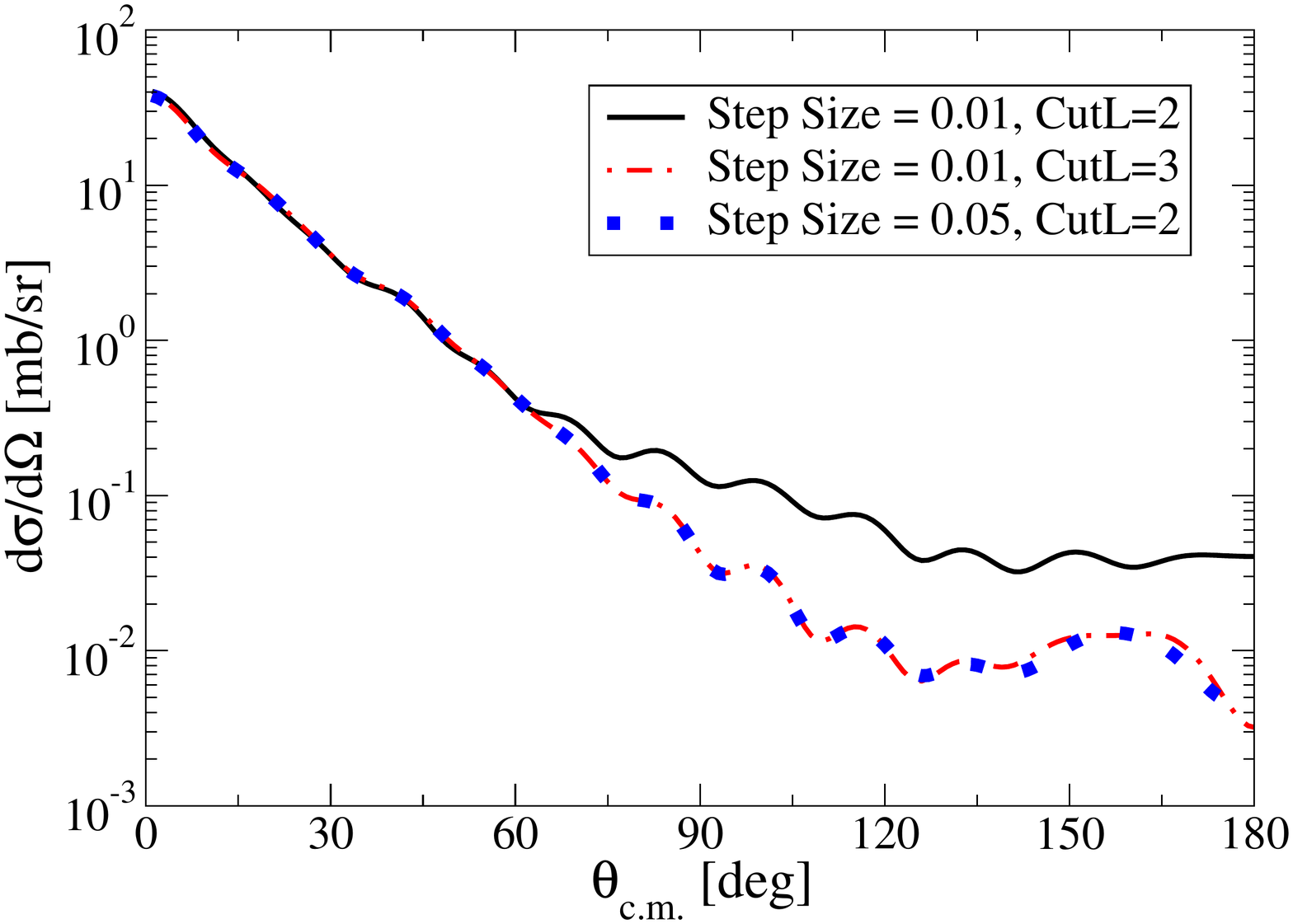}
\end{center}
\caption{Angular distributions for $^{208}$Pb$(d,p)^{209}$Pb at $E_d=50$ MeV obtained by using different step sizes and values of a cut parameter (CutL) to calculate the $rhs$ of Eq.(\ref{NL-Adiabatic}). The solid line uses a step size of $0.01$ fm with CutL=2, the dashed line a step size of $0.01$ fm with CutL=3, and the dotted line a step size of $0.05$ fm with CutL=2.}
\label{fig:CutL_Comparison}
\end{figure}

To figure out which calculation in Fig. \ref{fig:StepSizeComparison} is correct, we increased the CutL parameter for the $0.01$ fm step size calculation. The results are shown in Fig. \ref{fig:CutL_Comparison}. When we increased the CutL parameter of the $0.01$ fm calculation, from CutL=2 to CutL=3, the resulting angular distribution is now in agreement with the other two calculations. Therefore, the discrepancy was indeed due to an insufficient value for CutL. This problem was investigated for all of the reactions studied in this thesis. In all cases, a step size of $0.05$ fm for the $rhs$ of Eq.(\ref{NL-Adiabatic}) with CutL=2 was sufficient to obtain converged results.

\chapter{ Mirror Symmetry of ANCs}
\label{Mirror_Symmetry}

Direct proton capture at low relative energies needed for astrophysics are always peripheral due to the Coulomb barrier. At the limits of $E \rightarrow 0$ these reactions are uniquely determined by the  asymptotic normalization coefficient (ANC) of the single proton overlap function of the final nucleus \cite{Xu_prl1994}. It is for this reason that the ANC method \cite{Xu_prl1994} has been put forth as an indirect way of obtaining proton radiative-capture cross sections from the ANCs extracted from experiments, such as transfer. 

In some astrophysical environments, proton capture may occur on proton-rich nuclei. Obtaining the necessary ANC experimentally in order to understand these astrophysically important reactions may be difficult or impossible since the experiment would require proton-rich radioactive beams. However, an indirect technique has been proposed \cite{Timofeyuk_prl2003}  which uses information about the mirror system in order to extract the necessary ANC. The mirror nucleus is defined as the nucleus with interchanged numbers of protons and neutrons. While an experiment may not be able to be performed on the proton-rich nucleus of interest, experiments on the mirror system can sometimes be performed with stable beams, and thus, with much higher accuracy.

In \cite{Timofeyuk_prl2003, Timofeyuk_prc2005,Timofeyuk_epja2006}, the ratio, $\mathcal{R}$, of the proton to neutron ANC squared is determined for a wide range of light nuclei within a microscopic cluster model (MCM).  In \cite{Timofeyuk_prl2003} an analytic derivation of the ratio, $\mathcal{R}_o$, is presented. The ratio obtained from the MCM calculations is in fair agreement with the predictions of the analytic formula \cite{Timofeyuk_prc2005,Timofeyuk_epja2006}. In this work, we want to explore the effects of couplings induced by deformations of the core and core excitations. 

The reason for relying on charge symmetry arguments rather than just calculating the ANC directly is due to large uncertainties in the theoretical predictions of ANCs. The individual ANCs in \cite{Timofeyuk_prl2003, Timofeyuk_prc2005,Timofeyuk_epja2006} are strongly dependent on the $NN$ interaction used, but the ratio of ANCs was found to be independent of the choice of the $NN$ interaction, within a few percent.

\section*{Theoretical Considerations}

We consider the $A=B+x$ model used in \cite{Nunes_npa1996}, which starts from an effective Hamiltonian representing the motion of the valence nucleon ($x=n,p$) relative to the core, $B$:

\begin{eqnarray}
H_A=T_r+H_B+V_{Bx}(\textbf{r},\xi),
\end{eqnarray}

\noindent where $T_r$ is the relative kinetic energy operator, and $H_B$ is the internal Hamiltonian of the core. $V_{Bx}$ is the effective interaction between the core and the valence nucleon which depends on the $B-x$ relative coordinate, $\textbf{r}$, and the internal degrees of freedom of the core, $\xi$. In the model of \cite{Nunes_npa1996}, $V_{Bx}$ is taken to be a deformed Woods-Saxon potential

\begin{eqnarray}
V_{Bx}(\textbf{r})=-V_{WS}\left(1+\exp\left[\frac{r-R(\theta,\phi)}{a} \right] \right)^{-1}
\end{eqnarray}

\noindent where $V_{WS}$ is the depth, and may depend on the orbital angular momentum, $\ell$. The radius, $R$, is angle dependent, and given by:

\begin{eqnarray}
R(\theta,\phi)=R_{WS}\left[1+\sum_{q=2}^{Q}\beta_qY_{q0}(\theta,\phi) \right]
\end{eqnarray}

\noindent where $\beta_q$ characterized the deformation of the core, and thus, the strength of the coupling between the various $B+x$ configurations. As usual, the radius is given by $R_{WS}=r_{WS}A^{1/3}$ with $A$ the mass number of the $B+x$ system. We also include the typical undeformed spin-orbit potential described in Chapter \ref{Potentials}. 

The $B+x$ wave function is expanded in eigenstates of the core, $\Phi_{I^{\pi}B}$, with spin $I$, parity $\pi_B$, and eigenenergy $\epsilon_{I^{\pi}B}$:

\begin{eqnarray}
\Psi_{J^{\pi}}=\sum_{n \ell jI\pi_{B}}\psi_{n \ell j}(r)\mathcal{Y}_{\ell j}(\hat{\textbf{r}})\Phi_{I^{\pi}B}(\xi).
\end{eqnarray}

\noindent In this expansion, we factorize the radial part, $\psi_{n \ell j}$, and the spin-angular part, $\mathcal{Y}_{\ell j}$ for convenience. The quantum numbers $n$ and $j$ correspond to the principal quantum number and the angular momentum obtained from coupling the orbital angular momentum, $\ell$, with the spin, $s$, respectively. With this expansion, the coupled-channels equation for each $\psi$ is given by \cite{Nunes_npa1996}:

\begin{eqnarray}\label{coupled-eqn}
\left[T_r^\ell+V_{ii}(r) \right]\psi_i(r)+\sum_{j \ne i}V_{ij}\psi_j(r)=\left(\epsilon_{J^{\pi}}^{x}-\epsilon_i \right)\psi_i(r),
\end{eqnarray}

\noindent where $i$ represents all possible ($n \ell j I_{\pi_B}$) combinations, $\epsilon_{J^{\pi}}^{x}$ is the binding energy in the $A=B+x$ system, and the potential matrix elements $V_{ij}$ are given by

\begin{eqnarray}
V_{ij}(r)=\langle \Phi_i(\xi)\mathcal{Y}_i(\hat{\textbf{r}})|V_{Bx}(\textbf{r},\xi)|\mathcal{Y}_j(\hat{\textbf{r}})\Phi_j(\xi)\rangle.
\end{eqnarray}

We take $\Phi_i$ from the rotational model with parameters fixed phenomenologically. The solution of Eq.(\ref{coupled-eqn}) is found by imposing bound-state boundary conditions and normalizing $\Psi_{J^{\pi}}$ to unity. See \cite{Nunes_npa1996, Capel_prc2010} for more details.

In this model, the norm of $\psi_i$ relates directly to a spectroscopic factor, $S_i^x$:

\begin{eqnarray}\label{SF}
S_{i}^{x}=\int_{0}^{\infty} |\psi_i|^2r^2dr
\end{eqnarray}

\noindent while the ANC, $C_{i}^{x}$, is determined from the asymptotic behavior of $\psi_i$:

\begin{eqnarray}\label{eq:ANC}
\psi_{i}(r)\xrightarrow[r \rightarrow \infty]{}C_{i}^{x}W_{-\eta_{i}^{x},\ell+1/2}(2\kappa_ir)
\end{eqnarray}

\noindent with $\kappa_i=\sqrt{2\mu_{Bx}|\epsilon_{J^{\pi}}-\epsilon_i|/\hbar^2}$ and $\mu_{Bx}$ is the reduced mass. Here, $W$ is the Whittaker function with $\eta_{i}^{x}$ the Sommerfeld parameter \cite{Abramowitz_Book}.

As an example to illustrate the model, consider the mirror pair $^{17}$O and $^{17}$F. The core for both nuclei is $^{16}$O, which has a $0^+$ ground state, and two low-lying $2^+$ and $3^-$ states, which strongly couple to the ground state through $E2$ and $E3$ transitions, respectively. If we include the $0^+$ and $2^+$ states of $^{16}$O in our model space, then the ground, $5/2^+$ state of $^{17}$O and $^{17}$F would contain not only a $1d_{5/2}$ valence nucleon coupled to the ground state, but also, for example, a $2s_{1/2}$ nucleon coupled to the excited $2^+$ state. 

In this study, we compare the proton ANCs, $C_{i}^{p}$, with the neutron ANCs, $C_{i}^{n}$, through the ratio

\begin{eqnarray}\label{R}
\mathcal{R}=\left|\frac{C_{i}^{p}}{C_{i}^{n}} \right|^2.
\end{eqnarray}

\noindent The ratio of ANCs calculated in our model is then compared with the analytic formula derived in \cite{Timofeyuk_prl2003,Timofeyuk_prc2007}

\begin{eqnarray}\label{Ro}
\mathcal{R}_o=\left|\frac{F_{\ell}(i\kappa_i^pR_N)}{\kappa_i^p R_N j_{\ell}(i\kappa_i^nR_N)} \right|^2,
\end{eqnarray}

\noindent where $F_{\ell}$ and $j_{\ell}$ are regular Coulomb and spherical Bessel functions, respectively, \cite{Abramowitz_Book}, and $R_N=1.25A^{1/3}$ is the radius of the nuclear interior, of which $R_{o}$ is not strongly dependent. We will compare the ratio of the ANCs from our calculations with the value obtained from this relation.

\section*{Results}

\subsection*{Ratio for Specific Mirror Partners}
\label{Mirror_Ratios}

Since we are interested in the ANCs for each mirror nuclei, and these depend strongly on the energy of the system relative to threshold, it is important that we reproduce the experimental separation energies exactly. We do this by adjusting the depths of $V_{Bn}$ and $V_{Bp}$ to reproduce exactly the corresponding binding energies. All calculations are performed with the program \textsc{FACE} \cite{face}, and all details of the calculations for each mirror pair can be found in \cite{Titus_prc2011}. 

\begin{table*}[t]
\centering
\begin{tabular}{c c c c c c}
\hline
\hline
Nuclei                  & $I^{\pi_B}$ 		 & $n \ell j$	& $\mathcal{R}$       & $\mathcal{R}_o$&  $\mathcal{R}_{MCM}$\\
\hline
$^{8}$Li/$^{8}$B  	& $3/2^-,1/2^-$     & $1p3/2$	& $1.04 \pm 0.04$  & 1.12 & $1.08$	\\
$^{13}$C/$^{13}$N 	& $0^+,2^+$         & $1p1/2$	& $1.19 \pm 0.02$  & 1.20 & $1.14$ \\
$^{17}$O/$^{17}$F (g.s.)& $0^+,3^-$ 	& $1d5/2$	& $1.18 \pm 0.01$  & 1.22 & $1.19$ \\
$^{17}$O/$^{17}$F (e.s.)& $0^+,3^-$ 	& $2s1/2$	& $693  \pm 16$    & 799  & $736$  \\
$^{17}$O/$^{17}$F (g.s.)& $0^+,2^+$ 	& $1d5/2$	& $1.219 \pm 0.004$& 1.22 & $1.19$ \\
$^{17}$O/$^{17}$F (e.s.)& $0^+,2^+$ 	& $2s1/2$	& $756  \pm 23$    & 799  & $736$  \\
$^{23}$Ne/$^{23}$Al     & $0^+,2^+,4^+$   & $1d5/2$	
				& \;\; $(1.852  \pm 0.014)\times10^4$ \;   & \; $2.06\times10^4$ \; & \; $2.96\times10^4$  \\
$^{27}$Mg/$^{27}$P      & $0^+,2^+, 4^+$ 	& $2s1/2$	& $40.1  \pm 1.8$   & 43.7 & $44.3$ \\
\hline
\hline
\end{tabular}
\caption{Ratio of proton to neutron ANCs for the dominant component: Comparison of this work $\mathcal{R}$ with
the results of the analytic formula $\mathcal{R}_0$ Eq.(\ref{Ro}) and the results of
the microscopic two-cluster calculations  $\mathcal{R}_{MCM}$ \cite{Timofeyuk_prc2005,Timofeyuk_epja2006} including the Minnesota interaction. The uncertainty in $\mathcal{R}$ account for the sensitivity to the parameters of $V_{Bx}$.
}
\label{ratios}
\end{table*}

Our results are summarized in Table \ref{ratios}. From the proton and neutron wave functions calculated from Eq.(\ref{coupled-eqn}), we determine the ANCs and the ratio $\mathcal{R}$. For each case, $\mathcal{R}$ corresponds to $r_{WS}=1.25$ fm, $a=0.65$ fm, and $V_{so}=6$ MeV. The uncertainty reflects the range obtained with the geometry $r_{WS}=1.2$ fm, $a=0.5$ fm, and $V_{so}=8$ MeV. Our results for $\mathcal{R}$ are compared to the values obtained with the analytic formula $\mathcal{R}_o$ of Eq.(\ref{Ro}), and those obtained within the MCM, where they assumed two clusters and used the Minnesota interaction, $\mathcal{R}_{MCM}$ \cite{Timofeyuk_prc2005,Timofeyuk_epja2006}.

For nearly all cases, $\mathcal{R}$, $\mathcal{R}_o$, and $\mathcal{R}_{MCM}$ are all in fair agreement. However, there were a few cases where there were discrepancies. For $^{23}$Ne/$^{23}$Al, it is important to note that in our calculations we impose realistic binding energies, whereas in the MCM results, binding energies can sometimes differ significantly. Since $\mathcal{R}$ depends strongly on the binding energies, this can cause large differences between our values and those of \cite{Timofeyuk_epja2006}. The values for $\mathcal{R}_o$ presented in Table \ref{ratios} also assume the experimental binding energies, therefore, differences between $\mathcal{R}$ and $\mathcal{R}_o$ must be related to the failure of the simple analytic relations. Such is the case for $^{17}$O/$^{17}$F (e.s.) when the $3^-$ excited state is included in the model space. 

\subsection*{Testing Model independence}

The usefulness of the ratio method is that the ratio $\mathcal{R}$ should be model independent. This was demonstrated in Section \ref{Mirror_Ratios} where for most cases studied, the ratio obtained from this study, $\mathcal{R}$, using our simple model was in fair agreement with the ratio obtained with the much more sophisticated microscopic cluster model, $\mathcal{R}_{MCM}$. In this subsection, we use the deformation parameter as a free variable to tune the amount of coupling between the various configurations. With the configurations of $^{23}$Ne/$^{23}$Al and $^{27}$Mg/$^{27}$P pairs being very similar to the $^{17}$O/$^{17}$F systems in its ground and excited states, respectively, we concentrate of the three lighter cases.

We find no significant difference in the ratio $\mathcal{R}$ for both the $^{8}$Li/$^{8}$B and $^{13}$C/$^{13}$N mirror pairs. In these cases, the main components of the wave function are $p$ waves, even in the configurations including core excitation. For $|\beta_2|=0.0-0.7$, the resulting range of values for $\mathcal{R}$ are $1.038-1.044$ for $^{8}$Li/$^{8}$B and $1.201-1.251$ for $^{13}$C/$^{13}$N. This constancy is obtained even though the variation in $\beta_2$ leads to significant changes in the spectroscopic factor: $S^x_{1p_{3/2}}$ goes from $1$ to $0.75$ for $^{8}$Li/$^{8}$B, while $S_{1p_{1/2}}^x$ decreases down to $0.32$ for $^{13}$C/$^{13}$N. 

The situation for $^{17}$O/$^{17}$F differs. We consider the separate effects of including the $3^-$ state and the $2^+$ state. Let us first consider the inclusion of $^{16}$O($0^+,3^-$). Like for $^{8}$Li/$^{8}$B and $^{13}$C/$^{13}$N, the variation in $\mathcal{R}$ was small, even though over $30\%$ of the $5/2^+$ ground-state wave function is in a core-excited configuration at $\beta_3=0.7$. For this $\beta_3$, the $1/2^+$ excited-state wave function is almost exclusively in the $^{16}$O($0^+$)$\otimes 2s_{1/2}$ configuration ($S_{2s_{1/2}}^x \approx 95\%$). As a result, the change in the corresponding ratio was limited to less than $1\%$. 

\begin{figure}[h!]
\begin{center}
\includegraphics[width=0.4\textwidth]{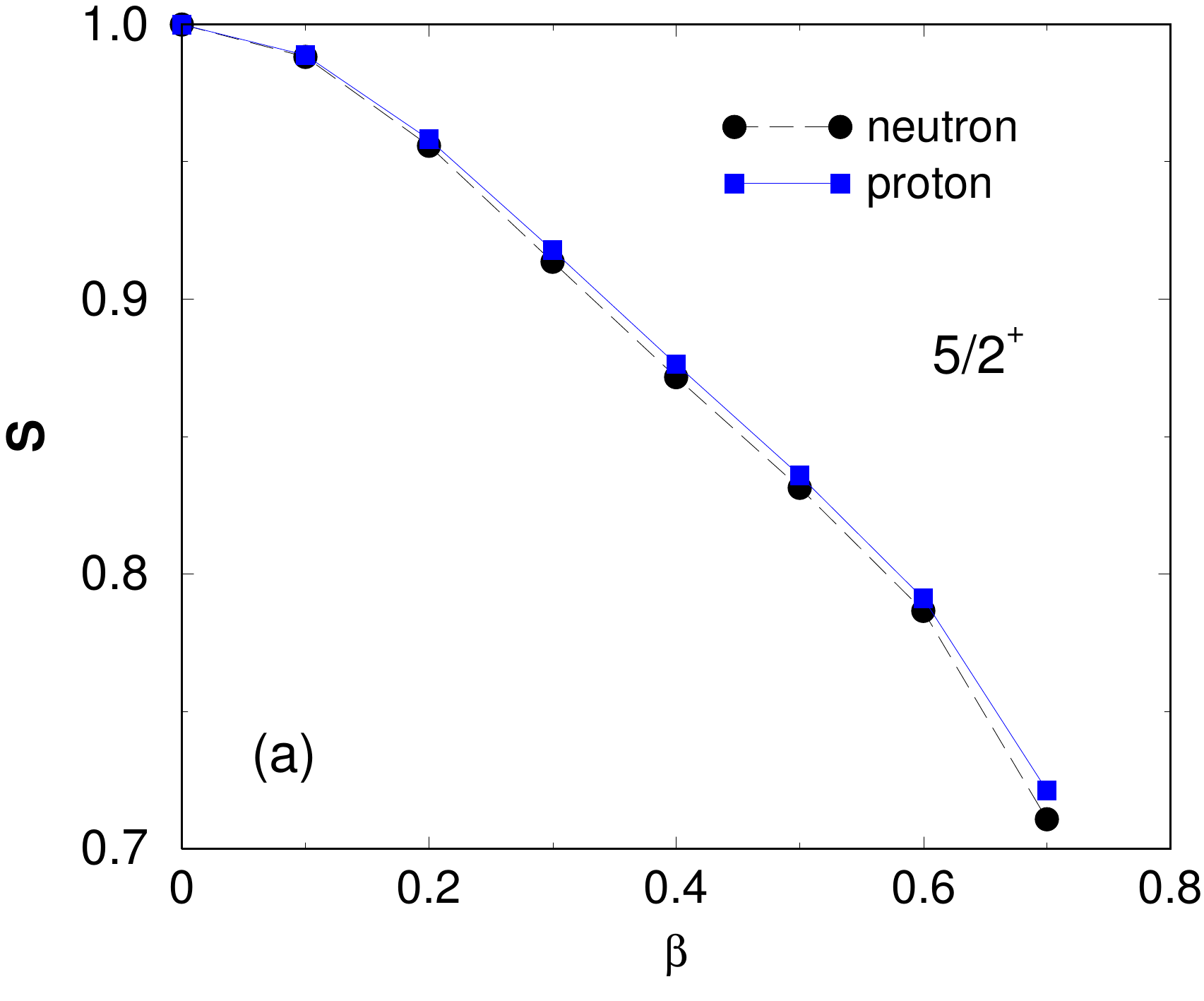}
\includegraphics[width=0.4\textwidth]{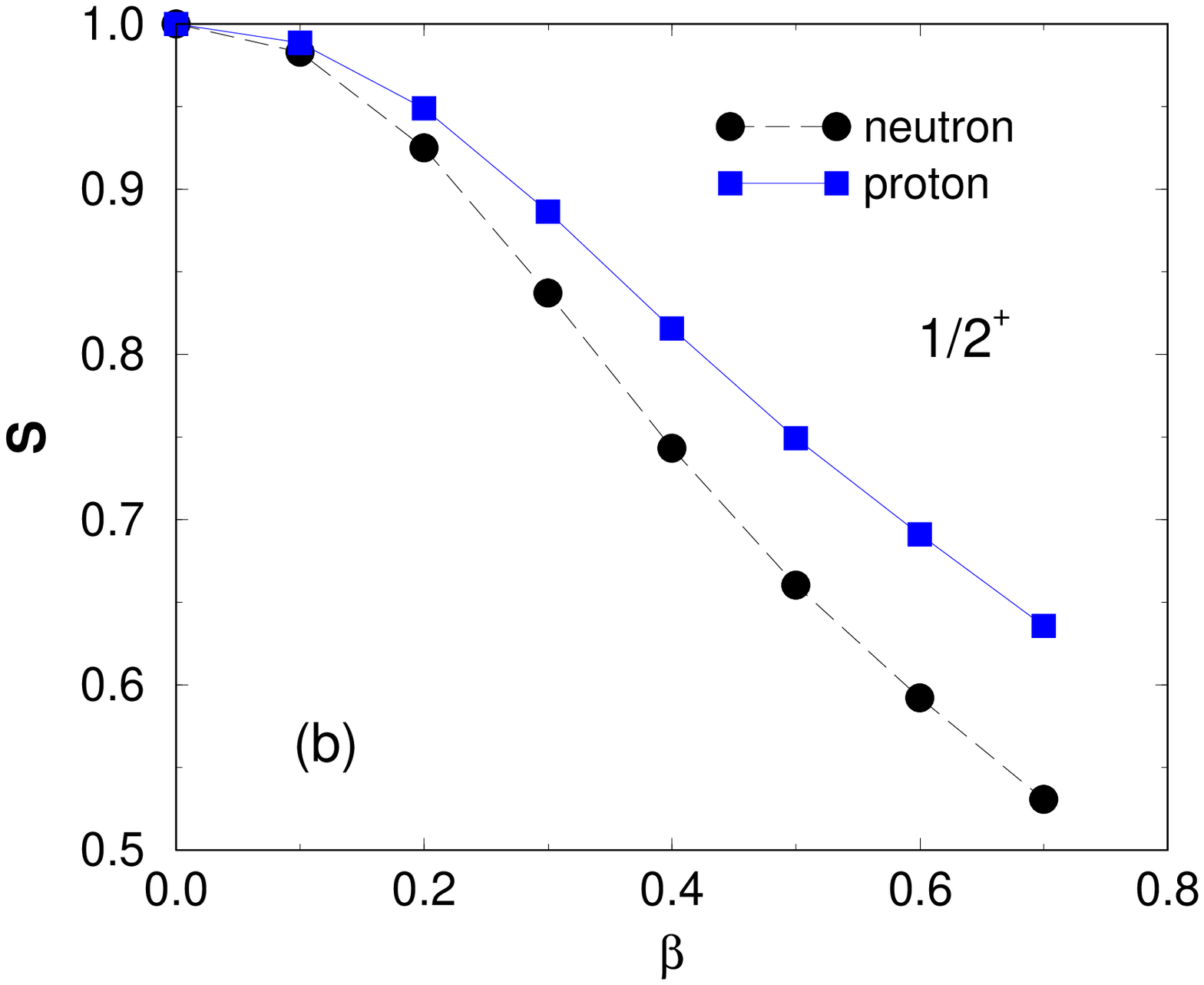}
\end{center}
\caption{Neutron and proton spectroscopic factors
for $^{17}$O and $^{17}$F, respectively, considering the $^{16}$O core in its
$0^+$ ground state and $2^+$ first excited state:
(a) $5/2^+$ ground state and
(b) $1/2^+$ first excited state. Figure reprinted from \cite{Titus_prc2011} with permission. }
\label{fig:sf}
\end{figure}

\begin{figure}[h!]
\begin{center}
\includegraphics[width=0.4\textwidth]{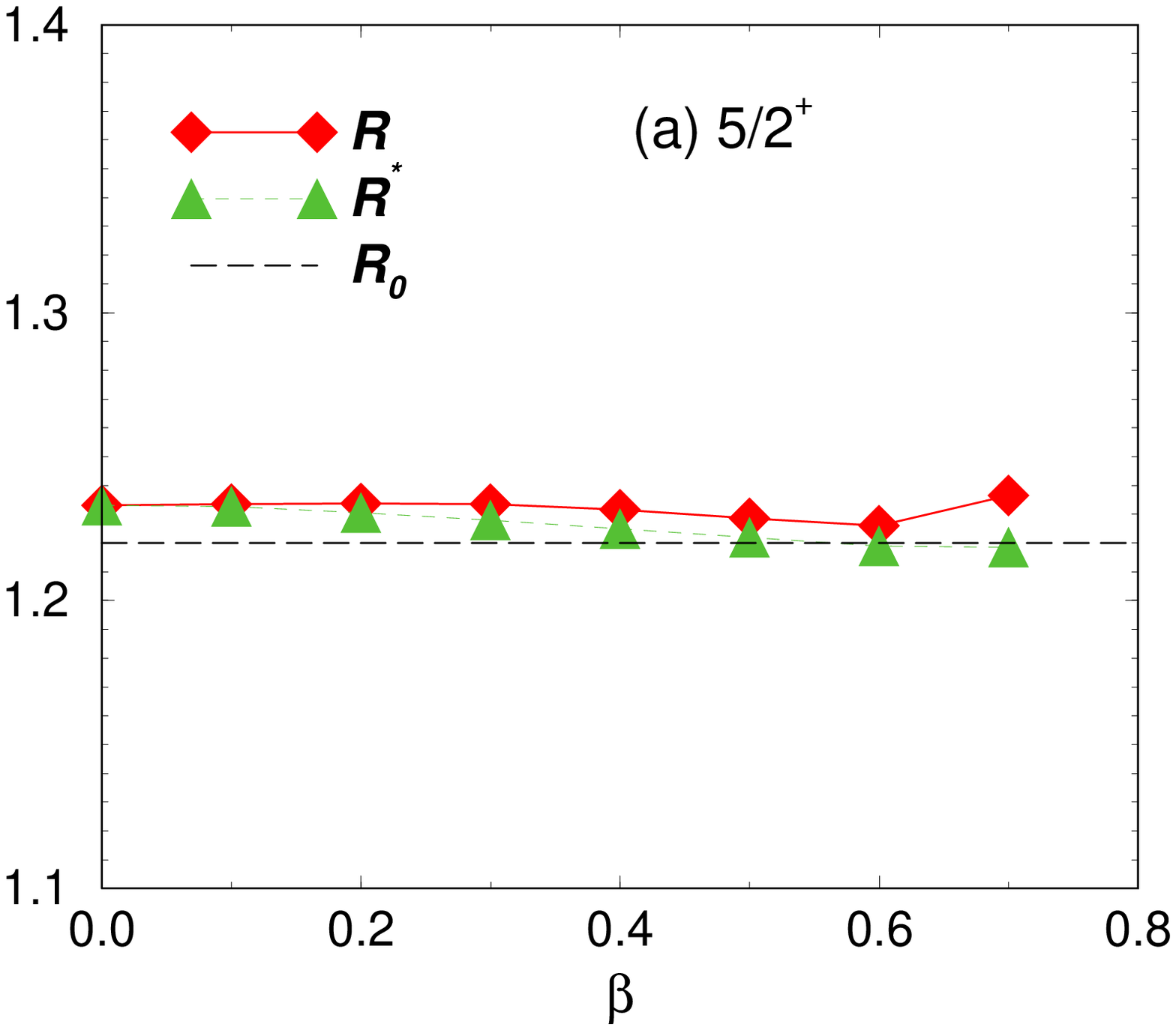}
\includegraphics[width=0.4\textwidth]{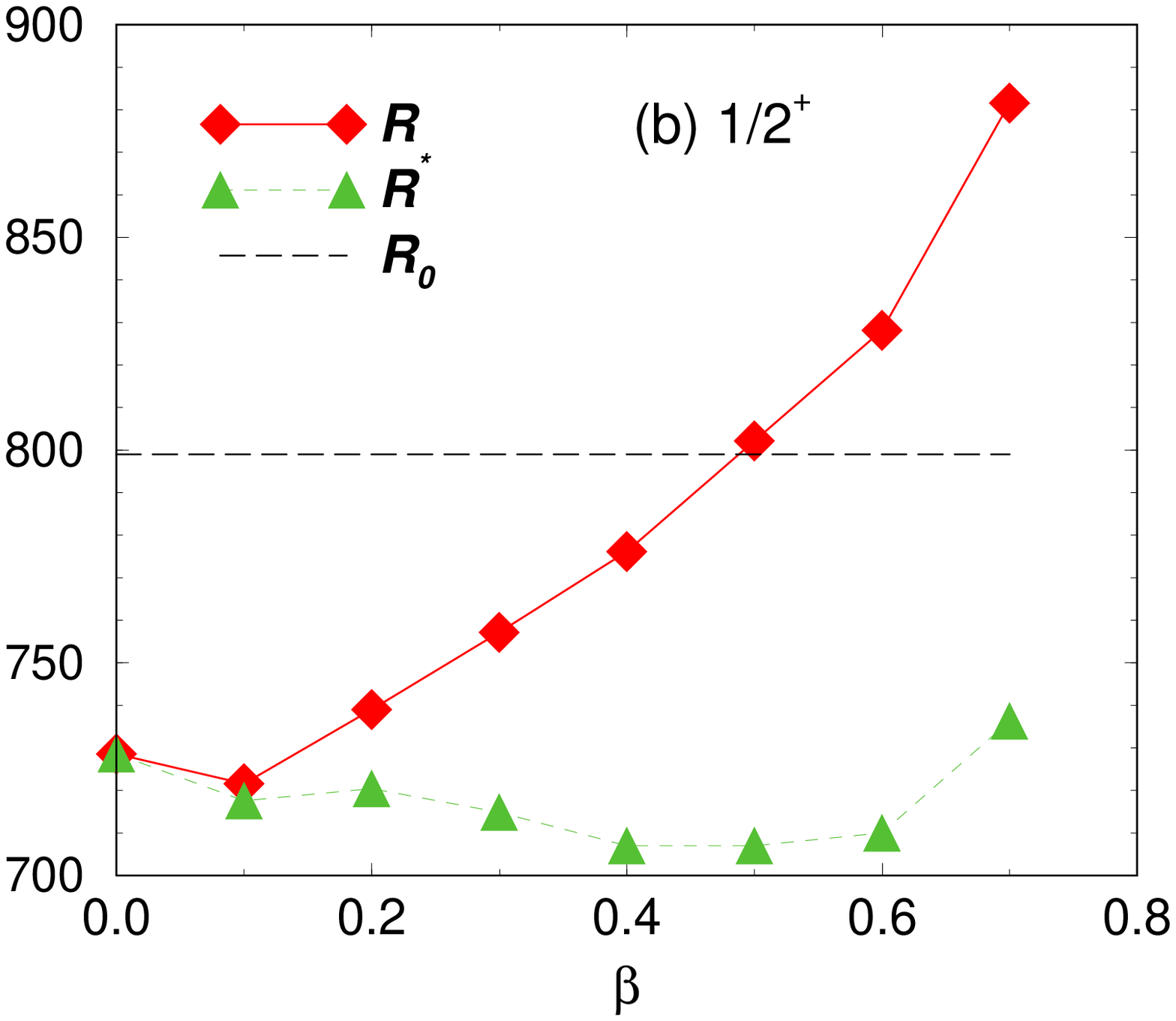}
\end{center}
\caption{Ratio of proton and neutron ANCs
for $^{17}$O and $^{17}$F, respectively, including $^{16}$O($0^+, 2^+$):
(a) $5/2^+$ ground state and
(b) $1/2^+$ first excited state. Figure reprinted from \cite{Titus_prc2011} with permission. }
\label{fig:ratio}
\end{figure}

We next consider the inclusion of $^{16}$O($0^+,2^+$). In this case, the $d_{5/2}$ ground state admixes with an $s_{1/2}$ component with the core in its excited state. For the $1/2^+$ excited state of $^{17}$O, the $s_{1/2}$ component coupled to the ground-state of the core admixes with $d$ components with the core in its $2^+$ excited state. Like in all cases, the energies of the two lowest states in $^{17}$O and $^{17}$F were refitted by simultaneously adjusting the depths of the potential for each $\beta_2$. For both the $5/2^+$ and $1/2^+$ states, the spectroscopic factors, of the component with the core in the ground state experiences a large reduction at large $\beta_2$, as is seen in Fig. \ref{fig:sf}. For the ground state ($5/2^+$), the proton and neutron spectroscopic factors vary together, while for the excited state ($1/2^+$), there is more admixture in the neutron system than for the proton system. This is reflected in a different behavior of the ANC ratios.

In Fig. \ref{fig:ratio} we present the ratio $\mathcal{R}$ as well as a modified ratio compensating for the changes in the spectroscopic factors $\mathcal{R}^*=\mathcal{R}S^n/S^p$. The analytic prediction, $\mathcal{R}_o$ is also shown by the horizontal dashed line. For the $5/2^+$ ground state, neither $\mathcal{R}$ nor $\mathcal{R}^*$ deviate much from the value at $\beta_2=0$, corresponding to the single-particle prediction, as seen in Fig. \ref{fig:ratio}a. Both of these ratios are close to the analytic prediction, $\mathcal{R}_o$. On the contrary, for the $1/2^+$ excited state, $\mathcal{R}$ shows a large variation partly caused by the difference between neutron and proton spectroscopic factors, as seen in Fig. \ref{fig:ratio}b. The features seen in Fig. \ref{fig:ratio} can be extrapolated to $^{23}$Al and $^{27}$P, since, as mentioned before, the former has a structure very similar to that of $^{17}$F(g.s.), while the latter exhibits the same components as $^{17}$F(e.s.). 

In \cite{Timofeyuk_prc2005,Timofeyuk_epja2006} core excitation is explored within the MCM. Even in these studies there was growing disagreement between ${\cal R}_{MCM}$ and ${\cal R}_0$ as more core states were explicitly included in the model space. This was understood in terms of the long range Coulomb quadruple term which was added to the Hamiltonian in the proton case, a term not considered in the derivation of ${\cal R}_0$, nor in our present calculations. Here, however, we not only see a deviation from ${\cal R}_0$, but also a strong dependence on the deformation parameter for some cases. Therefore we conclude the source for deviations from ${\cal R}_0$ and the breakdown of the constant ratio concept is induced by the nuclear quadruple term, which is present in both neutron and proton systems.

The surprising results for the $1/2^+$ mirror states led to several additional tests which isolated the cause for the large coupling dependence on ${\cal R}$. There are three essential ingredients: low binding, the existence of an $s$-wave component coupled to the ground state of the core, and a significant admixture with other configurations. It appears that when all three conditions are met, the differences between the neutron and proton wave functions increase around the surface, exactly where the nuclear quadruple interaction peaks. This results in a stronger effect of coupling on the neutron system compared to the proton system, inducing differences in $S^n$ relative to $S^p$, which are reflected in the coupling dependence on $\cal R$. Our tests show that the effect is independent of whether the wave functions have a node.

\subsection*{Conclusions}

The proposed indirect method for extracting proton capture rates from neutron mirror partners relies on the ratio between asymptotic normalization coefficients of the mirror state being model independent. In \cite{Titus_prc2011}, we tested this idea against core deformation and excitation. We considered a core + $N$ model where the core is deformed and allowed to excite, and applied it to a variety of mirror pairs ($^{8}$Li/$^{8}$B, $^{13}$C/$^{13}$N, $^{17}$O/$^{17}$F, $^{23}$Ne/$^{23}$Al, and $^{27}$Mg/$^{27}$P) and we explored how the mirror states evolve as a function of deformation. 

For most cases, the ratio of the ANC of mirror states was found to be independent of the deformation, and the calculated ratio of ANCs agreed well with the simple analytic formula. From our investigations we concluded that there are three conditions that need to be met for the idea of a model-independent ratio to break down with deformation or core excitation: (i) the proton system should have very low binding, (ii) the main configuration should be an $s$-wave component coupled to the ground state of the core, and (iii), there should be significant admixture with other configurations. This has implications for the application of the indirect method based on the ANC ratio of reactions relevant to novae, namely pertaining the direct capture component of $^{26}$Si($p,\gamma$)$^{27}$P. 


\chapter{ List of Acronyms}

\begin{table}[h]
\centering
\begin{tabular}{|c|l|}
\hline
ADWA   & adiabatic distorted wave approximation \\ \hline
ANC    & asymptotic normalization coefficient \\ \hline
CDCC   & continuum discretized coupled channel \\ \hline
DOM    & dispersive optical model \\ \hline
DWBA   & distorted wave Born approximation \\ \hline
LPE    & local phase equivalent potential \\ \hline
NLAT   & nonlocal adiabatic transfer \\ \hline
PCF    & Perey correction factor \\ \hline
\end{tabular}
\caption{List of acronyms used in this work.}
\end{table}



\end{doublespace}



\bibliography{thesis}
\bibliographystyle{ieeetr}

\end{document}